\documentclass[12pt,a4paper]{report}

\usepackage{geometry}
\usepackage{float}
\usepackage{titlesec}
\usepackage{amssymb,amsmath}
\usepackage[]{graphicx} \graphicspath{ {Fig/} }
\usepackage{caption}
\usepackage[]{xcolor}
\usepackage{booktabs, multirow}
\usepackage[colorlinks,linkcolor=black,urlcolor=blue,citecolor=gray]{hyperref}
\usepackage{natbib}
\usepackage{url}
\usepackage{subcaption}
\usepackage{tikz}
\usepackage{lineno}
\let\oldequation\equation
\let\oldendequation\endequation
\renewenvironment{equation}
{\linenomathNonumbers\oldequation}
{\oldendequation\endlinenomath}
\usepackage{makecell}   
\usepackage{lscape}     
\usepackage{longtable}  
\usepackage{arydshln}   
\usepackage{hyperref}   
\usepackage{comment}
\usepackage{caption}
\usepackage{xcolor}
\usepackage[italian, english]{babel}  
\usepackage{fancyhdr} 
\usepackage{datetime2} 
\usepackage{lastpage}  
\usepackage{caption}
\usepackage{threeparttable}
\usepackage{romannum}
\usepackage{txfonts}
\usepackage{multirow}
\usepackage{listings}

\newcommand\aap{A\&A}                
\newcommand\apj{ApJ}                 
\newcommand\mnras{MNRAS}             

\def\hxmt {\emph{Insight}-HXMT }

\def\lat{{\it Fermi}-LAT }
\def\hess{H.E.S.S. }
\def\fermi{{\it Fermi }}

\definecolor{MyGray}{rgb}{0.30,0.31,0.32} 
\definecolor{MyDarkBlue}{rgb}{0.,0.08,0.5} 
\definecolor{MyLightBlue}{rgb}{0.2,0.2,1.0} 
\definecolor{MyDarkRed}{rgb}{0.5,0.04,0} 
\definecolor{MyDarkGreen}{rgb}{0.0,0.4,0.08} 
\definecolor{ChapterBlack}{HTML}{000000}
\definecolor{SectionBlue}{HTML}{00008b}
\definecolor{SubSectionBlue}{HTML}{0000FF}
\definecolor{SubSubSectionBlue}{HTML}{4682B4}
\definecolor{SectionGray}{HTML}{4D4D4D}      
\definecolor{SubSectionGray}{HTML}{7F7F7F}   
\definecolor{SubSubSectionGray}{HTML}{A9A9A9} 

\titleformat{\chapter}[display]
  {\color{ChapterBlack}\normalfont\LARGE\bfseries}{\chaptertitlename\ \thechapter}{20pt}{\Huge}
\titleformat{\section}
{\color{SectionBlue}\normalfont\LARGE\bfseries}
{\color{SectionBlue}\thesection}{1em}{}
\titleformat{\subsection}
{\color{SubSectionBlue}\normalfont\Large\bfseries}
{\color{SubSectionBlue}\thesubsection}{1em}{}
\titleformat{\subsubsection}
{\color{SubSubSectionBlue}\normalfont\large\bfseries}
{\color{SubSubSectionBlue}\thesubsubsection}{1em}{}
\DeclareCaptionFont{myblue}{\color{SubSectionBlue}}
\newcommand{\myurl}[1]{{\color{MyDarkGreen}\url{#1}}} 

\begin{document}
\title{Gamma-ray Analysis of Pulsar Environments and Their Theoretical Explanation}
\author{Wei Zhang}
\date{31th January 2025}

\newgeometry{margin=2cm}
\begin{titlepage}
    \begin{center}
    
        \vspace*{0.2cm}
        
        \begin{figure}
            \centering
                \includegraphics[width=0.4\columnwidth]{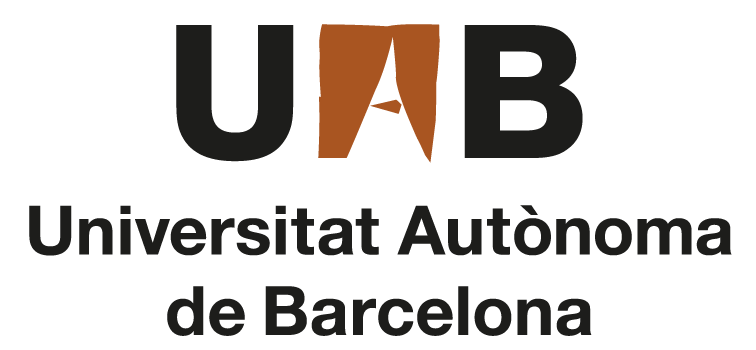}
                \hspace{0.6cm}
                \includegraphics[width=0.4\columnwidth]{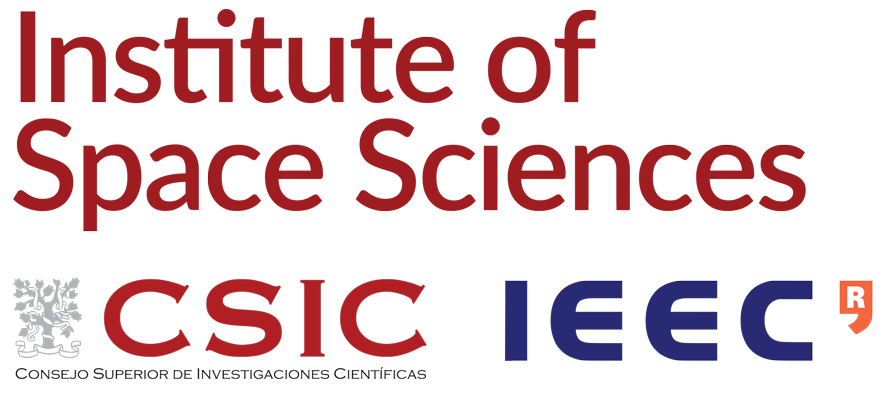}
                \hspace{0.6cm}
        \end{figure}
        
        \noindent\hrulefill
        \\
        \vspace{2cm}
        \Huge
    \textbf{Doctoral Thesis}
        \Huge
        \\
        \vspace{2cm}
        {\fontsize{22.5pt}{22.5}\textcolor{MyDarkBlue}{\textbf{Gamma-ray Analysis of Pulsar Environments and Their Theoretical Explanation}}}

        \vspace{3cm}
        \begin{minipage}[t]{0.34\textwidth}
        \begin{flushleft}
        {\fontsize{16pt}{16pt}Author: \\ \textbf{\textcolor{MyDarkBlue}{Wei Zhang}}}
        \end{flushleft}
        \end{minipage}
        \begin{minipage}[t]{0.64\textwidth}
        \begin{flushright} 
        {\fontsize{16pt}{16pt}Supervisor: \\ \textbf{\textcolor{MyDarkBlue}{Prof. Diego F. Torres}}}
        \end{flushright}
        \end{minipage}\\

        \vspace{1cm}
        \begin{minipage}[t]{0.48\textwidth}
        \begin{flushleft}
        {\fontsize{16pt}{16pt}Tutor: \\ \textbf{\textcolor{MyDarkBlue}{Prof. Llu\'{i}s Font Guiteras}}}
        \end{flushleft}
        \end{minipage}
        \begin{minipage}[t]{0.48\textwidth}
        \begin{flushright} 
        {\fontsize{16pt}{16pt}Mentor: \\ \textbf{\textcolor{MyDarkBlue}{Dr. Weiqiang Li}}}
        \end{flushright}
        \end{minipage}\\

         \vspace{2cm}
        \Large
    \textbf{Barcelona, September 2025}
        \Large
        
        \vfill
        \noindent\hrulefill
        \vspace{0.3cm}
        \Large
        
    \end{center}
\newpage
\thispagestyle{empty}
~
\newpage
\end{titlepage}

\restoregeometry
\vspace*{2cm}
\begin{flushleft}
  {\parbox{4cm}{\color{MyDarkBlue}{\textbf{\LARGE Acknowledgements}}}}
\end{flushleft}
\vspace{2cm}

First and foremost, I would like to sincerely thank my PhD supervisor, Prof. Diego F. Torres, for his insightful guidance, continuous support, and patience throughout my doctoral journey. His mentorship has been invaluable in helping me navigate the challenges of research and grow as a scientist. 

I am also very grateful to my second supervisor, Dr. Jonatan Martín, who provided essential guidance during my first year and helped me establish a strong foundation.

I would like to express my gratitude to my master’s supervisors, Prof. Jiancheng Wang and especially Prof. Xian Hou, whose encouragement and patient teaching have greatly influenced my academic path.

I deeply appreciate the collaborations and insightful discussions with my research partners, particularly Dr. Jean Ballet and Prof. Jian Li, whose expertise significantly contributed to my work.

Thanks to all my colleagues at UAB and ICE-CSIC for fostering a supportive and friendly research environment. A special thank you goes to Noemí Cortés, the assistant to the director at ICE, for her outstanding administrative assistance that made my daily work much easier. 

I am also thankful for the financial support from the China Scholarship Council (CSC), which made this study possible. 

Finally, my heartfelt thanks to my family and friends for their unwavering support, encouragement, and understanding throughout this journey. Their presence has been a constant source of strength.

This thesis is the result of the help and support of many wonderful people - I am truly grateful to all of you.

\newpage
\thispagestyle{empty}
~
\newpage

\vspace*{2cm}
\begin{flushleft}
  {\parbox{4cm}{\color{MyDarkBlue}{\textbf{\LARGE Abstract}}}}
\end{flushleft}
\vspace{2cm}
Pulsars and their surrounding pulsar wind nebulae (PWNe) serve as natural laboratories where extreme magnetic fields, relativistic particles, and shock dynamics converge. 
Understanding their high-energy emission is essential not only for tracing their evolutionary pathways but also for probing the origins of Galactic cosmic rays and the structure of the interstellar medium.

This dissertation presents a comprehensive investigation of the gamma-ray properties of pulsars and PWNe, integrating observational analysis with numerical modeling. 
A physically motivated, time-dependent leptonic model (TIDE) was used to simulate the spectral energy distribution (SED) evolution of PWNe. 
Validation against three representative sources — Crab Nebula, 3C 58, and G11.2$-$0.3 — demonstrated strong convergence and robust parameter recovery, even under limited data, by employing a multi-start fitting strategy. 

Building on this framework, a systematic search for MeV--GeV PWNe was performed using over 11 years of \lat data. 
Focusing on PWNe without known gamma-ray pulsars, the search identified several new candidates and enabled a population-wide characterization. 
Five sources were modeled in detail, with extensive consistency checks — such as epoch comparisons and likelihood weighting — confirming the reliability of the results.

The model was further applied to four potential TeV-emitting PWNe, selected via a pulsar clustering technique (“pulsar tree”). 
Their SEDs were predicted and compared with current and upcoming TeV instrument sensitivities. 
Additionally, several ultra-high-energy (UHE) gamma-ray sources (e.g., eHWC J2019+368, HESS J1427$-$608, LHAASO J2226+6057) were critically assessed within the PWN framework, revealing substantial challenges for standard leptonic interpretations.

Beyond PWNe, this work also explores other pulsar-related systems. A deep gamma-ray search for the high-mass X-ray binary pulsar 1A 0535+262 yielded the most stringent \lat upper limits to date. 
A separate study of the globular cluster M5 uncovered steady gamma-ray emission consistent with the cumulative output of internal millisecond pulsars.

These results deepen our understanding of the gamma-ray behavior of pulsar environments and underscore the utility of unified modeling frameworks in uncovering hidden source populations and constraining high-energy emission processes. 
Looking ahead, the methodologies and findings presented in this work provide a foundation for future multi-wavelength investigations and studies with the next-generation of gamma-ray observatories.

\subsection*{Publications related to this thesis}
The following publications form the main basis of this thesis: 
\begin{itemize}
  \item Zhang, W., Torres, D. F., García, C. R., Li, J., and Mestre, E., “Analysis of the possible detection of the pulsar wind nebulae of PSR J1208-6238, J1341-6220, J1838-0537, and J1844-0346”, A\&A, 691, A332 (2024).
  \item Hou, X., Zhang, W., Freire, P. C. C., Torres, D. F., Ballet, J., Smith, D. A., Johnson, T. J., Kerr, M., Cheung, C. C., et al., “Characterizing the Gamma-Ray Emission Properties of the Globular Cluster M5 with the Fermi-LAT”, ApJ, 964, 118 (2024).
  \item Hou, X., Zhang, W., Torres, D. F., Ji, L., and Li, J., “Deep Search for Gamma-Ray Emission from the Accreting X-Ray Pulsar 1A 0535+262”, ApJ, 944, 57 (2023).
  \item De Sarkar, A., Zhang, W., Martín, J., Torres, D. F., Li, J., and Hou, X., “LHAASO J2226+6057 as a pulsar wind nebula”, A\&A, 668, A23 (2022).
  \item J. Eagle, D. Castro, W. Zhang, D. Torres, J. Ballet, and The Fermi-LAT Collaboration, “A Systematic Search for
MeV-GeV Pulsar Wind Nebulae without Gamma-ray Detected Pulsars”, ApJ ({Corresponding author, in press}).
\end{itemize}

\tableofcontents{}

\newpage
\thispagestyle{empty}
~
\newpage
\listoffigures

\newpage
\thispagestyle{empty}
~
\newpage

\listoftables

\chapter{Introduction}
\label{introduction}
\vspace*{0.5cm}
\setcounter{page}{1}
\pagenumbering{arabic}

\pagestyle{fancy}
\fancyhf{} 
\setlength{\headheight}{14.5pt}
\renewcommand{\headrulewidth}{0.4pt} 
\renewcommand{\footrulewidth}{0.4pt} 
\renewcommand{\footruleskip}{4pt} 
\fancyhead[C]{\textcolor{gray}{WEI ZHANG · DOCTORAL THESIS}} 
\fancyfoot[C]{\thepage}


\section{Pulsars}
\label{psr}

\begin{figure}
    \centering
  	\includegraphics[width=\columnwidth]{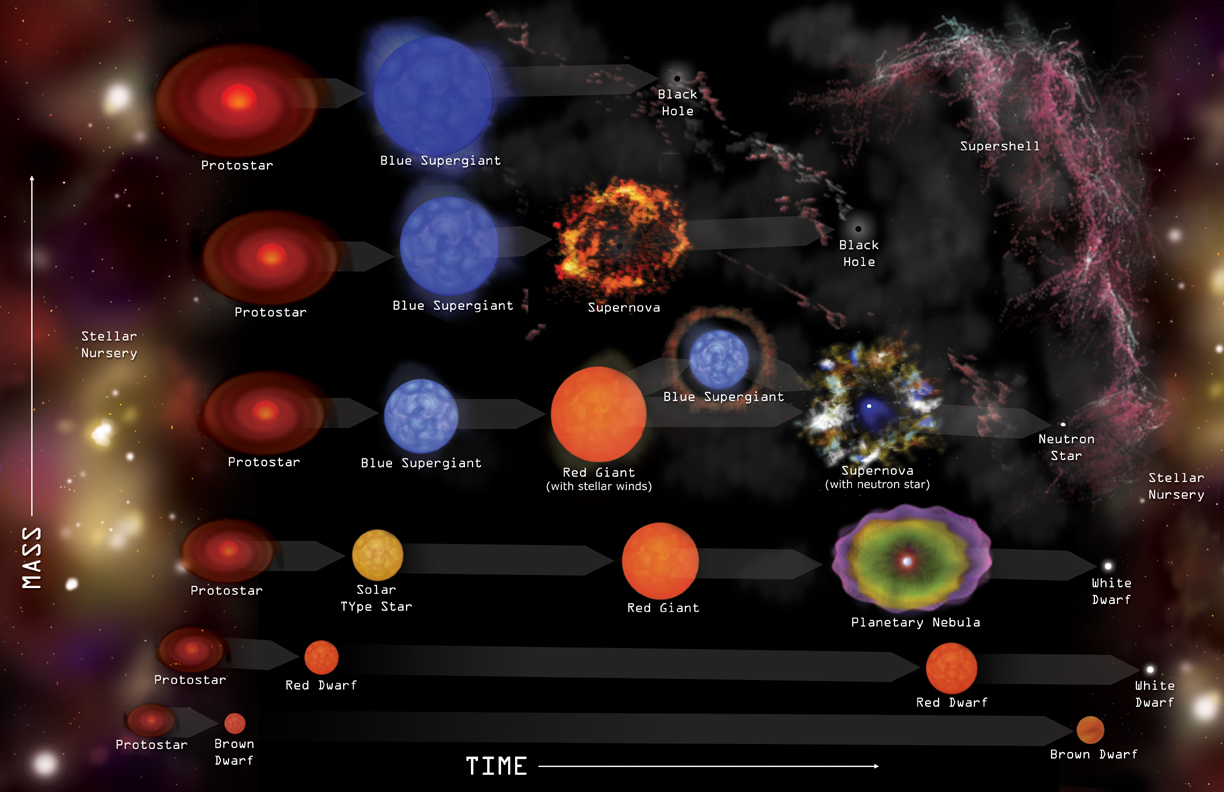}
    \caption{The evolution of the stars with different initial mass (adopted from Chandra X-ray Observatory website$^a$).}
    \raggedright\rule{0.4\linewidth}{0.4pt}  
    \caption*{\raggedright \scriptsize $^a$\url{https://chandra.si.edu/graphics/xray_sources/stellar_fate.jpg}}
    \label{C1.stellar_fate}
\end{figure}

\begin{figure}
    \centering
  	\includegraphics[width=0.5\columnwidth]{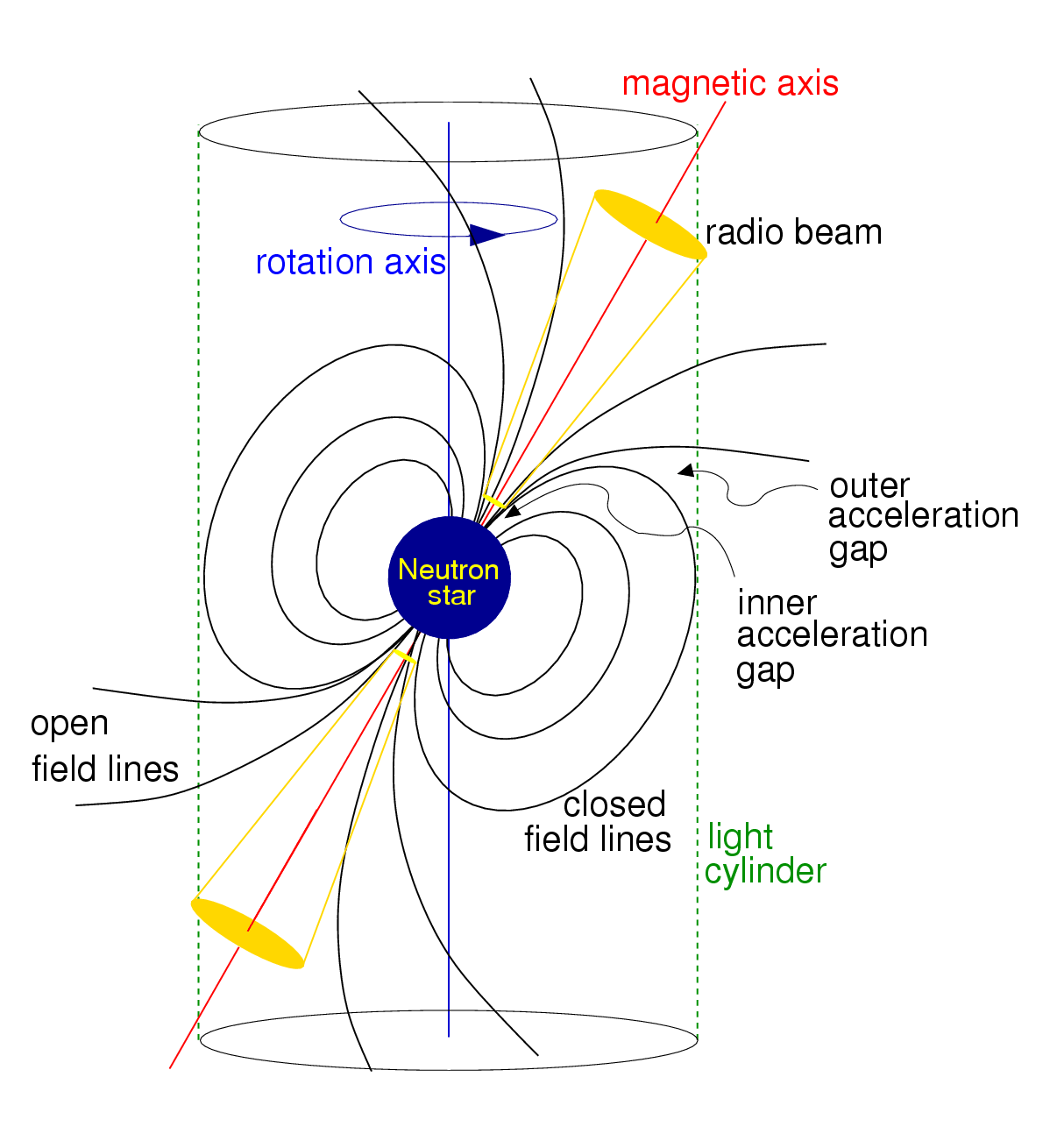}
    \caption{Radiation model of pulsars \citep{lorimer2005handbook}.}
    \label{fig.intro.PSR}
\end{figure}

Within stellar interiors, thermal pressure generated by ongoing thermonuclear fusion effectively counterbalances gravitational contraction. 
However, in the terminal phases of stellar evolution, the depletion of nuclear fuel leads to a decrease in thermal pressure, rendering it insufficient to counteract gravity. 
In low- to intermediate-mass stars, this yields a white dwarf stabilized by electron degeneracy pressure, with outer layers ejected as a planetary nebula \citep{gomez2020predicting}. 
For progenitor stars with more initial masses ($\gtrsim 10 M_\odot$), nuclear fusion progresses to the formation of an iron core. 
Subsequently, the iron core undergoes a catastrophic collapse, triggering a supernova explosion. 
The explosion expels the stellar envelope, forming a supernova remnant (SNR), while the core contracts into either a neutron star (for progenitor masses $\lesssim$ 30 M$\odot$) or a black hole (for masses $\gtrsim$ 30 M$_\odot$). 
Figure \ref{C1.stellar_fate} intuitively presents the evolutionary paths of stars with different initial masses. 
Unlike white dwarfs, neutron stars are supported by neutron degeneracy pressure, a consequence of the Pauli exclusion principle acting on closely packed neutrons \citep{Xiang2008}.

During the formation of a neutron star, most of the angular momentum of the progenitor is conserved, where the angular momentum is given by \( L = I\omega \), and the moment of inertia scales as \( I \propto MR^2 \), with \( M \) and \( R \) representing the stellar mass and radius, respectively. 
The core collapse leads to a drastic reduction in radius (typically to about 10 km) and partial mass ejection during the supernova explosion, resulting in a significant decrease in the moment of inertia and, consequently, a rapid increase in rotational frequency. 
As a result, neutron stars usually have quite short spin periods ranging from a few milliseconds to tens of seconds.

Similarly, the conservation of most magnetic flux generates intense surface magnetic fields, often reaching \(10^{12}\)--\(10^{15}\) G. 
Electrons and protons accelerated in such strong, varing fields emit high-energy radiation via synchrotron or curvature processes along magnetic poles. 
Due to the misalignment of the magnetic and rotational axes, the radiation beam sweeps the sky like a lighthouse (Figure \ref{fig.intro.PSR}). 
%
If Earth lies within the beam path, periodic pulses are observed. Such neutron stars are known as pulsars \citep{lorimer2005handbook, Xiang2008}.

\subsection{Categories}

Pulsars are usually classified into three different categories, depending on the main source of electromagnetic radiation \citep{Kohler2020}: 
\begin{enumerate}
    \item rotation-powered pulsars that is mainly powered by the loss of rotational energy;
    \item accretion-powered pulsars (including most X-ray pulsars, and usually with a companion star) that is mainly powered by the gravitational potential energy of the accreted matter from the companion stars; 
    \item magnetars are a subset of pulsars characterized by the presence of an exceedingly robust magnetic field, with the decay of this field serving as the primary source of their energy.
\end{enumerate}

Accretion-powered pulsars gain angular momentum from their companion stars, potentially shortening their spin periods to millisecond scales (called millisecond pulsars). 
These pulsars have magnetic fields much weaker than ordinary ones, likely because of field weakening during the accretion process. 
Millisecond pulsars are notable for their exceptional pulse period stability, comparable to that of atomic clocks, making them valuable as natural cosmic timekeepers for precise temporal measurements. 


\subsection{Radiation and Theoretical Models}

In general, rotational energy is an important source of electromagnetic radiation for all types of pulsars. 
According to the magnetic dipole model, the radiation power is given by \(\dot{E} \propto -B^2 R^6 \omega^4\), where \(R\) is the radius of the pulsar, \(B\) is the surface magnetic field, and \(\omega\) is the angular velocity. 
Assuming that the electromagnetic radiation originates solely from the loss of rotational energy, the radiation power can also be expressed as \citep{Xiang2008}:
\begin{equation}
 \dot{E}=\frac{d}{dt}(\frac{1}{2}I\omega^2)=I\omega \dot{\omega}{. }
\label{eq.intro.1}
\end{equation}
So we can obtain 
\begin{equation}
 \dot{\omega}=\frac{\dot{E}}{I\omega}\propto \frac{-B^2R^6\omega^4}{I\omega}=\frac{-B^2R^6\omega^3}{I}{. }
\label{eq.intro.2}
\end{equation}
The negative \(\dot{\omega}\) indicates an increasing spin period \(P\), which is commonly observed in most pulsars with a small and stable \(\dot{P}\).

Pulsars emit electromagnetic radiation across the spectrum, from radio to gamma rays. 
Understanding the electrodynamics of pulsar magnetospheres — the plasma-filled regions surrounding these stars — is critical to unraveling their emission mechanisms and evolutionary behavior. 
The modeling of pulsar magnetosphere and radiation has undergone a profound transformation over the past several decades, evolving from idealized vacuum approximations to sophisticated, plasma-filled, kinetic simulations \citep[see, e.g.][]{Cao2024psr,Philippov2022psr,Cerutti2025}. 
These models attempt to reproduce the complex interplay between electromagnetic fields, plasma dynamics, and observed multi-wavelength emission. 
%

To describe pulsar magnetospheres, a hierarchy of models has been developed, differing in physical assumptions and computational complexity: 
\begin{itemize}
    \item \textbf{The Vacuum Dipole Model} treats the pulsar as a rotating magnetic dipole in vacuum \citep{Deutsch1955}, forming the basis of early gap models (e.g., Polar Cap \citep[PC, see e.g.,][]{Ruderman1975,Daugherty1982}), Slot Gap / Two-Pole caustic \citep[SG / TPC, see e.g.,][]{Dyks2003,Muslimov2004}), and Outer Gap \citep[OG, see e.g.,][]{Cheng1986,Zhang1997psrmodel,Cheng2000}). While analytically tractable, it lacks plasma and current feedback, limiting its ability to reproduce the observed gamma-ray light curves seen by \lat. The interlay of assumptions underlying the gap models have been studied in \citep{Vigano2015p1,Vigano2015p2} 
    
    \item \textbf{The Force-Free Electrodynamics (FFE) Model} assumes a dense plasma that fully screens parallel electric fields ($\mathbf{E} \cdot \mathbf{B} = 0$), yielding realistic large-scale field structures and equatorial current sheets. FFE solutions support light curve and polarization modeling \citep[see e.g.,][]{Benli2021,Petri2021,Harding2017}, but cannot account for particle acceleration due to the absence of $E_{||}$. 

    \item \textbf{The Resistive Magnetosphere Model} introduces finite conductivity to allow non-zero $E_{||}$, enabling self-consistent acceleration regions and emission modeling \citep[see e.g.,][]{Kalapotharakos2012}. However, they remain incomplete due to the lack of kinetic microphysics, such as pair production or plasma feedback. 

    \item \textbf{Particle-in-Cell (PIC) Magnetosphere Model} solves for particle dynamics and electromagnetic fields self-consistently, capturing kinetic-scale physics and feedback processes, \citep[see e.g.,][]{Philippov2022psr,Cerutti2025}. However, they are computationally intensive and not yet feasible for realistic pulsar parameters. 

 
\end{itemize}

Regarding radiation/magnetospheric models, a remark goes to the Effective Synchro-curvature Model, which follows the dynamics of charged particles accelerated in the magnetosphere of a pulsar and computes their emission via synchro-curvature radiation \citep{Vigano2015}. The model has succeeded in fitting the gamma-ray spectra of the whole population of gamma-ray pulsars \cite{Vigano2015p3} and reproduces as well those pulsars that also have detected non-thermal X-ray pulsations \citep{Torres2018p,Torres2019p}, all with only three free effective parameters involved. Both general agreement on the global properties of the predicted spectra and light curves and specific fitting to all \lat pulsars have been presented \cite{2024Iniguez,2025Iniguez}.

%

\subsection{Remarks}

Almost six decades after the discovery of the first pulsar by Hewish and Bell in 1967 \citep{Hewish1969}, more than 4,000 pulsars have been identified. 
However, many key questions remain unresolved, including the origins of anomalous X-ray pulsars, nulling and giant pulses, glitches \citep{Wu2021}, and, more broadly, the lack of a comprehensive framework capable of describing the complex, multi-scale behavior of pulsar systems. 
%
%
With ultra-strong gravity, magnetic fields, and extreme densities, pulsars serve as natural laboratories where all four fundamental forces intersect under extreme conditions. 
Investigating their behavior is essential to deepening our understanding of fundamental physics and the universe itself. 

\section{Pulsar Wind Nebulae}
\label{pwn}

\begin{figure}[H]
    \centering
  	\includegraphics[width=\columnwidth]{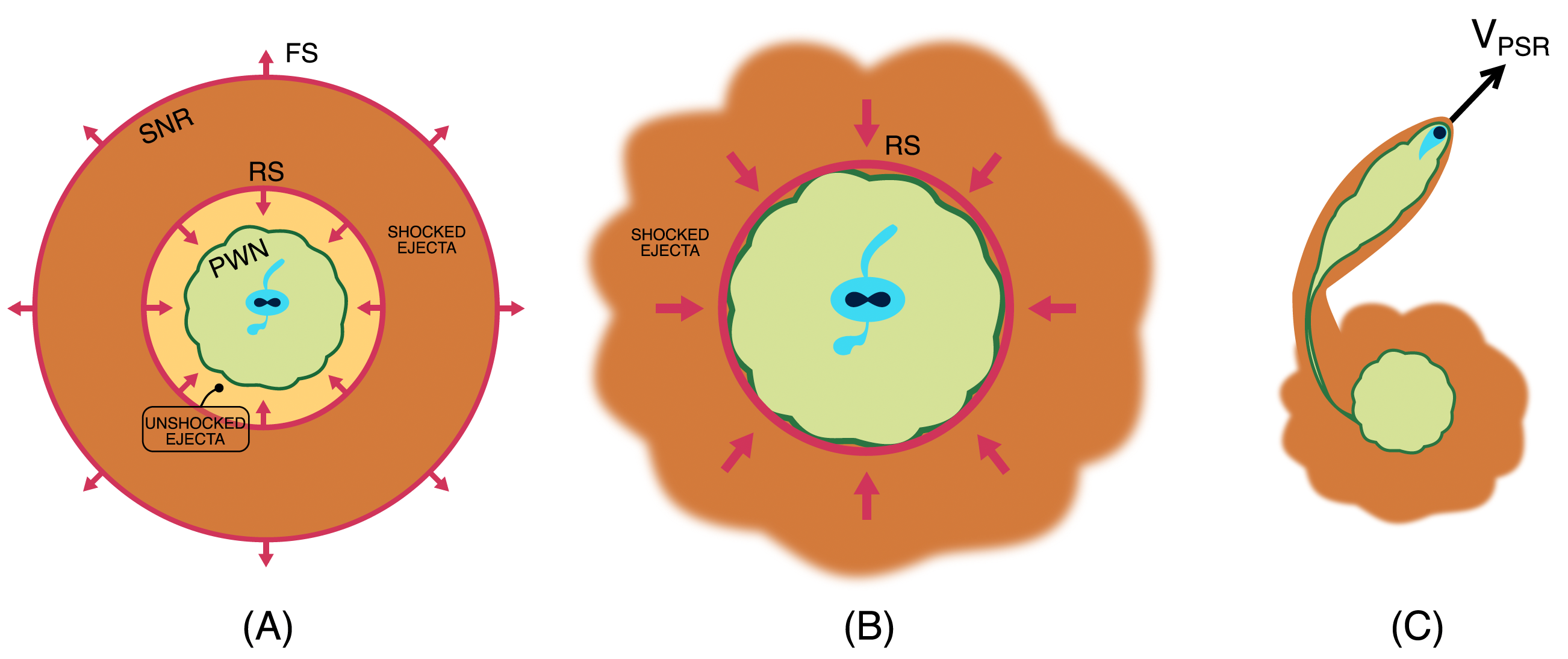}
    \caption{The evolution of PWNe \citep{Olmi2024}.}
    \label{fig.intro.PWN_evolution}
\end{figure}

\begin{figure}
    \centering
  	\includegraphics[width=0.485\columnwidth]{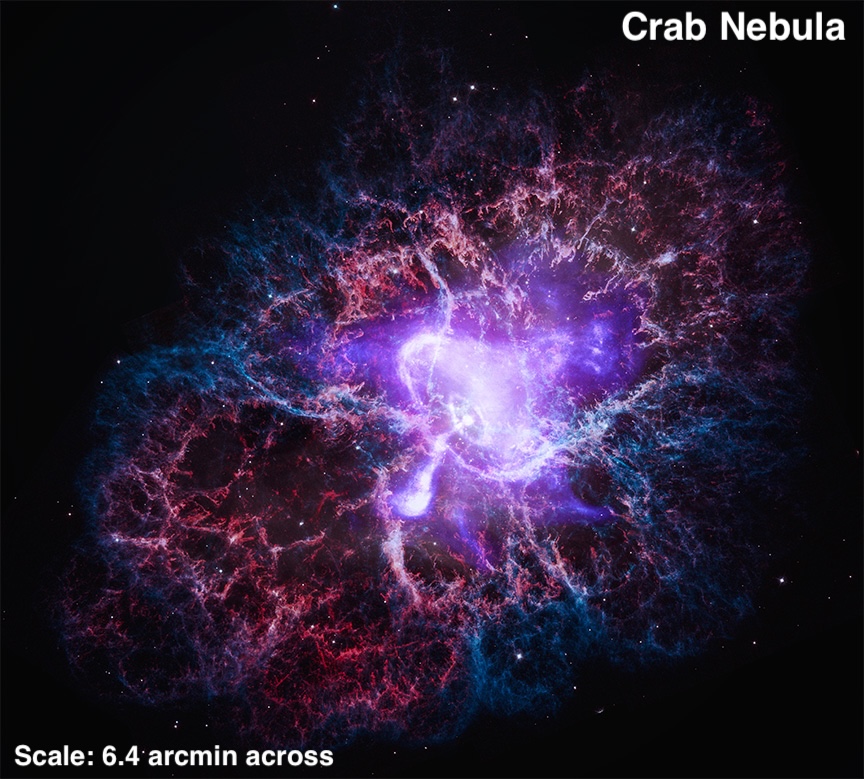}
        \includegraphics[width=0.502\columnwidth]{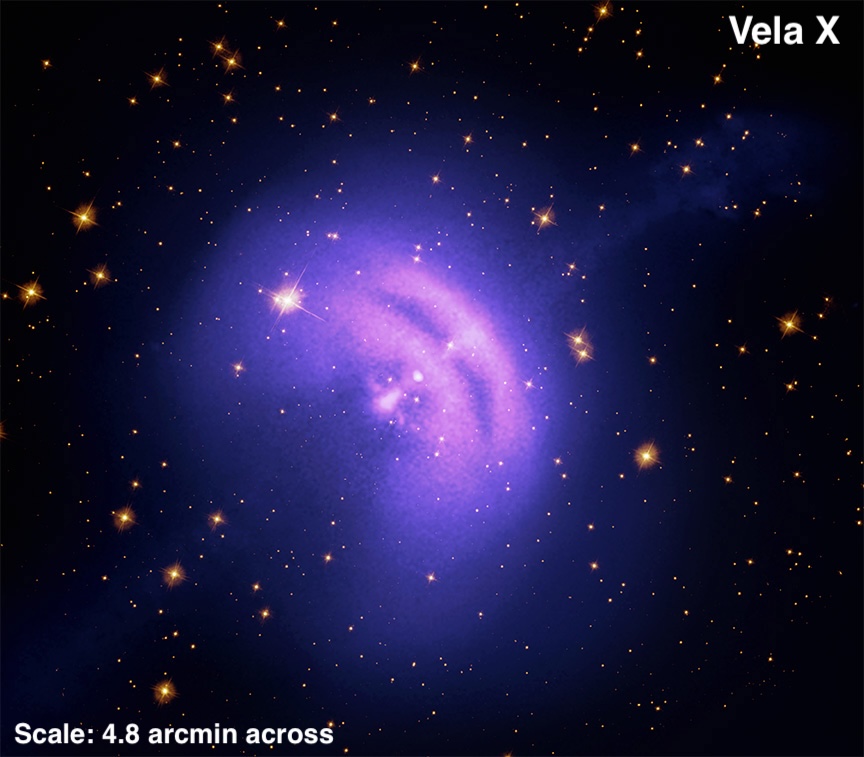}
        \includegraphics[width=0.431\columnwidth]{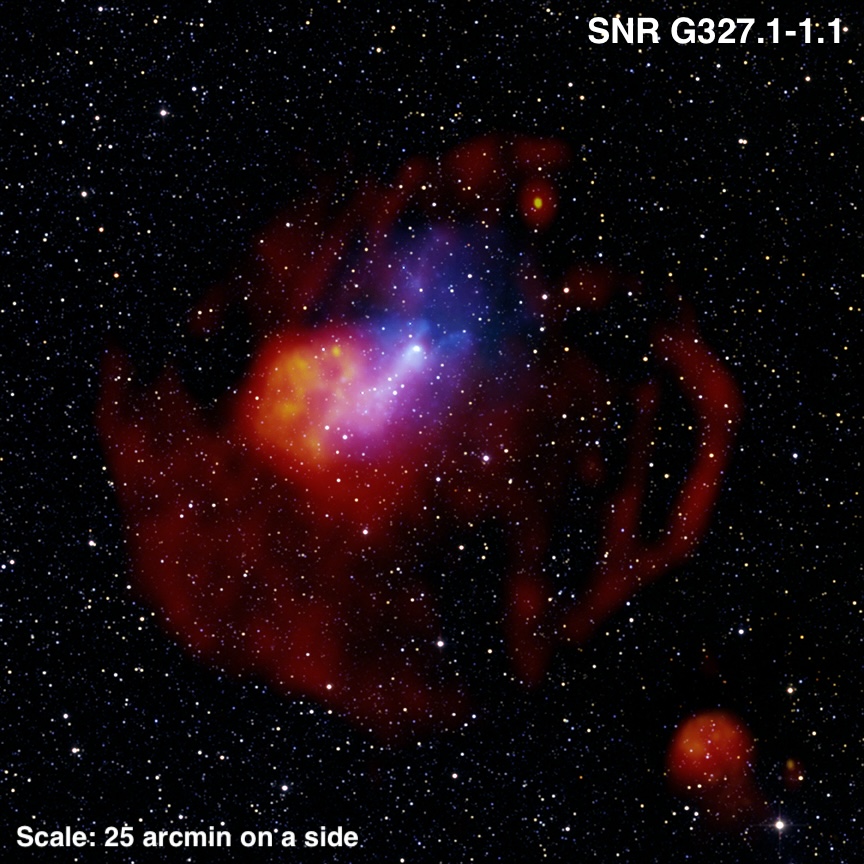}
        \includegraphics[width=0.559\columnwidth]{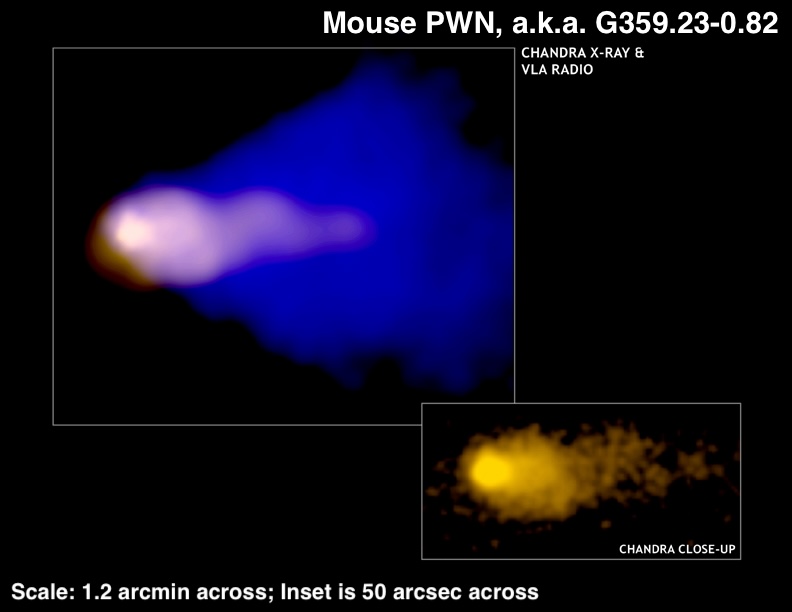}
    \caption{Several PWN examples during the different evolution phases.\\ 
        \textbf{Crab Nebula$^a$:} The composite X-ray (blue, white, and purple) and optical (red, green, and blue) image of the Crab Nebula. The rings around the pulsar and the jets blasting into space are clearly visible. \\ 
        \textbf{Vela X$^b$:} The composite X-ray (purple and light blue) and optical (yellow) image of PWN Vela X.  \\ 
        \textbf{SNR G327.1-1.1$^c$:} The composite X-ray (blue), radio (red and yellow), and infrared (RGB) image of SNR G327.1-1.1, encompassing a PWN, Snail. \\ 
        \textbf{Mouse$^d$:} The composite X-ray (gold) and radio (blue) image of Mouse PWN. A close-up of the head of the Mouse is inserted where a bow shock wave has formed as a pulsar plows through interstellar space. }
    \raggedright\rule{0.4\linewidth}{0.4pt}  
    \caption*{\raggedright \scriptsize     
        $^a$\textbf{Credit:}  X-ray: (Chandra) NASA/CXC/SAO, (IXPE) NASA/MSFC; Optical: NASA/ESA/STScI; Image \par {\hskip 4.5em}Processing: NASA/CXC/SAO/J. Schmidt, K. Arcand, and L. Frattare. \par
        $^b$\textbf{Credit:} X-ray: NASA/CXC/SAO; Optical: NASA/ESA/STScI; Image processing: NASA/CXC/SAO/J.  \par {\hskip 4.5em}Schmidt, K. Arcand. \par
        $^c$\textbf{Credit:} X-ray: NASA/CXC/SAO/T.Temim et al. and ESA/XMM-Newton; Radio: SIFA/MOST and 
        \par {\hskip 4.5em}CSIRO/ATNF/ATCA; Infrared: UMass/IPAC-Caltech/NASA/NSF/2MASS. \par
        $^d$\textbf{Credit:} X-ray: NASA/CXC/SAO/B.Gaensler et al.; Radio: NRAO/AUI/NSF.
}
    \label{fig.intro.PWN_examples}
\end{figure}

A pulsar is born in a supernova explosion, typically acquiring a natal kick velocity of the order of 100–500 km/s \citep{FG2006}. 
During the explosion, the cold ejecta from the explosion (the unshocked ejecta) expand into the surrounding interstellar medium (ISM) at speeds ranging from 300 to 5000 km/s \citep{Chevalier1976}, driving a forward shock (FS) that compresses and heats the ambient gas. 
In the early stages of the supernova remnant (SNR) evolution, the pulsar remains near the explosion center because of its relatively modest velocity compared to the rapidly expanding ejecta. 
As the FS propagates through the ISM, enhanced compression occurs in denser regions, giving rise to a reverse shock (RS). 
Initially moving outward behind the FS, the RS eventually reverses direction and propagates inward. 
Unlike the FS, the RS will decelerate the unshocked ejecta, also called the shocked ejecta.

Meanwhile, the pulsar continuously emits a cold plasma outflow (the pulsar wind), composed of charged particles accelerated to relativistic speeds within its magnetic field. 
This high-pressure pulsar wind drives a FS into the surrounding medium, which, during the early evolution (typically within the first few thousand years), is the unshocked ejecta. 
The PSR's FS defines the outer boundary of the pulsar wind nebula (PWN), compressing and heating the surrounding medium, potentially generating X-rays or lower-energy emissions. 
Consequently, the region dominated by the influence of a pulsar is also used to define its PWN (see, e.g., \citet{Giacinti2020}).

As the relativistic pulsar wind propagates through the expanding SNR shell, it gradually decelerates due to interaction with the ambient medium. This deceleration continues until a balance is achieved between the wind's ram pressure and the internal pressure, resulting in the formation of a termination shock (TS). 
In the TS, relativistic particles can be heated and re-accelerated, producing synchrotron radiation \citep{Gaensler_Slane06a}. 

PWNe exhibit three primary evolutionary stages, as illustrated in Figure \ref{fig.intro.PWN_evolution} \citep{Olmi2024}:
\begin{itemize}
    \item Free-expansion phase;
    \item Reverberation phase;
    \item Late phase, e.g., the bow-shock phase shown in panel C of Figure \ref{fig.intro.PWN_evolution}. 
\end{itemize}

\subsection{Free-expansion Phase}

During the free-expansion phase, a young PWN expands with a mild acceleration into the unshocked ejecta. 
The pulsar's kick velocity is much smaller than the expansion speeds of the PWN and SNR, allowing it to remain near the system's center and maintain the observed symmetry (Figure \ref{fig.intro.PWN_evolution}, panel A). 
A prime example at this stage is the Crab Nebula shown in Figure \ref{fig.intro.PWN_examples}. 
Its emission has been observed ranging from radio to TeV gamma rays, and the detection of PeV photons suggests the presence of a PeVatron within the nebula \citep{Cao2021}. 

In leptonic models, synchrotron radiation from relativistic electrons and positrons dominates the radio to X-ray band. 
Higher-energy gamma rays are primarily produced via inverse Compton scattering (ICS) of soft background photons, including synchrotron, infrared (NIR and FIR), and cosmic microwave background (CMB) photons \citep{martin2022unique,Olmi2023}. 

This phase typically lasts several thousand years, depending on the spin-down power of the pulsar and ambient density \citep{Swaluw2001}. 
When the RS reaches the outer edge of the PWN, the PWN moves to the reverberation phase.

\subsection{Reverberation Phase}

Following the collision between the SNR's RS and the PWN's FS, the expansion of the PWN is decelerated and subsequently undergoes compression unless the PWN is sufficiently energetic to resist it. 
This compression leads to an increase in the internal magnetic field, pressure, and energy of the PWN. 
Once the internal pressure becomes higher than the external pressure, the PWN re-expands. 
This cycle of compression and expansion can repeat, resulting in an oscillatory behavior that typically persists on timescales of a few thousand years, e.g., \cite{Gaensler_Slane06a,bandiera2023reverberation,Olmi2024}.

As shown in panel B of Figure \ref{fig.intro.PWN_evolution}, a simplified case is presented in which proper pulsar motion is neglected, and it is assumed that both the SNR ejecta and the ISM are isotropically distributed. 
Under these idealized conditions, the SNR+PWN system is highly symmetric. 

However, in a more realistic scenario, the pulsar is typically offset from the SNR center due to its proper motion. 
Furthermore, anisotropies in the ambient medium contribute to the non-uniform expansion of the SNR, causing the RS to interact with the PWN at different times along different directions. 
As a result, the SNR+PWN system exhibits significant asymmetry during this stage. 
A well-known example of this asymmetry is observed in Vela X (see Figure \ref{fig.intro.PWN_examples}) \citep{Blondin2001}.

Capturing such complex, anisotropic evolution necessitates multidimensional simulations. 
These are significantly more computationally demanding than simplified one-zone or 1D models and remain a major challenge in modeling the evolution of PWN during the reverberation phase (see e.g., \cite{Olmi2020} for an example of a hybrid approach that can perhaps alleviate such problematic). For a full discussion of the reveberation phase see the recent works by \citet{Bandiera2021,Bandiera:2020,Bandiera:2022,bandiera2023reverberation}.

\subsection{Late Phase}

Once the reverberations subside—typically after tens of thousands of years—the PWN resumes steady expansion within the hot, shocked ejecta. 
By this stage, the pulsar has often migrated far from its birthplace. 
If this distance exceeds the size of the original wind bubble, the pulsar escapes and forms a new PWN, leaving a relic nebula behind. 
Such systems commonly exhibit a radio and X-ray bridge linking the old and new PWNe \citep{Gaensler_Slane06a}, as exemplified by the composite SNR G327.1-1.1 (the "Snail" PWN; see Figure \ref{fig.intro.PWN_examples}). 

At later times (typically $>$ 40 kyr), the pulsar may exit the SNR shell and enter the ISM \citep{Gaensler_Slane06a}. 
In contrast to young PWNe, where X-ray and TeV emissions are similar in extent, older systems usually feature a compact X-ray core embedded within a much larger TeV halo \citep{gomez2020predicting}, such as the Mouse PWN (also in Figure \ref{fig.intro.PWN_examples}). 

Because of the reduced sound speed at the SNR edge or in the ISM, many pulsars move with supersonic velocities in this phase, generating bow shocks. 
The equilibrium between the pulsar wind and its surroundings along with the ram pressure constraints limits the spatial extent of the PWN and eventually halts its steady expansion. 
Observationally, if the pulsar's spin-down luminosity remains sufficiently high, the system may manifest itself as a compact, bright head at the leading edge of an elongated tail, as exemplified by the Mouse PWN \citep{Olmi2024}.

In the late phase, the asymmetry of a PWN becomes more pronounced. 
%
It is evident that modelling this phase (likely in 3D) should be able to replicate the PWN's earlier evolutionary stages. 
The construction of such a model remains a key challenge in the field.

\section{\lat and Data Analysis}
\label{lat}

\subsection{Instrument and Performance}

\begin{figure}
    \centering
  	\includegraphics[width=0.8\columnwidth]{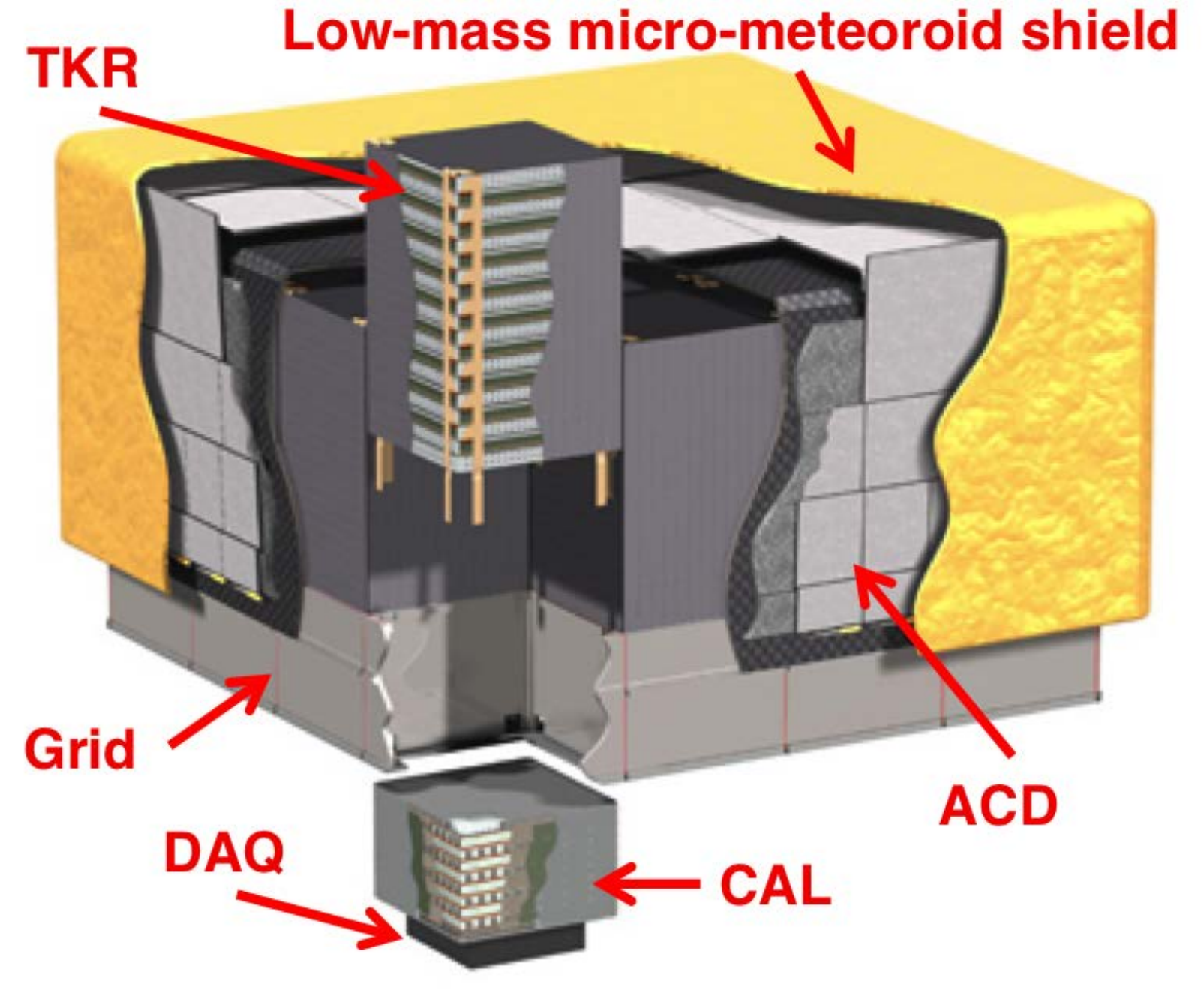}
    \caption{Structure of the \lat. Taken from \citep{Fermi2009ApJ,Nicolas2014}.} 
    \label{fig.intro.lat}
\end{figure}

The \fermi Gamma-ray Space Telescope is equipped with two primary instruments: the Large Area Telescope (LAT) and the Gamma-ray Burst Monitor (GBM). 
The GBM is primarily designed to detect gamma-ray bursts (GRBs) in the energy range of approximately 8 keV to 40 MeV \citep{GBM2009ApJ}, while the LAT is optimized to measure the direction, arrival time and energy of high-energy photons in the range from 20 MeV to 2 TeV \citep{Fermi2009ApJ}. 

For photons with energies exceeding 5 MeV, pair production becomes the dominant interaction mechanism, resulting in the generation of electron–positron pairs. 
Therefore, the LAT functions as a pair-conversion detector. 
Its architecture comprises four main components: 16 high-resolution converter-tracker (TKR) modules, 16 calorimeter (CAL) modules, an anti-coincidence detector (ACD), and a data acquisition system (DAQ), as illustrated in Figure \ref{fig.intro.lat}.

Compared to its predecessor, the Energetic Gamma Ray Experiment Telescope (EGRET) aboard the Compton Gamma Ray Observatory (CGRO), \lat exhibits significant improvements in key performance metrics, including energy range, energy resolution, effective area, field of view, source localization determination, and point source sensitivity\footnote{\url{https://fermi.gsfc.nasa.gov/ssc/data/analysis/documentation/Cicerone/Cicerone_Introduction/LAT_overview.html}.}.

\subsection{\lat Analysis}
\label{intro.lat_analysis}

Since its launch in 2008, the \lat has continuously accumulated over 16 years of observational data. 
The latest data release, Pass 8, and the source catalog, 4FGL-DR4 \citep{Abdollahi2022, Ballet2023}, provide the most up-to-date resources for analysis. 
The corresponding data processing tools, \texttt{Fermitools} package\footnote{\url{https://fermi.gsfc.nasa.gov/ssc/data/analysis/software/}} and \texttt{Fermipy} \citep{Wood2017} are publicly available on the \fermi website.

The Pass 8 data selections recommended\footnote{\url{https://fermi.gsfc.nasa.gov/ssc/data/analysis/documentation/Pass8_usage.html}; \url{https://fermi.gsfc.nasa.gov/ssc/data/analysis/documentation/Cicerone/Cicerone_Data_Exploration/Data_preparation.html}} for standard \lat analyses include: 
\begin{enumerate} 
\item Event class \texttt{P8 SOURCE} (\texttt{evclass}=128) and event type \texttt{FRONT+BACK} (\texttt{evtype}=3); 
\item Energy threshold above 100 MeV to mitigate low-energy IRF uncertainties; 
\item Zenith angle cut of $<90^\circ$ to suppress Earth limb contamination; 
\item Good time intervals selected with ``(\texttt{DATA\_QUAL}$>$0) \&\& (\texttt{LAT\_CONFIG==1})''. 
\end{enumerate}

The Test Statistic (TS) is used to quantify the detection significance of gamma-ray emission from a target source. 
It is defined as $TS = 2(\log \mathcal{L}_{1} - \log \mathcal{L}_{0})$, where $\log \mathcal{L}_{1}$ and $\log \mathcal{L}_{0}$ are the logarithms of the maximum likelihoods for models with and without the source (the ``null hypothesis"), respectively \citep{Mattox1996}. 
The TS approximately follows a $\chi^2$ distribution with degrees of freedom (DOF) equal to the number of free parameters. 
In the simplest case, when only flux normalization is free (that is, DOF = 1), the detection significance can be approximated by $\sqrt{TS}$. 
However, when DOF $>$ 1, the significance should instead be derived from the $p$-value of the $\chi^2_k$ distribution of $p = 1 - F_{\chi^2_k}(\mathrm{TS})$, where $F_{\chi^2_k}$ denotes the cumulative distribution function of the $\chi^2$ distribution with DOF = $k$.
In practice, a threshold of $\mathrm{TS} \geq 25$ is typically used to indicate a significant detection in the \fermi source catalogs. 
This threshold roughly corresponds to the significance of $4\sigma$ in a four-parameter model (e.g. coordinates, flux, and spectral index) \citep{Mattox1996}.

\citet{3FGL} found that adjustment of normalization in each energy band often improved the consistency of the spectral model, particularly at low energies. 
Thus, they introduced a novel approach to address systematics in the diffuse emission model by unweighting the likelihood when the statistical precision of the galactic diffuse emission exceeds the systematic variations in normalization between regions of interest (ROIs), typically at the level of 2–3\%. 
Building on this, Jean Ballet et al. further proposed a likelihood weighting method\footnote{See \url{https://fermi.gsfc.nasa.gov/ssc/data/analysis/scitools/weighted\_like.pdf}}. 
This method can effectively mitigate the risk of overfitting faint sources and reduces the TS overestimation caused by background mismodeling, and it was employed by \citet{Abdollahi2022}.

\section{Structure of the Thesis}
\label{intro.structure}

This thesis is organized to reflect a logical progression from its central focus — the searching, modeling and characterization of PWNe — toward a broader understanding of other pulsar-related gamma-ray systems.  
Accordingly, the chapters are arranged to follow a coherent thematic structure, beginning with foundational background, followed by detailed PWN modeling, population searches, and high-energy implications, and then expanding to studies of two additional pulsar-associated systems.

\textbf{Chapter~\ref{introduction}} provides an introduction to pulsars and PWNe, outlining their physical origin, radiation mechanisms, and evolutionary phases. 
It also introduces the \lat, outlining the gamma-ray data analysis techniques used throughout this work. 
The chapter concludes with the motivation and structure of the dissertation.

\textbf{Chapter~\ref{tide}} presents a time-dependent leptonic radiative model for PWNe, implemented in the TIDE simulation code. 
The model incorporates particle transport, magnetic field evolution, and nebular expansion, and is validated using benchmark sources such as the Crab Nebula, 3C~58, and G11.2$-$0.35. 

\textbf{Chapter~\ref{fermi_pwn}} presents a systematic search for MeV--GeV PWNe using \lat data, focusing on sources not associated with known gamma-ray pulsars. The results have been accepted for publication in \apj~\citep[corresponding author,][]{Eagle2025}. 
A complementary study targeting sources with associated gamma-ray pulsars, based on off-pulse data, is planned as a follow-up to this work. 
The chapter details sample selection, data processing, spatial and spectral analysis, PWN radiation modeling, and includes population-wide comparisons. 

\textbf{Chapter~\ref{tev.pwn}} investigates four individual pulsars (J1208$-$6238, J1341$-$6220, J1838$-$0537, and J1844$-$0346) as potential TeV PWN systems. A “pulsar tree” clustering method is introduced for source selection, and each system is modeled in detail. This work has been published in \aap~\citep{Zhang2024}. 

\textbf{Chapter~\ref{PeVatron}} critically evaluates whether PWNe can account for the multiwavelength emission observed from several ultra-high-energy (UHE) gamma-ray sources detected by LHAASO, HAWC, and HESS.  
Their potential role as Galactic PeVatrons is also examined. Part of this work has been published in \aap~\citep[second author;][]{Sarkar2022}. 

%

\textbf{Chapters~\ref{1A0535} and~\ref{M5}} extend the scope beyond PWNe to explore gamma-ray emission from other pulsar-related systems: the high-mass X-ray binary pulsar 1A~0535+262 and the globular cluster M5. 
We conducted deep gamma-ray spatial, spectral, and temporal analysis with the \lat data for each of them. Both of them have been published in \apj~\citep[as second author;][]{Hou2023,Hou2024}. 

\textbf{Chapter~\ref{Conclusion}} concludes the dissertation with a synthesis of the results and a discussion of future research directions.

\textbf{Appendix~\ref{Appendix}} includes the code scripts and technical documentation supporting the modeling and analysis described in the main chapters.

Additional work not included in this thesis includes a PWN-related study published in \mnras~\citep[co-author;][]{Kundu2024}, in which I contributed modeling results for Kes~75 and G21.5$-$0.9 using our PWN model to independently cross-check their findings. 
I'm also leading an ongoing study to establish the most plausible source model for the MeV--GeV emission from the region surrounding the magnetar Swift~J1818.0$-$1607. 

\newpage
\thispagestyle{empty}
~
\newpage


\chapter{PWN Model and Code Performance}
\label{tide}
\textbf{\color{SectionBlue}\normalfont\Large\bfseries Contents of This Chapter\\\\}

In this chapter, we present the time-dependent leptonic radiative model for PWNe implemented in the numerical simulation code \texttt{TIDE}.
References are provided below. 
The model self-consistently tracks the evolution of the particle energy distribution, magnetic field strength, and nebular size, allowing for robust predictions of multi-wavelength spectral energy distributions over the early stage of a PWN. 
The physical model incorporates a broken power-law injection spectrum, synchrotron and inverse Compton cooling, and magnetic field decay. 
Particle escape is treated with a simplified Bohm-like approximation.

In Section~\ref{performance}, the model is applied to three representative PWNe — Crab Nebula, 3C~58, and G11.2$-$0.3 —to evaluate the performance of TIDE. 
The script used for this analysis, developed by the author, is provided in Appendix~\ref{script2}.
These examples demonstrate that \texttt{TIDE} successfully reproduces the observed emission from PWNe. 
The consistency and flexibility of the results validate the theoretical formulation and the computational implementation, further establishing \texttt{TIDE} as a reliable tool for population-wide studies and future model extensions. 
In addition, this comparative analysis also demonstrates that the probability of reaching the global optimum strongly depends on the richness of the observational data. 
In particular, well-sampled sources such as the Crab Nebula exhibit a high success rate, while sparsely observed sources like G11.2-0.35 are more prone to local minima. 
These results highlight the importance of extensive data coverage and support the implementation of multi-start fitting strategies when modeling poorly constrained systems. 

\newpage

\section{Model Description}
\label{pwn_model}

TIDE is a numerical simulation tool based on a time-dependent leptonic PWN model, the details of which can be found in \citet{Martin2012, Torres2014, Martin2016}, and \citet{martin2022unique}.

\subsection{The Diffusion Equation}

The model describes the temporal evolution of the lepton distribution $N(\gamma, t)$ under the combined effects of particle injection, escape, and energy loss, governed by the following diffusion equation: 
\begin{equation}
 \frac{\partial N(\gamma, t)}{\partial t}=-\frac{\partial}{\partial \gamma}[\dot{\gamma}(\gamma, t) N(\gamma, t)]-\frac{N(\gamma, t)}{\tau(\gamma, t)}+Q(\gamma, t) \text {. }
\label{eq.tide.a}
\end{equation}
The first term on the right-hand side accounts for energy losses from different radiation mechanisms, including synchrotron, bremsstrahlung, inverse Compton (IC), and adiabatic processes here. 
The second term describes particle escape (assuming via Bohm diffusion in our model), with $\tau(\gamma, t)$ denoting the energy- and time-dependent escape timescale. 
The third term, $Q(\gamma, t)$, represents the injection rate per unit of energy and volume.

In principle,  we adopt a broken power-law injection spectrum: 
\begin{equation}
Q(\gamma, t)=Q_{0}(t)\left\{\begin{array}{lll}
\left(\frac{\gamma}{\gamma_{\mathrm{b}}}\right)^{-\alpha_{1}} & \text { for } & \gamma \leq \gamma_{\mathrm{b}}, \\
\left(\frac{\gamma}{\gamma_{\mathrm{b}}}\right)^{-\alpha_{2}} & \text { for } & \gamma>\gamma_{\mathrm{b}}, 
\end{array}\right.
\label{eq.tide.b}
\end{equation}
where $\gamma_b$ is the break energy and $\alpha_1$, $\alpha_2$ are the low- and high-energy spectral indices, respectively. 
The normalization $Q_0(t)$ is determined by the injected particle luminosity $L(t)$: 
\begin{equation}
\eta_p L(t)=\int_{\gamma \min }^{\gamma_{\max }} \gamma m_{e} c^{2} Q(\gamma, t) d \gamma . 
\label{eq.tide.c}
\end{equation}
The spin-down power is partitioned into contributions from particles, magnetic energy, and other forms: $\eta_p$, $\eta_B$, and $\eta_{others}$ (e.g., the multiband pulsar emission), respectively. 
For example, the magnetic fraction is defined as $\eta_B = L_B(t)/L(t)$. 
The spin-down luminosity of a pulsar evolves as: 
\begin{equation}
L(t)=L_{0}\left(1+\frac{t}{\tau_{0}}\right)^{-\frac{n+1}{n-1}}, 
\label{eq.tide.d}
\end{equation}
where $L_0$ is the initial spin-down luminosity, $n$ is the braking index (typically 3), and $\tau_0$ is the initial spin-down timescale: 
\begin{equation}
\tau_{0}=\frac{P_{0}}{(n-1) \dot{P}_{0}}=\frac{2 \tau_{\mathrm{c}}}{n-1}-t_{\mathrm{age}}. 
\label{eq.tide.e}
\end{equation}
\(P_0\) and \(\dot{P}_0\) refer to the initial period and the first period derivative of the pulsar, respectively. 
The parameter \(t_{age}\) represents the pulsar's true age , and \(\tau_c\) is the characteristic age, satisfying 
\begin{equation}
\tau_{\mathrm{c}}=\frac{P}{2 \dot{P}}, 
\label{eq.tide.f}
\end{equation}
where \(P\) and \(\dot{P}\) are the current period and the first period derivative, respectively.

%
The upper limit $\gamma_{\max}$ is restricted by synchrotron cooling and spatial confinement. 
The synchrotron-limited value is \citep{Jager2009, martin2022unique}: 
\begin{equation}
\gamma_{\max }^{\operatorname{sync}}(t)=\frac{3 m_{e} c^{2}}{4 e} \sqrt{\overline{e B(t)}}. 
\label{eq.tide.g}
\end{equation}
%
The gyro-radius constraint requires the Larmor radius $R_L$ smaller than the termination shock radius $R_S$ (i.e. $R_L=\epsilon R_S$, and containment factor $\epsilon < 1$), giving: 
\begin{equation}
\gamma_{\max }^{\operatorname{gyro}}(t)=\frac{\varepsilon e \kappa}{m_{e} c^{2}} \sqrt{\eta \frac{L(t)}{c}}, 
\label{eq.tide.h}
\end{equation}
where the magnetic compression ratio $\kappa \approx 3$ for strong shocks \citep{Becker2007, Holler2012}. 
The minimum of these two constraints is used in \texttt{TIDE}.

\subsection{Radius and Magnetic field evolution}

Before the interaction with the supernova reverse shock, the PWN expands freely. 
The radius evolution follows: 
\begin{equation}
R_{PWN}(t)=C\left(\frac{L_{0} t}{E_{0}}\right)^{1/5} V_{ej} t, 
\label{eq.tide.i}
\end{equation}
where C is a constant with a value of about 0.839 for a PWN made up of relativistic hot gases. 
$V_{ej} = \sqrt{10 E_0 / 3 M_{ej}}$ is the velocity of the ejecta, and \(E_0\) and \(M_{ej}\) denote the energy of the supernova explosion and the mass of the ejecta, respectively. 

The magnetic energy, defined as $W_B = \frac{B^2 R_{\mathrm{PWN}}^3}{6}$,
evolves under adiabatic losses as \citep[e.g.,][]{gelfand2009,Martin2016,martin2022unique}:
\begin{equation}
    \frac{dW_B(t)}{dt} = \eta_{\mathrm{B}} L(t) - \frac{W_B(t)}{R_{\mathrm{PWN}}(t)} \frac{dR_{\mathrm{PWN}}(t)}{dt},
\end{equation}
The resulting magnetic field is: 
\begin{equation}
    B(t) = \frac{1}{R_{\mathrm{PWN}}^2(t)} \sqrt{6 \eta_{\mathrm{B}} \int_0^t L(t') R_{\mathrm{PWN}}(t')\, dt'}.
\end{equation}

\section{Performance of TIDE}
\label{performance}


The core of \texttt{TIDE} is a numerical solution to the diffusion equation, optimized using the Nelder–Mead simplex algorithm \citep{Press1992}. 
This derivative-free method iteratively updates a simplex of $N+1$ vertices in an $N$-dimensional parameter space to approach the best-fit solution. 
Despite reduced efficiency in higher dimensions \citep{Gao2012}, an enhanced version integrated in \texttt{scipy.optimize.minimize} improves convergence. 
For typical cases with 6–8 free parameters and a reasonable initial guess, convergence is usually achieved within $\sim$1000 iterations \citep{martin2022unique}.

A systematic performance evaluation by \citet{martin2022unique} tested \texttt{TIDE} on the Crab Nebula and 3C 58, using 150 randomized initializations per source. 
Their results demonstrate:
\begin{itemize}
    \item stable convergence in over 70\% of trials for both PWNe, validating robustness against varied initial guesses; 
    \item strong constraints on physical parameters when multi-wavelength data are available, indicating uniqueness in model solutions;
    \item incorporation of an additional systematic uncertainty parameter to account for calibration differences across multi-instrument datasets, enhancing the robustness of statistical fitting;
    \item first-time estimation of PWN age purely from spectral fitting, achieving high precision (e.g., $<$ 50 yr error for the Crab Nebula).
\end{itemize}
These results illustrate that TIDE is an advanced tool for modeling PWNe, offering both improved computational efficiency and scientifically rigorous fitting capabilities. 
The model represents a significant step forward in constraining physical parameters of PWNe and in understanding their complex evolution across the electromagnetic spectrum.

To further assess the sensitivity of the fitting performance of \texttt{TIDE} to the quality and quantity of multiwavelength data, we applied it to three representative PWNe with varying degrees of observational coverage: the Crab Nebula (extensively observed across all wavelengths), 3C 58 (moderately observed), and G11.2-0.35 (sparsely observed). 
Although 3C~58 was considered as a less-observed case in \citet{martin2022unique}, its data coverage still exceeds that of most-usually observed PWNe. 
The inclusion of G11.2$-$0.35 thus enables a more representative evaluation of \texttt{TIDE}'s performance under typical observational conditions.

For each source, we conducted 200 independent fitting trials using randomly selected initial values for the free parameters, following the strategy of \citet{Mandal2022}. 
Fixed parameters include the far-infrared (FIR) and near-infrared (NIR) temperatures ($T_{fir}, T_{nir}$), ambient medium density ($n_{ism}$), braking index ($n$), containment factor ($\epsilon$), and explosion energy ($E_{sn}$), with their values listed in the corresponding summary tables. 
The free parameters, namely the break energy ($\gamma_b$), spectral indices ($\alpha_1, \alpha_2$), ejecta mass ($M_{ej}$), magnetic energy fraction ($\eta$), FIR and NIR energy densities ($U_{fir}, U_{nir}$), and the systematic uncertainty ($\sigma$), were allowed to vary within specified ranges.

Instead of only testing for convergence in \citet{martin2022unique}, we classified solutions into distinct model families by clustering both parameter values and their associated spectral energy distributions (SEDs). 
For each PWN, we identified the global best-fit model and compared it to local minima. We evaluated fits using statistical criteria such as the reduced $\chi^2$, parameter distributions, and SED quality.

\subsection{The Crab Nebula}

\begin{table}[H]
    \centering
    \caption{Summary of the physical magnitudes of the Crab Nebula}. \label{tab:mag.crab}
    \label{tab.tide.parameters.crab}
    \begin{tabular}{@{}llll@{}}
    \toprule
		Parameters & Symbol &Values &Fitting Range\\
		\hline
         Measured or assumed parameters: \\
        		\hline
		Age (kyr) &$t_{age}$  &0.968\\
		Period (ms) &P  &33.4 \\
        Period derivative $\mathrm{(s~s^{-1})}$ &$\dot{P}$ &$4.2\times10^{-13}$ &\\
		Characteristic age (kyr) &$\tau_c$ &1296\\
		Spin-down luminosity now $\mathrm{(erg~s^{-1})}$ &L &$4.5\times10^{38}$\\
		Braking index &n &2.509 &\\
    Initial spin-down luminosity $\mathrm{(erg~s^{-1})}$ &$L_0$ &$3.1\times10^{39}$\\
		Initial spin-down age (kyr) &$\tau_0$ &0.75\\
		Distance (kpc) &d &2\\
		SN explosion energy (erg) &$E_{sn}$ &$10^{51}$ &\\
		ISM density ($\mathrm{cm^{-3}}$) &$n_{ism}$ &0.5\\
		Minimum energy at injection &$\gamma_{min}$ &1 &\\
        Containment factor  &$\epsilon$ &0.3 & \\
        CMB temperature (K)  &$T_{cmb}$ &2.73 &\\
		CMB energy density $\mathrm{(eV~cm^{-3})}$  &$U_{cmb}$ &0.25 &\\
        FIR temperature (K) &$T_{fir}$ &70\\
        NIR temperature (K) &$T_{nir}$ &5000\\
                \hline
        Fitted parameters: \\
        		\hline
		Break energy ($10^5$)&$\gamma_b$ &6.08 (5.78, 7.40) &0.1 - 100\\
		Low energy index &$\alpha_1$ &1.51 (1.47, 1.53) &1 - 4\\
		High energy index &$\alpha_2$  &2.49 (2.46, 2.49) &1 - 4\\
		Ejected mass $(M_{\odot})$  &$M_{ej}$  &7.01 (7, 7.56) &7 - 12\\
		Magnetic energy fraction ($10^{-2}$)  &$\eta$ &2.042 (2.022, 2.435) &0.01 - 50\\
		FIR energy density $\mathrm{(eV~cm^{-3})}$ &$U_{fir}$ &0.16 (0.1, 0.26) &0.1 - 5\\
		NIR energy density $\mathrm{(eV~cm^{-3})}$ &$U_{nir}$ &2.64 (0.1, 4.65) &0.1 - 5\\
		PWN radius now (pc) &$R_{pwn}$ &1.86\\
		Magnetic field now ($\mu$G) &B &83.90\\
		Reduced $\chi^2$ &$\chi^2/D.O.F.$ &274.30/287 (0.96)\\
		Systematic uncertainty &$\sigma$ &0.19 &0.01 - 0.5\\
    \bottomrule
    \end{tabular}
\end{table}

\begin{figure}[H]
    \centering
    \includegraphics[width=0.32\columnwidth]{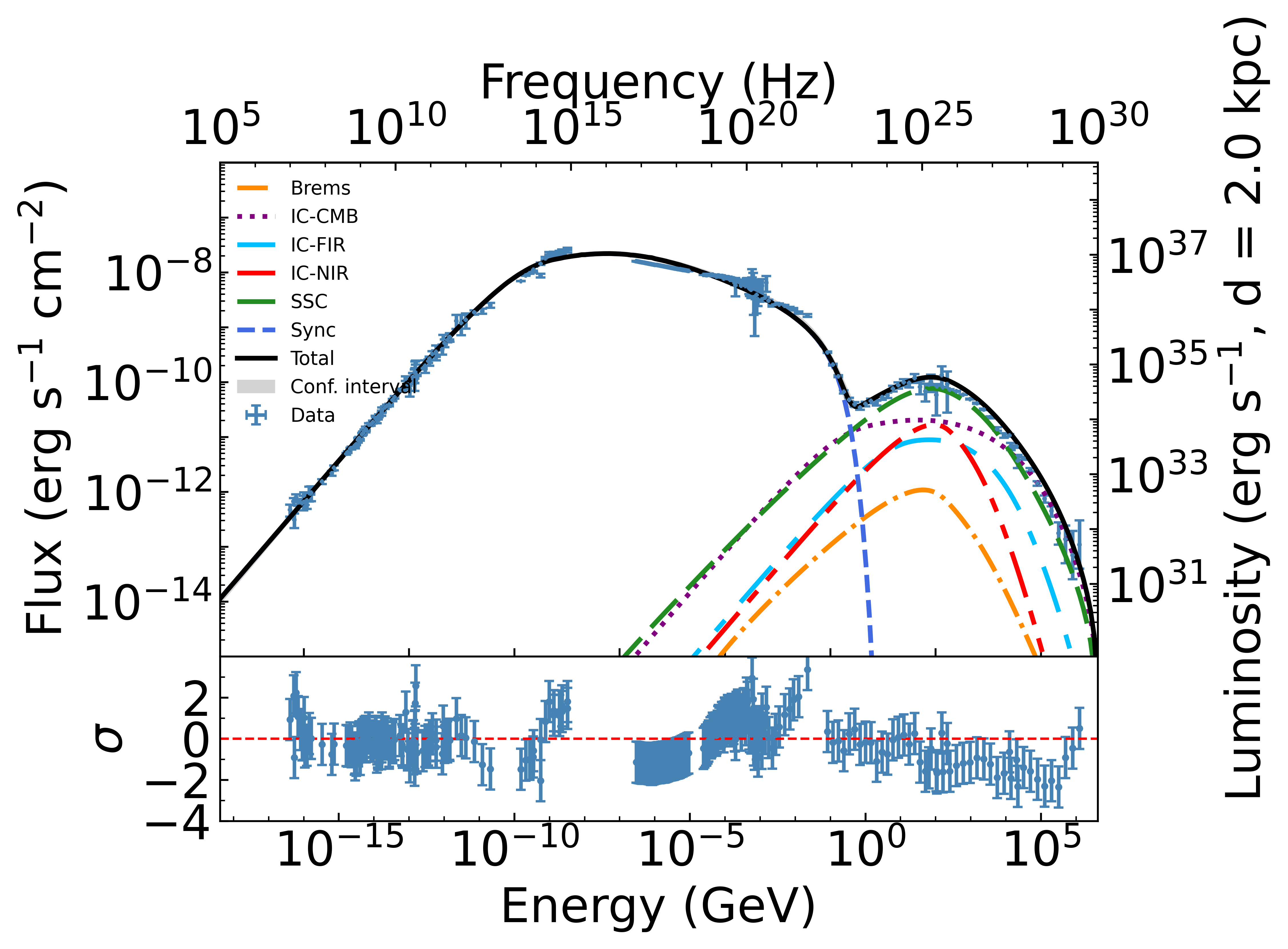}
    \includegraphics[width=0.32\columnwidth]{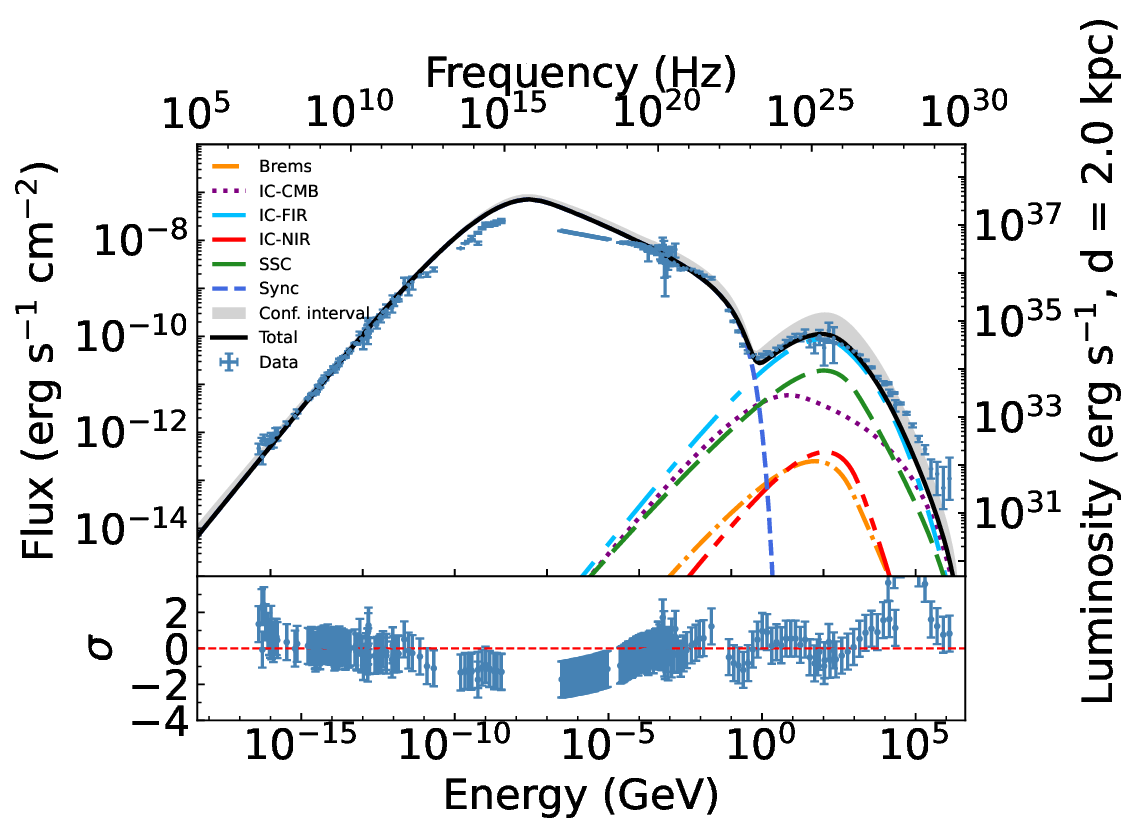}
    \includegraphics[width=0.32\columnwidth]{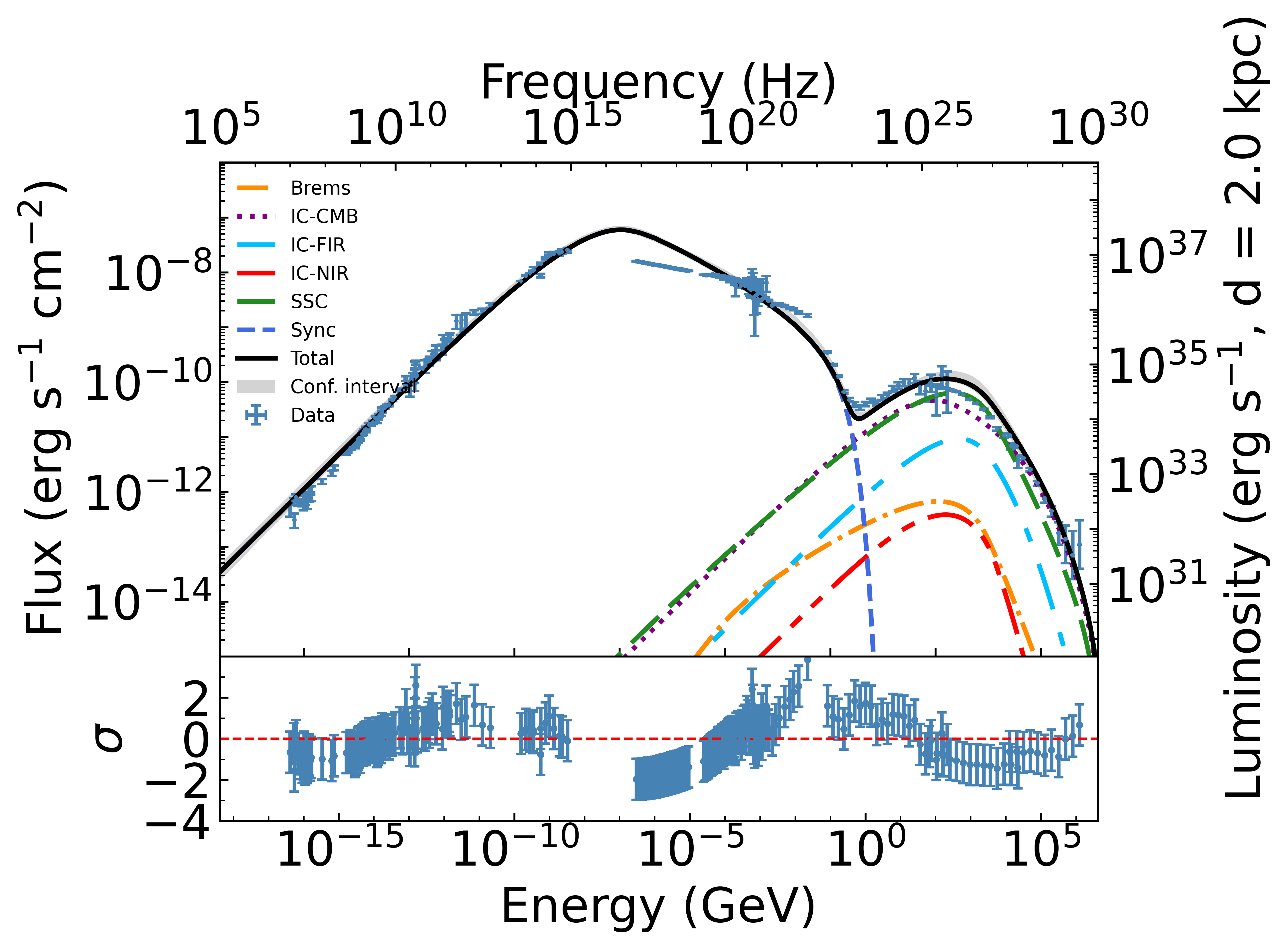}
    \caption{
 From left to right: Representative SEDs for Model 1, Model 2 and Model 3, respectively. The multi-band data in the SED are taken from \citet{Baldwin1971, Macias2010} (radio), \citet{Buehler2013} (from IR to TeV bands) and \citet{lhaaso2021} (LHAASO).
    }
    \label{pic.tide.sed.crab}
\end{figure}




The Crab Nebula is undoubtedly the most prominent member of the PWN family. 
Since its discovery by John Bevis in 1731, it has been extensively studied from radio to PeV energies. 
Because of the wealth of observational data available, it has become the benchmark for testing nearly all PWN models. 

The nebula is powered by PSR J0534+2200, a young and energetic pulsar discovered in 1968. 
With a true age of 968 years and a spin-down luminosity of \(4.5 \times 10^{38}~\mathrm{erg~s}^{-1}\), it serves as a key energy source for this bright extended emission. 
Detailed parameters for both the pulsar and the nebula are listed in Table \ref{tab.tide.parameters.crab}, and are adopted in this study. 
The fitting ranges and the best-fit values for the free parameters,\(\gamma_b\), \(\alpha_1\), \(\alpha_2\), \(M_\mathrm{ej}\), magnetic fraction \(\eta\), \(U_\mathrm{fir}\), \(U_\mathrm{nir}\), and the systematic uncertainty \(\sigma\) are also provided in the same table. 
The corresponding best-fit SED is shown in Figure \ref{pic.tide.sed.crab} (Model 1).

In general, these fits can be roughly categorized into the following models: 
\begin{itemize}
  \item \textbf{Model 1: 135 fits (67.5\%), global optimum.}  
  The representative values of the free parameters - \(\gamma_b\), \(\alpha_1\), \(\alpha_2\), \(M_\mathrm{ej}\), magnetic fraction \(\eta\), \(U_\mathrm{fir}\), \(U_\mathrm{nir}\), and systematic uncertainty \(\sigma\) - are approximately \(5 \times 10^5\), 1.49, 2.48, \(7.0~M_{\odot}\), 0.020, \(0.1~\mathrm{cm^{-3}}\), \(0.1~\mathrm{cm^{-3}}\), and 0.16, respectively. 
  The corresponding reduced \(\chi^2\), magnetic field \(B\), and PWN radius \(R_\mathrm{pwn}\) are about 1.05, \(84~\mu\mathrm{G}\), and 1.86 pc. 
  A representative SED is shown in the left panel of Figure \ref{pic.tide.sed.crab}.

  \item \textbf{Model 2: 58 fits (29\%), local optimum.}  
  The representative values of the free parameters are approximately \(2.7 \times 10^6\), 1.42, 2.58, \(7.0~M_{\odot}\), 0.43, \(5~\mathrm{cm^{-3}}\), \(0.1~\mathrm{cm^{-3}}\), and 0.39. 
  The corresponding reduced \(\chi^2\), magnetic field \(B\), and radius \(R_\mathrm{pwn}\) are about 1.18, \(284~\mu\mathrm{G}\), and 2.31 pc. 
  A representative SED is shown in the middle panel of Figure \ref{pic.tide.sed.crab}.

  \item \textbf{Model 3: 5 fits (2.5\%), local optimum.}  
  The representative values of the free parameters are approximately \(1 \times 10^7\), 1.73, 2.76, \(7.0~M_{\odot}\), 0.021, \(0.10~\mathrm{cm^{-3}}\), \(0.11~\mathrm{cm^{-3}}\), and 0.36. 
  The corresponding reduced \(\chi^2\), magnetic field \(B\), and radius \(R_\mathrm{pwn}\) are about 1.02, \(91~\mu\mathrm{G}\), and 1.80 pc. 
  A representative SED is shown in the right panel of Figure \ref{pic.tide.sed.crab}.

  \item \textbf{Others: 2 fits (1\%).}  
  These include a failed fit and a local solution that does not belong to any of the above categories. 
\end{itemize}

Among the 200 fits, only Model 1 corresponds to the global optimum, accounting for 67.5\% of the total. 
Model 2, while comprising a considerable fraction (29\%), clearly does not represent the best-fit solution. 
Model 3 and the remaining unclassified fits account for only 5 and 2 cases, respectively, and thus carry negligible statistical significance. 
In summary, more than 65\% of the trials converge to good fits, indicating that TIDE yields highly reliable results when applied to sources with rich multi-wavelength data. 
The likelihood of becoming trapped in a local optimum is low and can be further minimized by performing multiple fits with different initial values for the free parameters and selecting the best among them.

Furthermore, the non-global-optimal fits can be readily identified as suboptimal. 
They are often associated with excessively large systematic uncertainties (as in Models 2 and 3), parameter values approaching boundary limits (e.g. \(\gamma_b\) in Model 3), and poor SED matches.

\subsection{3C 58}

\begin{table}[H]
    \centering
    \caption{Summary of the physical magnitudes of 3C 58}. \label{tab:mag.3c58}
    \label{tab.tide.parameters.3c58}
    \begin{tabular}{@{}llll@{}}
    \toprule
		Parameters & Symbol &Values &Fitting Range\\
		\hline
         Measured or assumed parameters: \\
        		\hline
		Age (kyr) &$t_{age}$  &2.5\\
		Period (ms) &P  &65.7 \\
        Period derivative $\mathrm{(s~s^{-1})}$ &$\dot{P}$ &$1.93\times10^{-13}$ &\\
		Characteristic age (kyr) &$\tau_c$ &1296\\
		Spin-down luminosity now $\mathrm{(erg~s^{-1})}$ &L &$2.7\times10^{37}$\\
		Braking index &n &3 &\\
		Initial spin-down luminosity $\mathrm{(erg~s^{-1})}$ &$L_0$ &$9.3\times10^{37}$\\
		Initial spin-down age (kyr) &$\tau_0$ &2.878\\
		Distance (kpc) &d &2\\
		SN explosion energy (erg) &$E_{sn}$ &$10^{51}$ &\\
		ISM density ($\mathrm{cm^{-3}}$) &$n_{ism}$ &0.1\\
		Minimum energy at injection &$\gamma_{min}$ &1 &\\
        Containment factor  &$\epsilon$ &0.5 & \\
        CMB temperature (K)  &$T_{cmb}$ &2.73 &\\
		CMB energy density $\mathrm{(eV~cm^{-3})}$  &$U_{cmb}$ &0.25 &\\
        FIR temperature (K) &$T_{fir}$ &25\\
        NIR temperature (K) &$T_{nir}$ &2900\\
                \hline
        Fitted parameters: \\
        		\hline
		Break energy ($10^5$)&$\gamma_b$ &0.88 (0.85, 0.91) &0.1 - 100\\
		Low energy index &$\alpha_1$ &1.04 (1, 1.07) &1 - 4\\
		High energy index &$\alpha_2$  &3.01 (3.005, 3.011) &1 - 4\\
		Ejected mass $(M_{\odot})$  &$M_{ej}$  &12.82 (12.52, 13.30) &7 - 25\\
		Magnetic energy fraction ($10^{-2}$)  &$\eta$ &1.420 (1.252, 1.330) &0.01 - 50\\
		FIR energy density $\mathrm{(eV~cm^{-3})}$ &$U_{fir}$ &0.22 (0.1, 0.42) &0.1 - 5\\
		NIR energy density $\mathrm{(eV~cm^{-3})}$ &$U_{nir}$ &0.45 (0.1, 1.05) &0.1 - 5\\
		PWN radius now (pc) &$R_{pwn}$ &2.57\\
		Magnetic field now ($\mu$G) &B &15.33\\
		Reduced $\chi^2$ &$\chi^2/D.O.F.$ &234.43/225 (1.04)\\
		Systematic uncertainty &$\sigma$ &0.07 &0.01 - 0.5\\
    \bottomrule
    \end{tabular}
\end{table}

\begin{figure}[H]
    \centering
    \includegraphics[width=0.32\columnwidth]{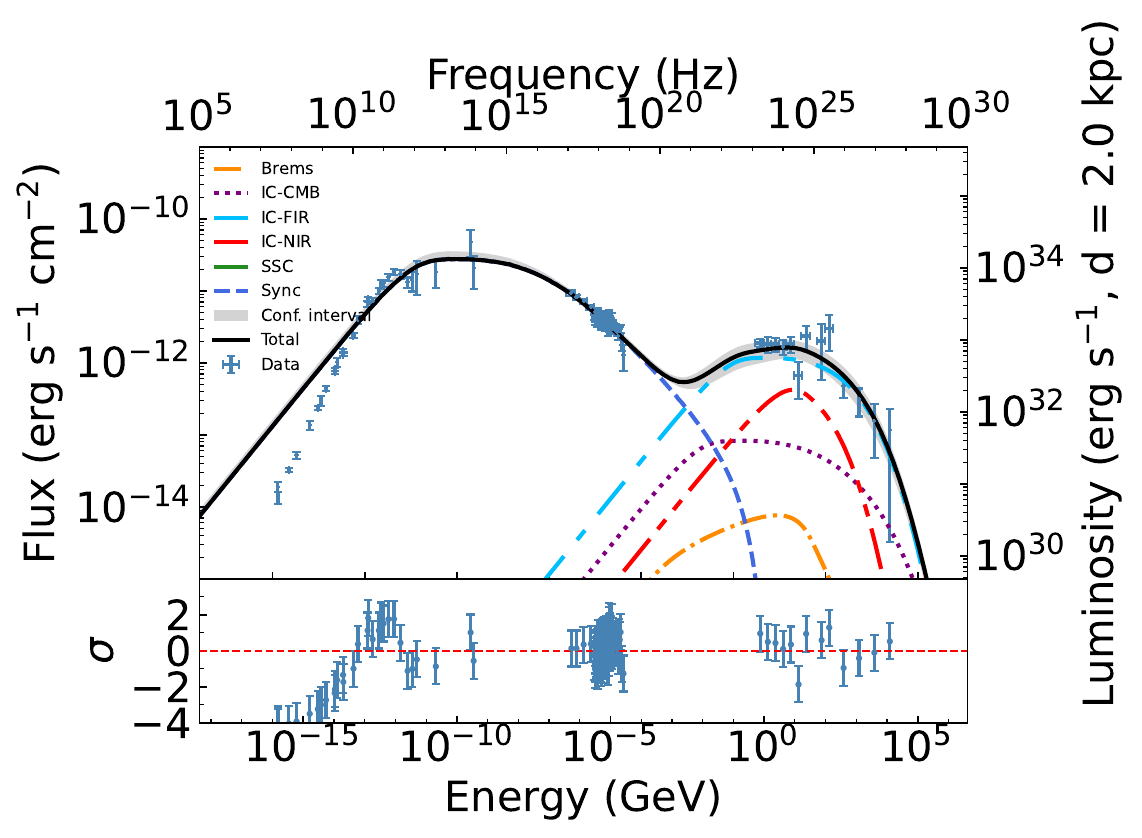}
    \includegraphics[width=0.32\columnwidth]{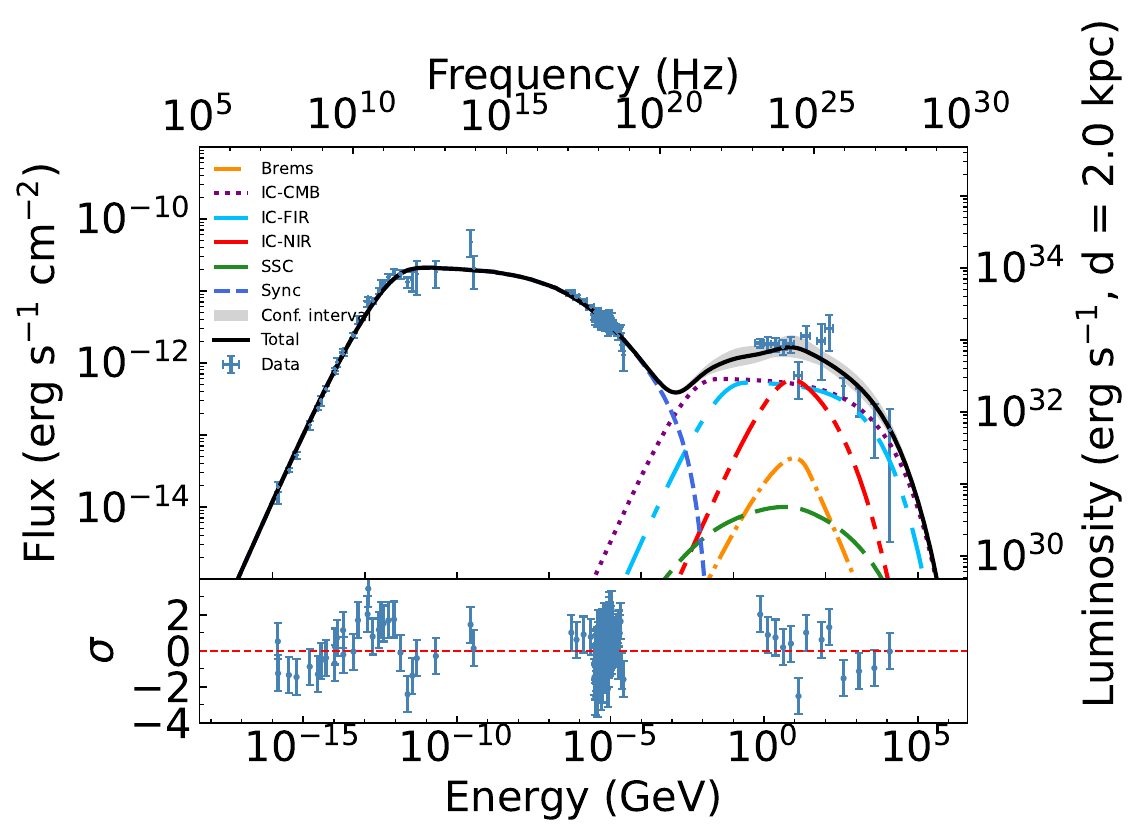}
    \includegraphics[width=0.32\columnwidth]{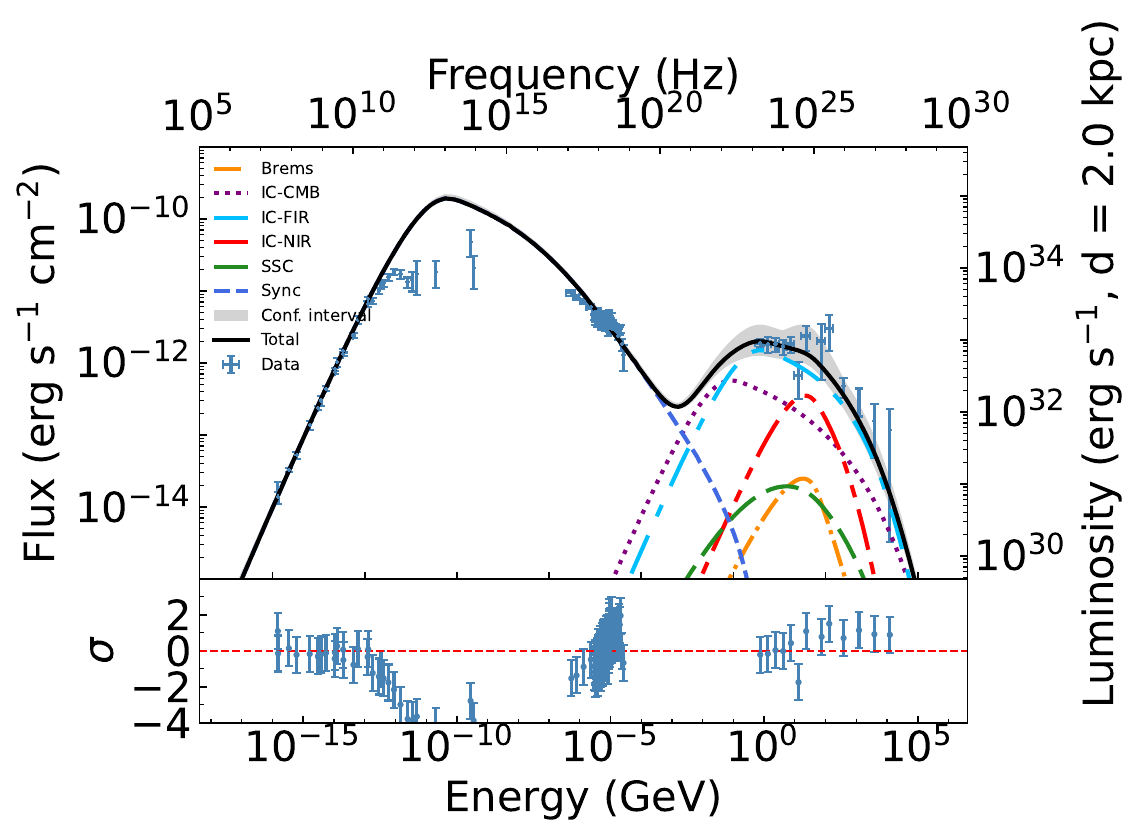}
    \includegraphics[width=0.32\columnwidth]{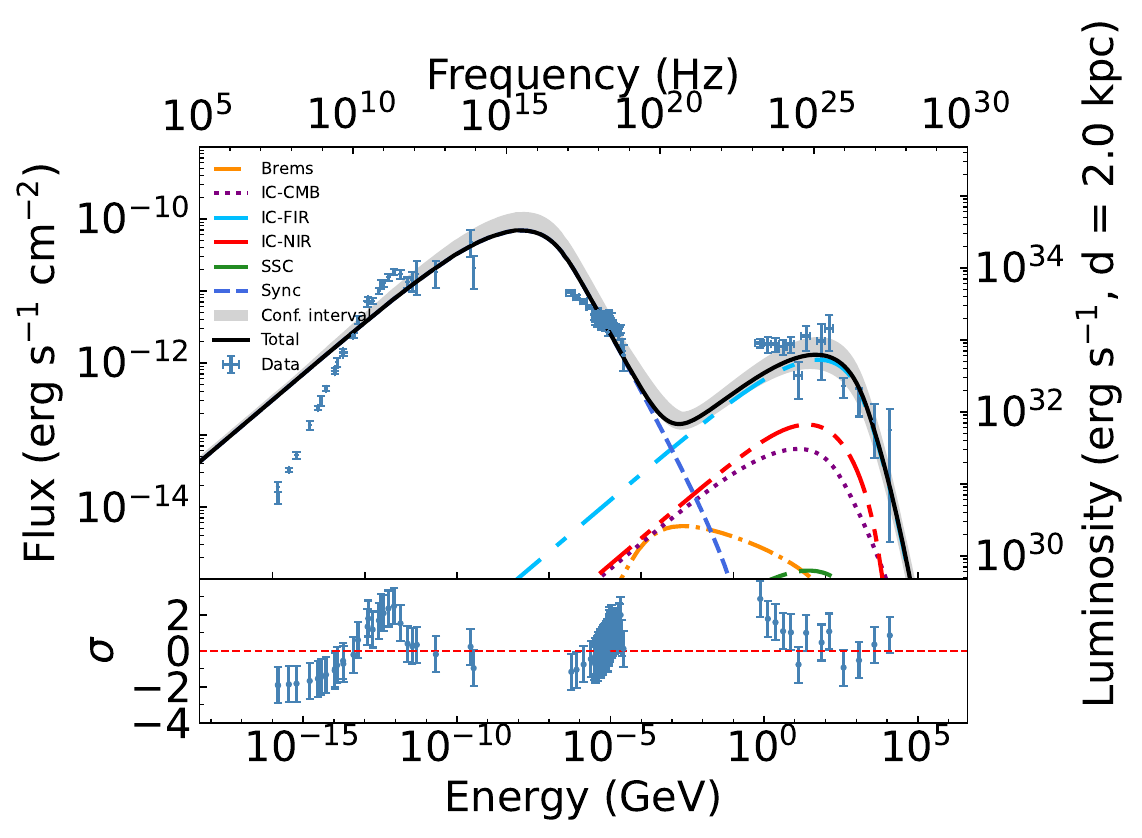}
    \includegraphics[width=0.32\columnwidth]{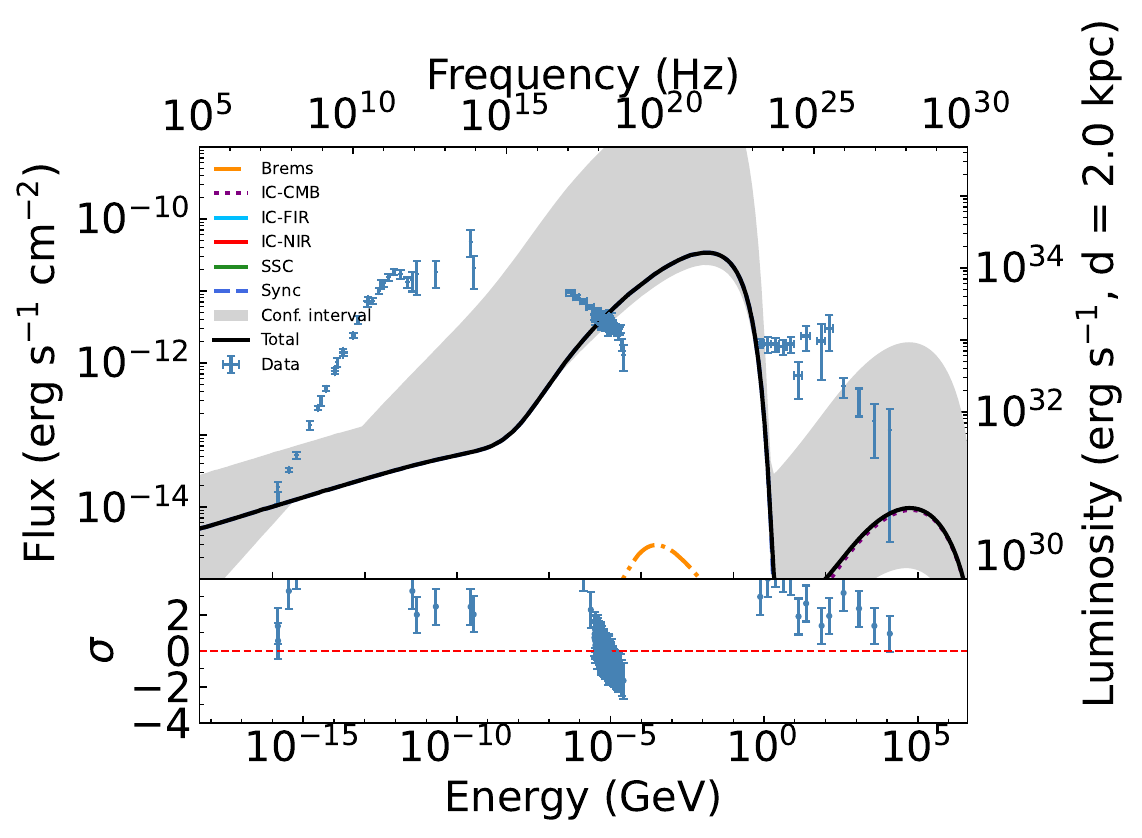}
    \includegraphics[width=0.32\columnwidth]{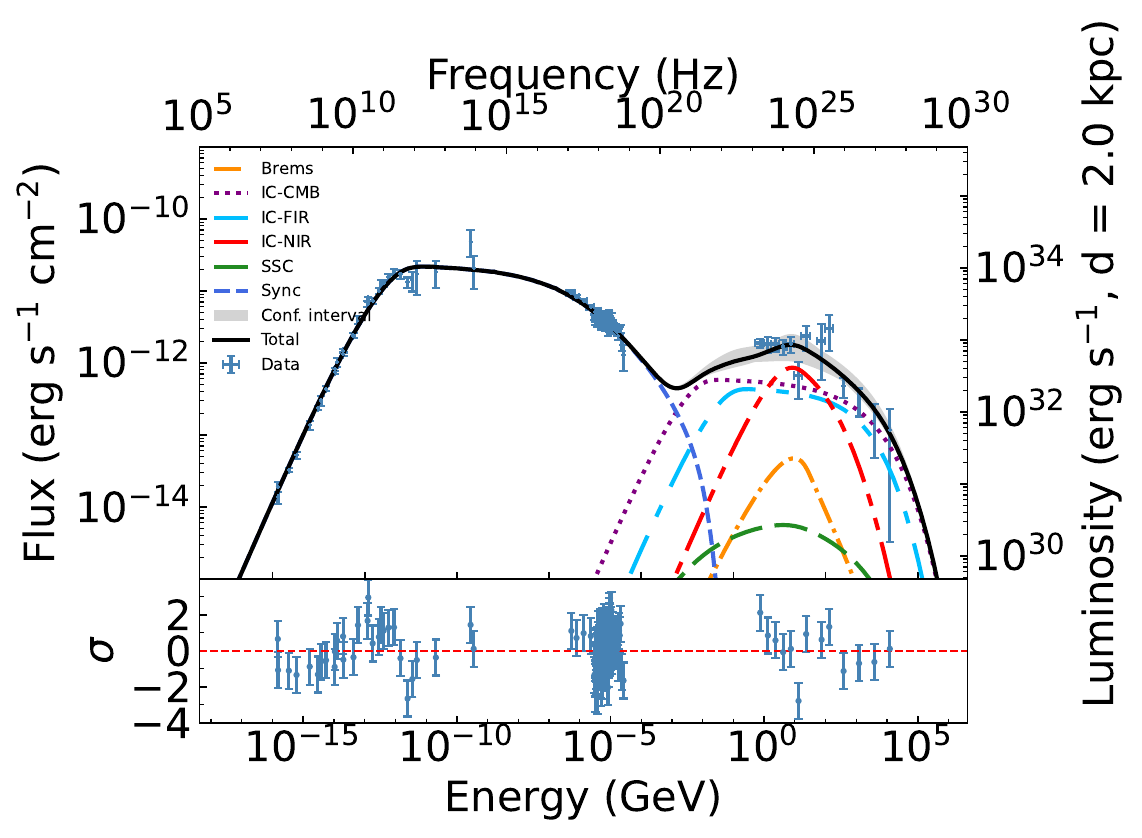}
    \includegraphics[width=0.32\columnwidth]{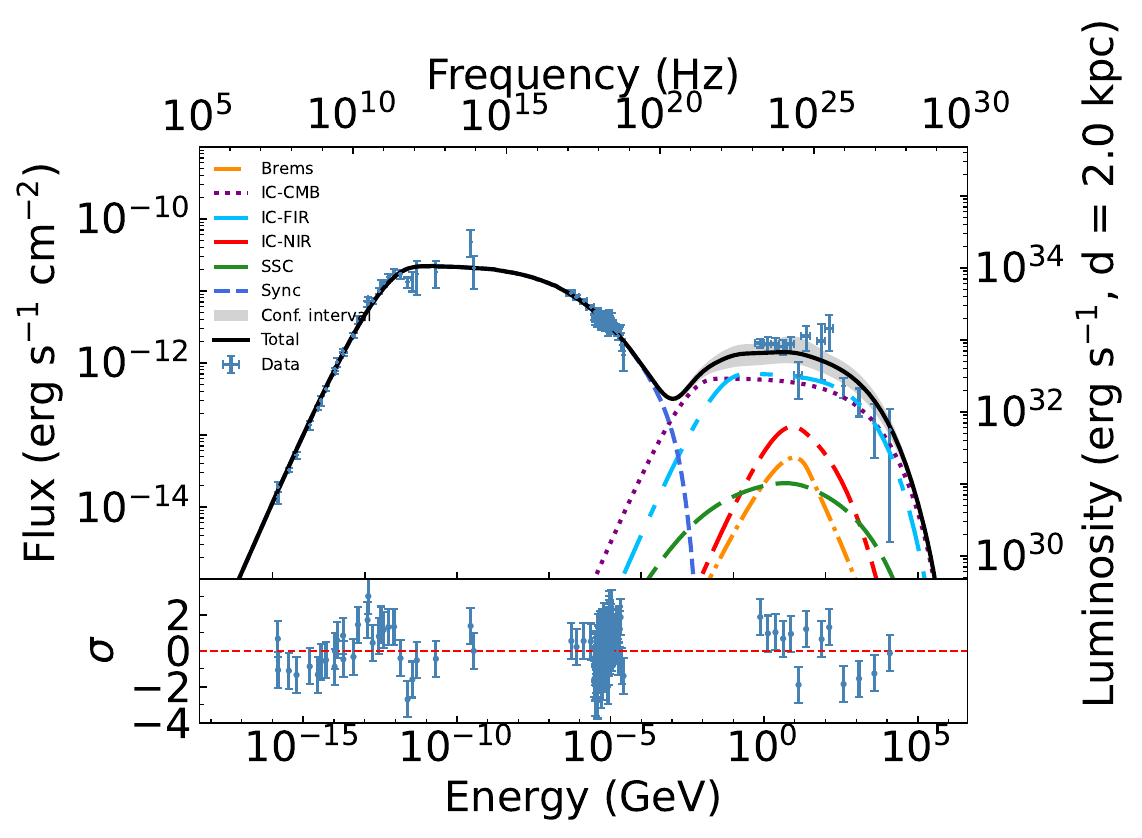}
    \caption{
 From top to bottom, left to right: Representative SEDs for Model 1, Model 2 Model 3, Model 4, Model 5, Model 6 and Model 7, respectively. The multi-band data in the SED are taken from \cite{Green1986, Morsi1987, Salter1989} (radio), \cite{Torii1999} (infrared), \cite{Green1994, Slane2008} (X-rays), \cite{lat2013, Ackermann2013, Li2018} (GeV) and \cite{Aleksic2014} (TeV).
    }
    \label{pic.tide.sed.3c58}
\end{figure}

3C 58 is another well-known PWN, powered by the pulsar PSR J0205+6449. 
Initially discovered in radio and classified as a supernova remnant (SNR) in 1971, it was later reclassified as a Crab-like PWN based on its flat radio spectrum and centrally filled morphology \citep{Weiler1978}. 
The powering pulsar, PSR J0205+6449, was discovered much later, in 2002 \cite{Murray2002}. 

To date, 3C 58 has been detected across a broad range of wavelengths, including radio, infrared, X-ray, GeV, and TeV bands, providing relatively rich multi-wavelength observations for modeling. 
Of particular note is the ongoing debate on the age and distance of this source. 
Following \cite{martin2022unique}, we adopt a dynamical age of 2.5 kyr \cite{Chevalier2004, Chevalier2005} and the latest distance measurement of 2 kpc \cite{Kothes2010} in this work.

Detailed information on PSR J0205+6449 and 3C 58 is provided in Table \ref{tab.tide.parameters.3c58}. 
This table also includes the best-fit parameters, and the corresponding SED is shown in Figure~\ref{pic.tide.sed.3c58} (Model 2). 
As in the Crab Nebula, 200 fits were performed using randomly selected initial values for the free parameters. 

All of these fits can be roughly grouped into eight types: 
\begin{itemize} 
\item \textbf{Model 1: 58 fits (29\%), local optimum.} 
The representative values of the free parameters - \(\gamma_b\), \(\alpha_1\), \(\alpha_2\), \(M_\mathrm{ej}\), \(\eta\), \(U_\mathrm{fir}\), \(U_\mathrm{nir}\), and systematic uncertainty \(\sigma\) - are approximately \(1.08 \times 10^5\), 1.95, 2.88, \(7.0~M_{\odot}\), 0.5, \(3.6~\mathrm{cm^{-3}}\), \(2.6~\mathrm{cm^{-3}}\), and 0.22, respectively. 
The corresponding reduced \(\chi^2\), magnetic field \(B\), and radius \(R_\mathrm{pwn}\) are about 0.93, \(47~\mu\mathrm{G}\), and 0.93~pc. 
A representative SED is shown in the top left panel of Figure~\ref{pic.tide.sed.3c58}. 

\item \textbf{Model 2: 78 fits (39\%), global optimum.}  
The representative values are approximately \(0.88 \times 10^5\), 1, 3, \(13~M_{\odot}\), 0.014, \(0.22~\mathrm{cm^{-3}}\), \(0.45~\mathrm{cm^{-3}}\), and 0.07. 
The corresponding reduced \(\chi^2\), magnetic field \(B\), and radius \(R_\mathrm{pwn}\) are about 1.05, \(16~\mu\mathrm{G}\), and 2.57~pc. 
A representative SED is shown in the upper middle panel of Figure~\ref{pic.tide.sed.3c58}. 

\item \textbf{Model 3: 11 fits (5.5\%), local optimum.}  
The representative values are approximately \(1.9 \times 10^5\), 1, 3.22, \(7.0~M_{\odot}\), 0.5, \(0.67~\mathrm{cm^{-3}}\), \(0.3~\mathrm{cm^{-3}}\), and 0.21. 
The corresponding reduced \(\chi^2\), magnetic field \(B\), and radius \(R_\mathrm{pwn}\) are about 0.99, \(48~\mu\mathrm{G}\), and 3.95~pc. 
A representative SED is shown in the top right panel of Figure~\ref{pic.tide.sed.3c58}. 

\item \textbf{Model 4: 7 fits (3.5\%), local optimum.}  
The representative values are approximately \(1 \times 10^7\), 2.27, 3.59, \(25~M_{\odot}\), 0.19, \(5~\mathrm{cm^{-3}}\), \(5~\mathrm{cm^{-3}}\), and 0.5. 
The corresponding reduced \(\chi^2\), magnetic field \(B\), and radius \(R_\mathrm{pwn}\) are about 0.7, \(90~\mu\mathrm{G}\), and 2~pc. 
A representative SED is shown in the middle-row left panel of Figure~\ref{pic.tide.sed.3c58}.

\item \textbf{Model 5: 4 fits (2\%), local optimum.}  
The representative values are approximately \(1 \times 10^7\), 2.74, 1, \(8~M_{\odot}\), 0.5, \(0.1~\mathrm{cm^{-3}}\), \(0.2~\mathrm{cm^{-3}}\), and 0.36. 
The corresponding reduced \(\chi^2\), magnetic field \(B\), and radius \(R_\mathrm{pwn}\) are about 14, \(49~\mu\mathrm{G}\), and 3.90~pc. 
A representative SED is shown in the middle-row middle panel of Figure~\ref{pic.tide.sed.3c58}. 

\item \textbf{Model 6: 12 fits (6\%), local optimum.}  
The representative values are approximately \(0.86 \times 10^5\), 1, 3, \(7.0~M_{\odot}\), 0.038, \(0.1~\mathrm{cm^{-3}}\), \(5~\mathrm{cm^{-3}}\), and 0.1. 
The corresponding reduced \(\chi^2\), magnetic field \(B\), and radius \(R_\mathrm{pwn}\) are about 0.97, \(16~\mu\mathrm{G}\), and 3.4~pc. 
A representative SED is shown in the right panel of the middle row of Figure~\ref{pic.tide.sed.3c58}.

\item \textbf{Model 7: 10 fits (5\%), local optimum.}  
The representative values are approximately \(0.86 \times 10^5\), 1, 3, \(25~M_{\odot}\), 0.005, \(0.2~\mathrm{cm^{-3}}\), \(0.5~\mathrm{cm^{-3}}\), and 0.07. 
The corresponding reduced \(\chi^2\), magnetic field \(B\), and radius \(R_\mathrm{pwn}\) are about 1.05, \(16~\mu\mathrm{G}\), and 1.83~pc. 
A representative SED is shown in the bottom panel of Figure~\ref{pic.tide.sed.3c58}. 

\item \textbf{Others: 10 fits (5\%).}  
This category includes several different local optimum models that do not belong to any of the above categories. 
\end{itemize}




Among the 200 fits, only Model 2 represents the global optimum, accounting for 39\% of all trials. 
Although this is the largest share, it is significantly lower than the 67.5\% observed for the Crab Nebula, reflecting the impact of sparse data coverage. 
This highlights the need for multiple trials with carefully chosen initial values to avoid local minima.
Model 1, which comprises 29\% of the fits, clearly cannot reproduce the radio observations, as is evident from its SED.

Moreover, non-global-optimal fits (e.g., Model 1, Model 3, Model 4, and Model 5) are often characterized by large systematic uncertainties, parameters close to their boundaries (e.g., $\gamma_b$ for Model 4 and Model 5, $\eta$ for Model 1, Model 3, and Model 5), and poor SED performance.

\subsection{PWN G11.2-0.35}

\begin{table}[H]
    \centering
    \caption{Summary of the physical magnitudes of PWN G11.2-0.35}. \label{tab:mag.G11}
    \label{tab.tide.parameters.G11}
    \begin{tabular}{@{}llll@{}}
    \toprule
		Parameters & Symbol &Values &Fitting Range\\
		\hline
         Measured or assumed parameters: \\
        		\hline
		Age (kyr) &$t_{age}$  &1.6\\
		Period (ms) &P  &64.698 \\
        Period derivative $\mathrm{(s~s^{-1})}$ &$\dot{P}$ &$3.432\times10^{-14}$ &\\
		Characteristic age (kyr) &$\tau_c$ &29.885\\
		Spin-down luminosity now $\mathrm{(erg~s^{-1})}$ &L &$6.4\times10^{36}$\\
		Braking index &n &3 &\\
		Initial spin-down luminosity $\mathrm{(erg~s^{-1})}$ &$L_0$ &$7.14\times10^{36}$\\
		Initial spin-down age (kyr) &$\tau_0$ &28.285\\
		Distance (kpc) &d &3.7\\
		SN explosion energy (erg) &$E_{sn}$ &$10^{51}$ &\\
		ISM density ($\mathrm{cm^{-3}}$) &$n_{ism}$ &1.7\\
		Minimum energy at injection &$\gamma_{min}$ &1 &\\
        Containment factor  &$\epsilon$ &0.5 & \\
        CMB temperature (K)  &$T_{cmb}$ &2.73 &\\
		CMB energy density $\mathrm{(eV~cm^{-3})}$  &$U_{cmb}$ &0.25 &\\
        FIR temperature (K) &$T_{fir}$ &36\\
        NIR temperature (K) &$T_{nir}$ &3300\\
                \hline
        Fitted parameters: \\
        		\hline
		Break energy ($10^5$)&$\gamma_b$ &6.08 (5.78, 7.40) &0.1 - 100\\
		Low energy index &$\alpha_1$ &1.51 (1.47, 1.53) &1 - 4\\
		High energy index &$\alpha_2$  &2.49 (2.46, 2.49) &1 - 4\\
		Ejected mass $(M_{\odot})$  &$M_{ej}$  &7.01 (7, 7.56) &8 - 20\\
		Magnetic energy fraction ($10^{-2}$)  &$\eta$ &2.042 (2.022, 2.435) &0.1 - 50\\
		FIR energy density $\mathrm{(eV~cm^{-3})}$ &$U_{fir}$ &7.65 (1.46, 10) &0.01 - 10\\
		NIR energy density $\mathrm{(eV~cm^{-3})}$ &$U_{nir}$ &0.08 (0.01, 5) &0.01 - 5\\
		PWN radius now (pc) &$R_{pwn}$ &0.80\\
		Magnetic field now ($\mu$G) &B &17.35\\
		Reduced $\chi^2$ &$\chi^2/D.O.F.$ &6.39/2 (3.19)\\
		Systematic uncertainty &$\sigma$ &0.28 &0.01-0.5\\
    \bottomrule
    \end{tabular}
\end{table}

\begin{figure}[H]
    \centering
        \includegraphics[width=0.45\columnwidth]{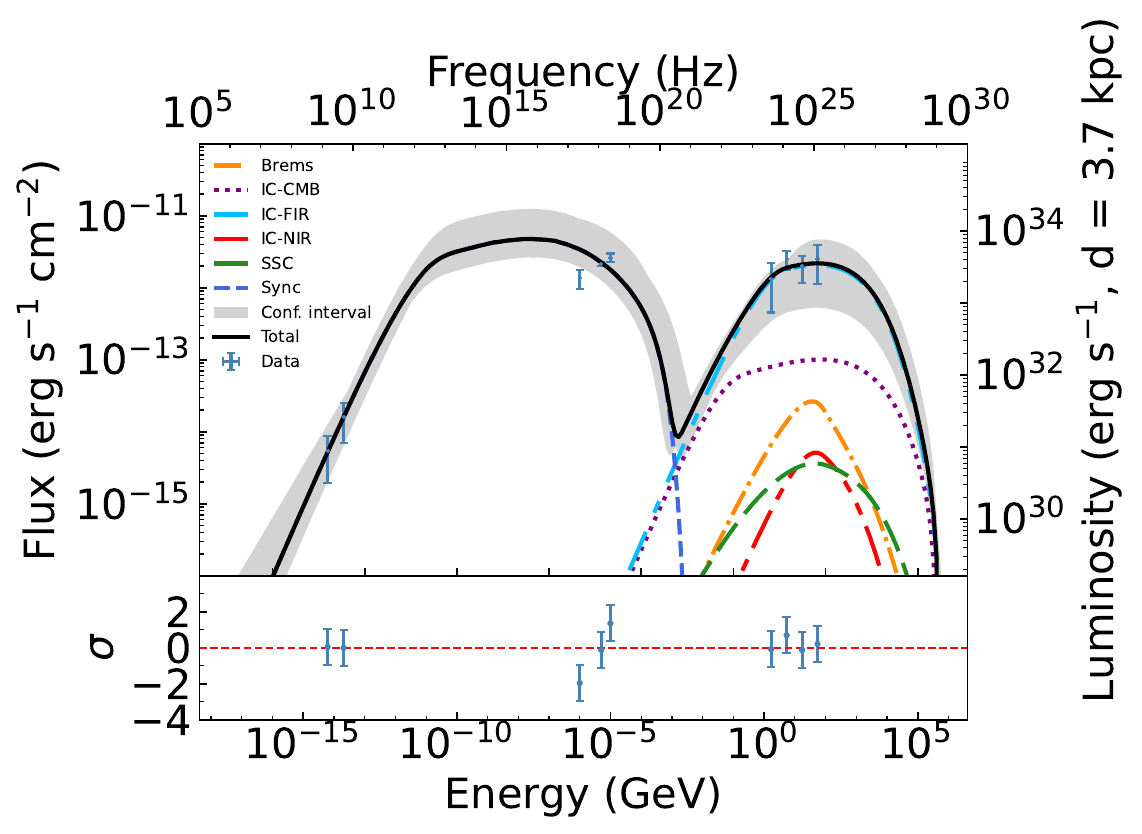}
        \includegraphics[width=0.45\columnwidth]{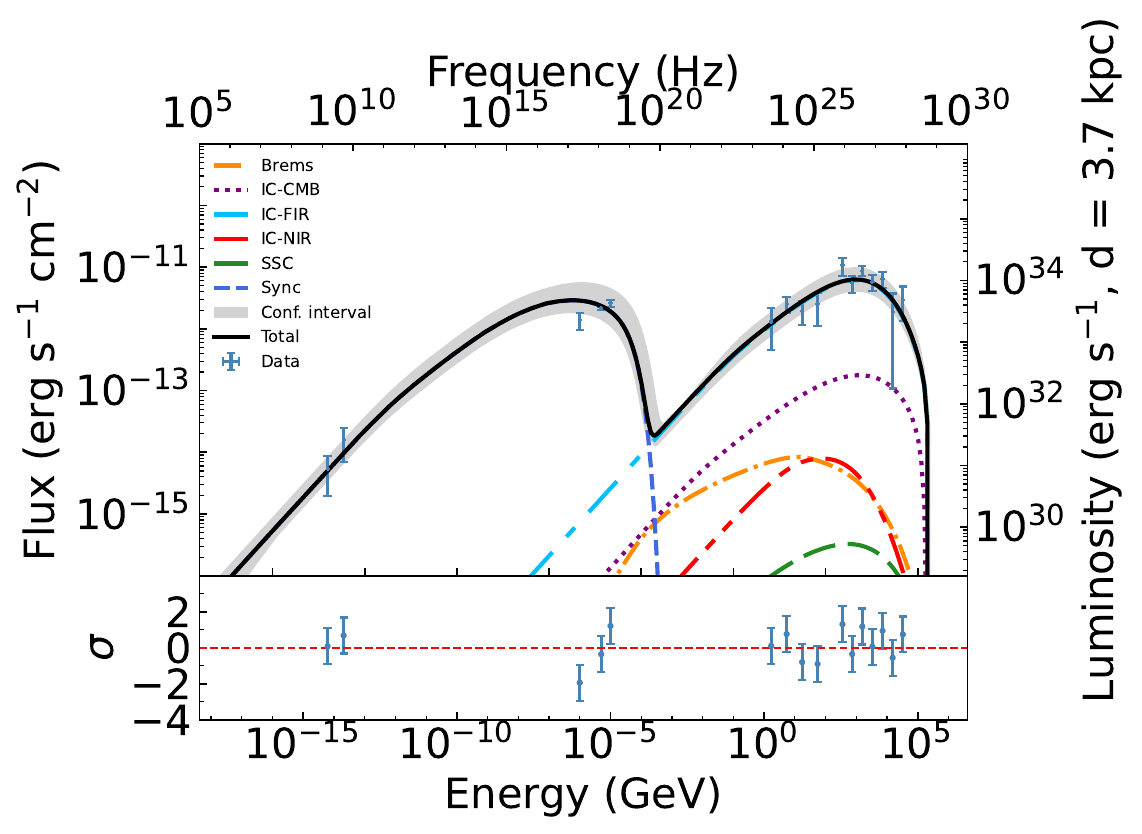}
    \caption{
 From left to right: SED from our model for PWN G11.2-0.35, and the resulting joint SED for PWN G11.2-0.35 and HESS J1809-193. The radio, X-ray and $\gamma$-ray data (including the upper limits indicated by the short blue horizontal line) for G11.2-0.35 are from \cite{Tam2002}, \cite{Roberts2003} and \cite{Eagle2022} respectively. TeV data are from HESS J1809-193 \citep{Aharonian2007}. 
    }
    \label{pic.tide.fit.G11}
\end{figure}

\begin{figure}[H]
    \centering
    \includegraphics[width=0.32\columnwidth]{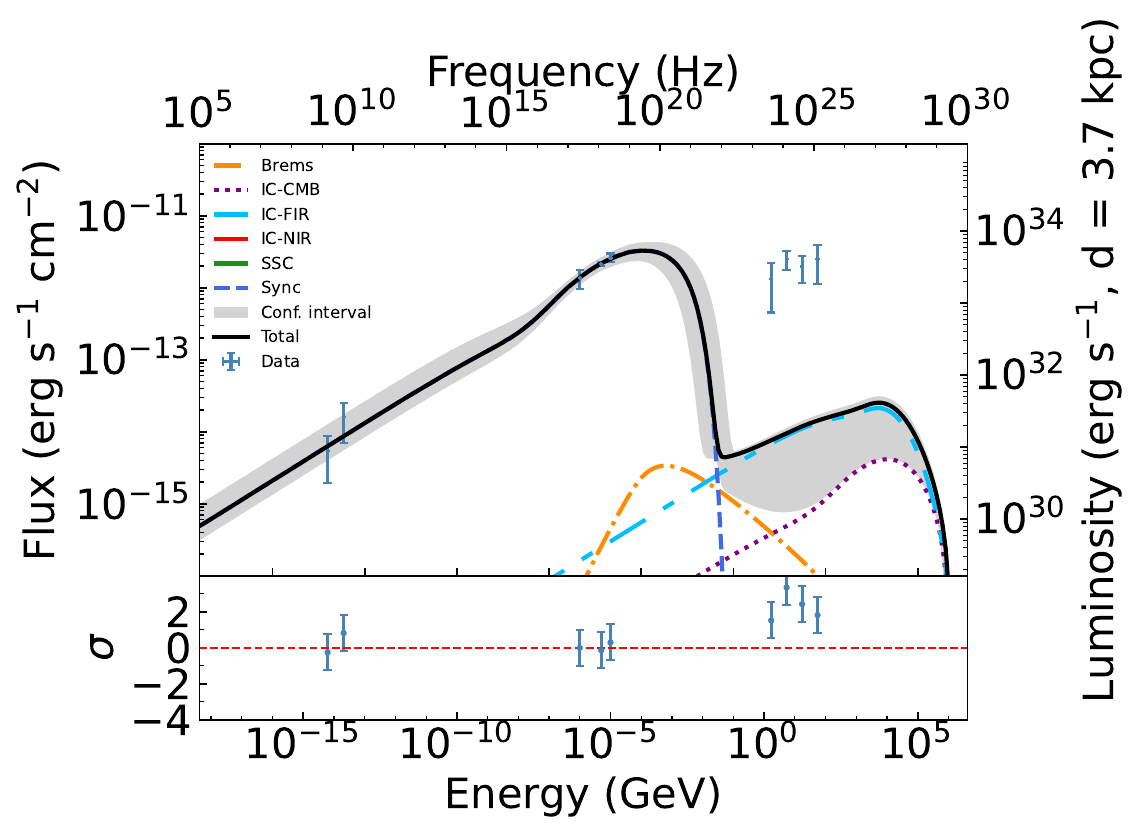}
    \includegraphics[width=0.32\columnwidth]{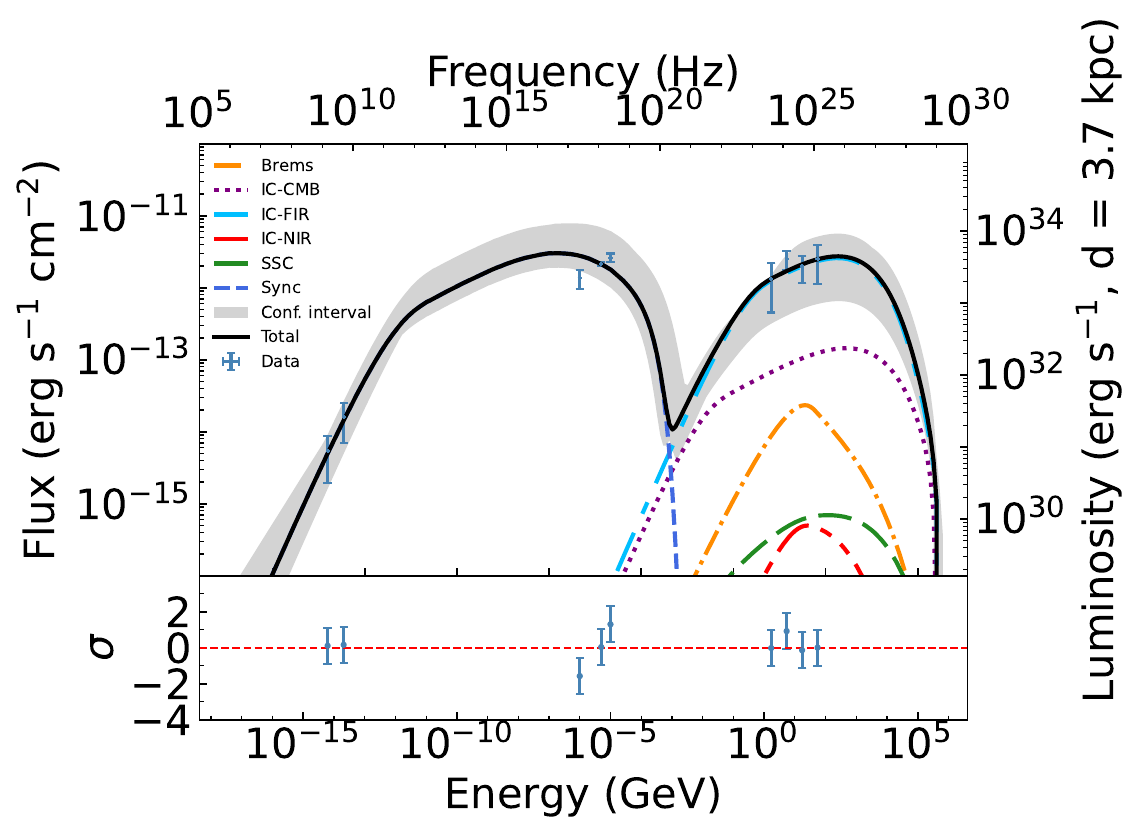}
    \includegraphics[width=0.32\columnwidth]{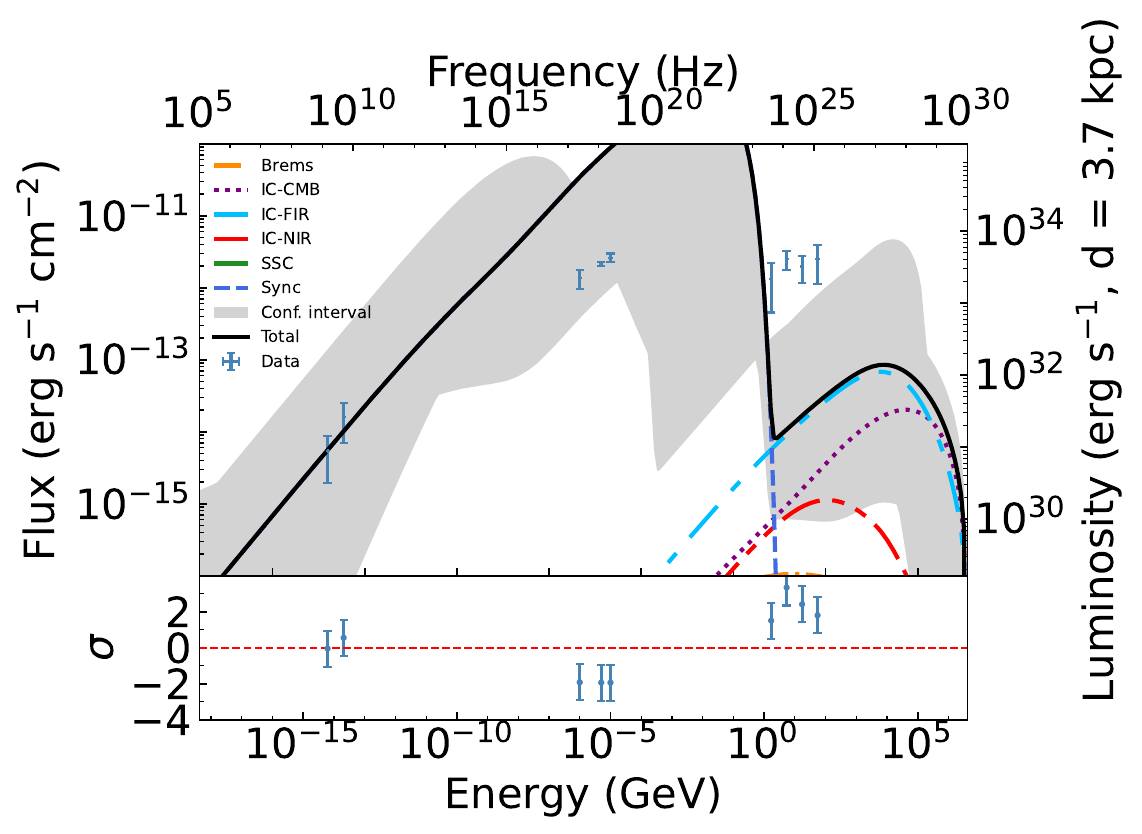}
    \includegraphics[width=0.32\columnwidth]{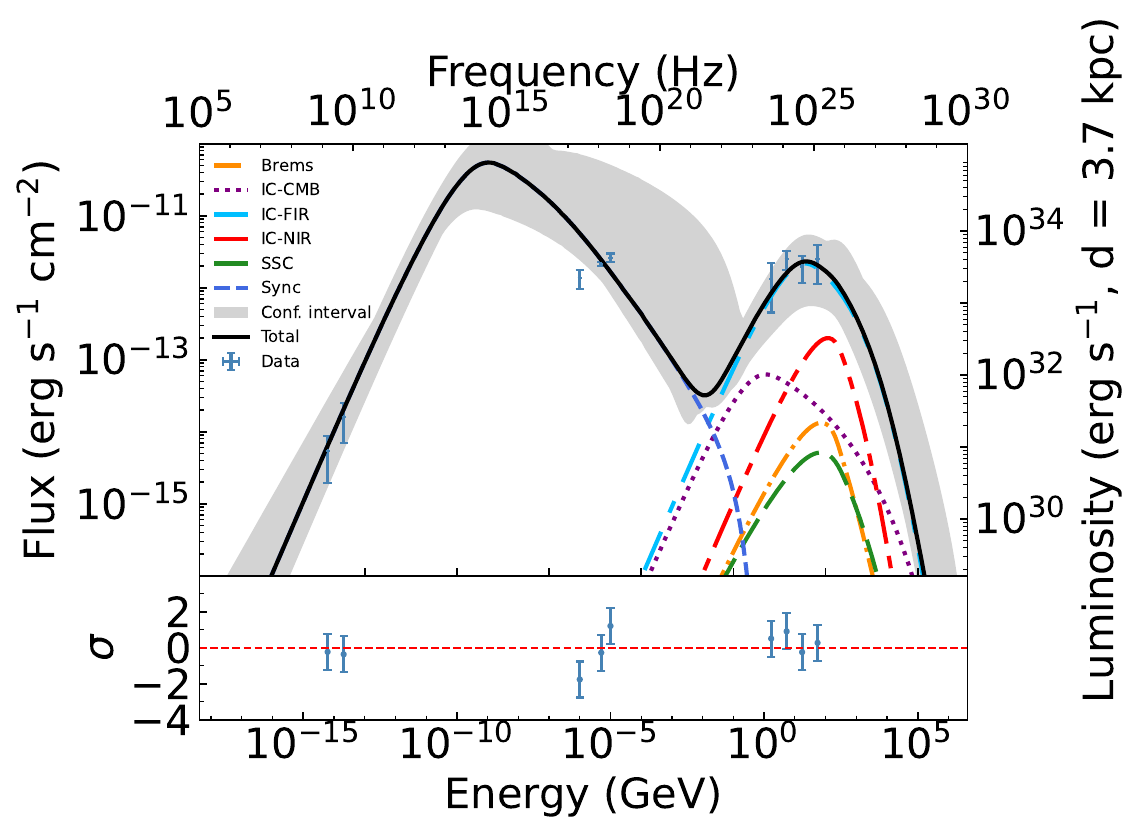}
    \includegraphics[width=0.32\columnwidth]{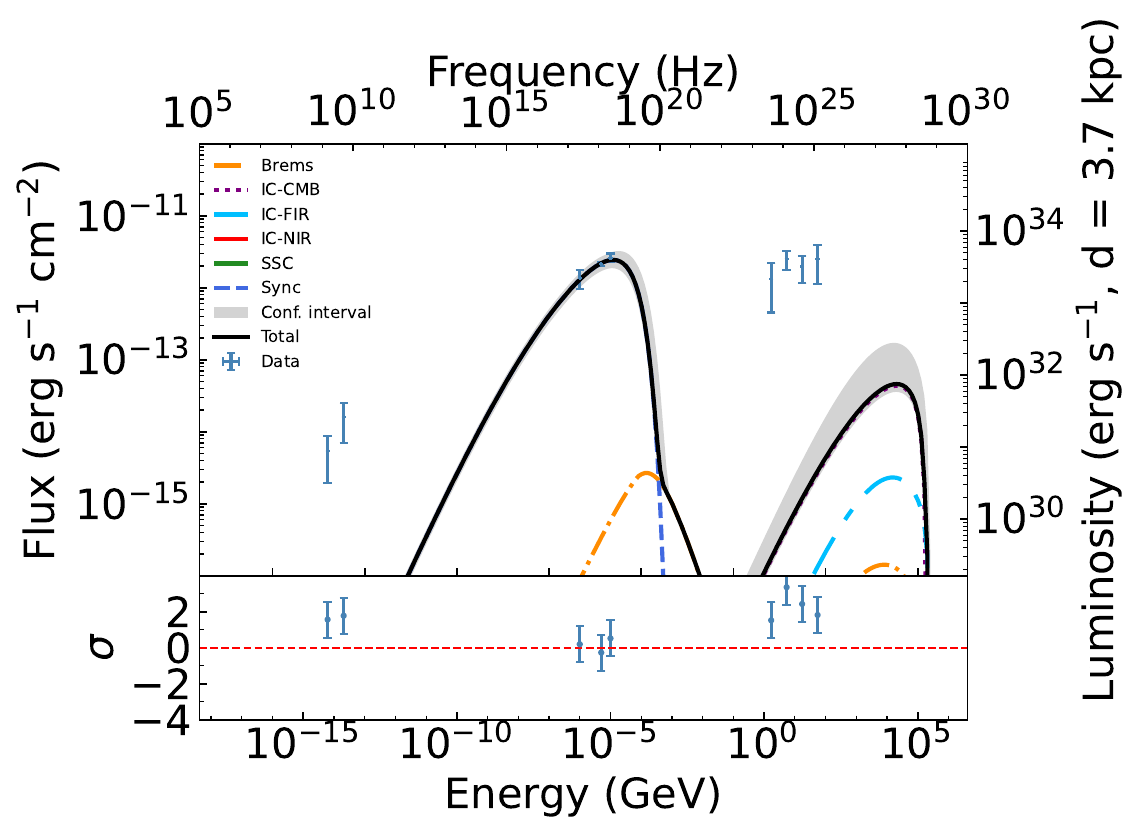}
    \caption{
 From top to bottom, left to right: Representative SEDs for Model 1, Model 2 Model 3, Model 4 and Model 5, respectively. The radio, X-ray and \(\gamma\)-ray data in the SED are from \cite{Tam2002}, \cite{Roberts2003} and \cite{Eagle2022} respectively.
    }
    \label{pic.tide.sed.G11}
\end{figure}


SNR G11.2-0.3 is a young composite SNR with a central PWN (PWN G11.2-0.35) likely associated with a supernova explosion recorded in 386 A.D. by Chinese astronomers \citep{Clark1977}. 
It hosts a $\sim 65$ ms X-ray pulsar, PSR J1811-1925, with a spin-down luminosity of $\dot{E} = 6.4\times10^{36}~\text{erg s}^{-1}$ \citep{Torii1997, Torii1999}. 
Distance estimates for this system vary significantly, ranging from 10 kpc \citep{Pavlovic2014} to as close as 4.4 kpc \citep{Kilpatrick2015, Green2004}. 
\cite{Roberts2003} performed a detailed X-ray study of the shell and PWN, suggesting an SNR age of $\sim$2 kyr, while the supernova explosion association implies an age of 1.6 kyr, which we adopt here. 

A nearby TeV $\gamma$-ray source, HESS J1809-193, was reported by \cite{Aharonian2007}, but its association with G11.2-0.3 remains uncertain. 
The angular separation (about 0.4$^\circ$ between PSR J1811-1925 and the center of HESS J1809-1903) and generally smaller distance estimates for HESS J1809-193 (ranging from 3.3 to 3.7 kpc, \cite{Yao2017, Cordes2002, Morris2002}) argue against a connection. 
This view is also supported by several studies \citep{Kargaltsev2007, Dean2008, Tanaka2013} that favor alternative associations with this TeV source, such as with PSR J1809-1917, SNR G11.0-0.0, and G11.18+0.01. 

However, the recent estimate of the kinematic distance of $3.7 \pm 0.2$ kpc by \cite{Ranasinghe2022}, which incorporates both the measurements of H\uppercase\expandafter{\romannumeral1} and the velocity-broadened molecular line data \citep{Kilpatrick2015}, brings the distance of G11.2-0.3 to $3.7 \pm 0.2$ kpc.

\citet{Eagle2022} searched for $\gamma$-ray emission from various PWNe, including PWN G11.2-0.35, using \lat data. 
Earlier, \cite{Tanaka2013} analyzed the radio and X-ray emission from PWN G11.2-0.35 using their spectral evolution model for young TeV PWNe. 
Their model uses the distance, size, pulsar period, spin-down rate and braking index \(n\) as input, and fits parameters such as age (\(t_{age}\)), magnetization fraction (\(\eta\)) and particle distribution (\(\gamma_{max}\), \(\gamma_{min}\), \(\gamma_b\), \(p_1\), \(p_2\)). 
These parameters allow calculation of the expansion velocity (\(\nu_{pwn}\)), current magnetic field (\(B_{now}\)), initial spin-down timescale (\(\tau_0\)), and initial rotational energy (\(L_0 \tau_0\)). 
Assuming a distance of 5 kpc, they obtained \(t_{age} = 2.0~\text{kyr}\), \(\gamma_b = 1.0\times10^5\), \(p_1 = 1.5\), \(p_2 = 2.6\), and \(B_{now} = 17~\mu G\). 

Incorporating the \lat data from \citet{Eagle2022}, we performed a detailed multi-band spectral fit for PWN G11.2-0.35. 
The parameters used for our model of PWN G11.2-0.35 are summarized in Tab. \ref{tab.tide.parameters.G11}. 
Based on its possible association with the supernova explosion in 386 A.D., we adopt an age \(t_{age} = 1.6\) kyr. 
The ISM density is set to \(n_{ism} = 1.7\) as in \cite{Priestley2022}, and the distance is fixed at the latest 3.7 kpc \citep{Ranasinghe2022}. 
The \(\gamma\)-ray data used in the fit are from \cite{Eagle2022}, with flux errors representing 1\(\sigma\) statistical uncertainties.

The SED on the left panel of Figure \ref{pic.tide.fit.G11} shows that our fit yields a \(\chi^2/\text{D.O.F.} = 6.39/2\), largely due to the limited available data. 
Despite this, the model captures the overall spectral shape well, with the dominant high-energy component arising from IC scattering off FIR photons. The fitted FIR energy density is relatively high at 7.65 \(eV~cm^{-3}\). 
The current PWN radius is estimated at 0.80 pc, consistent with the observed size of 0.97 pc, and the present magnetic field is 17.35 \(\mu G\), aligning well with the 17 \(\mu G\) reported in \cite{Tanaka2013}, typical for young PWNe. 
The model also predicts a maximum photon energy of \(\sim 0.4\) PeV. 

We also attempted a joint fit using radio, X-ray, and GeV $\gamma$-ray data from PWN G11.2-0.35 and the TeV data of HESS J1809-193. The resulting energy spectrum is also shown in Figure \ref{pic.tide.fit.G11}. 
Although the overall fit appears reasonable, several issues exist. 
First, achieving a good fit required an unusually high FIR density of 20 $cm^{-3}$. 
Second, the angular offset between PWN G11.2-0.35 and the center of HESS J1809-193 is $\sim0.4^\circ$, significantly greater than the $\sim0.1^\circ$ offset for PSR J1809-1917, which is considered to be a more likely counterpart. 
Lastly, replacing the Fermi data for PWN G11.2-0.35 with that for the entire HESS J1809-193 region leads to a pretty poor multi-band fit.

Given the complexity of this region, it is more plausible that HESS J1809-193 originates from a combination of multiple sources, including SNR G11.2-0.3 (possibly including PWN G11.2-0.35), G11.18+0.01 (likely powered by PSR J1809-1917) and G11.0-0.0 (near the peak of the TeV emission). 
This probably explains why the GeV data from the extended TeV source cannot be matched solely with the radio and X-ray emission from PWN G11.2-0.35. 
%

Similarly, a performance evaluation was performed based on 200 fits. 

\begin{itemize}
\item \textbf{Model 1: 112 fits (56\%), local optimum.} 
The representative values of the free parameters - \(\gamma_b\), \(\alpha_1\), \(\alpha_2\), \(M_\mathrm{ej}\), \(\eta\), \(U_\mathrm{fir}\), \(U_\mathrm{nir}\), and systematic uncertainty \(\sigma\) - are approximately \(1 \times 10^7\), 2.46, 1.58, \(20~M_{\odot}\), 0.032, \(10~\mathrm{cm^{-3}}\), \(5~\mathrm{cm^{-3}}\), and 0.22, respectively. 
The corresponding reduced \(\chi^2\), magnetic field \(B\), and radius \(R_\mathrm{pwn}\) are about 12.5, \(45~\mu\mathrm{G}\), and 0.77~pc. 
A representative SED is shown in the top left panel of Figure~\ref{pic.tide.sed.G11}.

\item \textbf{Model 2: 33 fits (16.5\%), global optimum.} 
The representative values of the free parameters are approximately \(1.5 \times 10^5\), 1.0, 2.52, \(10~M_{\odot}\), 0.03, \(9~\mathrm{cm^{-3}}\), 0.01-5, and 0.26, respectively. 
The corresponding reduced \(\chi^2\), magnetic field \(B\), and radius \(R_\mathrm{pwn}\) are about 3.2, \(15~\mu\mathrm{G}\), and 0.76-1.1~pc. 
A representative SED is shown in the upper middle panel of Figure~\ref{pic.tide.sed.G11}.

\item \textbf{Model 3: 7 fits (3.5\%), local optimum.} 
The representative values of the free parameters are approximately \(0.1 \times 10^5\), 1.97, 1.0, \(20~M_{\odot}\), 0.5, \(10~\mathrm{cm^{-3}}\), \(5~\mathrm{cm^{-3}}\), and 0.5, respectively. 
The corresponding reduced \(\chi^2\), magnetic field \(B\), and radius \(R_\mathrm{pwn}\) are about 17.06, \(163~\mu\mathrm{G}\), and 0.84~pc. 
A representative SED is shown in the top right panel of Figure~\ref{pic.tide.sed.G11}.

\item \textbf{Model 4: 10 fits (5\%), local optimum.} 
The representative values of the free parameters are approximately \(7.5 \times 10^5\), 1.0, 3.22, \(8~M_{\odot}\), 0.5, \(10~\mathrm{cm^{-3}}\), \(5~\mathrm{cm^{-3}}\), and 0.42, respectively. 
The corresponding reduced \(\chi^2\), magnetic field \(B\), and radius \(R_\mathrm{pwn}\) are about 3.65, \(85~\mu\mathrm{G}\), and 1.3~pc. 
A representative SED is shown in the bottom left panel of Figure~\ref{pic.tide.sed.G11}.

\item \textbf{Model 5: 21 fits (10.5\%), local optimum.} 
The representative values of the free parameters are approximately \(1 \times 10^4\), 3.4, 1-1.45, \(20~M_{\odot}\), 0.01, \(10~\mathrm{cm^{-3}}\), \(5~\mathrm{cm^{-3}}\), and 0.01, respectively. 
The corresponding reduced \(\chi^2\), magnetic field \(B\), and radius \(R_\mathrm{pwn}\) are about 15, \(10-40~\mu\mathrm{G}\), and 0.72~pc. 
A representative SED is shown in the bottom right panel of Figure~\ref{pic.tide.sed.G11}.

\item \textbf{Others: 17 fits (8.5\%).} This includes a failed fit and several different local optimal models that do not belong to any of the above models. 
\end{itemize}

Of the 200 fits, only Model 2 corresponds to the global optimal solution, accounting for only 16.5\%, which is significantly lower than that for the Crab Nebula and 3C 58. 
The majority of the fits (Model 1) and all remaining models (Models 3 to 5) are trapped in local minima and thus fail to capture the best-fit results.
As with the Crab Nebula and 3C 58, these non-global-optimal solutions can be readily identified by their unusually large reduced $\chi^2$ values (e.g., Model 1, Model 3, and Model 5), systematic uncertainties pinned to their lower (0.01) or upper (0.5) limits (e.g., Model 1, Model 3, and Model 5), and overall poor SED fits.

\subsection{Conclusions}

In conclusion, the risk of falling to a local minimum increases when the available observational data are sparse. Thus, it is particularly important to perform multiple fits with varied initial parameter values in such cases to ensure a more reliable global optimal solution. 

\chapter{A Systematic Search for MeV–GeV Pulsar Wind Nebulae without Gamma-ray Detected Pulsars}
\label{fermi_pwn}
\textbf{\color{SectionBlue}\normalfont\Large\bfseries Contents of This Chapter\\\\}

This chapter presents the results of a collaborative project accepted for publication in \apj \citep[in press,][]{Eagle2025}, in which I served as one of the corresponding authors, focusing on the identification and modeling of MeV-GeV PWNe using \lat observations. 
After a brief overview of the source selection criteria and data analysis pipeline, two sections are emphasized: 
\begin{itemize}
    \item Comprehensive system checks to validate the spectral analysis results, based on extended \lat datasets and alternative analysis assumptions (e.g., with and without weighting).  
    \item Detailed radiative modeling of five representative PWN candidates using a time-dependent leptonic emission model. The corresponding script, developed by the author, is provided in Appendix~\ref{script1}.  
\end{itemize}

These two components constitute my primary contributions to this work and are presented in Sections~\ref{FP.rad_modeling} and~\ref{FP.system_check}.

\newpage

\section{Introduction}
\label{FP_intro}

%
To date, approximately 125 PWNe and candidates have been identified from radio to TeV energies, with PWNe constituting the dominant TeV source class in the Galaxy \citep{Abdalla2018}. 
In contrast, only 21 of them are associated with \lat sources in the 4FGL-DR4 catalog \citep{Ballet2023}, reflecting challenges in the GeV band due to the strong diffuse Galactic background and potential confusion with pulsar magnetospheric emission.

This study presents a systematic search for $\gamma$-ray emitting PWNe using 11.5 years of \lat Pass 8 data in the 300\,MeV to 2\,TeV range. 
We focus on 58 regions of interest (ROIs) associated with previously known PWN candidates that lack MeV-GeV $\gamma$-ray-detected pulsars, thereby minimizing pulsar contamination and improving sensitivity to nebular emission. 
Each ROI is analyzed using standard it{FermiPy} tools, and sources are classified based on spectral and spatial characteristics. 
The resulting catalog includes 36 GeV-detected sources, of which many are newly reported PWNe or strong candidates. 
A significant portion of this work includes the broadband spectral modeling and cross-validation of these detections, which are the focus of Sections~\ref{FP.rad_modeling} and \ref{FP.system_check}, where I contributed the most.

\section{Analysis Workflow}
\label{FP.workflow}

\subsection{Sample Selection and \lat Data Reduction}
\label{FP.selection}

The initial step in this study involved building a comprehensive list of PWNe or PWN candidates previously identified in radio, X-ray, and TeV observations. 
From this list, we selected [$125-63-4=$] 58 ROIs \citep[for the full list, see Table 1 in][]{Eagle2025} based on the following criteria: 
\begin{itemize}
  \item Exclusion of systems with \lat detected pulsars to reduce contamination from pulsed emission (63 PWNe / PWN candidates). 
  \item Removal of sources within $1^\circ$ of the Galactic Center due to complex background modeling (4 PWNe / PWN candidates).  
  \item Selection of PWNe from well-established TeV catalogs (e.g., \hess, MAGIC, VERITAS) and X-ray/radio SNR/PWN catalogs \citep[see e.g.,][]{ferrand2012census}.
\end{itemize}
\noindent
We analyzed each ROI using Pass~8 SOURCE class data (zenith angle $<90^\circ$) collected from August 2008 to January 2020, and performed spectral and spatial modeling with the ittt{FermiPy} package \citep{Wood2017}. 
Spatial and energy bin sizes were settled to 0.1$^\circ$ and 10 bins per decade, respectively.  
For each ROI, an initial global binned likelihood analysis was performed by combining four binned likelihood analyses on the four PSF event types (PSF0--PSF3)\footnote{PSF0--PSF3 correspond to increasing levels of angular reconstruction quality, with PSF3 providing the best resolution. See \url{https://www.slac.stanford.edu/exp/glast/groups/canda/lat_Performance.htm} for details.}, respectively. 
To refine spatial characterization, we subsequently repeated the likelihood analysis using only PSF3 events, which offer the best angular resolution. 
For ROIs with faint (TS $\lesssim$ 25 using PSF3 events only) and point-like residuals, we included all PSF types to enhance photon statistics. 
This applies to four ROIs: G54.10+0.27, B0453-685, G327.15-1.04, and G318.90+0.40. 
The remaining 54 ROIs, which show significant or extended residual emission, were analyzed using PSF3 events exclusively. 
%

\subsection{Spatial and Spectral Analyses}
\label{FP.analysis}

Each ROI was modeled following this procedure: 
\begin{itemize}
  \item Definition of ROI size: A $10^\circ \times 10^\circ$ region is used for most ROIs, with a $15^\circ$ radius for the source model. For two ROIs overlapping large SNRs, we adopt larger regions: $15^\circ$ [$20^\circ$ model radius] for G74.00-8.50, and $20^\circ$ [$25^\circ$] for G179.72-1.69.
  \item Load background components: 
    \begin{itemize}
      \item Galactic diffuse model: \texttt{gll\_iem\_v07.fits};
      \item Isotropic component: \texttt{iso\_P8R3\_SOURCE\_V3\_v1.txt} (for the four ROIs using all events) or \texttt{iso\_P8R3\_SOURCE\_V3\_PSF3\_v1.txt} (for others);
      \item Known sources from 4FGL-DR2 (or DR4 when necessary). 
      \item Allow all sources with TS~$\geq$~25 and within $3^\circ$ of the ROI center, to vary in both normalization and spectral index (and the normalization for isotropic and Galactic diffuse components) during the fit. Faint sources (TS~$<25$) are removed. 
    \end{itemize}
  \item Source detection: Use TS and count maps in multiple energy bands (300~MeV–2~TeV, 1–10\,GeV, 10–100\,GeV and 100~GeV-2~TeV) to identify $\gamma$-ray residuals. For each ROI with significant residual (TS~$>25$, ), add or select (from 4FGL sources) only one source as the corresponding PWN counterpart candidate. 
  \item Morphology fitting: Test both point-like and extended templates (RadialDisk and RadualGaussian) for these selected candidates. The resulting extended model with TS$_\mathrm{ext}>16$ would be adopted. The results of the spatial analyses for these candidates are listed in Table 2, 3, and 4 of \citet{Eagle2025}.  
  \item Spectral modeling: Fit with power-law or log-parabola (when $\mathrm{TS}_{\log \mathrm{P}}=2 \log \left(\frac{\mathcal{L}_{\log P}}{\mathcal{L}_{P L}}\right)$ $> 4$) functions. The results of the spatial analyses for these candidates are listed in Table 5 of \citet{Eagle2025}. 
  \item SED: For detected candidates, i.e. TS~$>25$, generate the SED in energy range of 300~MeV-1~TeV with seven energy bins, and measure the systematic errors for each energy bin \citep[see the full results in Table 6 and C1 of][]{Eagle2025}. For non-detections, calculate their 95\% confidence level (C.L.) upper limit fluxes for the energy range of 300~MeV-2TeV \citep[see Figure 6 in][]{Eagle2025}. 
\end{itemize}

\subsection{Source Classification}
\label{FP.classify}

\begin{table}[htbp]
\setlength\tabcolsep{1pt}
\scriptsize
\begin{center}
\caption{Summary of likely PWNe and PWN candidates. Adapted from \citet{Eagle2025}.}
\label{tab.FP.pwne}
\scalebox{1.0}{
\begin{tabular}{lccccccc}
\hline\hline
Galactic PWN Name & 4FGL Name & R.A. (deg) & Dec. (deg) & TS & TS$_{\rm ext}$ & $r$ ($^\circ$) & 95\% U.L. $r$ ($^\circ$) \\
\hline
\multicolumn{8}{c}{\textbf{Likely Point-like PWNe}} \\
G29.70-0.30 &J1846.4$-$0258 (DR4) &281.60  &-2.96 &20.7 &none &none &0.09 \\
G54.10+0.27 &J1930.5$+$1853 (DR3) &292.65 &18.90 &31.1 &none &none &0.10 \\
G279.60-31.70 &J0537.8$-$6909 &84.40  &-69.18 &168.3 &none &none &0.04 \\
G279.80-35.80 &* &73.46  &-68.47 &18.4 &none &none &0.15 \\
G315.78-0.23 &J1435.8$-$6018 &219.36  &-60.11 &37.4 &none &none &0.14 \\
G327.15-1.04 &J1554.4$-$5506 (DR3) &238.59  &-55.11 &34.1 &none &none &0.08 \\
\hline
\multicolumn{8}{c}{\textbf{Point-like PWN Candidates}} \\
G11.18-0.35 &J1811.5$-$1925 &272.88  &-19.44 &53.9 &none &none &0.05 \\
G16.73+0.08 &J1821.1$-$1422 &275.29  &-14.35 &142.5 &none &none &0.05 \\
G18.90-1.10 &J1829.4$-$1256 &277.34  &-12.88 &101.8 &none &none &0.06 \\
G20.20-0.20 &J1828.0$-$1133 &277.02  &-11.57 &153.9 &none &none &0.06 \\
G27.80+0.60 &J1840.0$-$0411 &279.97  &-4.27 &131 &none &none &0.04 \\
G39.22-0.32 &J1903.8$+$0531 &285.97  &+5.50 &129.7 &none &none &0.05 \\
G49.20-0.30 &J1922.7$+$1428c &290.70  &14.27 &157.2 &none &none &- \\
G49.20-0.70 &* &290.74  &14.09 &28.0 &none &none &0.03 \\
G63.70+1.10 &J1947.7$+$2744 &296.96  &27.73 &93.9 &none &none &0.07 \\
G65.73+1.18 &J1952.8$+$2924 &298.13  &29.46 &123.3 &none &none &0.05 \\
G74.94+1.11 &J2016.2$+$3712 &304.04  &37.20 &66.3 &none &none &0.04 \\
G318.90+0.40 &J1459.0$-$5819 &224.69  &-58.38 &55.6 &none &none &0.10 \\
G337.20+0.10 &* &248.88  &-47.19 &18.4 &none &none &0.06 \\
G337.50-0.10 &J1638.4$-$4715c (DR3) &249.67  &-47.28 &42.7 &none &none &0.09 \\
G338.20-0.00 &J1640.6-4632 (DR1) J1640.7-4631e (DR3) &250.19  &-46.57 &168.3 &none &none &0.19 \\
\hline
\multicolumn{8}{c}{\textbf{Likely Extended PWNe}} \\
G8.40+0.15 & J1804.7$-$2144e & 271.11 & $-$21.74 & 104.8 & 59.65 & $0.29 \pm 0.02 \pm 0.11$ & 0.34 \\
G25.10+0.02 & J1836.5$-$0651e & 279.24 & $-$6.91 & 1174 & 616.2 & $0.53 \pm 0.02 \pm 0.06$ & 0.56 \\
G25.24$-$0.19 & J1836.5$-$0651e & - & - & - & - & - & - \\
G336.40+0.10 & J1631.6$-$4756e & 248.14 & $-$47.91 & 94.5 & 24.56 & $0.19 \pm 0.03 \pm 0.83$ & 0.23 \\
\hline
\multicolumn{8}{c}{\textbf{Extended PWN Candidates}} \\
G11.03$-$0.05 & J1810.3$-$1925e & 272.39 & $-$19.42 & 84.0 & 25.88 & $0.41 \pm 0.05 \pm 0.05$ & 0.49 \\
G11.09+0.08 & J1810.3$-$1925e & - & - & - & - & - & - \\
G12.82$-$0.02 & J1813.1$-$1737e & 273.47 & $-$17.65 & 854.6 & 195.6 & $0.41 \pm 0.02 \pm 0.03$ & 0.45 \\
G15.40+0.10 & J1818.6$-$1533 & 274.60 & $-$15.10 & 394.5 & 20.60 & $0.19 \pm 0.03 \pm 0.10$ & 0.24 \\
G24.70+0.60 & J1834.1$-$0706e & 278.53 & $-$7.32 & 290.9 & 72.57 & $0.19 \pm 0.02 \pm 0.02$ & 0.22 \\
G29.40+0.10 & J1844.4$-$0306 & 281.15 & $-$3.10 & 195.7 & 71.05 & $0.27 \pm 0.04 \pm 0.03$ & 0.34 \\
G328.40+0.20 & J1553.8$-$5325e & 238.61 & $-$53.38 & 1070 & 436.5 & $0.43 \pm 0.02 \pm 0.03$ & 0.46 \\
\hline
\end{tabular}}
\end{center}
\begin{tablenotes}
\item \textbf{Note.} Columns include the PWN name, its associated 4FGL source (entries marked with `*' indicate no association; sources J1836.5$-$0651e, J1616.2$-$5054e, and J1810.3$-$1925e each correspond to two PWNe), J2000 coordinates, TS values, and the 95\% upper limit on $r$.  
For extended sources, the table also lists TS$_{\rm ext}$ values for sources modeled with a radial Gaussian model, and the best-fit extension radius $r$ along with its 1$\sigma$ statistical and systematic uncertainties. \\
\end{tablenotes}
\end{table}

\begin{table}[htbp]
\setlength\tabcolsep{4pt}
\scriptsize
\begin{center}
\caption{Summary of PWN Classification Criteria. }
\label{tab.FP.classify}
\scalebox{1.0}{
\begin{tabular}{l|c|c}
\hline
\textbf{Classification Criteria} & \textbf{Likely PWNe (Total = 9)} & \textbf{Candidate PWNe (Total = 21)} \\
\hline
TeV counterpart present & 7 & 8 \\
TeV source identified as PWN & 6 & 2 \\
Multiwavelength study supports PWN origin & 9 & $\sim$3 \\
Confirmed Multiwavelength PWN counterpart & 9 & 8 \\
Location in LAT sky & Generally in non-complex regions & Some in complex or uncertain regions \\
\hline
\end{tabular}}
\end{center}
\begin{tablenotes}
\item \textbf{Note.} The numbers indicate how many sources in each group satisfy the respective criterion. Reproduced from Table 7 of \citet{Eagle2025}. 
\end{tablenotes}
\end{table}

\begin{figure}[htbp]
\centering
\includegraphics[width=\textwidth]{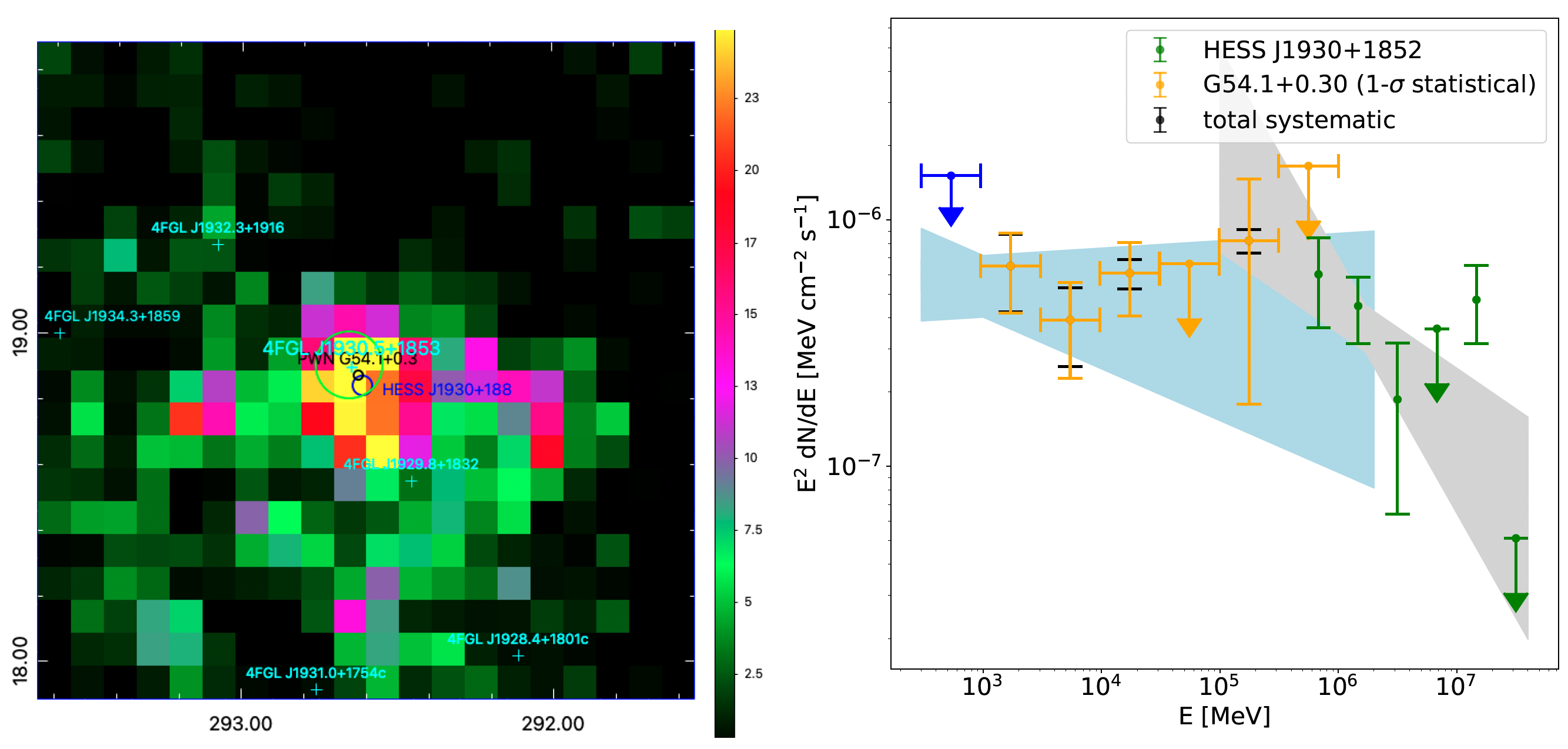}
\includegraphics[width=\textwidth]{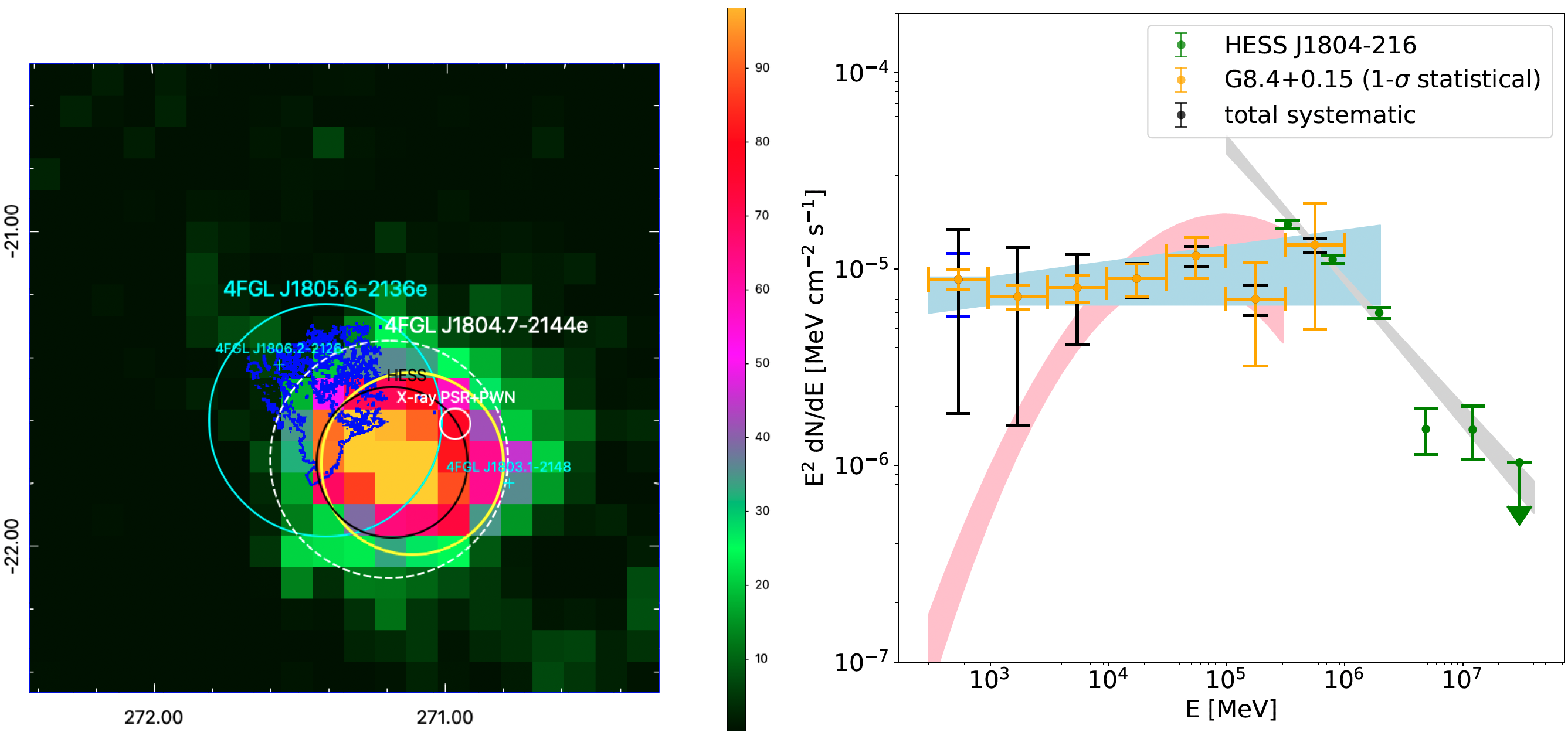}
\caption{
    \textbf{Top:} \textit{Left:} TS map of the likely point-like PWN G54.1+0.30 (J2000 coordinates). The green circle denotes the 95\% localization of 4FGL\,J1930.5+1853; black and blue circles indicate the X-ray PWN and the 68\% extension of HESS\,J1930+188, respectively. Unrelated 4FGL sources are shown in cyan. Maximum TS $\approx 32$. 
    \textit{Right:} \lat spectrum of 4FGL\,J1930.5+1853 (blue band, yellow points), overlaid with the HESS spectrum (green points and grey band; \citet{hess2018population}). The blue upper limit below 1\,GeV indicates additional systematic error, as discussed in Section 3.6 of \citet{Eagle2025}. \\
    \textbf{Bottom:} \textit{Left:} TS map of the likely extended PWN G8.40+0.15. HESS\,J1804$-$216 is shown in black; PSR\,J1803$-$2137 and its X-ray nebula in white. Radio contours of SNR G8.7$-$1.4 are overlaid in blue. The dashed white circle marks the unmodeled 4FGL\,J1804.7$-$2144e. The best-fit LAT extension is shown in yellow. Maximum TS $\approx 122$. All 4FGL sources are labeled in cyan.
    \textit{Right:} \lat SED (yellow points with blue band) and best-fit model compared with the HESS data (grey band and points; \citet{hess2018population}). The Log-Parabola model from \citet{Liu2019G8} is shown in pink. Additional systematic error below 1\,GeV is indicated as a blue flux error. \\
    Reproduced from Figure 3 and 6 of \citet{Eagle2025}. 
 }
\label{FP.fig.PWNexamples}
\end{figure}

To identify a GeV detection as a likely PWN counterpart, we applied the following criteria:
\begin{itemize}
  \item spatial coincidence with a known PWN detected at radio, X-ray, or TeV wavelengths; 
  \item consistent spatial extension between the \lat detection and that observed in other wavebands; 
  \item the energetics of the associated pulsar, PWN, and the host SNR. 
\end{itemize}

While positional overlap provides an initial clue, it is insufficient on its own due to source confusion and the limited angular resolution of \lat. 
The extent of an evolved PWN, often comparable in radio and GeV but smaller in X-ray/TeV, can vary significantly with evolutionary phase. 
In particular, PWNe in the reverberation phase may develop distinct high- and low-energy components with different morphologies, complicating interpretation \citep[see e.g.,][]{Bandiera:2022,bandiera2023reverberation,Torres2017}. 
Energetic diagnostics offer additional insight but rely on detecting the associated pulsar and SNR shell, which often absent or uncertain. 
Pulsars are the most numerous Galactic $\gamma$-ray sources \citep{Smith2023,Ballet2023}, and SNRs outnumber confirmed PWNe by a factor of two, making contamination likely. 
At $E < 10$\,GeV, several sources show spectral signatures suggestive of pulsar emission, further blurring classification.

With these considerations, among the 58 ROIs analyzed, we identified 36 $\gamma$-ray sources, including 9 likely PWNe (6 point-like and 3 extended) and 21 PWN candidates (15 point-like and 6 extended). 
The full list is provided in Table~\ref{tab.FP.pwne}, with classification criteria summarized in Table~\ref{tab.FP.classify}. 
Most of these sources are also discussed in detail in Section~4.3 of \citet{Eagle2025}.

%
The nine likely PWNe exhibit significant emission above 10~GeV, suggestive of a PWN origin. 
Figure~\ref{FP.fig.PWNexamples} shows the TS maps and best-fit SEDs for G54.1+0.30 and G8.4+0.15, representative of point-like and extended likely PWNe, respectively. 
However, multiwavelength follow-up observations are necessary to confirm their classification and to clarify the origin of the observed $\gamma$-ray emission. 
Notably, three extended sources among them would increase the number of \textit{Fermi}-detected extended PWNe without known $\gamma$-ray pulsars from six to nine, and expand the total extended LAT-detected PWNe from 12 to 15. 
No point-like PWNe are currently identified in the 4FGL-DR4 catalog \citep{Ballet2023}.

An additional 21 sources are classified as PWN candidates. 
These lack firm TeV associations or reside in complex PSR/PWN/SNR environments where multiple high-energy emission mechanisms may contribute. 
%
%
Our results (see Section~4.2 of \citet{Eagle2025}) indicate that pulsars may contribute in 11 cases. 
Disentangling these contributions requires further multiwavelength observational constraints.

In general, reliable PWN identification demands either morphological consistency with known PWNe (e.g., \citealt{Li2018}) or correlated variability, the latter so far observed only in the flaring Crab Nebula.

\subsection{Systematic Checks and Validation}
\label{FP.system_check}

\begin{table}[htbp]
\centering
\setlength\tabcolsep{1pt}
\caption{Data configuration details for the 11.5-year and 14-year \lat analyses used in system checks. Reproduced from Table A1 of \citet{Eagle2025}. }
\label{FP.tab.lat_config}
\scalebox{0.8}{
\begin{tabular}{l|ccc}
\hline
\lat Data Configuration & 14 years (Weighted) & 11.5 years (Weighted) & 11.5 years (Unweighted) \\
\hline
Time range & 4 Aug. 2008--15 Aug. 2023 & 4 Aug. 2008--1 Jan. 2020 & 4 Aug. 2008--1 Jan. 2020 \\
Energy range & 300 MeV--1 TeV & 300 MeV--1 TeV & 300 MeV--2 TeV \\
Catalog & 4FGL-DR4 & 4FGL-DR2 & 4FGL-DR2 \\
\multirow{2}{*}{Event Type} & 8/16/32 (0.3--1 GeV) & 8/16/32 (0.3--1 GeV) & \multirow{2}{*}{32} \\
           & 4/8/16/32 (1--1000 GeV) & 4/8/16/32 (1--1000 GeV) & \\
\multirow{2}{*}{Zmax} & 100$^\circ$ (0.3--1 GeV) & 100$^\circ$ (0.3--1 GeV) & \multirow{2}{*}{100$^\circ$} \\
            &105$^\circ$ (1--1000 GeV) & 105$^\circ$ (1--1000 GeV) & \\
Extended Source Template & Extended\_14years & Extended\_8years & Extended\_8years \\
Weighted (Y/N) & Y & Y & N \\
Fermitools version & 2.2.0 & 2.2.0 & 2.0.8 \\
\hline
\end{tabular}}
\end{table}

\begin{figure}[htbp]
\centering
\includegraphics[width=0.48\textwidth]{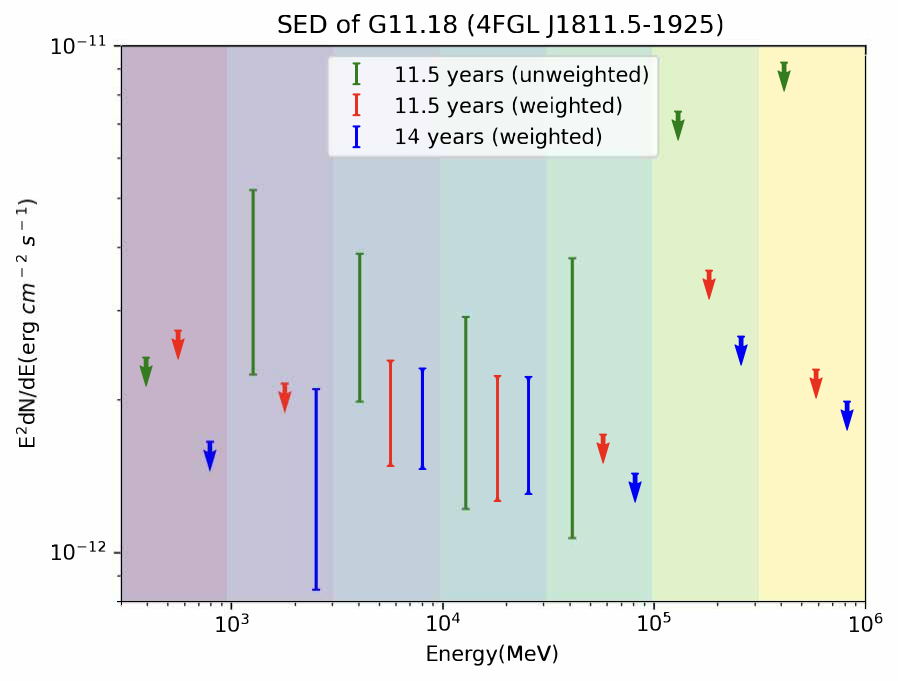}
\includegraphics[width=0.48\textwidth]{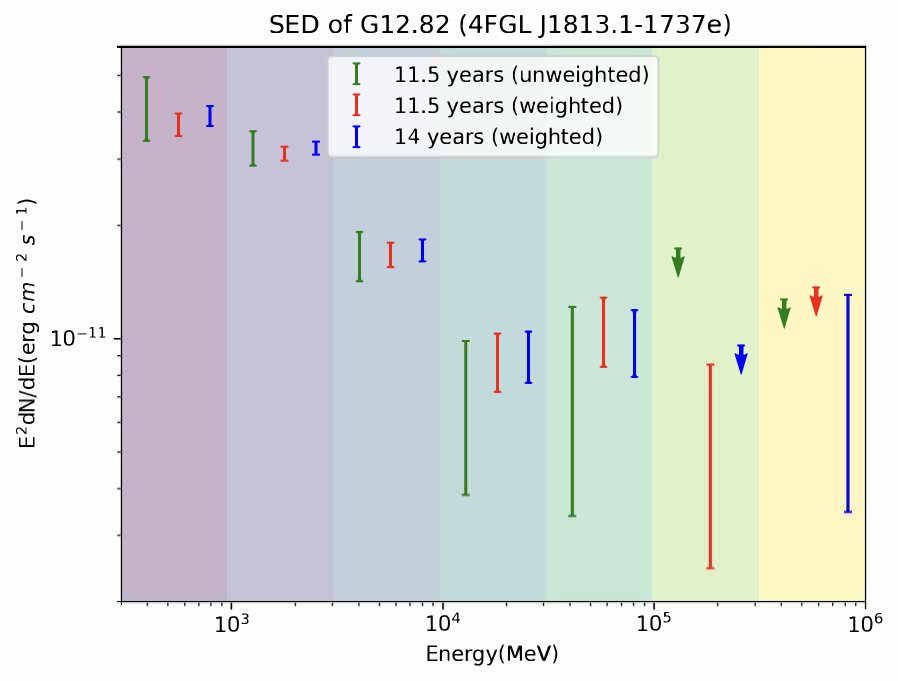}
\includegraphics[width=0.48\textwidth]{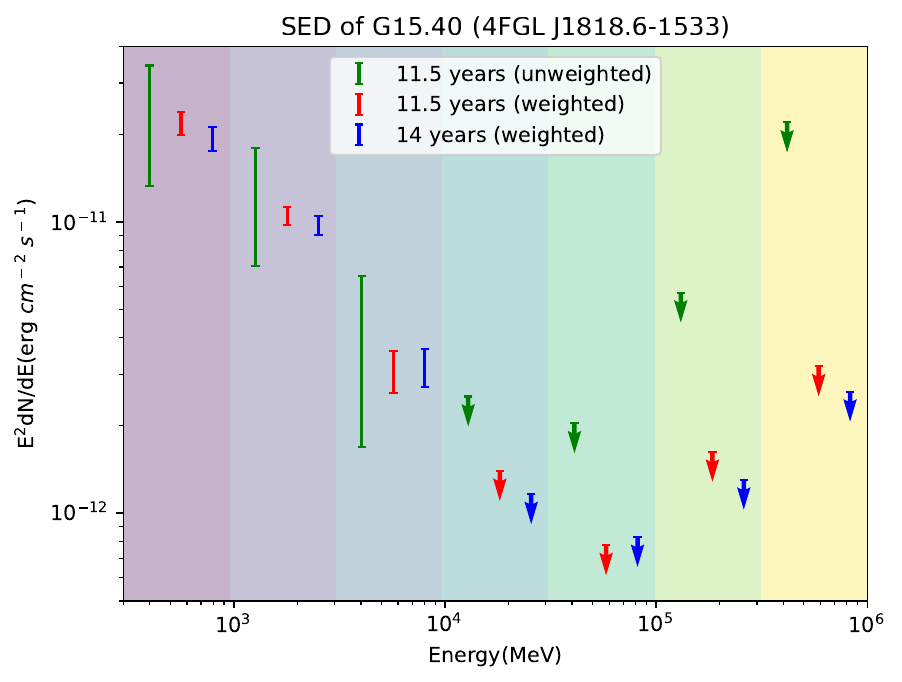}
\includegraphics[width=0.48\textwidth]{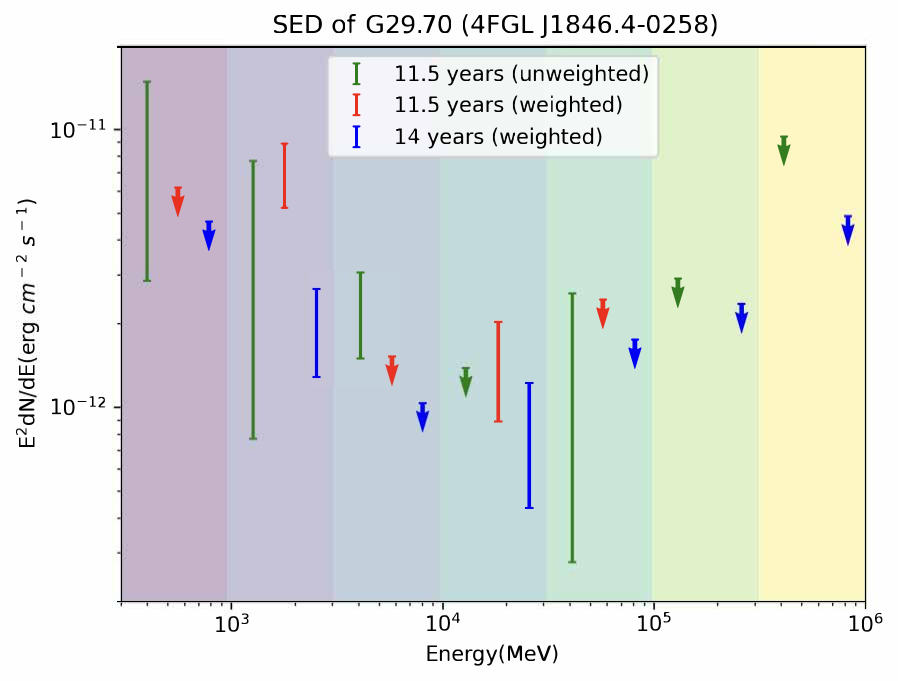}
\includegraphics[width=0.48\textwidth]{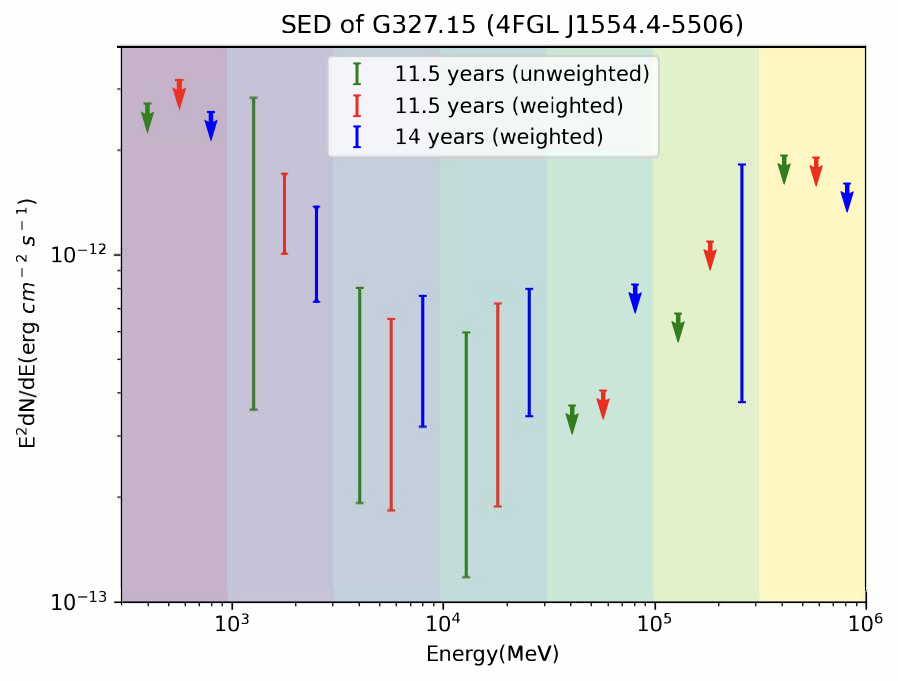}
\caption{Comparison of SEDs derived from the original (11.5 yr) and updated (14 yr) \lat data for five selected sources.}
\label{FP.fig.syscheck_seds}
\end{figure}

To evaluate the robustness of the primary results, we conducted a systematic validation using updated \lat data (14 years) and the latest 4FGL-DR4 catalog. 
The impact of incorporating the likelihood weighting method described in Section~\ref{intro.lat_analysis} was also assessed. 
Configuration details are summarized in Table~\ref{FP.tab.lat_config}, and the same five representative sources from Section~\ref{FP.rad_modeling} were re-analyzed. 

The best-fit spectral parameters and SEDs remained consistent across all test cases. 
Figure~\ref{FP.fig.syscheck_seds} compares the resulting SEDs, confirming that the main conclusions are robust against changes in data volume, catalog version, and analysis method. 
These results support the internal consistency of the methodology and reinforce the persistent nature of the identified PWN candidates.

\section{Radiative Modeling of PWNe}
\label{FP.rad_modeling}

\begin{table}[htbp]
\setlength\tabcolsep{4pt}
\scriptsize
\begin{center}
\caption{Parameters used in radiative modeling of five PWN candidates. Adapted from Table 8 of \citet{Eagle2025}. }
\label{FP.radmod_targets}
\scalebox{1.0}{
\begin{tabular}{l|ccccc}
\hline
PWN & $\dot{E}$ (erg\,s$^{-1}$) & $\tau_c$ (kyr) & d (kpc) & PWN Radius (deg) & Reference \\
\hline
G11.18-0.35 & $6.4 \times 10^{36}$ & 29.88 & 3.7 & 0.011 & \citet{Madsen2020,Ranasinghe2022} \\
G12.82-0.02 & $5.6 \times 10^{37}$ & 5.6 & 6.2 & 0.05 & \citet{hess2018population,Joshi2023} \\
G15.40+0.10 & $7.0 \times 10^{36}$ & 17 & 9.3 & 0.14 & \citet{Su2017,HESS2014} \\
G29.70-0.30 & $8.1 \times 10^{36}$ & 0.723 & 5.8 & 0.0083 & \citet{Straal2023} \\
G327.15-1.04 & $3.1 \times 10^{36}$ & 17.4 & 9.0 & 0.02 & \citet{Temim2015} \\
\hline
\end{tabular}}
\end{center}
\begin{flushleft}
\footnotesize
\textbf{Note.} For G11.18$-$0.35, we adopt an age of 1.6\,kyr based on its historical association \citep{Clark1977}. 
For G327.15$-$1.04 and G15.40+0.10, with no identified pulsars, $\dot{E}$, $\tau_c$, and distance follow estimates from the listed references. 
G15.40+0.10 is assigned a typical $\dot{E}$ for pulsars powering PWNe.
\end{flushleft}
\end{table}

\begin{figure}[htbp]
\centering
\includegraphics[width=0.48\textwidth]{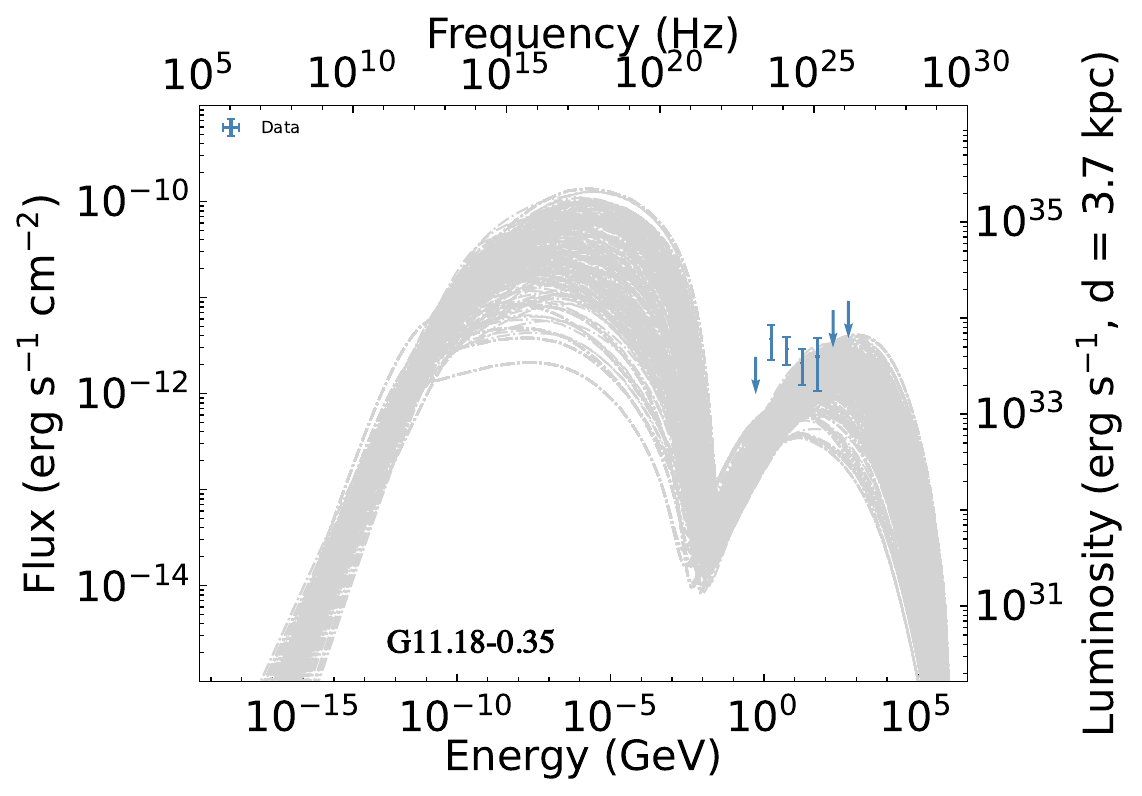}
\includegraphics[width=0.48\textwidth]{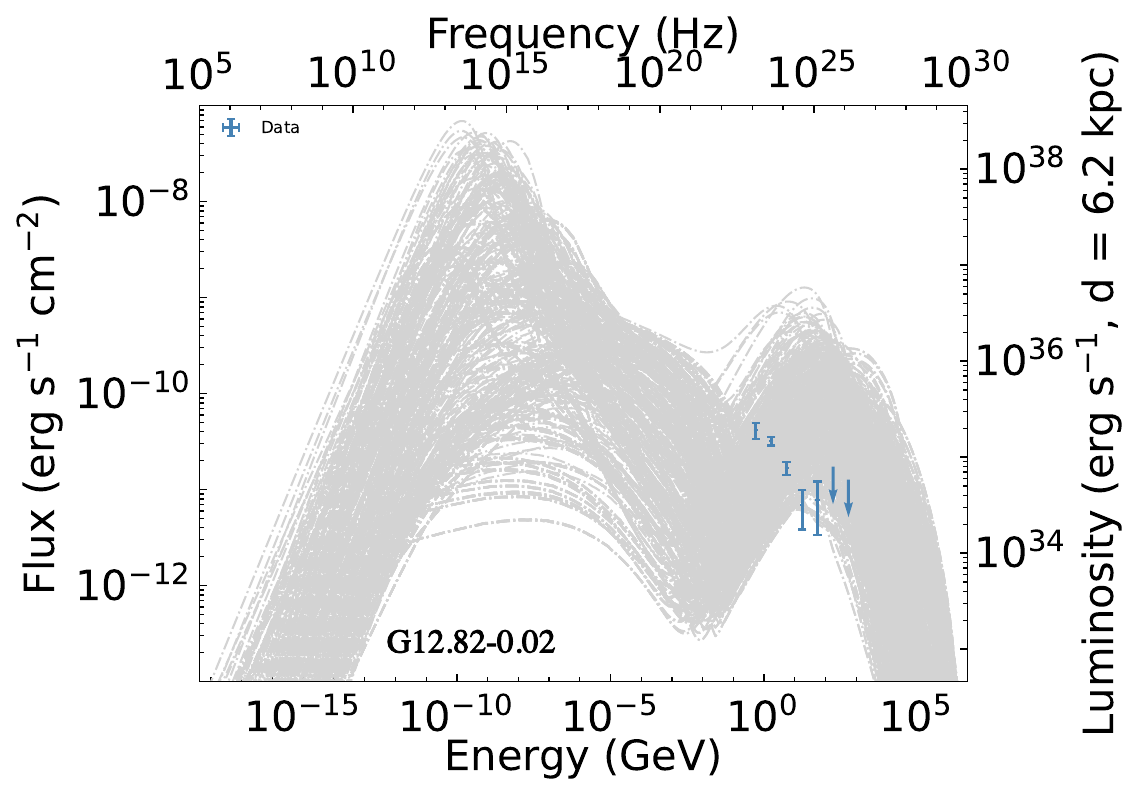}
\includegraphics[width=0.48\textwidth]{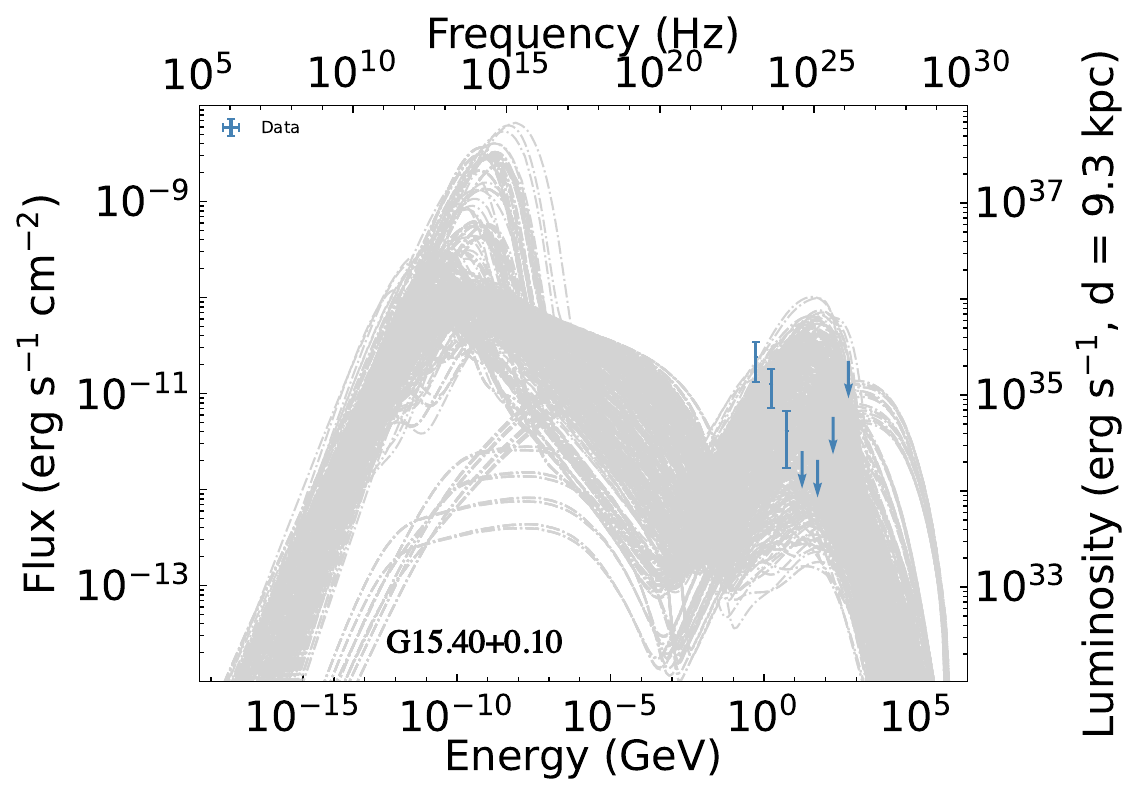}
\includegraphics[width=0.48\textwidth]{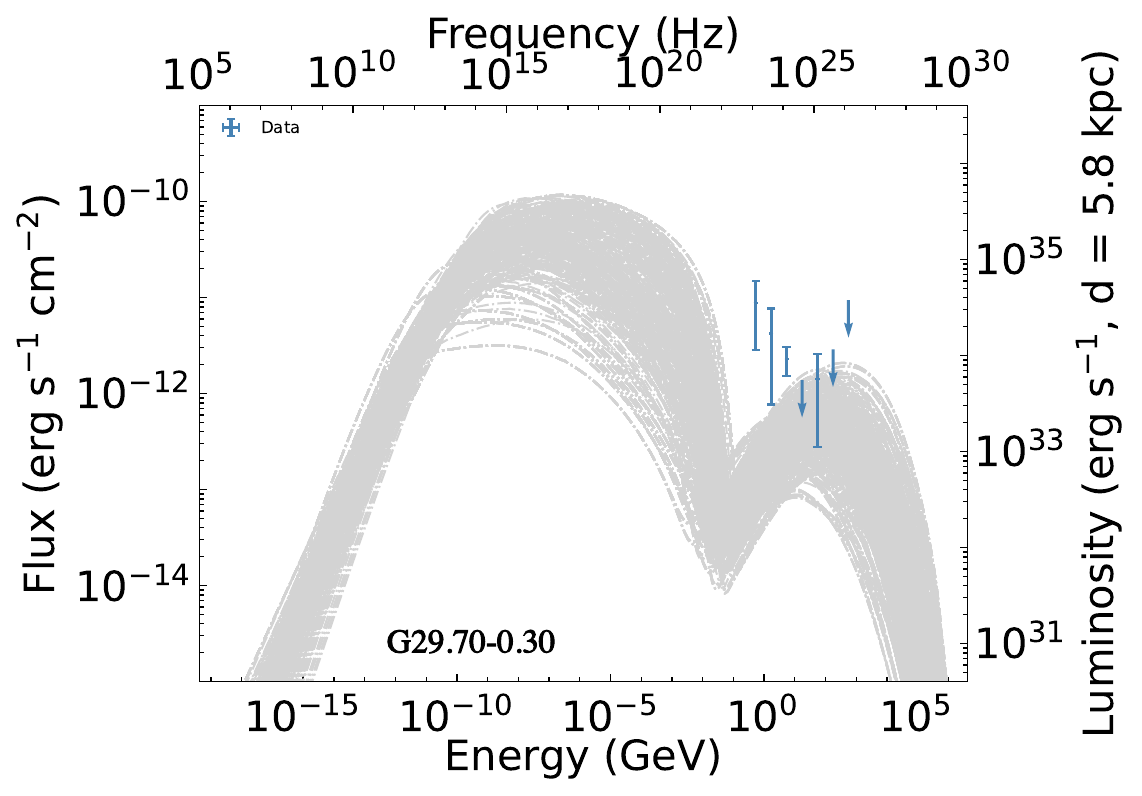}
\includegraphics[width=0.48\textwidth]{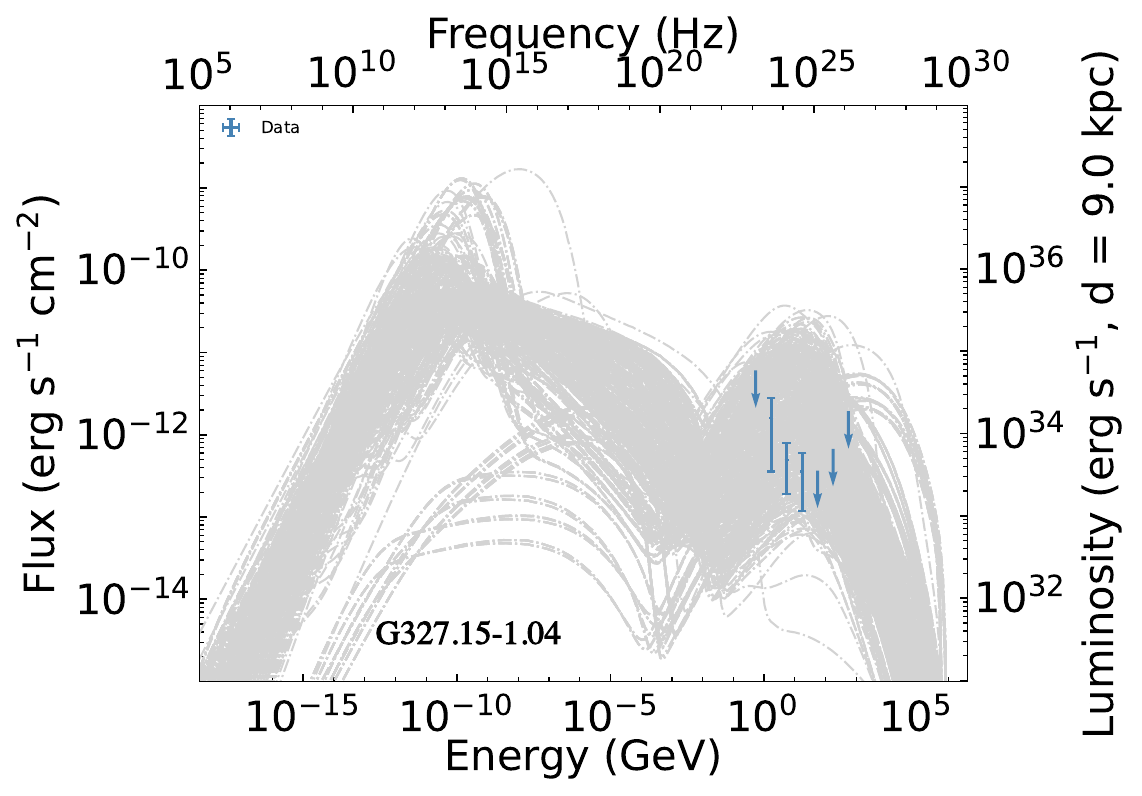}
\caption{Example predicted SEDs (gray) compared with \lat data (blue data points) for five selected sources. Adapted from Figure 21 of \citet{Eagle2025}. }
\label{FP.fig.rad_models}
\end{figure}

To assess the physical plausibility of the $\gamma$-ray emission in selected cases, we applied a one-zone, time-dependent PWN evolutionary model \citep{Torres2014, martin2022unique}. 
Five representative sources were modeled, including three PWN candidates (G11.18$-$0.35, G12.82$-$0.02, G15.40+0.10) and two likely PWNe (G29.70$-$0.30, G327.15$-$1.04), whose adopted pulsar and nebular properties are listed in Table~\ref{FP.radmod_targets}.

We varied nine physical parameters over conservative ranges:
\begin{itemize}
    \item $M_{\rm ej} \in [8, 15]\,M_\odot$;
    \item $\alpha_1 \in [1.0, 1.6]$, $\alpha_2 \in [2.2, 2.8]$;
    \item $\gamma_b \in [10^5, 10^6]$;
    \item $n = 2.0$, 2.5, 3.0 (2.16 for G29.70$-$0.30 post-burst; \citealt{Livingston2011});
    \item $\eta_B \in [0.02, 0.04]$;
    \item $n_{\rm ISM} \in [0.1, 1.0]\,{\rm cm}^{-3}$;
    \item $U_{\rm FIR/NIR} \in [1, 3] \times U_{\rm GALPROP}$;
    \item $t_{\rm age} \in [0.7, 1.3]\,\tau_c$ (fixed at 1.6 kyr for G11.2$-$0.35; \citet{Clark1977}).
\end{itemize}

For each target, we generated 712 models sampling the parameter space: 512 from binary combinations of extrema ($2^9$; or $2^8 = 256$ for G11.2$-$0.35 due to fixed age), plus 200 randomly sampled models. 
The resulting SEDs were compared with \lat spectra derived in this work (Figure~\ref{FP.fig.rad_models}). 
While the modeled flux levels generally match the \lat data, the spectral shapes often diverge, suggesting possible contamination from nearby sources (e.g., pulsars or SNRs), environmental variations, or unaccounted physical processes beyond the assumed parameter space.

\section{The \lat PWN Population}
\label{FP.pwn_population}

\begin{figure}[htbp]
    \centering
    \includegraphics[width=\textwidth]{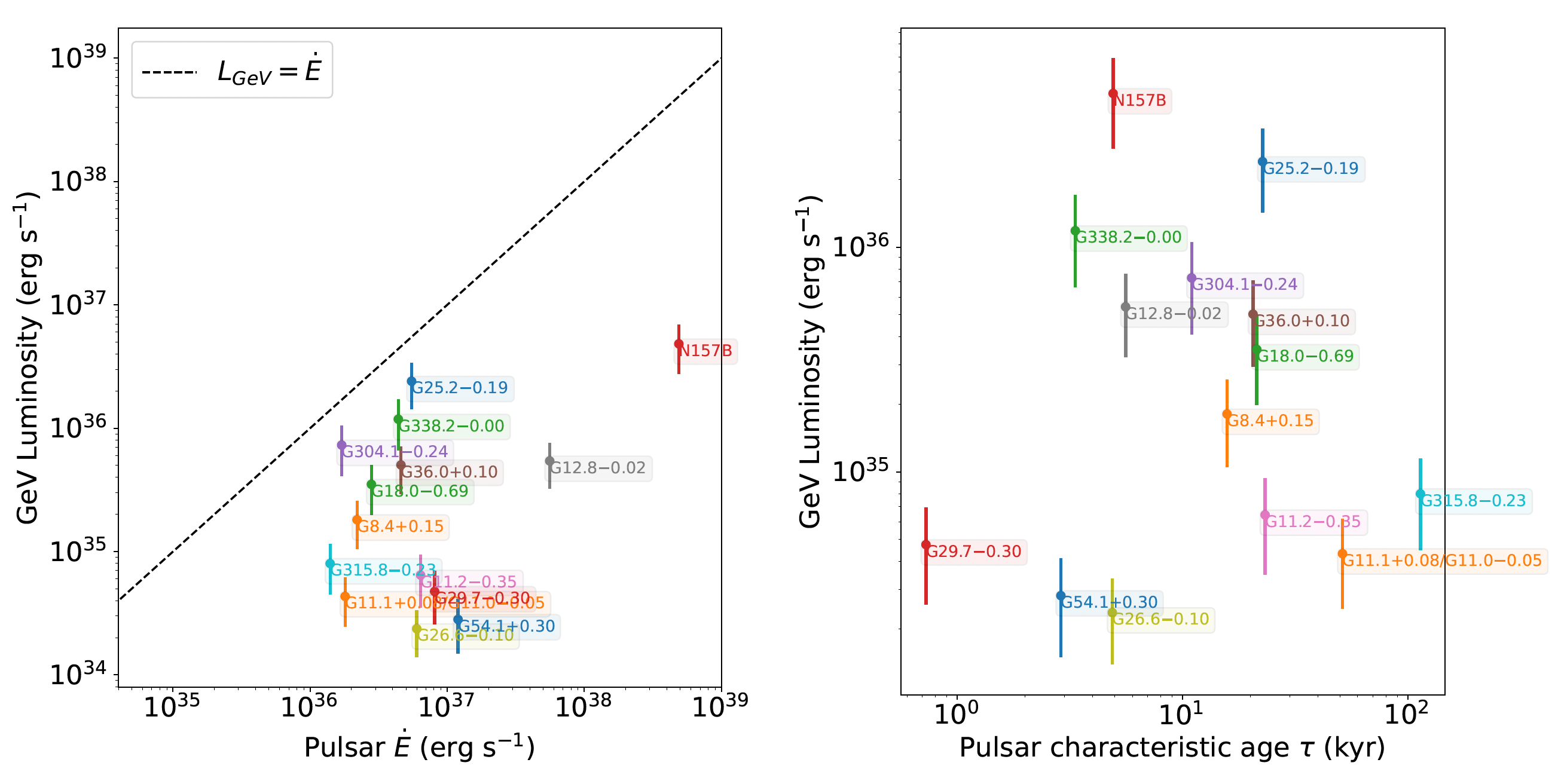}
    \caption{Left: GeV luminosity (300~MeV–2~TeV) versus pulsar spin-down power $\dot{E}$. 
    Right: GeV luminosity versus pulsar characteristic age $\tau_c$. Both panels include 30\% uncertainties to account for flux and distance uncertainties. No significant correlation is observed in either relation.}
    \label{FP.fig.pwn_population}
\end{figure}

\begin{table}[htbp]
\centering
\setlength\tabcolsep{2pt}
\scriptsize
\caption{Key properties of pulsars associated with the analyzed PWNe. The upper portion lists firmly associated pulsars for which \fermi counterparts were detected, while the lower portion includes sources without \fermi detections. Reproduced from Table B1 of \citet{Eagle2025}. }
\label{FP.tab.pwn.population}
\scalebox{1.0}{
\begin{tabular}{cccccccc}
\hline
ROI & Galactic PWN Name & 4FGL ID & Associated Pulsar & $d$ (kpc) & $\tau_c$ (kyr) & $\dot{E}$ ($10^{36}$ erg s$^{-1}$) \\
\hline
1  & G8.40$+$0.15   & J1804.7$-$2144e & J1803$-$2137 & 3.42 & 15.8  & 2.2  \\
2  & G11.09$+$0.08  & J1810.3$-$1925e & J1809$-$1917 & 3.27 & 51.4  & 1.8  \\
3  & G11.18$-$0.35  & J1811$-$1925e   & J1811$-$1925 & 5.00 & 23.3  & 6.4  \\
4  & G12.82$-$0.02  & J1813.1$-$1737e & J1813$-$1749 & 6.15 & 5.6   & 56   \\
5  & G18.00$-$0.69  & J1824.5$-$1351e & J1826$-$1334 & 3.61 & 21.4  & 2.8  \\
6  & G25.24$-$0.19  & J1836.5$-$0651e & J1838$-$0655 & 6.60 & 22.7  & 5.5  \\
7  & G26.60$-$0.09  & J1839.0$-$0532e & J1838$-$0537 & 1.3  & 4.89  & 6.0  \\
8  & G29.70$-$0.30  & J1846.4$-$0258 (DR4) & J1846$-$0258 & 5.8  & 0.728 & 8.1  \\
9  & G36.01$+$0.10  & J1857.7$+$0246e & J1856$+$0245 & 6.32 & 20.6  & 5.0  \\
10 & G54.10$+$0.27  & J1930.5$+$1853 (DR3) & J1930$+$1852 & 6.17 & 2.89  & 12   \\
11 & G279.60$-$31.70 (N 157B) & J0537.3$-$6909 & J0537$-$6910 & 49.7 & 4.93  & 490  \\
12 & G304.10$-$0.24 & J1303.0$-$6312e & J1301$-$6305 & 10.7 & 11    & 1.7  \\
13 & G315.78$-$0.23 & J1435.8$-$6018  & J1437$-$5959 & 8.54 & 114   & 1.4  \\
14 & G322.50$-$0.28 & J1616.2$-$5054e & J1617$-$5055 & 4.74 & 8.13  & 16   \\
15 & G338.20$-$0.00 & J1640.6$-$4632 (DR1), J1640.7$-$4631e (DR3) & J1640$-$4631 & 12.7 & 3.35  & 4.4  \\
\hline
1  & G0.87$+$0.08   & -- & J1747$-$2809 & 8.1  & 5.3   & 43.0  \\
2  & G23.50$+$0.10  & -- & J1833$-$0827 & 4.4  & 147   & 0.58  \\
3  & G32.64$+$0.53  & -- & J1849$-$0001 & 7.0  & 43.1  & 9.8   \\
4  & G34.56$-$0.50  & -- & J1856$+$0113 & 2.81 & 20.3  & 0.43  \\
5  & G47.38$-$3.88  & -- & J1932$+$1059 & 0.2  & 3100  & 0.004 \\
6  & G108.60$+$6.80 & -- & J2225$+$6535 & 1.9  & 1120  & 0.001 \\
7  & G179.72$-$1.69 & -- & J0538$+$2817 & 0.95 & 618   & 0.05  \\
8  & G266.97$-$1.00 & -- & J0855$-$4644 & 5.6  & 141   & 1.1   \\
9  & G290.00$-$0.93 & -- & J1101$-$6101 & 8$^\dag$  & 116   & 1.4   \\
10 & G310.60$-$1.60 & -- & J1400$-$6325 & 9.1  & 12.7  & 51    \\
11 & G341.20$+$0.90 & -- & J1646$-$4346 & 6.2  & 32.5  & 0.36  \\
12 & G358.29$+$0.24 & -- & J1740$-$3015 & 2.9  & 20.6  & 0.008 \\
\hline
\end{tabular}}
\begin{tablenotes}
\item \textbf{Note.} Pulsar distances, characteristic ages, and spin-down powers are taken from the ATNF pulsar catalog \citep{manchester2005australia}.
\item $^\dag$ The distance for PSR J1101$-$6101 is adopted from \citet{Halpern2014}. 
\end{tablenotes}
\end{table}

Figure~\ref{FP.fig.pwn_population} explores the relationship between GeV luminosity and pulsar properties for firmly associated sources (see Table~\ref{FP.tab.pwn.population}). 
The left panel shows the 300~MeV--2~TeV luminosity as a function of the pulsar spin-down power $\dot{E}$, while the right panel plots luminosity versus characteristic age $\tau_c$. 
Luminosities were computed using flux uncertainties and a 20\% distance error, following the method of \citet{Acero2013}. 
Pulsar distances and characteristic ages are taken from the Australia Telescope National Facility (ATNF) pulsar catalog\footnote{\url{https://www.atnf.csiro.au/research/pulsar/psrcat}} based on dispersion measures.

Consistent with earlier studies \citep[e.g.,][]{Acero2013}, no significant correlation is found between GeV luminosity and either $\dot{E}$ or $\tau_c$. 
This suggests that gamma-ray output in this energy range does not scale simply with these basic pulsar parameters.

\section{Discussion and Concluding Remarks}
\label{FP.conclusion}

In this work, we presented a part of a large systematic study of $\gamma$-ray emission from known and candidate PWNe that lack of detected \fermi pulsars using 11.5 years of \lat data. 
Through a targeted search across 58 ROIs, we identified 36 significant sources, including 9 likely PWNe and 21 PWN candidates, classified based on morphological and spectral characteristics, as well as energetic consistency. 

These results significantly expand the sample of potential $\gamma$-ray PWNe, suggesting that the Galactic PWN population detected by \lat is likely underrepresented in current catalog. 
We demonstrate that radial Gaussian models effectively describe the morphology of most extended sources, including those with ambiguous associations. 
Multiwavelength evidence remains critical for classifying and understanding many of the sources, particularly in complex regions where confusion with SNRs or pulsars is possible. 
Radiative modeling for five representative sources reinforces this conclusion. 
While model fluxes generally align with \lat measurements, mismatches in spectral shape point to possible contamination from nearby objects, environmental inhomogeneities, or unmodeled physical processes.

The expanded sample serves as a valuable reference for future studies and presents a prime target list for multiwavelength follow-up observations. 
A forthcoming study will examine the off-pulse emission from LAT-detected pulsars to uncover additional PWNe hidden beneath bright magnetospheric emission. 
\chapter{Analysis of the possible detection of the pulsar wind nebulae of PSR J1208-6238, J1341-6220, J1838-0537, and J1844-0346}
\label{tev.pwn}
\textbf{\color{SectionBlue}\normalfont\Large\bfseries Contents of This Chapter\\\\}

In this chapter, we investigate the detectability of PWNe using the pulsar tree - a graph-theoretical framework recently developed in my research group to classify pulsars based on their intrinsic properties. 
By applying this method, we identified promising TeV PWN candidates and assessed their potential for detection by current and future gamma-ray observatories.

Specifically, we selected four candidate pulsars - PSR~J1208$-$6238, J1341$-$6220, J1838$-$0537, and J1844$-$0346 - based on their positions in the pulsar tree. 
Using our time-dependent leptonic model and the script described in Appendix \ref{script1}, we simulate the SEDs for these candidates. 
We then evaluate their detectability by comparing model predictions with the sensitivities of major TeV observatories.

This study provides insight into the effectiveness and limitations of using pulsar tree topology as a predictive tool, revealing its utility in isolating pulsars that may host TeV-emitting nebulae, even when only intrinsic pulsar parameters are considered.

This work has been published in \texttt{Astronomy \& Astrophysics (A\&A)} \citep{Zhang2024}. 

\newpage

\section{Introduction}
\label{sec1}


Most of the radiative models used to describe pulsar wind nebulae (PWNe) are applicable only to relatively young PWNe (where PWNe are not yet in contact with the reverse shock, so reverberation does not need to be considered in detail; see \citet{Bandiera:2022, bandiera2023reverberation} and references therein for discussion). According to existing models found to be in agreement with observations, the low-energy radiation of a PWN is mainly from synchrotron radiation generated by relativistic charges moving in a magnetic field, while the high-energy radiation is due to inverse Compton (IC) scattering from environmental photons -- mainly far infrared (FIR) photons (see, e.g., \citet{Atoyan1996,Aharonian1997,zhang2008, gelfand2009, dejager2009, tanaka2011, Martin2012, Bucciantini2011, Torres2014, Vorster2013,torres2018,zhu2021}. Only Crab is known to be a self-synchrotron Compton (SSC) source. The SSC becomes relevant only for highly energetic (around 70\% of Crab) particle-dominated nebulae at low ages (of less than a few thousand years) located in a FIR background with a relatively low energy density \citep{Torres2013}. 

The model used in this work to describe the evolution of PWNe is based on TIDE, a one-zone leptonic time-dependent code capable of calculating both the energy spectrum and the dynamical evolution of young and middle-aged PWNe. Here, a simple thin-shell model is taken into account for the evolution of a PWN during the reverberation phase (see the incorporation of model details with time in \citealt{Torres2014,Martin2016,martin2022unique}). At the core of this model is seeking the solution of the equation \citep{ginzburg1964}
\begin{equation}
 \frac{\partial N(\gamma, t)}{\partial t}=-\frac{\partial}{\partial \gamma}[\dot{\gamma}(\gamma, t) N(\gamma, t)]-\frac{N(\gamma, t)}{\tau(\gamma, t)}+Q(\gamma, t) \text {. }
\end{equation}
The left side of this equation shows the evolution of the lepton distribution with time. The right side represents the variation of the particle energy with time due to the energy losses of different processes, including synchrotron, bremsstrahlung, IC, and adiabatic losses. Additionally, it accounts for the effect of particle escape on the lepton distribution, where $\tau(\gamma, t)$ is the escape time (assuming through Bohm diffusion) with $Q(\gamma, t)$ being the particle injection per unit energy per unit volume at a certain time $t$ and energy $\gamma$. 

The pulsar tree (see \citet{MST-I,MST-II,Garcia2024frb,Garcia2024msp} for full details) is a new way of visualizing the pulsar population.
In this paper, we search for potential TeV PWNe based on their locations in the pulsar tree and use our one-zone PWN leptonic model to predict their possible energy spectrum characteristics. In Sect. \ref {method}, we introduce an analysis of MST where most known TeV PWNe are located in the lower-left region and choose candidates for further study, and subsequent sections show our results for the four objects chosen. Conclusions are given in Sect. \ref{conclusion}.

\section{Methodology}
\label{method}
\subsection{Global properties and candidate selection}
\label{MST}

\begin{figure}
\centering
    \includegraphics[width=0.32\columnwidth]{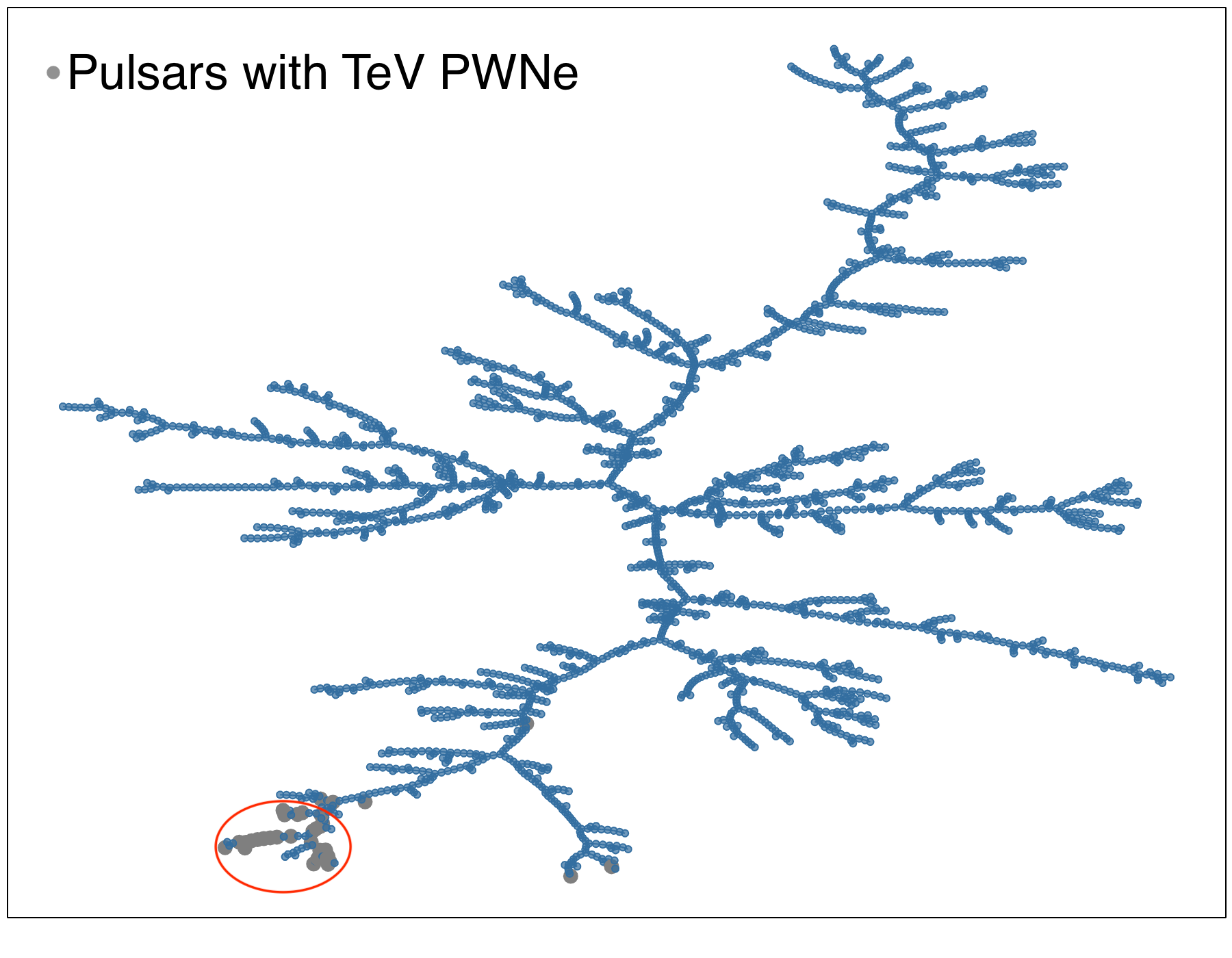}
    \includegraphics[width=0.32\columnwidth]{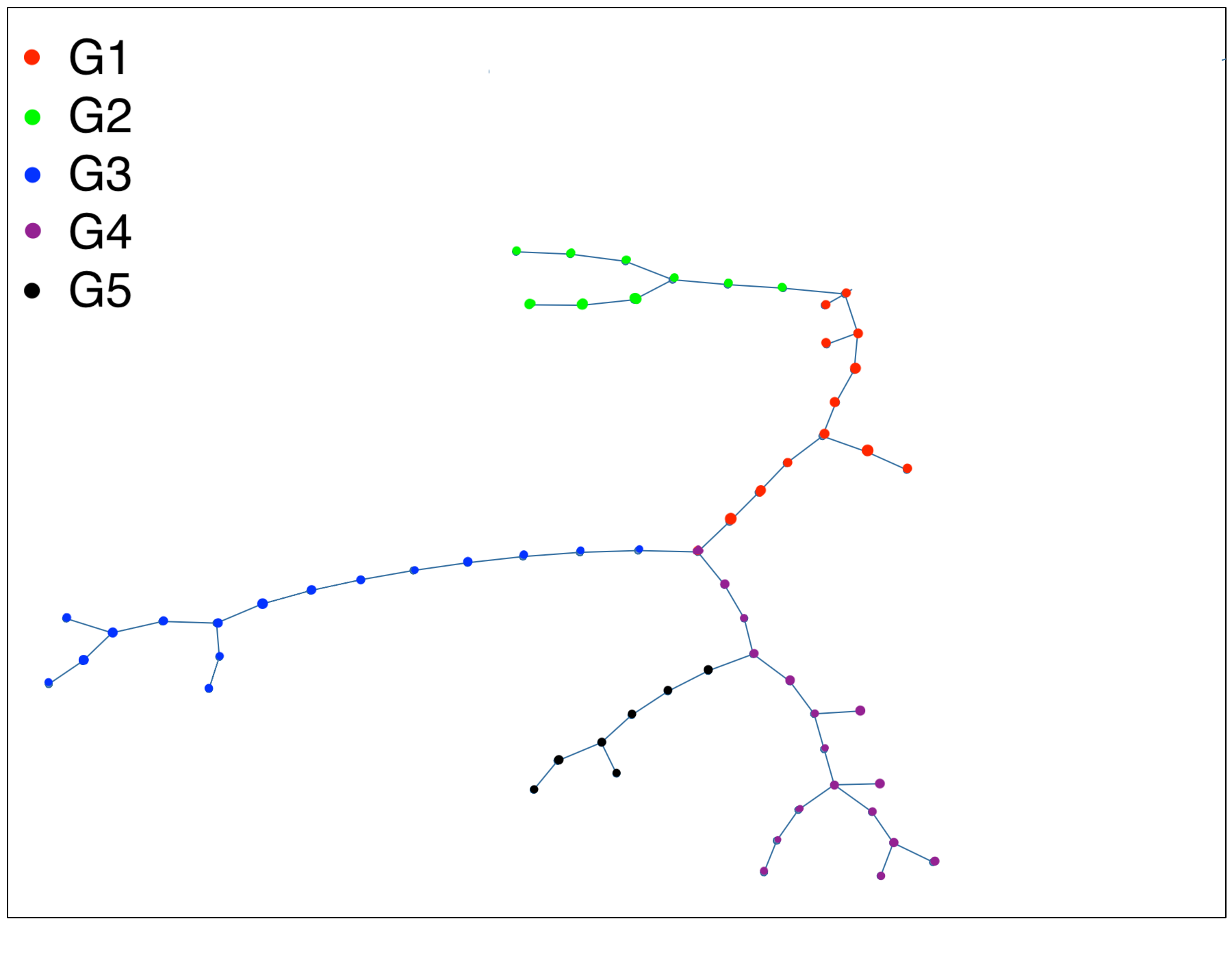}
    \includegraphics[width=0.32\columnwidth]{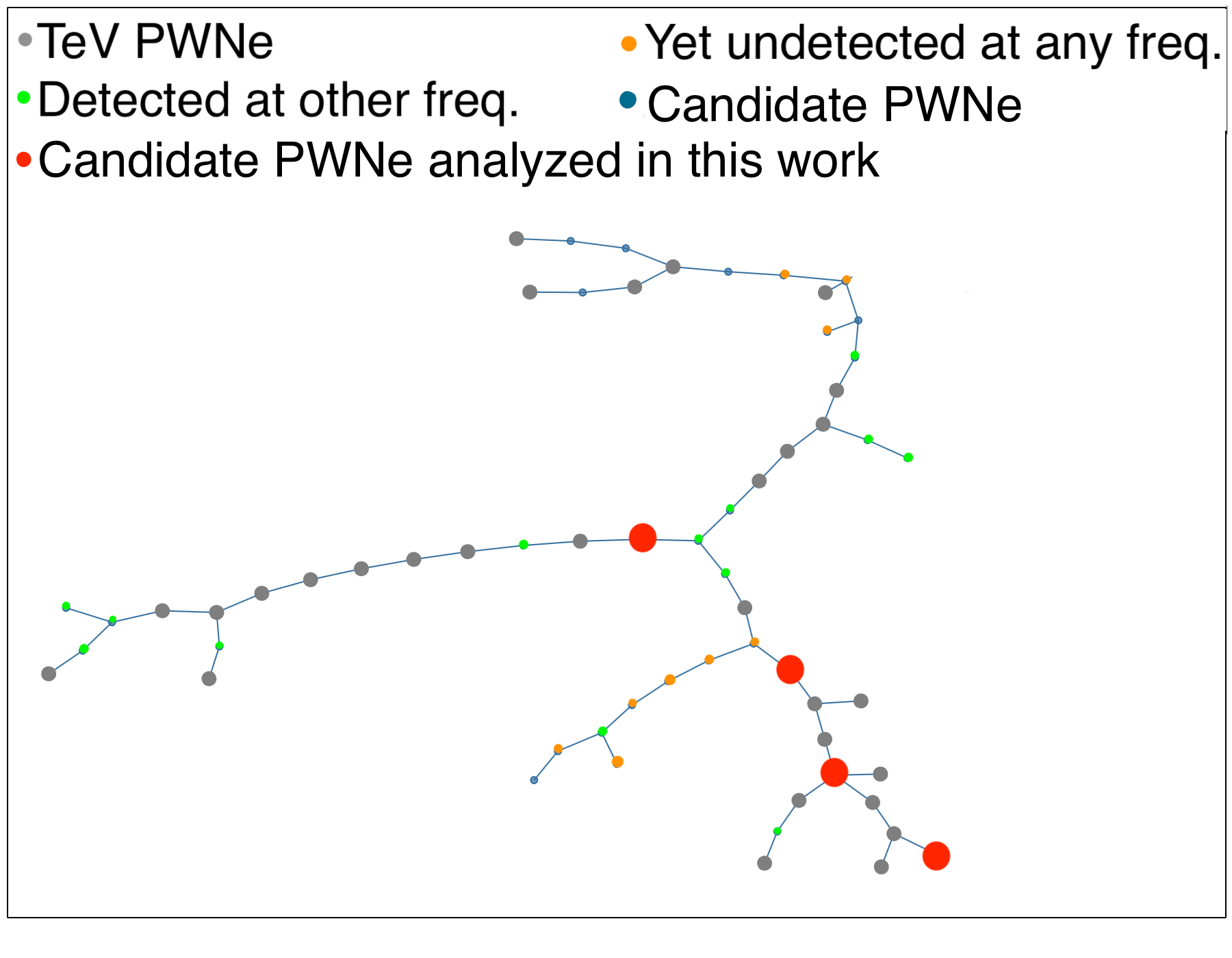}
\caption{The pulsar tree and the positions of the selected pulsars in the pulsar tree. Left panel: Pulsars with detected TeV PWNe (gray dots) noted in the pulsar tree, the minimal spanning tree of the pulsar population.
Middle panel: Grouping of pulsars. The pulsars have been grouped in a way that allows us to comment on their different features (see text). This is a zoom-in of the bottom-leftmost part of the MST {(the area enclosed by the red ellipse)}, which is where most of the TeV PWNe are located. 
Right panel: Same zoom-in as in the middle panel, but we note the 29 pulsars with confirmed TeV PWNe, the 13 pulsars with confirmed PWNe at other frequencies, the nine pulsars without a PWN detected at any frequencies yet, and the ten remaining pulsars for which there are only unconfirmed candidate PWNe claimed. 
Four selected candidates analyzed in this work are noted in red, and these also belong to the group of ten pulsars. 
From top to bottom, the red points represent PSR J1844-0346, PSR J1341-6220, PSR J1838-0537, and PSR J1208-6238, respectively.}
    \label{tev.fig1}
\end{figure}

\subsection{Computing the pulsar tree}

The first panel of Fig.~\ref{tev.fig1} shows the pulsar tree \citep{MST-I}. Online access to this figure (and tools to zoom and label each node) is also available.\footnote{\url{https://sites.google.com/view/thepulsartree}} 
The pulsar tree uses a graph theory tool known as a minimum spanning tree (MST) to represent the $P\dot P$ diagram. 
The MST is a graph that connects points in a multi-dimensional space. Each point (pulsar) is linked to at least another by an edge whose length is associated with a given distance. The latter is taken to be the Euclidean distance over the principal components of magnitudes representing the intrinsic characteristics of pulsars. The pulsar properties considered are the spin period ($P$), spin period derivative ($\dot{P}$), characteristic age ($\tau_c$), spin-down luminosity ($\dot{E}$), surface magnetic field ($B_s$), light cylinder magnetic field ($B_{lc}$), Goldreich-Julian charge density ($\eta_{GJ}$), and surface electric voltage ($\Phi_s$). The edges of an MST are chosen so that the sum of their lengths (equivalently, the sum of the total distance) when all nodes are linked and the whole population is connected is minimal. Graph theory shows that as long as distances are distinct, the MST is unique. Its definition intuitively implies that the MST is a classification technique. 

\subsection{The pulsar tree and the PWNe}

As can be seen in the first panel of Fig. \ref{tev.fig1}, most of the pulsars with detected TeV PWNe \citep{hess2018population} are clustered in the lower-left region of the pulsar tree. There are 61 pulsars in this region, shown in the second panel of Fig. \ref{tev.fig1}. Twenty-nine of them have confirmed TeV PWNe, 13 pulsars have confirmed PWNe detected in other bands, and another ten pulsars have been noted to have possible PWN counterparts at some frequencies that have not yet been confirmed. We did not find any reports regarding possible PWNe in any bands for the remaining nine pulsars. The overdensity of TeV-detected PWNe in a small part of the MST is quite notorious. If TeV PWNe were placed randomly in the MST, the probability that this overdensity existed would be negligible. The MST is joining similarly energetic and relatively young sources that are prone to have TeV PWN behavior. However, we recall that no connection with the environment or with the progenitor is part of the distances underlying the MST.

Table \ref{tabA1} lists the (61-29=) 32 pulsars without confirmed TeV PWNe. We used the visually obvious branches appearing in the pulsar tree in this region to divide the pulsars into five groups for further study (see the middle panel of Fig. \ref{tev.fig1}). The sources located at the junctions of these branches are included in the group to which they are the most similar. For example, PSR J1826-1256, which is at the junction of G1, G3, and G4, has six (five) [three] of the eight magnitudes considered that are larger or smaller than the corresponding values of the sources in G1 (G3) [G4]. So we include it in G4. None of the sources in G5 have a detected TeV PWNe. Also, we observed that the values of $\dot{E}$ and $\Phi_s$ (proportional to $\dot P/P^3$) in G5 are smaller than those in any other group. Of the six pulsars in G1 with the smallest values in $\dot{E}$ and $\Phi_s$, only PSR J1809-1917 has a detected TeV PWN. Similarly, the G4 pulsars
PSR J1341-6220, PSR J2111+4606, and PSR J1208-6238, which have the smallest $\dot{E}$ and $\Phi_s$ (PSR J2111+4606 does not have a PWN detected in any band so far), and the other two sources have only a possible PWN in X-ray and radio, respectively. The tree seems to group together pulsars that are less likely to have detectable PWNe. 

From the sources that do not have a confirmed TeV PWN already, we selected candidates for further study from possibly young sources with $\tau_c<15$ kyr, which still lack a detailed radiative analysis from which predictions can already be derived. We chose PSR J1838-0537, PSR J1208-6238, PSR J1844-0346, and PSR J1341-6220 under these conditions. The positions in the pulsar tree of these four sources are also shown in Fig. \ref{tev.fig1}. Our subsequent investigation sought to determine whether these four pulsars could be producing plausibly observable PWNe, or if models similar to those used to describe other well-known PWNe predict that they should be undetectable.

\begin{landscape}
\begin{table}
\setlength\tabcolsep{2pt}
        \centering
  \scriptsize
 \caption{Pulsars without confirmed TeV PWNe. The parameters in columns 2, 3, 4, and 5 are from the ATNF catalog \citep[see \url{http://www.atnf.csiro.au/research/pulsar/psrcat}][]{manchester2005australia}. The ``Y" in column 6 indicates that its PWN has been confirmed in one or several bands; an ``N" means that the PWN hasn't been observed yet to our knowledge; and a question mark means that there is a possible PWN counterpart in some band or bands but that it has not yet been confirmed. In the eighth column (PWN energy band), ``R, O, X, G, T" stand for radio, X-rays, visible, GeV, and TeV energy ranges. }  

 \label{tabA1}
        \begin{tabular}{lcccccccc} 
                \hline
        \hline
                    PSR Name & $P$  &$\dot{P}$  &$\dot{E}$  &$\tau_c$  &PWN? &PWN / Composite  &Energy &Refs.\\
             & ($10^{-2}$ s) & ($10^{-14}$ s s$^{-1}$) & ($10^{36}$ erg s$^{-1}$) & (kyr) & &SNR Name &Band & (see also other refs. therein)\\
                        
  \hline
            J0540-6919 &5.06 &47.9 &146 &1.67 &Y &G279.7-31.5 &R, X, O, G$^?$ &\cite{manchester1993,mignani2012,bamba2022spectral}\\
            J0940-5428 &8.75 &3.29 &1.93 &42.2 &N & & &\\
            J1015-5719 &14.0 &5.74 &0.828 &38.6 &Y &G283.1-0.59 &R &\cite{ng2017discovery}\\
            J1044-5737 &13.9 &5.46 &0.803 &40.3 &N & & &\\
            J1048-5832 &12.4 &9.61 &2.00 &20.4 &Y & &X &\cite{gonzalez2006}\\
            J1111-6039 &10.7 &19.5 &6.35 &8.66 &Y &G291.0-0.1 &R, X, G &\cite{slane2012broadband}\\
            J1112-6103 &6.50 &3.15 &4.53 &32.7 &$?$ & &X$^?$, G$^?$ &\cite{townsley2011integrated,ackermann2016}\\
            J1124-5916 &13.5 &75.3 &11.9 &2.85 &Y &G292.04+1.75 &R, X, G$^?$ &\cite{ajello2017,park2007,gaensler2003}\\
            J1135-6055 &11.5 &7.93 &2.06 &23.0 &Y &G293.8+0.6 &X, R$^?$ &\cite{zhang2019chandra,bordas2020}\\
            J1203-6242 &10.1 &4.41 &1.71 &36.1 &N & & &\\
            J1208-6238 &44.1 &327 &1.51 &2.14 &? & &X$^?$ &\cite{bamba2020low}\\
            J1341-6220 &19.3 &25.3 &1.38 &12.1 &? &G308.8-0.1 &R$^?$, G$^?$ &\cite{3FGL,green1997continuation}\\  
            J1400-6325 &3.12 &3.89 &50.7 &12.7 &Y & &X, R &\cite{bamba2022spectral,renaud2010}\\
            J1410-6132 &5.01 &3.20 &10.1 &24.8 &? &G312.4-0.4 &X$^?$ &\cite{doherty2003}\\
            J1524-5625 &7.82 &3.90 &3.21 &31.8 &N & & &\\
            J1637-4642 &15.4 &5.93 &0.640 &41.2 &? &HESS J1640-465 &X$^?$,G$^?$, T$^?$ &\cite{slane2010fermi,lemiere2009,aharonian2006}\\
            J1730-3350 &14.0 &8.48 &1.24 &26.1 &N & & &\\
            J1747-2958 &9.88 &6.13 &2.51 &25.5 &Y &Mouse &R,X &\cite{klingler2018}\\
            J1801-2451 &12.5 &12.8 &2.59 &15.5 &Y &Duck &R,X &\cite{kaspi2001,gaensler2000}\\
            J1813-1246 &4.81 &1.76 &6.24 &43.4 &? &HESS J1813-126 &X$^?$, T$^?$ &\cite{marelli2014puzzling,tibolla2022pulsar}\\
            J1826-1256 &11.0 &12.1 &3.58 &14.4 &Y &Eel &X, G$^?$, T$^?$ &\cite{burgess22,roberts2007,breuhaus2022}\\
            J1837-0604 &9.63 &4.52 &2.00 &33.8 &? &LHAASO J1839-0545 &T$^?$ &\cite{wilhelmi22,Cao2021}\\
            J1838-0537 &14.6 &47.2 &6.02 &4.89 &? &HESS J1841-055 &T$^?$ &\cite{magic2020studying,Aharonian2008}\\
            J1844-0346 &11.3 &15.5 &4.25 &11.6 &? &HESS J1843-033 &T$^?$ &\cite{sudoh2021highest,Devin2021,amenomori2022measurement}\\
            J1932+2220 &14.5 &5.76 &0.754 &39.8 &N & & &\\
            J1934+2352 &17.8 &13.1 &0.908 &21.6 &N & & &\\
            J1935+2025 &8.01 &6.08 &4.66 &20.9 &? &G054.1+00.3 &X$^?$, T$^?$ &\cite{temim2010,xia2023}\\
            J2021+3651 &10.4 &9.57 &3.38 &17.2 &Y &G75.2+0.1 &R, X, T$^?$ &\cite{roberts2008,woo2023,Abeysekara2020}\\
            J2022+3842 &4.86 &8.61 &29.6 &8.94 &Y &G076.9+1.0 &R, X &\cite{arzoumanian2011,marthi2011}\\
            J2111+4606 &15.8 &14.3 &1.44 &17.5 &N & & &\\
            J2229+6114 &5.16 &7.83 &22.5 &10.5 &Y &Boomerang &R, X, G, T$^?$ &\citet{magic2023,pope2024}\\
            J2238+5903 &16.3 &9.70 &0.889 &26.6 &N & & &\\
                \hline
        \end{tabular}
\end{table}
\end{landscape}

\begin{landscape}
\begin{table}
        \centering
  \scriptsize
        \caption{Physical magnitudes. The symbol ``-" in the table means the value or range is the same as noted in the previous column. To make sure the $\tau_0$ will not be negative, the braking index for the sources except J1208-6238 was assumed to be 2/2.5/3, which means that with the increasing of the assumed value for $t_{age}$ from 0.7 to three times $\tau_c$, the braking index would be assumed as 3, 2.5, and 2, respectively.} 
        \label{tab2}
        \begin{tabular}{llllll} 
                \hline
        \hline
                Parameters & Symbol & J1208-6238 &J1341-6220 &J1838-0537 &J1844-0346 \\
                \hline
         Measured, assumed, or derived parameters: \\
                        \hline
                
                Period (s) & $P$ &0.441 &0.193 &0.146 &0.113 \\
        Period derivative $(s~s^{-1})$ &$\dot{P}$ &$3.27\times10^{-12}$ &$2.53\times10^{-13}$ &$4.72\times10^{-13}$ &$1.55\times10^{-13}$\\
                Characteristic age (kyr) &$\tau_c$ &2.14 &12.1 &4.89 &11.6\\
                Spin-down luminosity now (erg s$^{-1})$ &$\dot{E}$ &$1.51\times10^{36}$ &$1.38\times10^{36}$ &$6.02\times10^{36}$ &$4.25\times10^{36}$ \\
                Distance (kpc) & $d$ &3, 6, 10  &12.6 &2.0  &2.4, 4.3, 10 \\
        Initial spin-down age &$\tau_0$ &$({2\tau_c})/({n-1}) - t_{age}$ & & &\\
        Initial spin-down luminosity (erg s$^{-1})$ &$L_0$ &$\dot{E}/(1+{t_{age}}/{\tau_0})^{-(n+1)/(n-1)}$ & & &\\
   \hline               
        Parameter ranges: \\
                        \hline
        Age ($\tau_c$) &$t_{age}$ &[0.7, 1.2] &[0.7, 1.3] &- &- \\
        ISM density (cm$^{-3}$) &$n_{ism}$ &[0.1, 1] &- &- &- \\
        Break energy &$\gamma_b$ &[$10^5, 10^7$] &- &- &-\\
        Low energy index &$\alpha_1$ &[1.0, 1.6] &- &- &-\\
                High energy index &$\alpha_2$ &[2.2, 2.8] &- &- &-\\
                Ejected mass $(M_{\odot})$  &$M_{ej}$ & {\bf [5, 15]} &- &- &-  \\
                Magnetic energy fraction  &$\eta$ &[0.02, 0.04] &- &- &-\\
                FIR energy density (eV cm$^{-3}$) &$U_{fir}$ &[0.55, 1.65] &[0.29, 0.87] &[0.82, 2.46] &[0.82, 2.46]\\
                NIR energy density (eV cm$^{-3}$) &$U_{nir}$ &[0.82, 2.46] &[0.51, 1.53] &[1.06, 3.18] &[1.06, 3.18] \\
        Breaking index & $n$ &2.598 (measured) &2/2.5/3 &- &- \\
        Minimum energy at injection &$\gamma_{min}$ &1 &- &- &- \\
        Containment factor  &$\epsilon$ &0.5 &- &- &- \\
        SN explosion energy (erg) &$E_{sn}$ &$10^{51}$ &- &- &- \\
        CMB temperature (K)  &$T_{cmb}$ &2.73 &- &- &-\\
                CMB energy density (eV cm$^{-3}$)  &$U_{cmb}$ &0.25 &- &- &-\\
        FIR temperature (K)  &$T_{fir}$ &70 &- &- &-\\
                NIR temperature (K)  &$T_{nir}$ &3000 &- &- &-\\
                \hline
        \end{tabular}
\end{table}
\end{landscape}

\subsection{Prior TeV observations and PWN models}
\label{analyse}

The positions of all four candidates are covered in the H.E.S.S. Galactic Plane Survey (HGPS; \citet{Abdalla2018}), and no PWNe was claimed at its sensitivity for the regions of PSR J1208-6238 and PSR J1341-6220. HESS J1841-055 and HESS J1843-033 are reported near PSR J1838-0537 and PSR J1844-0346, respectively, and both of them are regarded as one of a few plausible counterparts, based on being spatially coincident with the very high energy (VHE) source, albeit without certainty. Thus, we note the sensitivity of the HGPS at the position of the pulsars as an upper limit in their corresponding spectral energy distributions (SEDs) below, when appropriate. 

We simulated the possible SEDs of the putative PWNe of all four candidates using our one-zone leptonic model. The relevant physical parameters involved are shown in Table \ref{tab2}. The true age, $t_{age}$, of the pulsar was assumed to range from either 0.7 to 1.3 times $\tau_c$ or as wide a range as possible in order to secure a positive initial spin-down age $\tau_0$ given the adopted braking index. The base energy densities of soft photon fields, FIR and near infrared (NIR) photons, were adopted from the GALPROP code according to the coordinates and distances of these sources \citep{porter2022galprop}. However, considering that the GALPROP model estimates could underpredict the densities at the scale relevant for PWNe (see the discussion in \cite{Torres2014} and references therein), the energy densities were assumed to range from one to three times these values. 

We constructed models that range over nine parameters, as described in Table \ref{tab2}. We combined the minimum and maximum values of each of the nine parameters, which gave us 2$^{9}$=512 models, and then we added a group of 1000 model realizations by taking random values within all of these intervals, yielding a total of 1512 possible models for each source. These models span the range of possible SEDs given the uncertainties, and we used them to assess how likely it is to detect these PWNe in case they behave in a manner similar to the ones that are already known. We note that we considered a wide span of ejecta masses.
A lower $M_{ej}$ would result from a zero-age main sequence eight solar mass progenitor and could also be possible for windy progenitors such as luminous blue variables and Wolf-Rayet stars but would not be the usual case. Variations in the explosion energy are expected to be mild for such young pulsars, where approximately, for instance, $R_{pwn}(t) \propto \sqrt[5]{L_0t/E_{sn}}V_{ej}t$, and $V_{ej}=\sqrt{10E_{sn}/(3M_{ej})}$ while in free expansion.

\section{Results}
\label{results}

\subsection{PSR J1208-6238}
\label{secJ1208}

A very young, energetic, and highly magnetized radio-quiet $\gamma$-ray pulsar, PSR J1208-6238 has a spin-down luminosity of $1.51 \times 10^{36}$ erg s$^{-1}$. Its spin frequency and its first-order derivative are 2.2697 Hz and $-16.8427\times10^{-12}$ Hz/s, respectively, resulting in a characteristic age $\tau_c$ of 2672 years, and its braking index is $n = 2.598$, based on five years of Fermi-LAT observations \citep{clark2016braking}. The inferred dipolar magnetic field is relatively high, $3.8 \times 10^{13}$ G. The distance ($D$) to this pulsar is unknown, and we adopted 3, 6, and 10 kpc as different assumed values. Since the discovery of the pulsar, no evidence of a PWN has been observed at any wavelength until \cite{bamba2020low} reported X-ray emission from a possible PWN. The confidence level of this detection is 4.4 $\sigma$, but the association with a PWN is unclear (see the discussion of these authors). Their results would imply a conversion factor of $\dot{E}$ to the X-ray luminosity, which is unusually small ($< 10^{-4}\, D_3^2$, where $D_3$ represents the distance of the pulsar in units of 3 kpc). 
%

\begin{figure}
    \includegraphics[width=.32\columnwidth]{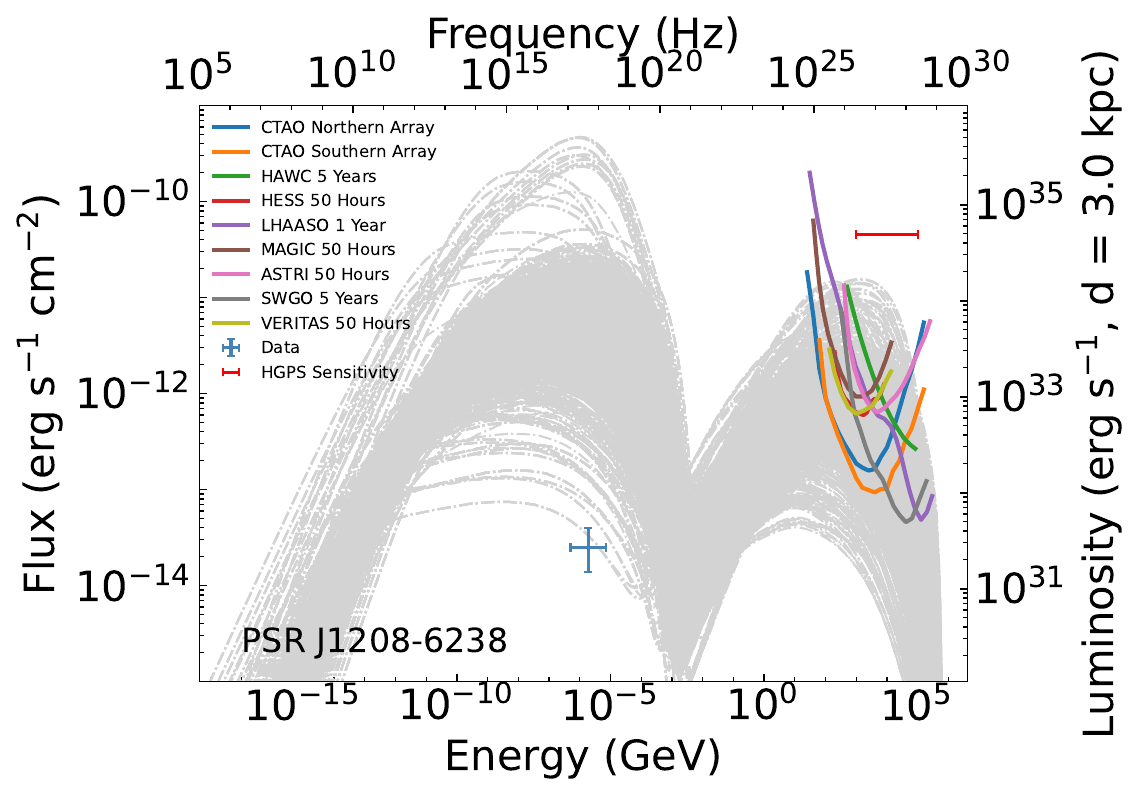}
    \includegraphics[width=.32\columnwidth]{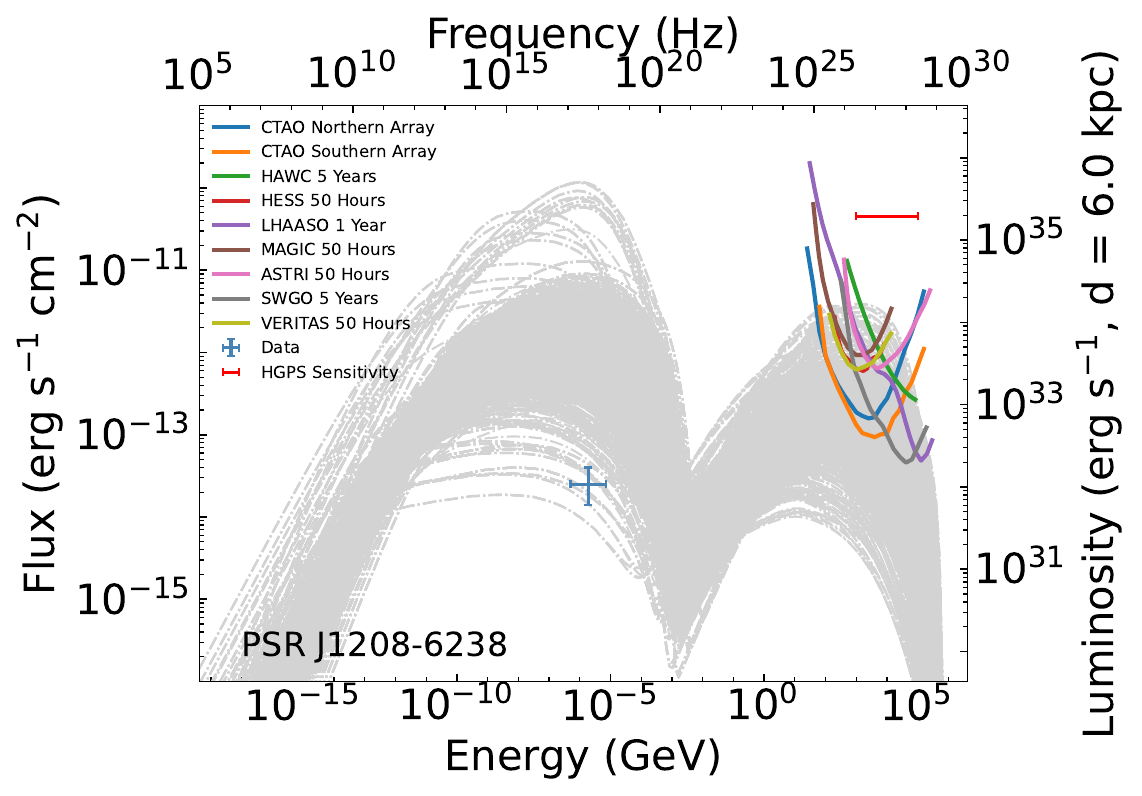}
    \includegraphics[width=.32\columnwidth]{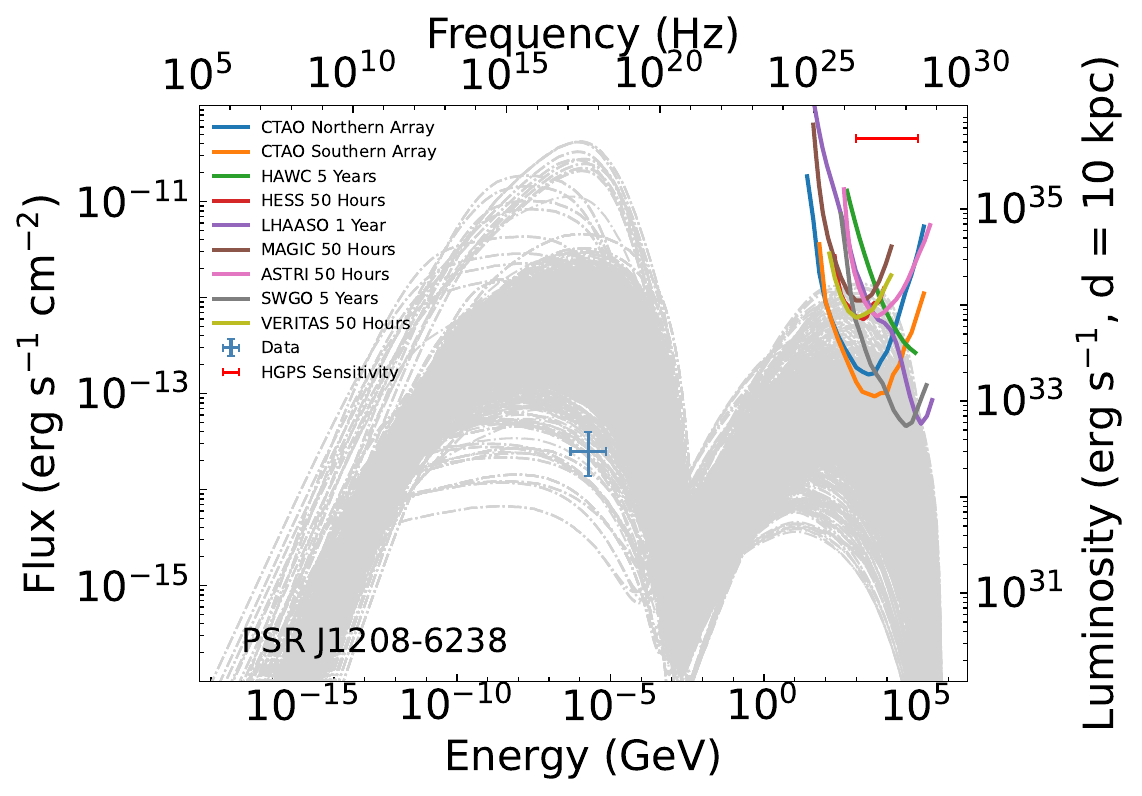}
    \includegraphics[width=.32\columnwidth]{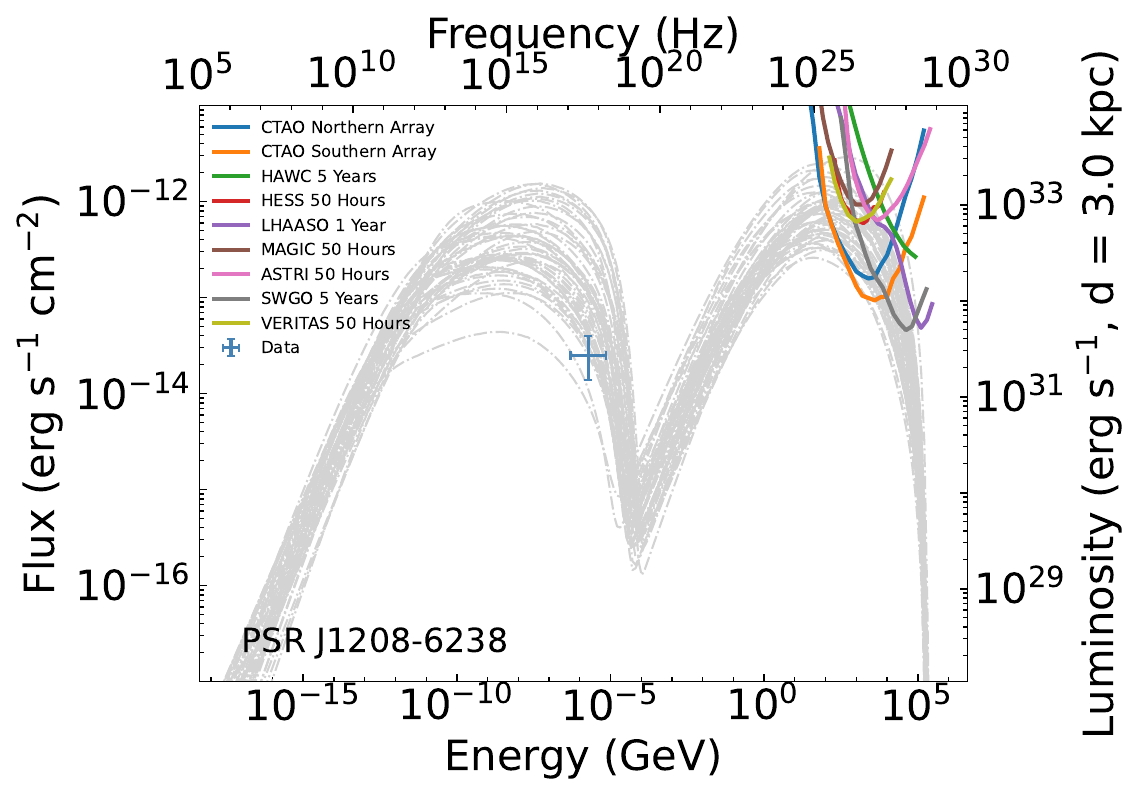}
    \includegraphics[width=.32\columnwidth]{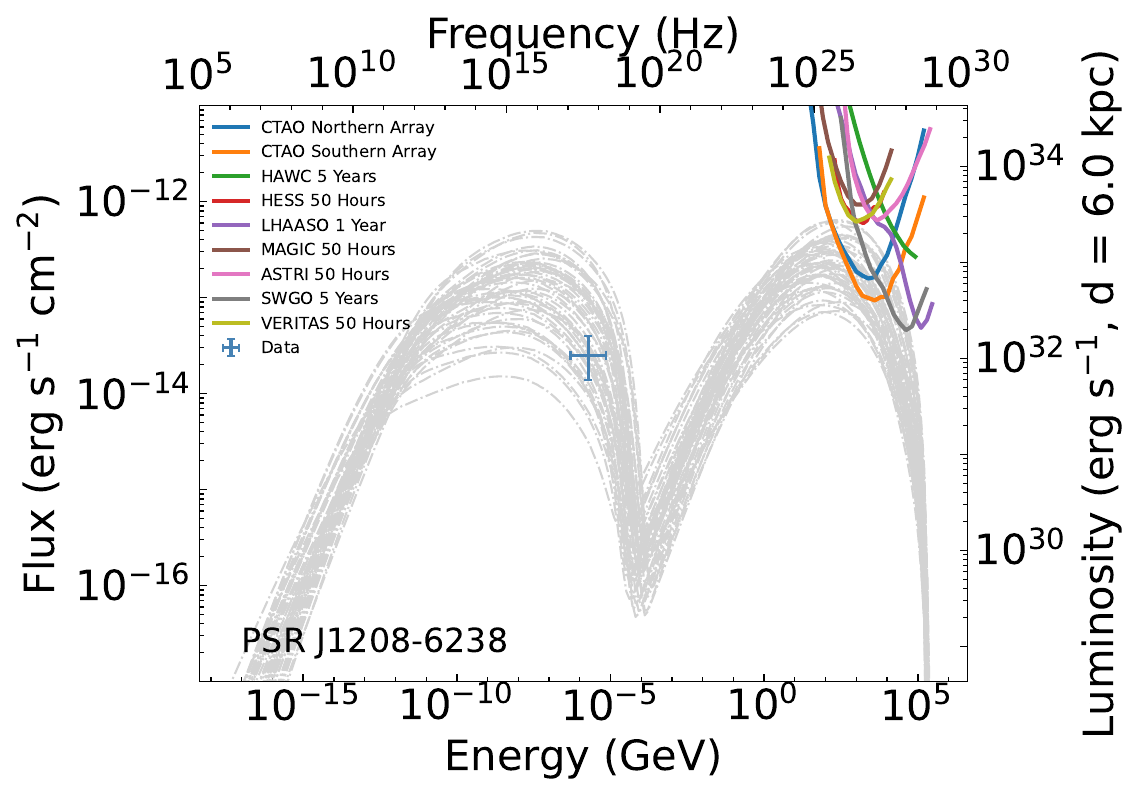}
    \includegraphics[width=.32\columnwidth]{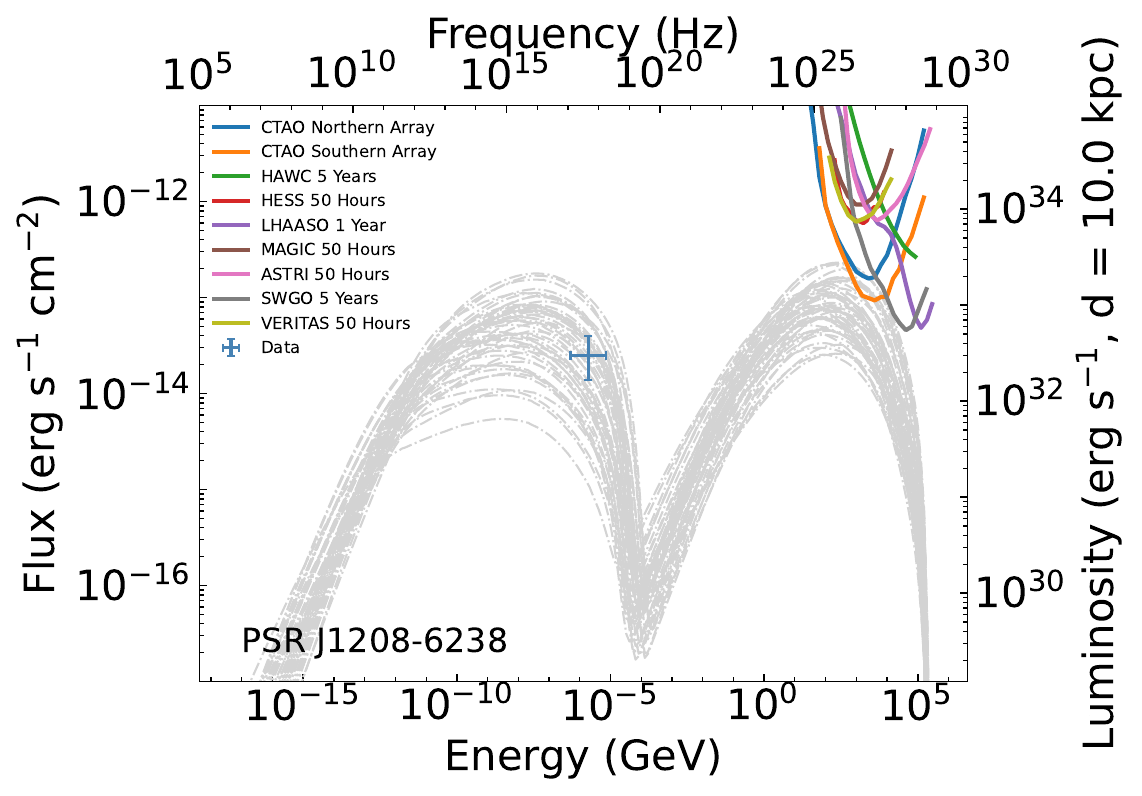}
   \caption{
Predicted set of SEDs for the PWNe of PSR J1208-6238. First three panels: Predicted SEDs for the PWN of PSR J1208-6238 with distances assumed to be 3 kpc, 6 kpc, and 10 kpc, respectively. 
The sensitivities of several instruments are from the CTA public website\protect\footnotemark. 
Last three panels: Fifty predicted SEDs for the PWN of PSR J1208-6238. These predictions were obtained by {fixing $\eta$ to 0.003 (for 3 kpc) or 0.004 (for 6 and 10 kpc)} and randomly selecting the other eight parameters in the ranges shown in Table \ref{tab2}. 
}
    \label{sed_J1208}
\end{figure}
\footnotetext{\url{https://www.ctao.org/for-scientists/performance/}}

Figure \ref{sed_J1208} shows the model realizations under different distance assumptions. At 3 kpc (this distance is also adopted in \citealt{clark2016braking} and \citealt{bamba2020low}), most of the model realizations would lead to a detectable PWN surrounding PSR J1208-6238, despite all of the models concluding that it would be undetectable in the HGPS. However, the realizations would all conflict with the X-ray observations of the region unless the size of the PWN is different from the X-ray region that is covered. Indeed, in the analysis of \cite{bamba2020low}, the authors used a ring centered on the pulsar as the possible PWN region, which had an inner radius of $\sim 5^{\prime\prime}$ and an outer radius of $\sim 9^{\prime\prime}$ (i.e., 0.13 pc outer radius at an assumed distance of 3 kpc). However, in our models, $R_{pwn}$ ranges from 0.3 to 4 pc at the same distance. If the X-ray size of this PWN is underestimated, the X-ray flux may also be underrated. 

In the case of 6 kpc, only eight models do not exceed the X-ray upper limit, but all of them would be below the sensitivity of any available TeV observational equipment, rendering the PWN undetectable. Even at 10 kpc (which is still within the maximum distance of 18.9 kpc reported in \cite{clark2016braking} by assuming that this pulsar is at the Galaxy edge for the given line of sight), this number rises to 41, representing 2.71\% of the total 1512 models, all of which are also undetectable in the TeV band. This result would be consistent with the fact that no TeV sources have been observed in this region so far. If this PWN behaved similarly to others in the known sample, we would likely find no TeV counterpart from it.

If the X-ray flux is indeed to be taken as an upper limit of the PWN, we also considered whether such a low flux could be just the result of a more extreme value of the magnetic fraction. Such low values were also needed in the modeling of some other PWNe (see the compilation provided by  \cite{Abdelmaguid2023}). Here, the value of $\eta$ needs to be less than 0.004 for 3 kpc and less than 0.005 for 6 and 10 kpc (as shown in the bottom panels of Fig. \ref{sed_J1208}) to respect the X-ray upper limit and make the TeV fluxes reach the sensitivity of the southern array of the Cherenkov Telescope Array (CTA) (which we refer to as S-CTA). These values of $\eta$ are much lower than the typical value found in other nebulae, which is typically around a few percent. This does not seem to offer a solution to the fact that it seems unlikely that PSR J1208-6238 would produce a detectable TeV PWN. Only the S-CTA, or H.E.S.S., can make a dedicated observation of this source, which we nevertheless promote as a way of testing these conclusions. 

Finally, if we consider the PWN to be a diffuse source of uncertain size and the X-ray data for a smaller PWN in \cite{bamba2020low} to provide an upper limit on its surface brightness, our models produce a surface brightness consistent with the X-ray constraint in about 45\% of all 1512 trials, and most of them are visible with S-CTA. However, we note the caveat that this assumes that the source is extended, dim, and uniformly diffuse, which indeed promotes fewer constraints onto the models and does not appear to be the case in X-ray observations of younger PWNe, where the X-ray emission is rather peaked.

\subsection{PSR J1341-6220}  
\label{secJ1341}

PSR J1341-6230 is a Vela-like radio pulsar discovered by the Parkes radio telescope with a spin period of 0.19 s, a characteristic age of $\tau_c = 12.1$ kyr, and a spin-down luminosity of $\dot{E} = 1.38 \times 10^{36}$ erg s$^{-1}$ \citep{manchester1985search,manchester2005australia}. Its distance is taken as 12.6 kpc from the ATNF catalog, based on the YMW16 electron density model. The weak X-ray counterpart of this pulsar was also detected in X-rays \citep{kuiper2015soft}. Frequent glitch phenomena were detected for this source; for instance, \cite{lower2021impact} reported 15 glitches. 

\cite{wilson1986x} gave an upper limit for the X-ray flux of PSR J1341-6220 on the order of $2.5 \times 10^{-13}$ erg s$^{-1}$ cm$^{-2}$ ($0.2-3.5$ keV). This pulsar is also listed as a radio pulsar with an X-ray counterpart in table 3 of \cite{Kaplan2004} with an about one order of magnitude smaller flux than quoted by \cite{wilson1986x}; however, it is unclear where this limit comes from, as they provide no analysis and quote an unpublished reference. \cite{Prinz2015} reported that the pulsar was in the field of view of the XMM-Newton telescope for about 42 ksec during two different epochs with different off-axis angles. Their analysis revealed two close sources (barely distinguishable and just above 2 keV; see their Fig. 1). They provided an upper limit for the harder of these two sources only. Although it is clear that there is no bright X-ray PWN around this pulsar, assigning a specific upper limit in order to discard models with it is, at this time, risky.
We promote the undertaking of a dedicated X-ray observation of this pulsar with Chandra to finally determine its nebular characteristics, if any.

\begin{figure}
\centering
    \includegraphics[width=.99\columnwidth]{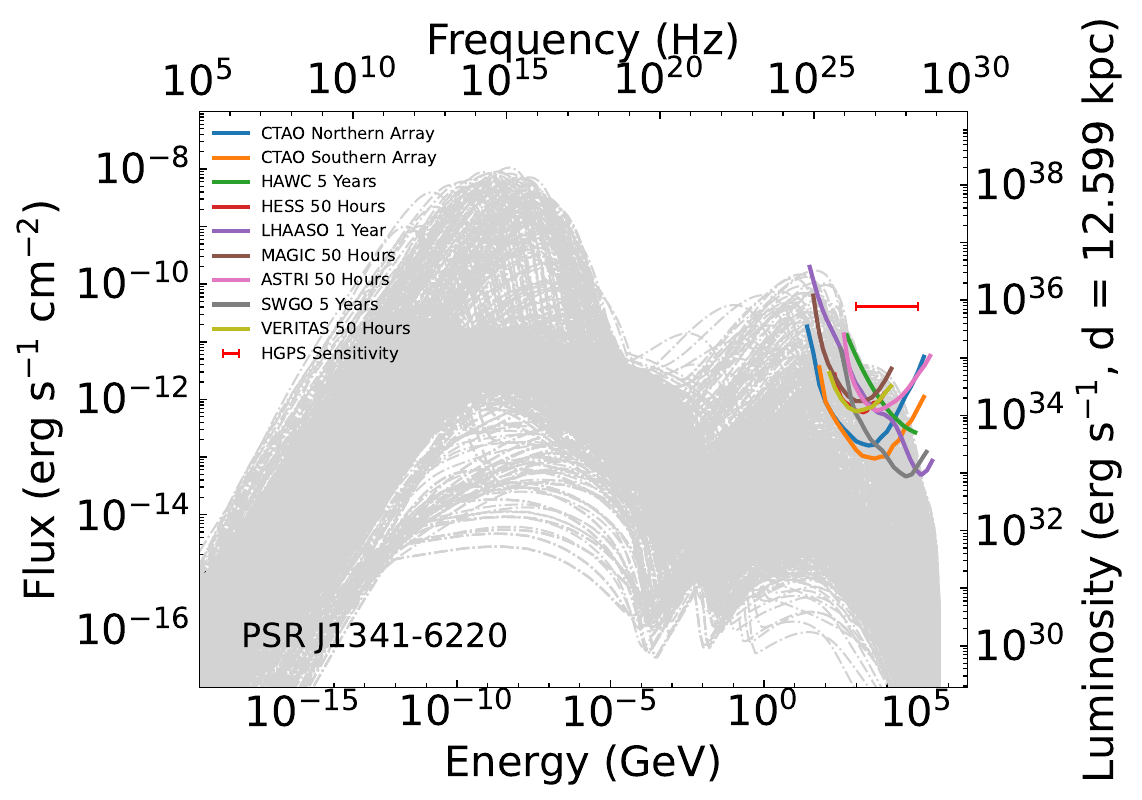}
   \caption{
Predicted set of SEDs for the PWNe of PSR J1341-6220. 
}
    \label{sed_J1341}
\end{figure}

The 1512 modeled SEDs for the putative PWN of PSR J1341-6220 are presented in Fig. \ref{sed_J1341}, and the relevant physical parameters involved are also shown in Table \ref{tab2}. This pulsar has the largest characteristic age among the four candidates, and from its SEDs, we observed that the PWN of PSR J1341-6220 might have already entered the reverberation phase in some model realizations. The TeV fluxes of the potential PWN in most models cannot be detected except by CTA. In addition, because of the location of this source, only S-CTA is suitable for a dedicated observation. We find that none of the 1512 model realizations would lead to a detectable TeV PWN in the HGPS, in agreement with the absence of detection. 

We classified all the models into three categories according to their visibility on different TeV telescopes:
    \begin{enumerate}
        \item {[$<$ S-CTA]}: Lower than the sensitivity of the CTA Southern Array. 
        \item {[$<$  H.E.S.S.]} \& [$>$ S-CTA]: Between H.E.S.S. and the CTA Southern Array sensitivities. 
        \item {[$>$ H.E.S.S.]}: Higher than the H.E.S.S. sensitivity. 
    \end{enumerate}
The number of models in different categories is listed in Table \ref{visibility}. One can see that in the majority of the models (close to 83\% of all those studied), the TeV emission from the PWN is undetectable. The TeV fluxes predicted for the remaining 17\% can be detected by S-CTA, and 6.55\% yield a TeV flux that could be discovered in a 50-hour dedicated observation by H.E.S.S.. 

\begin{table}
\setlength\tabcolsep{4pt}
        \centering
 \scriptsize
        \caption{Tera-electron volt visibility of the models for potential PWNe of selected pulsars that respect the X-ray upper limits when available. }
        \label{visibility}
 \begin{tabular}{llll} 
                \hline
        \hline
        PSR   & [$<$ S-CTA] & [$<$ H.E.S.S.] \& [$>$ S-CTA] & [$>$ H.E.S.S.]  \\
        \hline
        J1341-6220 (3 kpc)  &1132 (74.87\%)  &102 (6.75\%) &191 (12.63\%) \\ 
        J1341-6220 (6 kpc) &1206 (79.76\%)  &99 (6.55\%) &203 (13.43\%) \\
        J1341-6220 (12.6 kpc)  &1254 (82.94\%)  &159 (10.51\%) &99 (6.55\%) \\
        \hline
        J1838-0537 (2 kpc) &0   &16 (1.06\%) &73 (4.83\%) \\
        J1838-0537 (4 kpc) &0   &93 (6.15\%) &200 (13.23\%) \\ 
        J1838-0537 (6 kpc) &34 (2.25\%)  &229 (15.15\%) &202 (13.36\%) \\ 
        \hline
        J1844-0346 (2.4 kpc)  &0  &0 &0 \\ 
        J1844-0346 (4.3 kpc)  &0  &0 &0 \\ 
        J1844-0346 (10 kpc)  &8 (0.53\%)  &17 (1.12\%) &0 \\ 
\hline
        \end{tabular}
\end{table}

To understand which model parameters lead the separation in the different TeV visibility categories, we plotted the distributions of the nine parameters mentioned above and the radius $R_{pwn}$ and magnetic field $B_{pwn}$ for the 1000 random models under the detectable ([$>$ S-CTA]) and undetectable ([$<$ S-CTA]) categories. A Kolmogorov-Smirnov test (KS test) was also done to compare these two categories. Since the parameter values are randomly selected from their assumed ranges, their distributions are uniform. In contrast, a nonuniform distribution of a parameter in a particular category would show its influence over the PWNe TeV emission. 

\begin{figure}
        \includegraphics[width=\columnwidth]{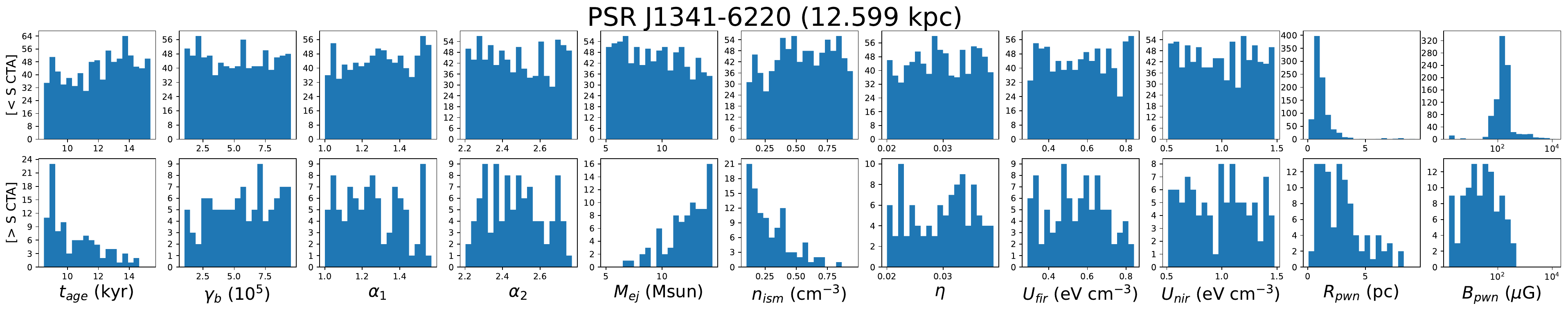}
   \caption{
Parameter distributions for the models analyzed for PSR J1341-6220 grouped into different categories. }
    \label{pars_distribution}
    \end{figure}

As shown in Fig. \ref{pars_distribution} and Table \ref{KStest}, for the 897 models with a relatively low TeV emission ([$<$ S-CTA]),  their $t_{age}$ and $n_{ism}$ are concentrated in the high end of the selected range, while $M_{ej}$ mostly appears in the lower end. The distributions of $R_{pwn}$ and $B_{pwn}$ are also different from models in the detectable category ([$>$ S-CTA]), with more models having smaller radii and bigger magnetic fields. The distributions of the other parameters appear to be uniformly distributed, implying that their values have a secondary effect on the TeV radiation of this potential PWN. Most of the values of $R_{pwn}$ and $B_{pwn}$ are less than 4 pc and between 40 and 300 $\mu$G, respectively. Models with a high TeV emission ([$>$ S-CTA] ) show the opposite of these distribution concentrations.

\begin{table}
    \setlength\tabcolsep{4pt}
        \centering
 \scriptsize
        \caption{Kolmogorov-Smirnov test for the distributions of 11 parameters in different categories. In the table, Cat. 0 relates to random models with a flux [$<$ S-CTA], and Cat. 1 relates to random models with a flux [$>$ S-CTA], and we refer to the p-value for a common parent population of the two quoted distributions. }
        \label{KStest}
        \scalebox{1.0}{
        \begin{tabular}{lccccccccccc} 
                \hline
        \hline
                    \textbf{Parameters} &$t_{age}$ &$\gamma_b$ &$\alpha_1$ &$\alpha_2$ &$M_{ej}$  &$n_{ism}$ &$\eta$ &$U_{fir}$ &$U_{nir}$ &$R_{pwn}$ &$B_{pwn}$\\
                \hline
  \multicolumn{3}{l}{\textbf{PSR J1341-6220}} & & & & & & & & &\\
            p-value Cat. 0 and Cat. 1: &$<10^{-10}$ &0.43 &0.06 &0.31 &$<10^{-10}$ &$<10^{-10}$ &0.54 &0.41 &0.90 &$<10^{-10}$ &$<10^{-10}$\\
                \hline
        \end{tabular}}
\end{table}

Table \ref{KStest} shows that the results of the KS test for the distributions of each parameter support that the distributions of $t_{age}$, $n_{ism}$, $M_{ej}$, $R_{pwn}$, and $B_{pwn}$ for the different TeV visibility categories are not consistent with them being born from the same parent population, ruling out the null hypothesis with a p-value of smaller than $10^{-10}$. The [$<$ S-CTA] category has older realizations than [$>$ S-CTA], and if only free expansion is considered, then the PWN should have a larger radius and a weaker magnetic field. Figure \ref{pars_distribution} shows that it is the other way around, which means that the PWN has indeed entered a compression process in the framework of these models. Moreover, $M_{ej}$ and $n_{ism}$ will affect the age of the PWN from the free expansion phase to the reverberation phase, and the magnetic field grows enough to burn off energetic electrons when the PWN is compressed, which in turn affects the IC radiation and reduces the TeV radiation. All of this promotes further theoretical studies with models capable of coping better with the reverberation processes (see \citet{bandiera2023reverberation}) as well as a dedicated observation using S-CTA, which will help further constrain our models and test the above conclusions. 

We also investigated scenarios with varying distance assumptions, similar to our approach for PSR J1208-6238. In cases of 3 and 6 kpc, nearly all models predict a PWN that remains undetectable in the HGPS. After excluding models with fluxes exceeding the HGPS sensitivity (only four models at 6 kpc and 87 models at 3 kpc), we obtained the number of models in the various categories specified in Table \ref{visibility}. Despite the larger flux with decreasing distance, the majority of the models indicate that this PWN remains undetectable by currently observing facilities ($>$74\%) even in the case that the uncertainty results in being in favor of increasing the flux on Earth. Thus, the level of distance uncertainty will have a minimal effect on the generic conclusions obtained.

\subsection{PSR J1838-0537}
\label{secJ1838}

PSR J1838-0537 is also a young and energetic radio-quiet $\gamma$-ray pulsar with a characteristic age of 4.89 kyr and a spin-down luminosity of $6.02 \times 10^{36}$ erg/s \citep{manchester2005australia}. The distance to this pulsar is also uncertain, and as in \cite{pletsch2012psr, albert21}, we chose 2.0 kpc. This value is based on the observed correlation between the $\gamma$-ray luminosity and $\dot{E}$ of a pulsar, and its uncertainty brings an obvious caveat, too. 

In the region where this pulsar is located, there is a TeV source, HESS J1841-055, that was also observed by HAWC as eHWC J1839-057 \citep{Abeysekara2020}. The potential PWNe of PSR J1838-0537 and PSR J1841-0456, as well as the supernova remnant (SNR) Kes 73, are suspected to be contributors to this VHE source \citep{gomez2020predicting}. The \cite{magic2020studying} tentatively investigated the physical nature and origin of the $\gamma$-ray emission from HESS J1841-055 and proposed leptonic and hadronic multi-source models. In their leptonic scenario, rather than IC radiation, bremsstrahlung is dominant, which is inconsistent with the usual results of PWN models. In their model, the TeV source is powered by one or several PWNe relics, so the IC emission efficiency is considered to be significantly higher than synchrotron emission, which can explain why there is no bright synchrotron nebula. 

\begin{figure}
\centering
    \includegraphics[width=.99\columnwidth]{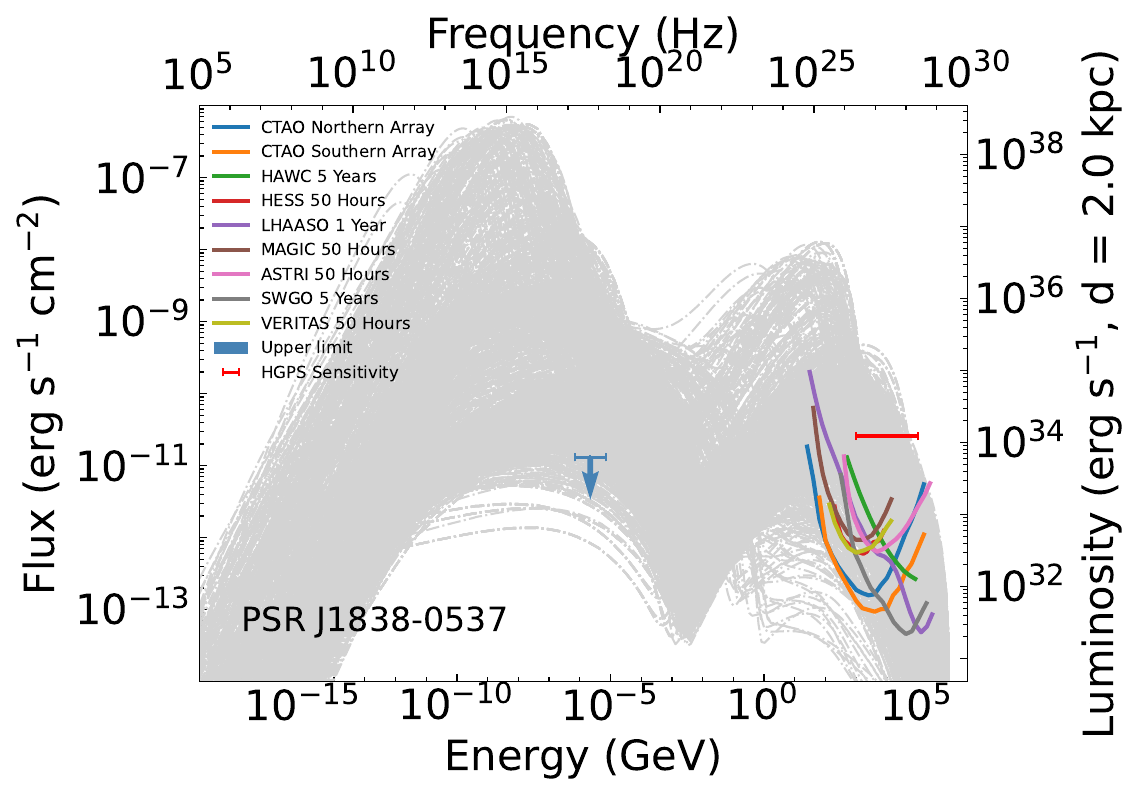}
   \caption{
Predicted set of SEDs for the PWN of PSR J1838-0537.
}
    \label{sed_J1838}
\end{figure}

However, PSR J1838-0537 is unlikely to leave such a PWN relic at such a young characteristic age. In addition, \cite{magic2020studying} considered the X-ray emission of the possible contributors (including nonpulsation X-ray flux from PSR J1838-0537) to HESS J1841-055 and used their total X-ray flux as the X-ray upper limit of this VHE extended source. We adopted this X-ray upper limit for the potential PWN of PSR J1838-0537 in this work as well. Possible SEDs and related physical parameters are shown in Table \ref{tab2} and Fig. \ref{sed_J1838}. 

The TeV radiation in the majority of the 1512 different models can be detected by H.E.S.S. (it might even have been detected already in HGPS as HESS J1841-055), and almost all of them are detectable with CTA. However, only in a small percentage of models does the X-ray emission respect the corresponding upper limit. In addition, in a small portion of models, this PWN has also entered the reverberation phase according to their SEDs. 

After excluding all the models that exceeded the X-ray upper limit, we divided the remaining 89 of them into the same three categories according to their detectability by different facilities, as described above. We found that most of them belong to [$>$ H.E.S.S.], as displayed in Table \ref{visibility}. Only 17 of these 89 models respecting the X-ray upper limit come from the 1000 random model realizations, and all of them belong to the very luminous class [$>$ H.E.S.S.], although none exceed the HGPS sensitivity. A dedicated deep observation of this source using CTA will be necessary in order to try to disentangle the different contributors in the region. 

Similar to our previous analyses, we also examined the impact of the distance uncertainty. For distances of 4 kpc and 6 kpc, many of the 1512 models studied (293 models for 4 kpc and 431 models for 6 kpc) comply with the X-ray upper limit. Table \ref{visibility} refers to their distributions in the various categories considered according to their detectability in the different observing facilities. Out of the 293 (431) models, none (only a small number) resulted in an undetectable PWN for S-CTA. However, there are still no models that can predict a detectable PWN in the HGPS as the assumed distance increases. In conclusion, the putative PWN powered by PSR J1838-0537 can likely be identified by S-CTA (and even possibly by H.E.S.S.) despite the considerable uncertainty regarding its distance. If the real distance exceeds 2 kpc, it is highly unlikely that HESS J1841-055 is the TeV counterpart of the putative PWN; instead, this PWN may at most be one of several contributors to this VHE source.

\subsection{PSR J1844-0346}
\label{secJ1844}

PSR J1844-0346 is also a $\gamma$-ray pulsar. It was found in the {\it Fermi}-LAT blind search survey and has a characteristic age of 11.6 kyr and a spin-down luminosity of 4.25$\times10^{36}$ erg s$^{-1}$ \citep{clark2017einstein}. Its distance is also unknown. 

HESS J1843-033, eHWC J1842-035, LHAASO J1843-0388, and TASG J1844-038 are VHE sources detected in this region. \cite{amenomori2022measurement} found that the energy spectrum of these VHE sources can be well fitted with a power-law function with an exponential cutoff. In the HGPS, \cite{Abdalla2018} found that HESS J1843-033 consists of two merged offset components (HGPSC 83 and HGPSC 84), which seems to imply that this TeV source may have multiple origins. Because of their spatial proximity, PSR J1844-0346 and the radio SNR G28.6-0.1 were suspected of being the origin of the VHE emission. However, the association between PSR J1844-0346 and SNR G28.6-0.1 is unlikely. The estimated distance of SNR G28.6-0.1 is 6 – 8 kpc \citep{Devin2021}, and for it, the resulting transverse velocity of the pulsar needed to reach from the center of the SNR to its current position would be over 1400 km/s, which is much larger than the typical value of a few hundred kilometers per second (e.g., \citet{Devin2021}). The same transverse velocity argument was made for a more likely association of PSR J1844-0346 with the star-forming region N49 at 5.1 kpc. Alternative empirical estimates of the distance of PSR J1844-0346 are 2.4 kpc \citep{wu2018einstein}, obtained by assuming that the $\gamma$-ray luminosity scales as $\sqrt{\dot{E}}$, and 
4.3 kpc, based on the empirical relation obtained for $\gamma$-ray pulsars \citep{parkinson2010eight}. None of these are certain, however, and they are herein considered to span the possible results.

\cite{zyuzin2018x} used nearly 100 ks exposure data from the Swift X-Ray Telescope (XRT) to discover an X-ray counterpart candidate of this $\gamma$-ray pulsar. The candidate has an estimated unabsorbed flux of $2.2^{+1.3}_{-0.4}\times10^{-13}$ erg cm$^{-2}$ s$^{-1}$ (0.3 - 10 keV). \citet{Devin2021} searched for radio or X-ray counterparts around PSR J1844-0346 that could indicate a possible PWN but did not find any, and they do not provide any diffuse upper limit to compare our results with. We took the X-ray flux of \cite{zyuzin2018x} as an X-ray upper limit for this pulsar's potential PWN in case the latter can be regarded as point-like. Although if the latter is more extended and diluted, which is not expected at this age, the flux can be larger.

\begin{figure}
\centering
    \includegraphics[width=0.32\columnwidth]{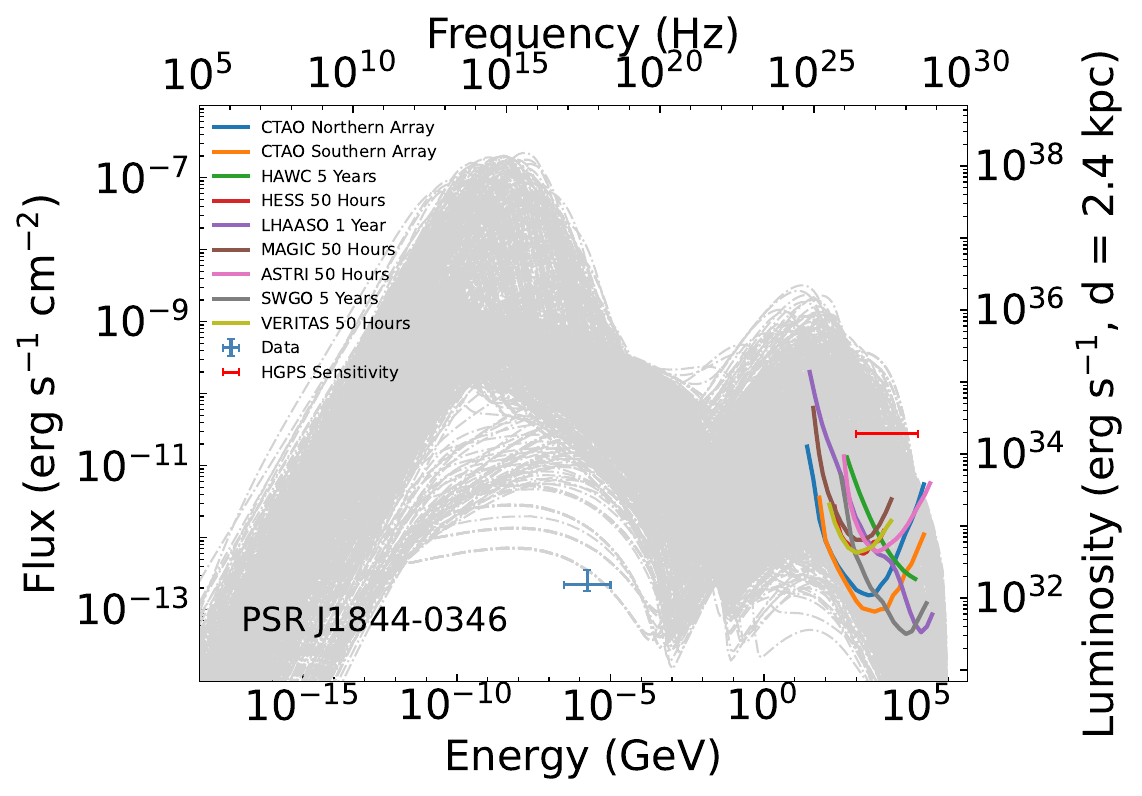}
    \includegraphics[width=0.32\columnwidth]{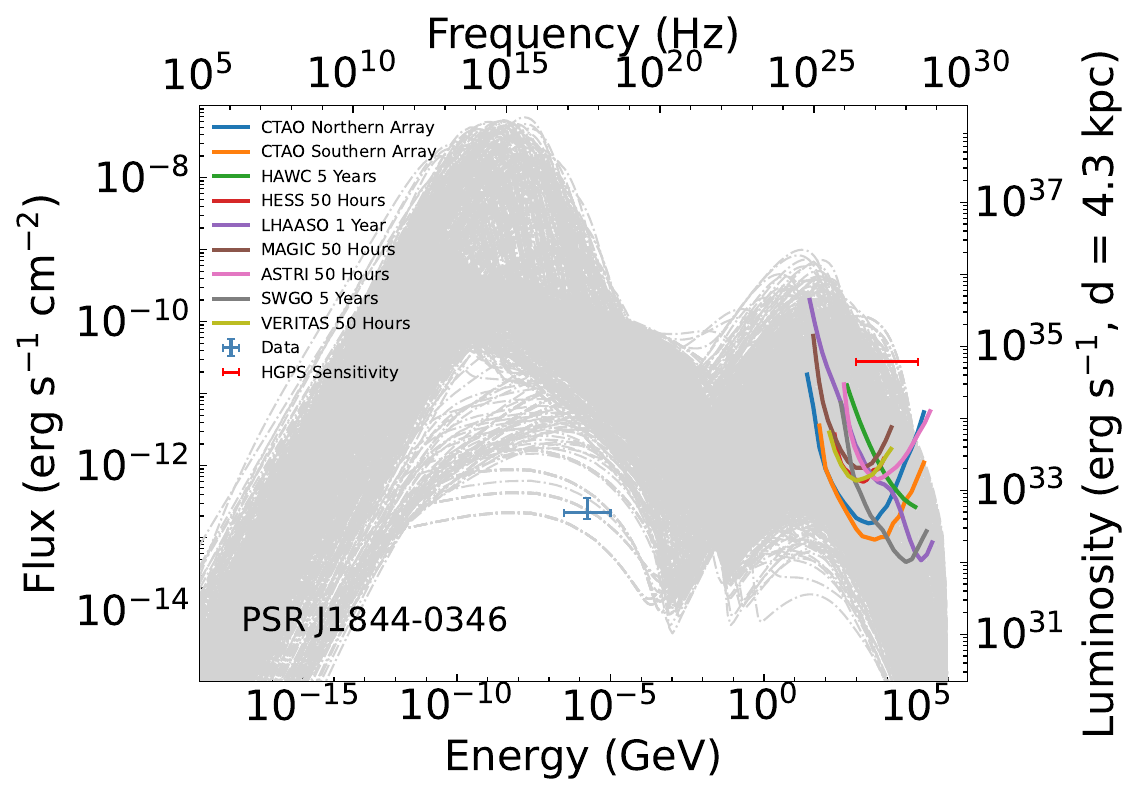}
    \includegraphics[width=0.32\columnwidth]{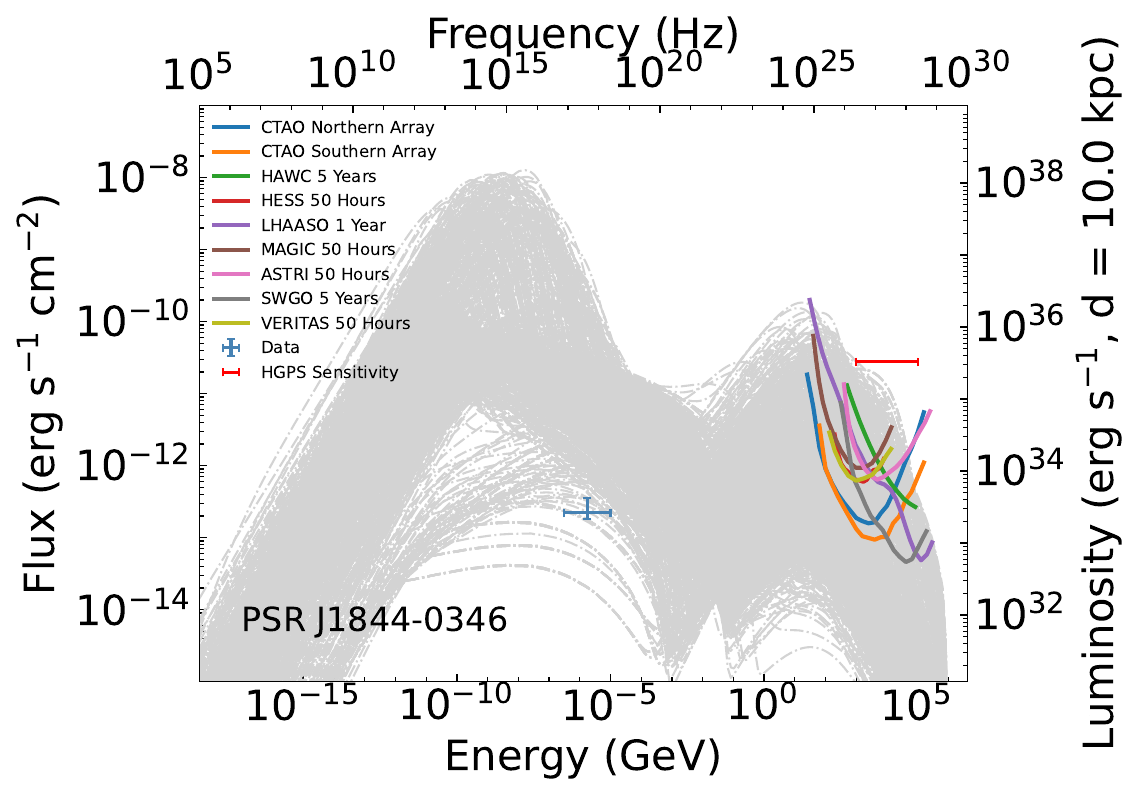}
   \caption{
Predicted set of SEDs for the PWN of PSR J1844-0346 located at 2.4, 4.3, and 10 kpc, respectively. 
}
    \label{sed_J1844}
\end{figure}

Similar to the other three candidates, the 1512 different model realizations with distances assumed to be 2.4 kpc can also be seen in Table \ref{tab2} and in the top panel of Fig. \ref{sed_J1844}. The characteristic age of 11.6 kyr allowed us to see that many of the possible PWNe have entered the reverberation phase. Due to the closer distance and the more energetic $\dot{E}$, the overall energy flux of PSR J1844-0346's PWN is higher compared to PSR J1341-6220's PWN, and this leads to the TeV emission being detectable by CTA in the most ($>$ 85\%) models and by H.E.S.S. in close to 30\%. However, if we are to assume the pulsar's X-ray flux as an upper limit for the PWN (which would only make sense for small PWN), none of the SEDs predicted by our models lead to detectable TeV emission except with an assumed distance of $\gtrsim 10$ kpc (see Fig. \ref{sed_J1844} and Table \ref{visibility}).

Ultimately, this PWN would not be an obvious candidate for S-CTA or H.E.S.S.. However, this conclusion is weakened when the true distance of this PWN is much larger than 2.4 kpc or its X-ray diffuse emission is much larger than the X-ray flux of the pulsar that we adopt here.

\section{Conclusions}
\label{conclusion}

In this work, we aimed to assess the detectability of PWNe when interested candidates are identified in the PWNe. With this in mind, we selected four pulsars and studied them as candidates to produce a detectable TeV PWN based on their locations on the pulsar tree, the MST of the pulsar population, as we have noted that the majority of the known TeV PWNe are closely grouped together.

Using the observational results of the four candidates and assuming within reasonable ranges some model parameters when unknown, we used our leptonic model to predict a large set of SEDs for the putative PWNe of each of these four pulsars. We then proceeded to analyze the set, and by comparing the TeV fluxes predicted by the models that respect the X-ray observations when existing with the sensitivities of current and future instruments, we concluded how likely it is for each of the PWNe to be detectable. 

We find that the TeV emission of the potential PWNe of PSR J1208-6238 and PSR J1341-6220 is expected to be relatively low, which is consistent with the fact that we have not detected their TeV counterparts yet. This is also the case if the pulsars are located at larger distances than assumed, which can happen within the associated uncertainty of this parameter. The TeV emission predicted by the models for PSR J1208-6238 at different distances is even lower than the sensitivity of S-CTA. The other two pulsars, PSR J1838-0537 and PSR J1844-0346, have a relatively higher TeV emission from their putative PWNe, and our results tend to suggest that they could be detectable by S-CTA or even by H.E.S.S. in sufficiently long dedicated observations. However, only a reduced number of model instances could lead to detectable TeV emission without violating X-ray constraints if they are used as PWN upper limits.

The above results also allowed us to understand the convenience and caveats of the pulsar tree in pulsar population analysis and pinpointing detectable PWNe. The pulsar tree that we used to select candidates only contains intrinsic information about the pulsars themselves, with no reference to the environment or the distance to them. Even with just this kind of information, the grouping of all PWNe is impressive (see Fig. \ref{tev.fig1}). However, we find that the location of a yet nondetected (and in principle likely undetectable) PWN, among others that are observed, does not necessarily imply a different behavior for the system. Finally, we note that most of the nondetected PWNe pertain to the same branch within the region of the pulsar tree where most of the PWNe are located, which is formed by the less energetic pulsars of the set. 

Given that we had a thousand models for each PWN, we could also test the general correlations without imposing any constraints. We used Pearson’s and Spearman’s tests to analyze whether $L_{TeV}$ is correlated with any of the model parameters. The largest correlation coefficients were obtained in the case of J1208-6238 for the pairs ($\alpha_2$, $L_{TeV}$), with $r=-0.61$ (Pearson) and $r=-0.69$ (Spearman). This represents a moderate to strong correlation. All other model parameters present milder correlation coefficients (typically well below $0.3$). We encountered a number of factors when searching for larger correlation coefficients. On the one hand, the cross-influence of all parameters is a well-known effect. For instance, if we keep the electron spectrum fixed, the higher the energy density is, the higher the TeV luminosity will be. However, if at the same time the magnetization is increased (so that the synchrotron losses are higher, affecting the electron population) or the age is increased, the electrons are more or less cooled as a result, and even for a higher photon density, there could be a lower TeV luminosity. As we have several parameters and all vary at once, correlations are correspondingly less clear. This is a widely known effect. On the other hand, the appearance of models entering into reverberation also affects correlations, as this phase gives rise to new phenomenology when the medium compresses the PWN shell. Our models have also been tested by comparing the parameters of those models that generate a large $L_{TeV}$ (in regard to the observational sensitivities) with those models that do not. We did this by testing via a KS test the null hypothesis that states that the distribution of parameters of both TeV luminous and dim PWNe are consistent with having the same parent population. The case of PSR J1341-6220 can be considered an example of this investigation. As for the distribution of $M_{ej}$, most models predicting a high TeV emission have a relatively larger $M_{ej}$ (and a smaller $n_{ism}$), which means a larger Sedov time ($t_{Sed} \propto M_{ej}^{5/6}\cdot n_{ism}^{-1/3}$). This is not obvious in the Pearson's and Spearman's tests (which in this case have a Pearson coefficient of 0.24 and a Spearman coefficient of 0.11), but it is captured by the KS test (see, e.g., Fig.~\ref{pars_distribution} and Table~\ref{KStest}).

In conclusion, this study provides a case study of pulsar population analysis using the pulsar tree, as well as two promising PWNe that could be detected with CTA and even H.E.S.S.. Therefore, they are also worthy of further observation in the TeV and other energy bands. Furthermore, we examined the impact of varying intrinsic pulsar parameters on the TeV radiation of their young PWNe. This work underscores the potential for advancing our understanding of pulsar characteristics and PWN evolution through high-energy observations. 
\chapter{Caveats of considering normal PWNe as Potential PeVatrons}
\label{PeVatron}
\textbf{\color{SectionBlue}\normalfont\Large\bfseries Contents of This Chapter\\\\}

In this chapter, we explore the possibility that PWNe are the origin of several ultra-high energy (UHE) sources and their potential as PeVatrons. 
Three representative UHE sources - LHAASO J2226+6057, eHWC J2019+368, and HESS J1427-608 - are systematically analyzed using our time-dependent leptonic model to interpret their broadband spectra and constrain their physical parameters. 
Key results include magnetic field estimates, PWN sizes, and implications for their evolutionary stages and gamma-ray production mechanisms. 

The first section presents a detailed modeling of LHAASO J2226+6057. 
This work has been published in \citet{Sarkar2022}, where I am listed as the second author. 
My main contributions include participating in the multi-wavelength spectral modeling, the scientific discussion, and 
assisting with manuscript review and revision. 

Subsequent sections focus on two other UHE sources - eHWC J2019+368 and HESS J1427-608 - applying similar modeling frameworks, which we decided to left unpublished since conclusionrs are similar as in the former case, in general terms. The script described in Appendix~\ref{script2} is employed in these analyses. 
We evaluate the plausibility of a PWN origin, assess their ability to accelerate particles to PeV energies, and highlight the challenges arising from our resulting models. 
%

\newpage

\section{LHAASO J2226+6057}
\label{J2226}

\subsection{Introduction}
\label{J2226.intro}

Recent advances in ultra-high energy (UHE, $E_\gamma \ge 100$ TeV) gamma-ray astronomy have revealed numerous Galactic sources, driven by observatories like LHAASO, Tibet AS$\gamma$, and HAWC \citep{Abeysekara2020, amenomori19, amenomori21, Cao2021}. 
Future facilities like the Cherenkov Telescope Array (CTA) and the Southern Wide-field Gamma-ray Observatory (SWGO) will further enhance our understanding of these extreme accelerators, known as PeVatrons, capable of reaching particle energies up to PeV scales. 

LHAASO, a cutting-edge facility in China at 4410 m altitude, has recently reported 12 sources with significant ($>7\sigma$) UHE gamma-ray emission, extending up to $\sim$1$^\circ$ \citep{Cao2021}. 
Some very high energy ((VHE, $100 \mathrm{GeV} \le E_\gamma \le 100$ TeV)) counterparts of these UHE sources have been linked to pulsar wind nebulae (PWNe) based on spatial alignment with powerful pulsars and their extended morphologies \citep{Abdalla2018}. 
For instance, the Crab nebula, powered by PSR B0531+21, is a confirmed PeVatron \citep{Cao2021}, supporting the notion that PWNe associated with high spin-down luminosity pulsars ($\Dot{E} > 10^{36}$ erg s$^{-1}$) may be a common source of UHE gamma rays \citep{albert21}.

PWNe are among the most efficient lepton accelerators in the Galaxy, powered by the rapid rotational energy loss of young, highly energetic pulsars. 
These systems release ultra-relativistic electron-positron pairs, forming a wind that generates a termination shock where particles can be further accelerated. 
The resulting high-energy leptons then produce broadband emission via synchrotron radiation, IC scattering, and Bremsstrahlung, creating spectra that span from radio to gamma-ray energies.

LHAASO J2226+6057, detected at RA = 336.75$^\circ$ and Decl. = 60.95$^\circ$ with a significance of 13.6$\sigma$ above 100 TeV and a spectrum with maximum energy of around 0.57 $\pm$ 0.19 PeV \citep{Cao2021}. 
This area is quite crowded and complex. 
The head region, containing PSR J2229+6114, and its wind nebula, the Boomerang PWN, likely emits via leptonic processes, while the tail region, potentially powered by SNR G106.3+2.7 and molecular clouds (MCs), may have a significant hadronic component.

PSR J2229+6114 is a bright gamma-ray pulsar with a spin period of 51.6 ms, characteristic age of 10.5 kyr, and a high spin-down luminosity of $2.2 \times 10^{37}$ erg s$^{-1}$ \citep{halpern01}. 
The region has also been detected in GeV gamma rays \citep{abdo09, xin19}, diffuse non-thermal X-rays \citep{fujita21}, and radio emission \citep{pineault2000}. 
However, its distance remains uncertain, with estimates ranging from 0.8 kpc (based on H\Romannum{1} and molecular line measurements; \citet{kothes01}) to 3 kpc (based on X-ray absorption; \citet{halpern01}) and 7.5 kpc (from pulsar dispersion measure; \citet{abdo09}). 
In this work, we adopt a distance of 3 kpc, consistent with recent studies \citep{joshi22, yu22}.

The physical connection between LHAASO J2226+6057 and PSR J2229+6114 has been explored in several works. 
\citet{breuhaus2022} considered a steady-state, one-zone leptonic model but neglected both the multi-wavelength (MWL) data and the evolutionary history of the PWN, focusing only on explaining the highest energy gamma-ray observations. 
\cite{joshi22} adopted a time-dependent, one-zone scenario, but did not account for the effects of particle escape, the evolution of the injected leptonic population, or the potential influence of the SNR reverse shock on the PWN's expansion, which may be important for a PWN with a age of older than 10 kyr\citep[see e.g.,][]{Martin2016}. 
\cite{yu22} used a similar approach, suggesting that a distorted PWN created by the reverse shock could explain the observed GeV gamma-ray emission \citep{xin19}.

A possible hadronic scenario has also been proposed by \cite{tibet21}, linking the observed UHE emission to interactions between the SNR and nearby MCs. 
However, this model faces challenges, as the associated SNR appears too old to produce the hard gamma-ray spectrum seen at PeV energies, potentially requiring a more complex or hybrid model to fully capture the observed spectrum \citep{breuhaus2022, desarkar22}.

In this work, we build on these previous studies, incorporating a more comprehensive treatment of PWN evolution and particle escape, and examining the potential impacts of SNR reverse shocks. 
We aim to reassess the nature of LHAASO J2226+6057 by relaxing some of the simplifying assumptions made in earlier models, providing a more complete picture of this intriguing UHE source.

\subsection{Results}
\label{J2226.results}

\subsubsection{Braking Index and True Age Estimation}
\label{j2226.sub1}

\begin{figure}[htp]
\centering
\includegraphics[width=.32\textwidth]{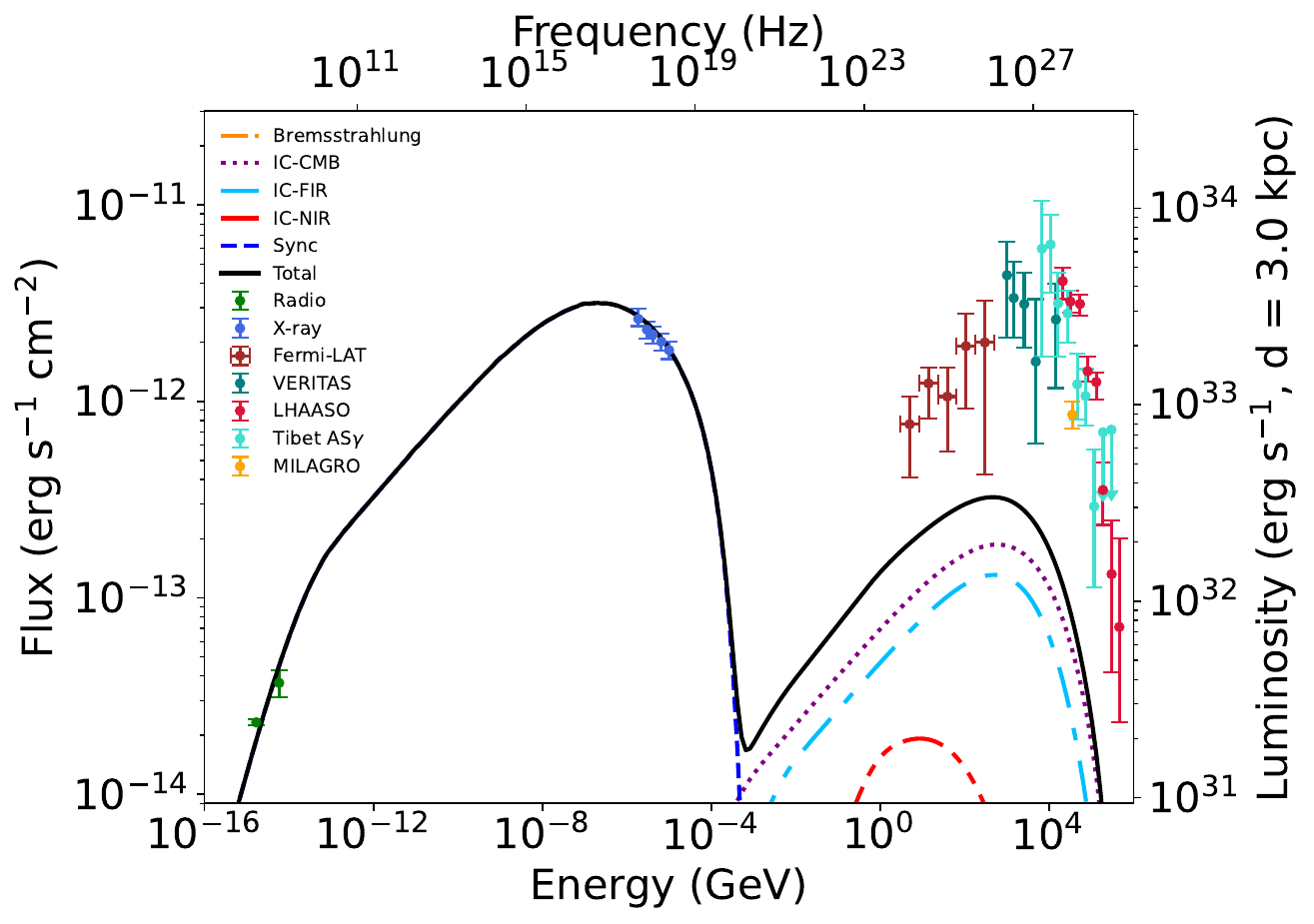}
\includegraphics[width=.32\textwidth]{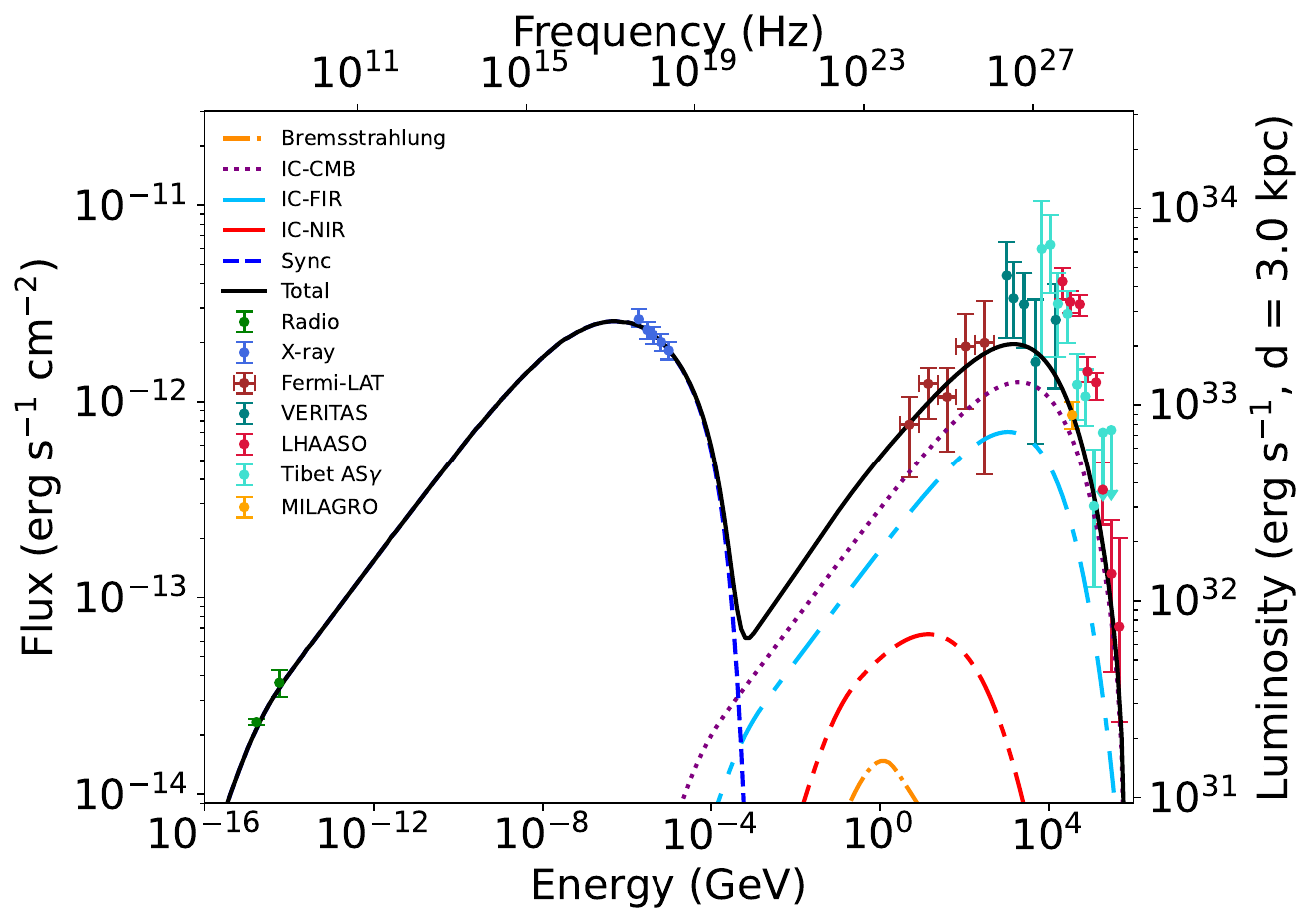}
\includegraphics[width=.32\textwidth]{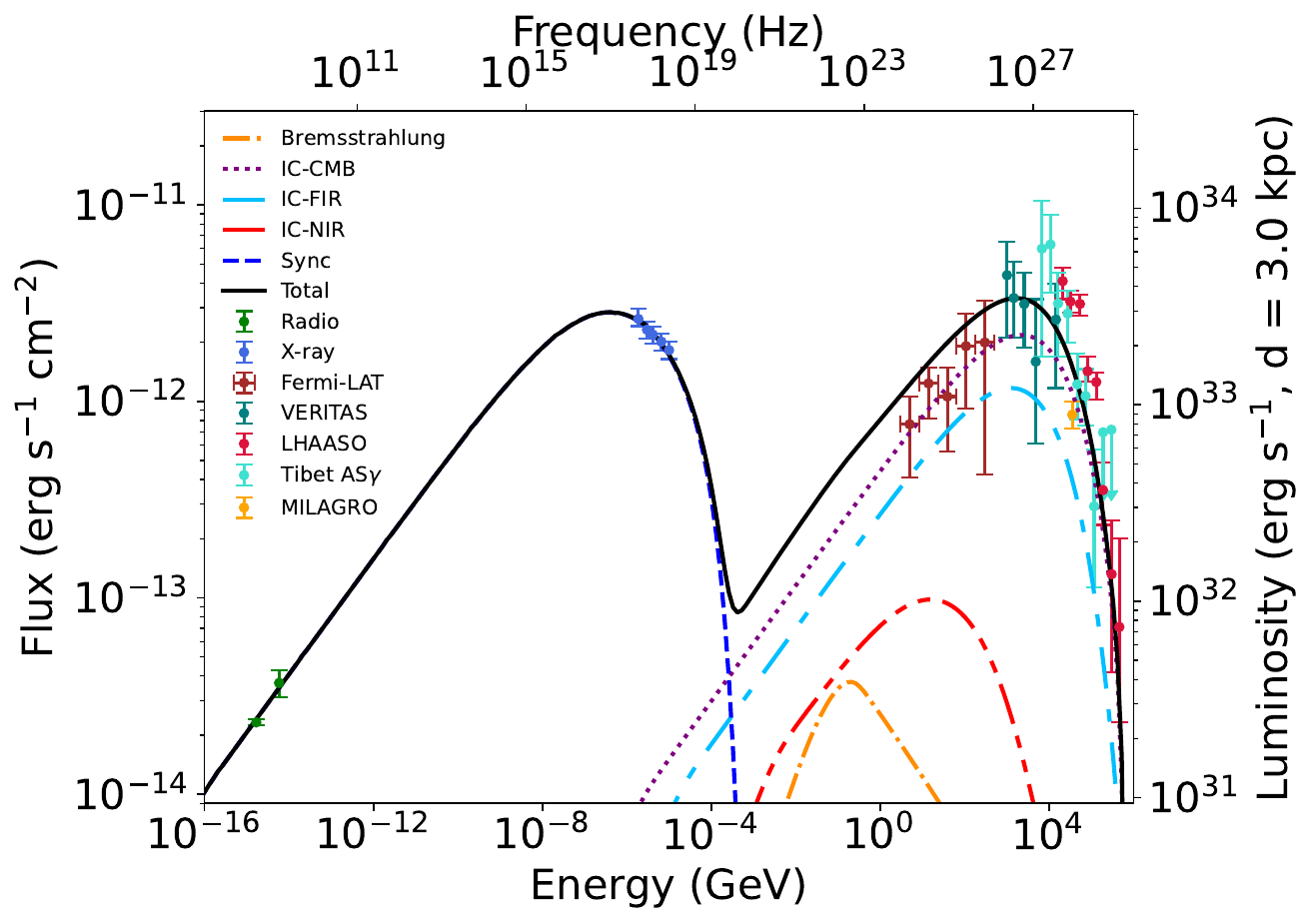}
\includegraphics[width=.32\textwidth]{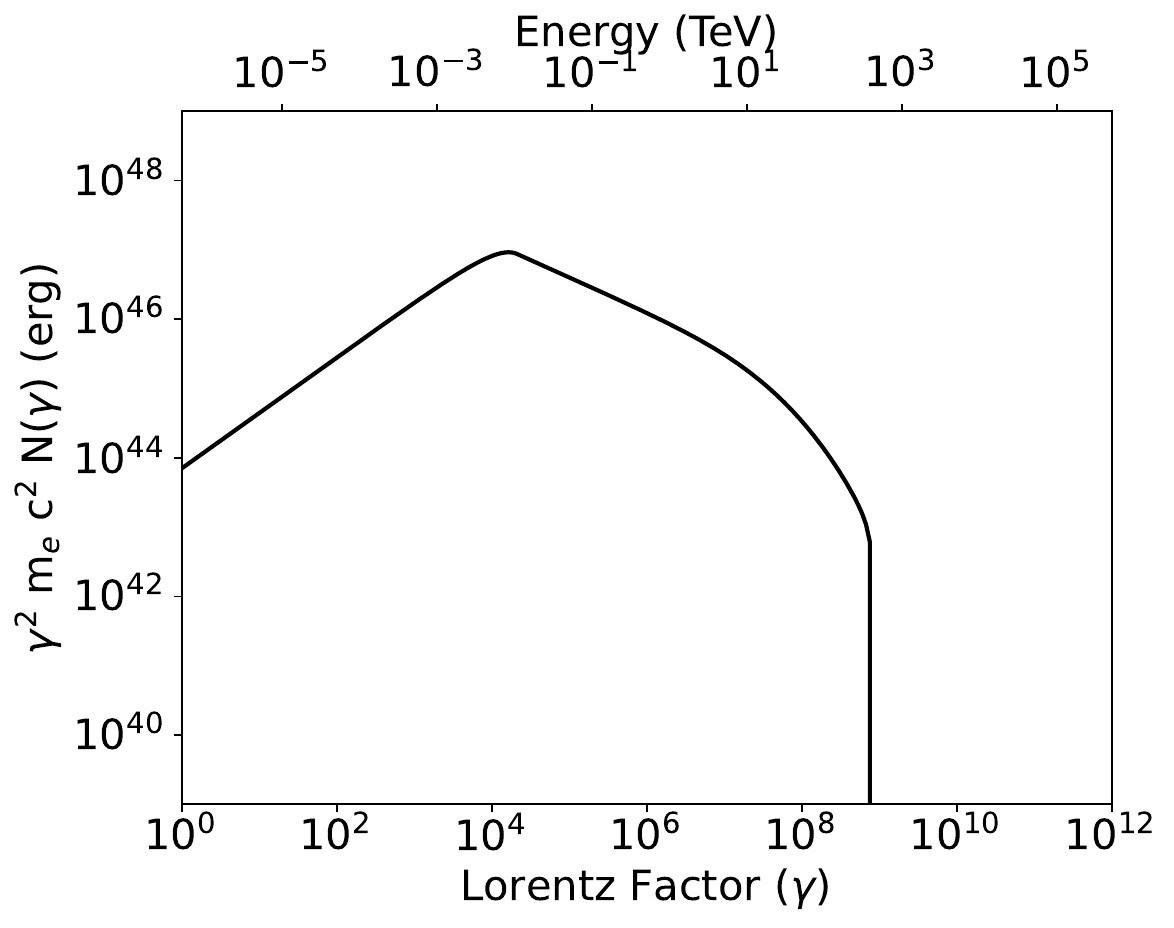}
\includegraphics[width=.32\textwidth]{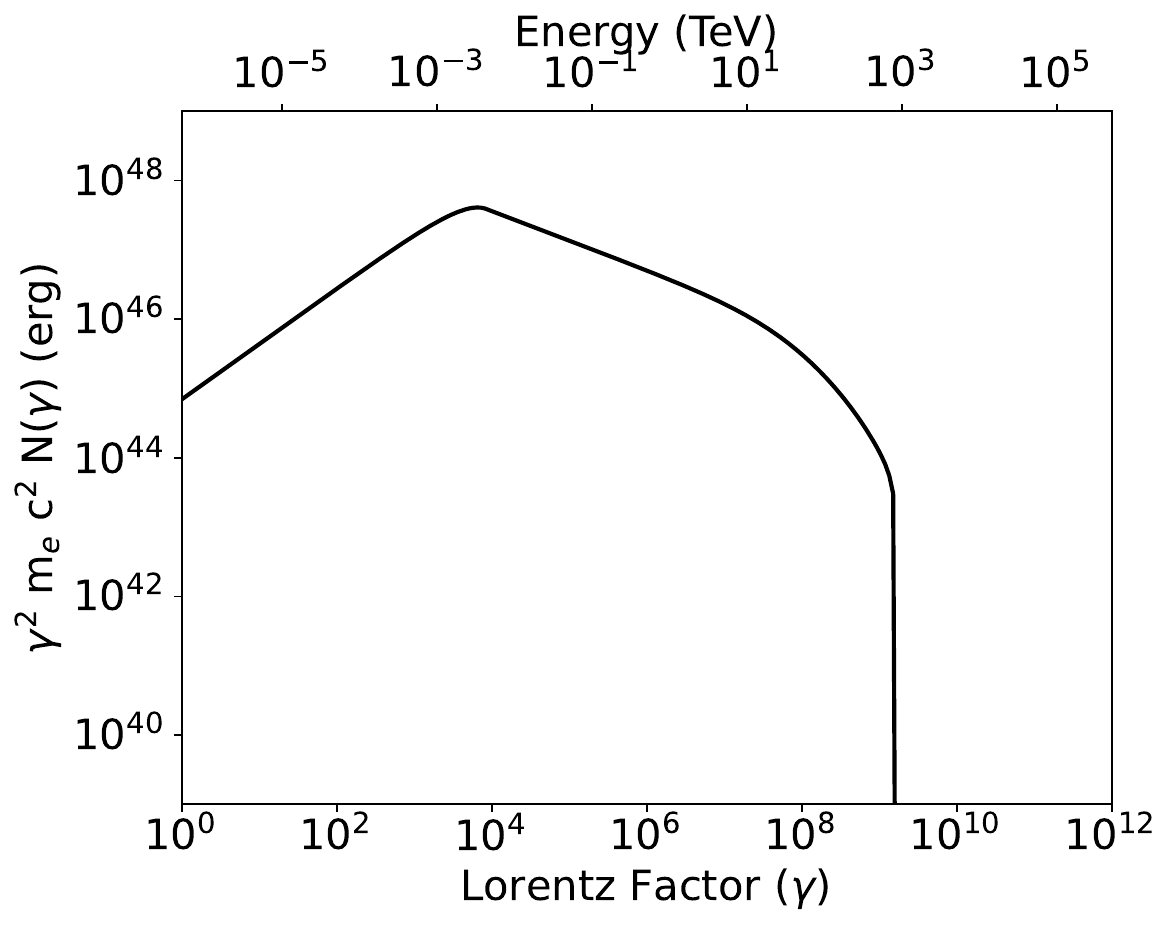}
\includegraphics[width=.32\textwidth]{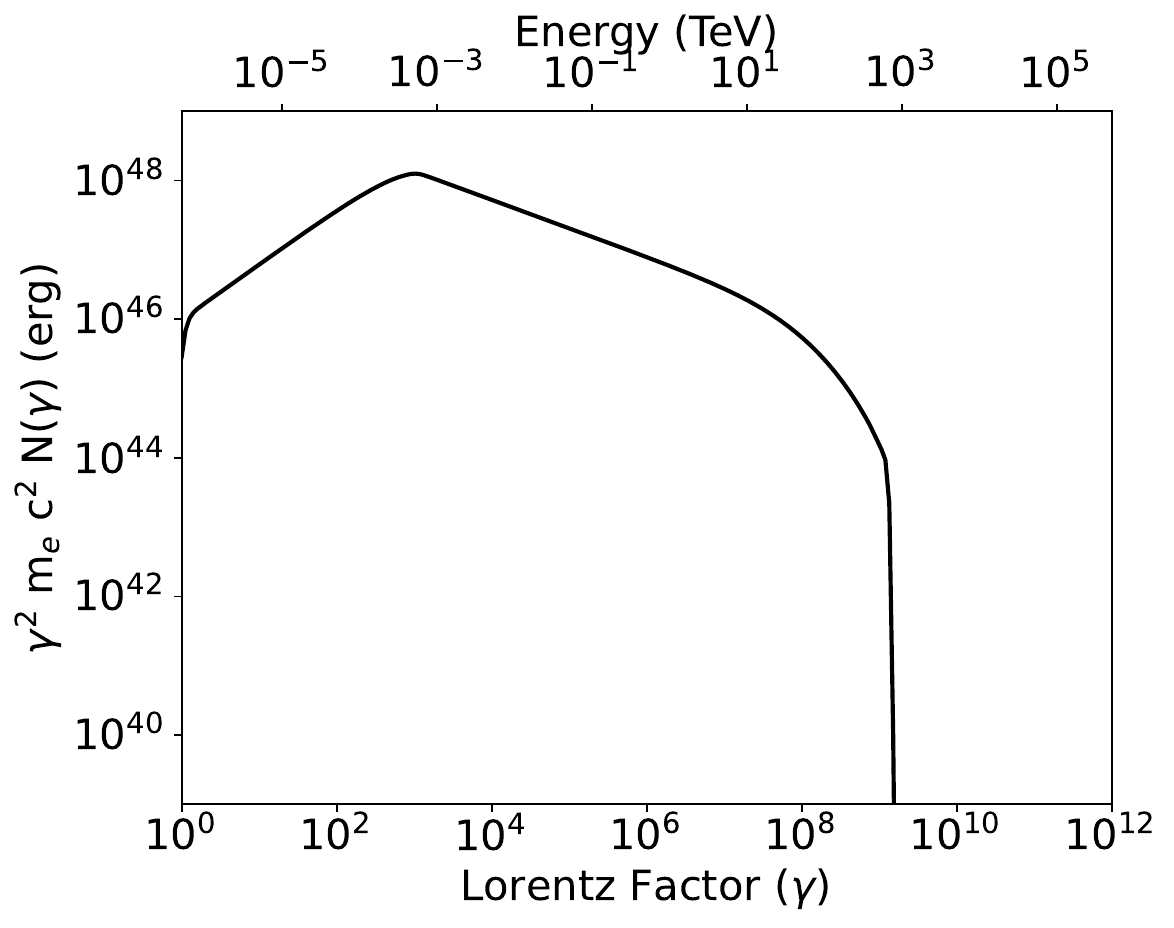}
\includegraphics[width=.32\textwidth]{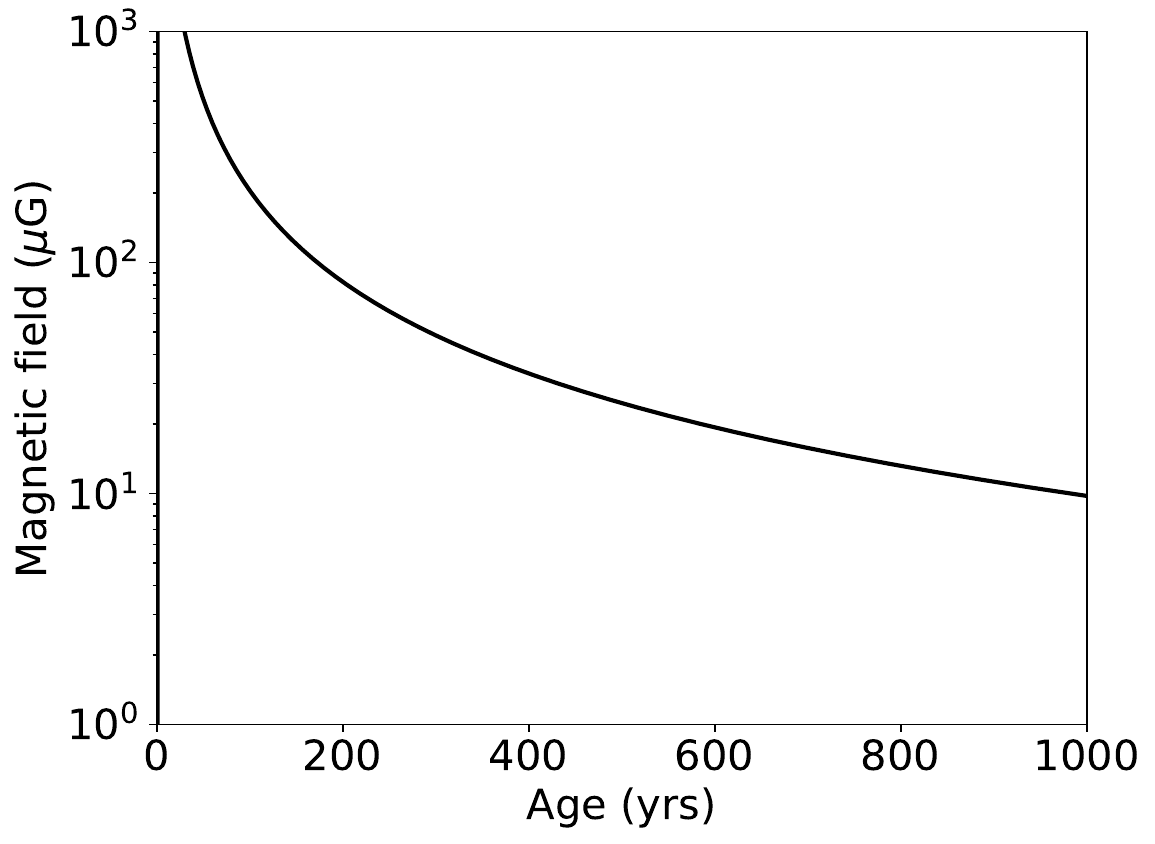}
\includegraphics[width=.32\textwidth]{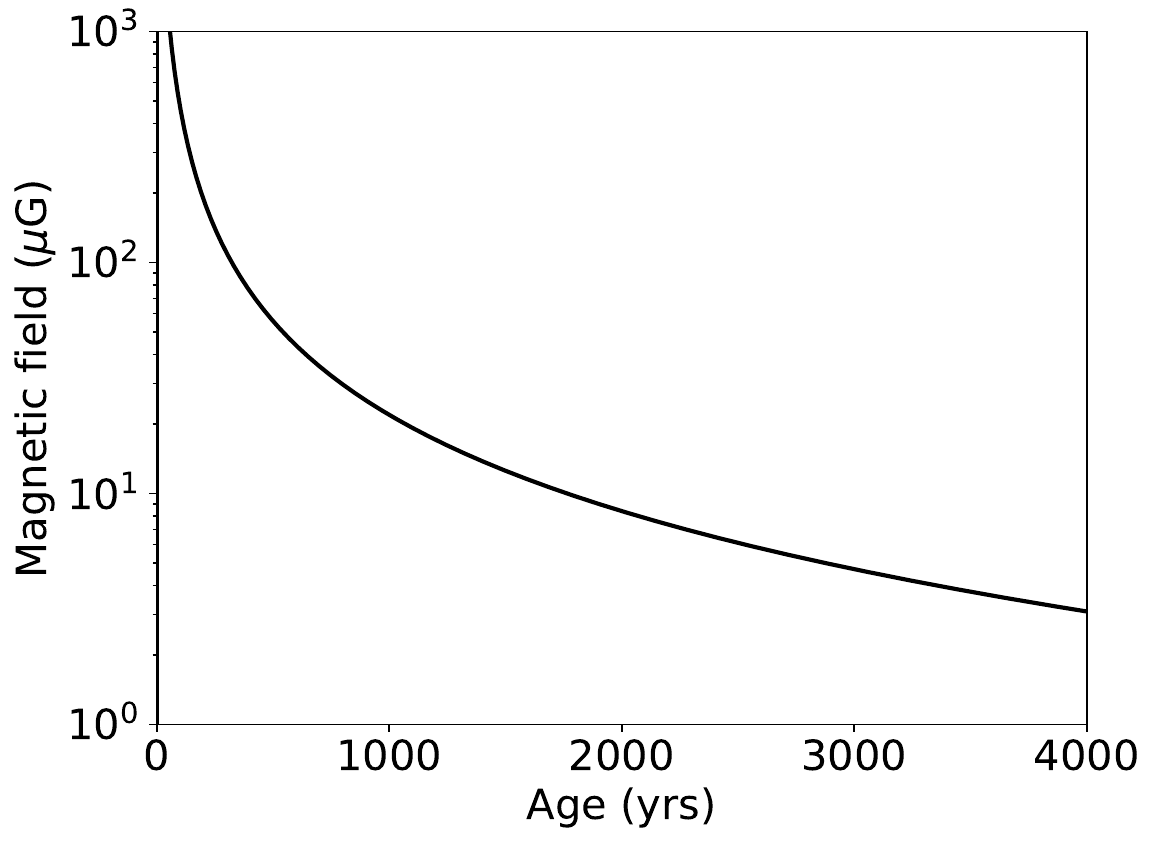}
\includegraphics[width=.32\textwidth]{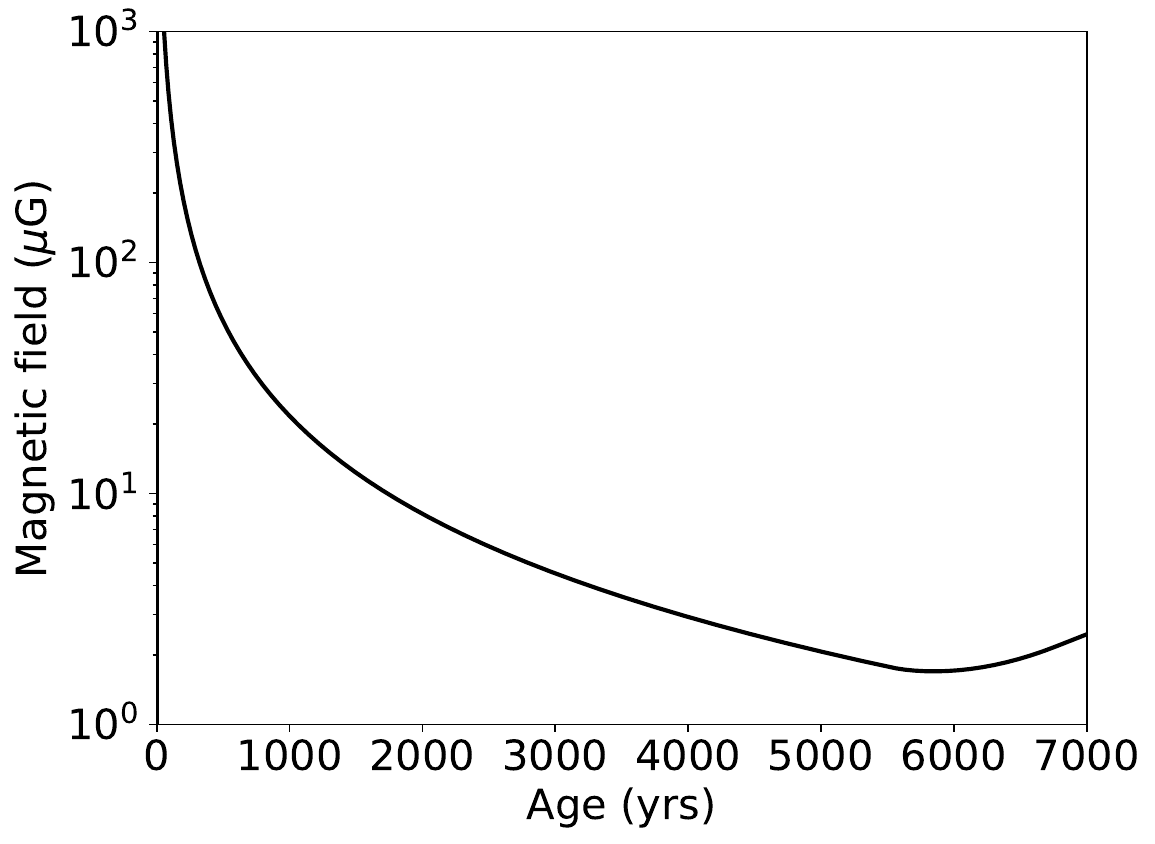}
\caption{Modeled MWL SEDs (top row), present-day lepton spectra (middle row), and magnetic field evolution (bottom row) for $t_{age}$ = 1000, 4000, and 7000 years, each with $n = 2.5$. 
The data sources are: radio (green, \citet{pineault2000}), X-ray (royalblue, \citet{fujita21}), Fermi-LAT (brown, \citet{xin19}), VERITAS (teal, \citet{acciari09}), Tibet AS$\gamma$ (turquoise, \citet{tibet21}), MILAGRO (orange, \citet{milagro09}), and LHAASO (crimson, \citet{Cao2021}). Copy from Figure 1 in \citet{Sarkar2022}. }
\label{fig1}
\end{figure}

\begin{figure}[htp]
\centering
\includegraphics[width=0.75\textwidth]{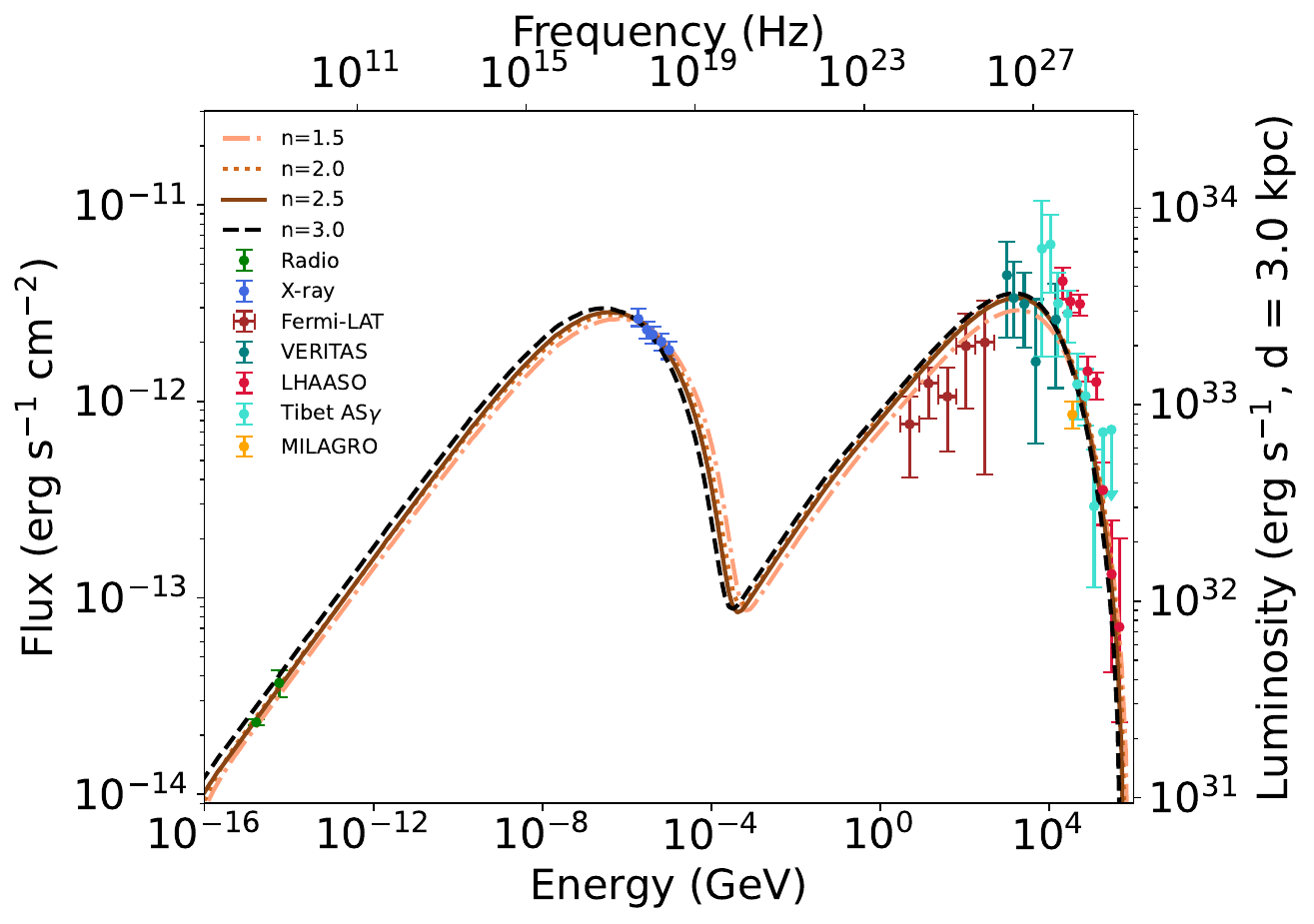}
\caption{Modeled MWL SEDs for different braking indices ($n = 1.5, 2.0, 2.5, 3.0$) at fixed $t_{age} = 7000$ years. The color scheme follows Figure \ref{fig1}. Copy from Figure 2 in \citet{Sarkar2022}. }
\label{fig2}
\end{figure}

The accurate determination of the true age ($t_{age}$) and braking index ($n$) of a pulsar is critical for modeling the long-term evolution of PWNe. 
In previous studies, $t_{age}$ and $n$ were subjectively assumed to be 7 kyr and 3, respectively \citep{joshi22, yu22}. 
However, such choices can significantly influence the derived physical properties of the PWN, and thus, a more comprehensive exploration of these parameters is warranted.

For an isolated pulsar, the braking index is defined by the nature of its spin-down process. 
Under the assumption of pure magnetic dipole radiation (MDR) in vacuum, the braking index is exactly 3 \citep{manchester77}. 
However, if the pulsar's spin-down is dominated by a particle wind, the index approaches 1 \citep{michel69, manchester85}. 
In practice, the braking index often falls between these limits due to a combination of MDR and particle wind braking, typically yielding values between 1 and 3, as observed for most of pulsars \citep{archibald16, espinoza11, pons12}. 
Despite the existence of a few exceptions, such as PSR J1640-4631, with a measured braking index of $n = 3.15 \pm 0.03$ \citep{archibald16}, and PSR J1734-3333, with a lower index of $n = 0.9 \pm 0.2$ \citep{espinoza11b}, the overall pattern remains consistent. 
Thus, we restrict our analysis to $1 < n < 3$.

To investigate the impact of $n$ and $t_{age}$ on the observed MWL SED of LHAASO J2226+6057, we considered $t_{age} = 1000, 4000,$ and $7000$ years, and $n = 1.5, 2.0, 2.5$ and $3.0$. 
The initial spin-down luminosity ($L_0$) and initial spin-down age ($\tau_0$) for each case were derived from the characteristic age ($\tau_c \approx 10.5$ kyr) and present-day spin-down luminosity ($L(t_{age}) \approx 2.25 \times 10^{37}$ erg/s) using equations~\ref{eq.tide.d} and \ref{eq.tide.e}. 

The other physical parameters were held constant across the models to isolate the effects of $n$ and $t_{age}$. 
These include the minimum Lorentz factor of the injected particles ($\gamma_{min} = 1$), supernova explosion energy ($E_{SN} = 10^{51}$ erg), ambient ISM density ($\rho_{ISM} = 0.1$ cm$^{-3}$), and SNR ejecta mass ($M_{ej} = 8 M_{\odot}$). 
The soft photon fields for IC scattering were fixed based on \cite{porter06}, assuming CMB, FIR, and NIR components with respective temperatures and energy densities of $T_{CMB} = 2.73$ K, $\omega_{CMB} = 0.25$ eV cm$^{-3}$, $T_{FIR} = 25$ K, $\omega_{FIR} = 0.29$ eV cm$^{-3}$, $T_{NIR} = 5000$ K, and $\omega_{NIR} = 0.45$ eV cm$^{-3}$. 
The parameters describing the injection function (i.e. low energy index $\alpha_1$, high energy index $\alpha_2$, energy break $\gamma_b$) and the magnetic fraction $\eta$ are adjusted to align with the MWL observations. 
These choices minimize the influence of uncertain environmental parameters, allowing a focused assessment of $n$ and $t_{age}$.

Fixing n to 2.5 and assuming three different values for $t_{age}$, we generated the MWL SEDs with our model, shown in Figure \ref{fig1}. 
For $t_{age} = 1000$ years, the IC emission is insufficient to account for the high-energy data, while for $t_{age} = 4000$ years, the model still underestimates the observed gamma-ray flux. 
In contrast, the model with $t_{age} = 7000$ years provides a significantly better fit across the entire MWL range, consistent with the results in \cite{joshi22} and \cite{yu22}. 
%

Next, the influence of varying $n$ on the SED was examined (Figure \ref{fig2}), fixing $t_{age}$ at 7000 years. 
The resulting spectra indicate that within the considered range, changes in $n$ have a relatively modest impact on the overall SED shape. 
Consequently, we adopted $t_{age} = 7000$ years and $n = 2.5$ for further modeling.

\subsubsection{MWL SED Fitting}
\label{j2226.sub2}

\begin{table}
\caption{Physical parameters used by and resulting from the fit. The bracketed terms in the fitted parameters section signify the lower and  upper bounds of 1$\sigma$ confidence interval respectively.}             
\label{tab1}      
\centering          
\begin{tabular}{l l l}      
\hline\hline       
                      
Definition & Parameter & Value\\ 
\hline
Measured or assumed parameters: & & \\
\hline
   Age & $t_{age}$ [kyr] & 7\\  
   Characteristic age & $\tau_c$ [kyr] & 10.5 \\
   Braking index & n & 2.5\\
   Present day spin down luminosity & L($t_{age}$) [erg s$^{-1}$] & 2.25 $\times$ 10$^{37}$ \\
   Distance & D [kpc] & 3 \\
   Minimum energy at injection & $\gamma_{min}$ & 1 \\
   SN explosion energy & E$_{SN}$ [erg] & 10$^{51}$\\
   ISM density & $\rho_{ISM}$ [cm$^{-3}$] & 0.1 \\
   SNR core density index & w$_{core}$ & 0\\
   SNR envelope density index & w$_{env}$ & 9\\
   PWN adiabatic index & $\gamma_{PWN}$ & 1.333\\
   SNR adiabatic index & $\gamma_{SNR}$ & 1.667\\
   Containment factor & $\epsilon$ & 0.5\\
   Magnetic compression ratio & $\kappa$ & 3\\
   CMB temperature & T$_{CMB}$ [K] & 2.73\\
   CMB energy density & $\omega_{CMB}$ [eV cm$^{-3}$] & 0.25\\
   FIR temperature & T$_{FIR}$ [K] & 25\\
   NIR temperature & T$_{NIR}$ [K] & 5000\\
   \hline
   Derived parameters: & & \\
   \hline
   Initial spin down luminosity & L$_0$ [erg s$^{-1}$] & 1.13 $\times$ 10$^{38}$\\
   Initial spin down age & $\tau_0$ [kyr] & 7\\
   \hline
   Fitted parameters: & & \\
   \hline
   Energy break at injection & $\gamma_b$ & 3338.00 (2082.91, 10597.30)\\
   Low energy index at injection & $\alpha_1$ & 1.4522 (1.0000, 1.6432)\\
   High energy index at injection & $\alpha_2$ & 2.3727 (2.3316, 2.3890)\\
   Ejected mass & $M_{ej}$ [$M_{\odot}$] & 8.8927 (8.1735, 9.3202)\\
   Magnetic fraction & $\eta$ & 0.0033 (0.0026, 0.0060)\\
   FIR energy density & $\omega_{FIR}$ [eV cm$^{-3}$] & 0.0100 (0.0100, 0.4611)\\
   NIR energy density & $\omega_{NIR}$ [eV cm$^{-3}$] & 0.0100 (0.0100, 5.0000)\\
   \hline
   Resulting features: & &\\
   \hline
   PWN radius & R$_{PWN}$ (t$_{age}$) [pc] & 9.33\\
   SNR forward shock radius & R$_{FS}$ (t$_{age}$) [pc] & 16.23\\
   SNR reverse shock radius & R$_{RS}$ (t$_{age}$) [pc] & 8.98\\
   PWN magnetic field & B$_{PWN}$ (t$_{age}$) [$\mu$G] & 1.91\\
\hline
\hline
\end{tabular}
\end{table}

\begin{figure*}[htp]
\centering
\includegraphics[width=\textwidth]{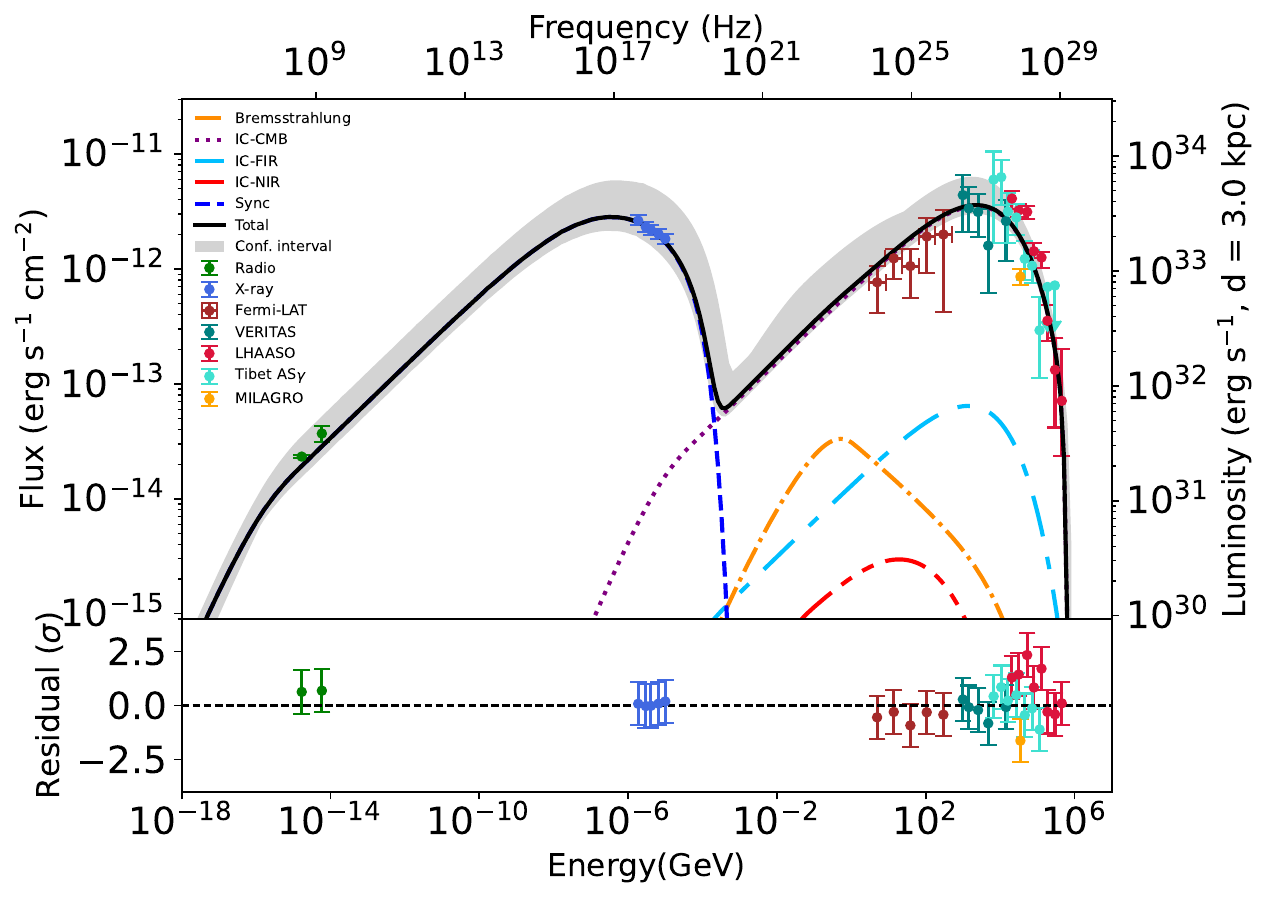}
\caption{Best-fit SED of LHAASO J2226+6057, using the same color scheme as in Figure \ref{fig1}. Adopted from 
 Figure 3 in \citet{Sarkar2022}. }
\label{fig3}
\end{figure*}

We modeled the MWL SED of the PWN using our TIDE code \citep{martin2022unique}, fitting the observed data by varying key physical parameters. 
$L_0$ and $\tau_0$ were calculated using equations \ref{eq.tide.d} and \ref{eq.tide.e}, while the pulsar age ($t_{\rm age} = 7000$ years) and braking index ($n = 2.5$) were chosen based on the estimates in Section \ref{j2226.sub1}. 
The parameters like the ejected mass ($M_{\rm ej}$) and the FIR/NIR energy densities were left free within the typical value ranges. 
Table~\ref{tab1} summarizes all these parameters.

Figure~\ref{fig3} presents the resulting MWL spectrum along with the 1$\sigma$ confidence interval. The model accurately captures the observed MWL data, as indicated by the small residuals and the reduced $\chi^2$/D.O.F. of 35.65/30. 
Notably, the model suggests that the IC scattering responsible for the VHE-UHE gamma rays primarily involves CMB photons, while the contributions from FIR and NIR fields are negligible. 
Although this result differs from that of the photon fields, which range from 1.5 to 3.0 times the CMB, in \citet{joshi22}, it aligns with recent findings by \citet{wilhelmi22}, which also highlight the dominance of CMB photons for IC scattering.

However, the best-fit model predicts a highly extended PWN with a radius of 9.33 pc and a relatively weak magnetic field of 1.91 $\mu$G. 
This size is consistent with the large extension observed by LHAASO but conflicts with the smaller radio and X-ray sizes reported for the Boomerang PWN \citep{halpern01, Halpern_2001b}. 
The low magnetic field also raises questions about the confinement of high-energy particles within such a dilute medium, a challenge we discuss further in Section~\ref{j2226.discussion}. 

Overall, the current model effectively reproduces the MWL spectrum although points to a physically large and weakly magnetized PWN.

\subsubsection{Impact of Reverberation}

\begin{figure}[htp]
\centering
\includegraphics[width=.49\textwidth]{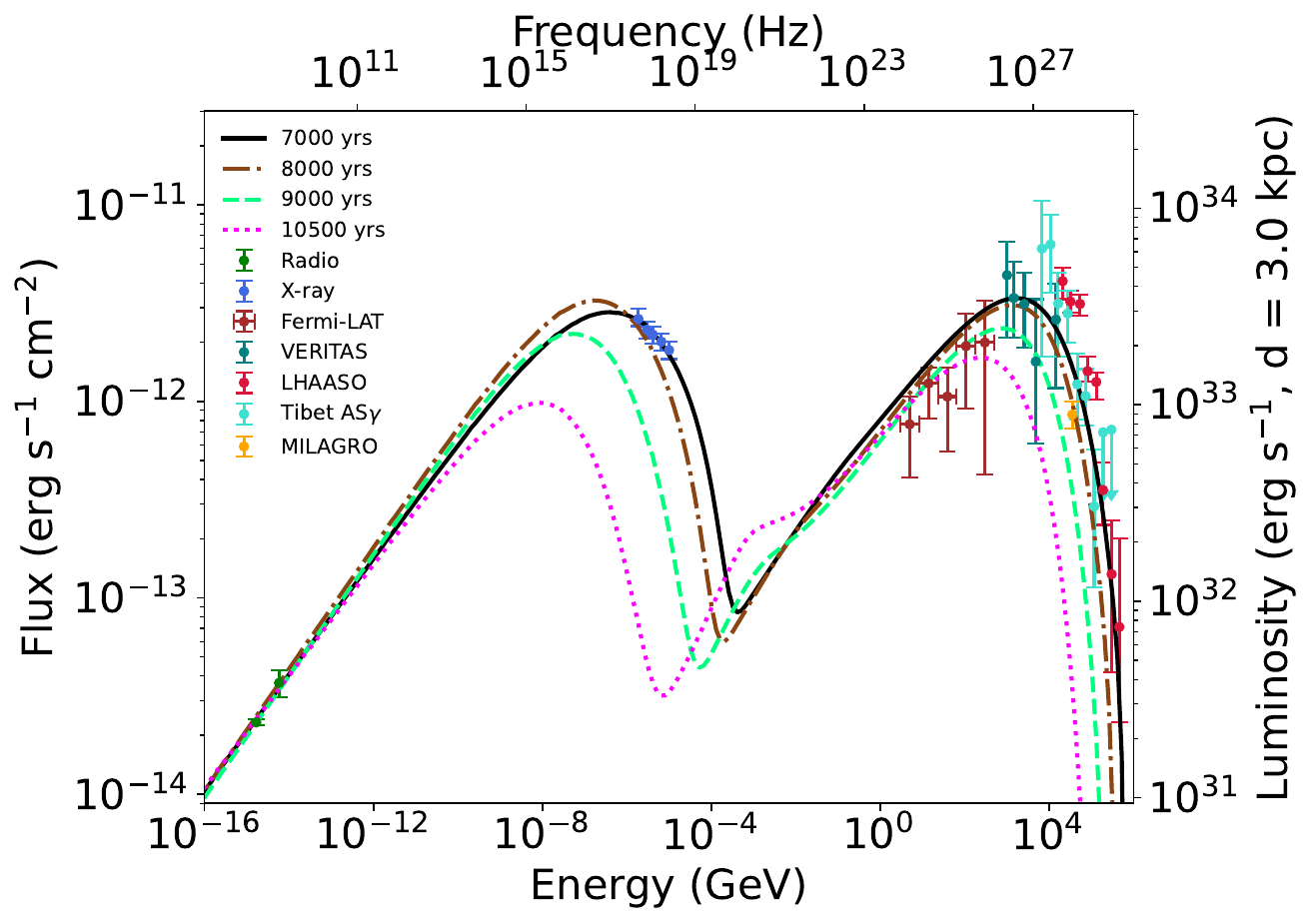}
\includegraphics[width=.49\textwidth]{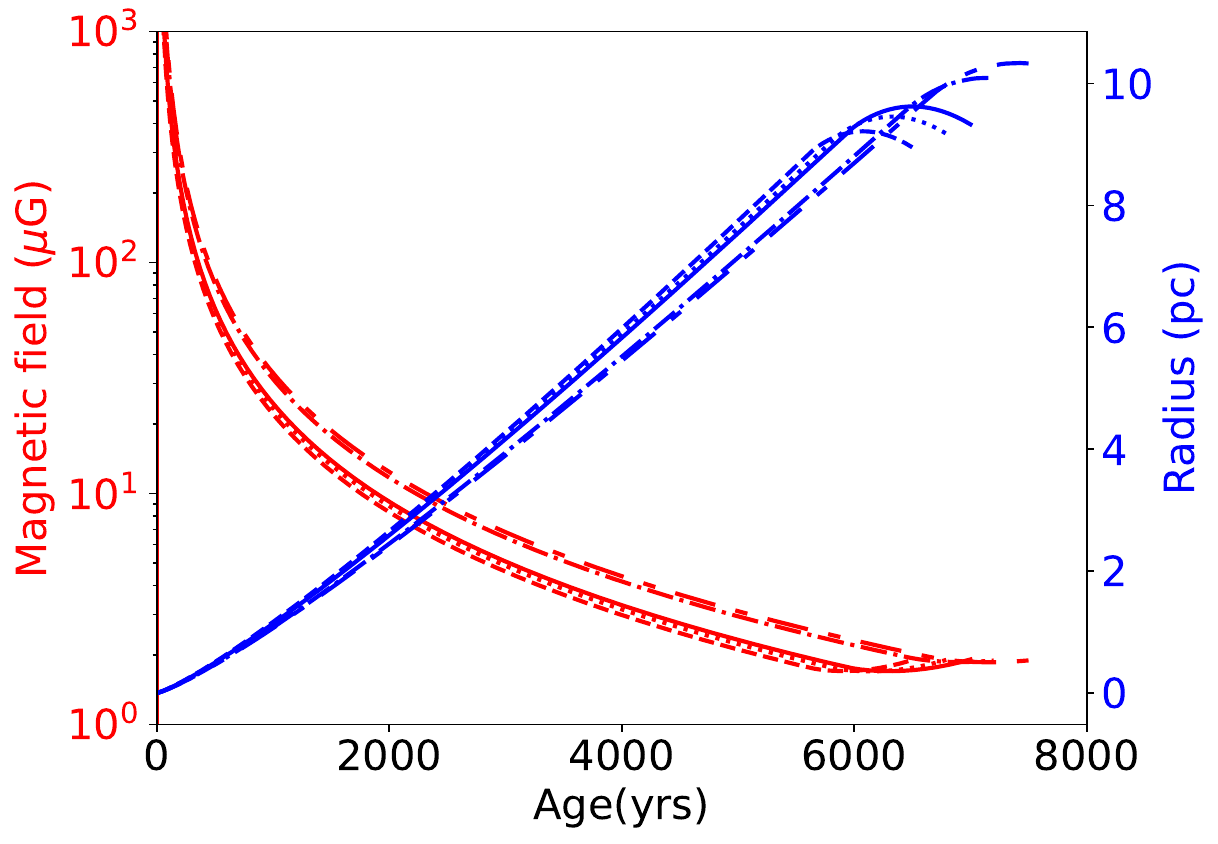}
\caption{Left: Impact of reverberation on the PWN SED for different ages. 
Right: Evolution of PWN magnetic field (red) and radius (blue) as a function of age, with $t_{age}$ = 6500 years (dashed), 6800 years (dotted), 7000 years (solid), 7200 years (dot-dashed), and 7500 years (long-dashed). Copy from Figure 5 in \citet{Sarkar2022}. }
\label{fig6}
\end{figure}

As shown in Table~\ref{tab1}, the radius of the SNR reverse shock is currently smaller than the PWN radius, indicating the onset of the reverberation phase, where the reverse shock begins to interact with the PWN shell. 
At the assumed age of 7000 years, this interaction is just beginning, implying only minor distortions to the PWN structure.

To assess the influence of reverberation on the MWL spectrum, we explored four possible ages: 7000, 8000, 9000, and 10500 years. 
For each case, the braking index was set to 2.5. 
We adopted the FIR/NIR energy densities from \cite{porter06} and varied the parameters $\alpha_1$, $\alpha_2$, $\gamma_b$, and $\eta$ to obtain the best fits for the observational data. 
The resulting SED are shown in the left panel of Figure~\ref{fig6}. 

As the age increases, the reverberation phase compresses the PWN, increasing its magnetic field strength. 
To account for this, the magnetic fraction $\eta$ was progressively reduced to limit the fraction of the spin-down luminosity that powers the magnetic field. 
Despite this adjustment, the high-energy component of the MWL spectrum softens significantly with age, reflecting increased synchrotron cooling of high-energy electrons, and the overall fit to the observed data deteriorates, particularly in the X-ray and VHE-UHE bands.

In addition, we also considered a narrower age range of 6500, 6800, 7200, and 7500 years, examining the corresponding changes in magnetic field and PWN radius. 
The results, presented in the right panel of Figure~\ref{fig6}, confirm that the PWN is in the early stages of reverberation phase in all cases, with the radius commencing to decrease and the magnetic field remaining close to or below the Galactic average.

\subsubsection{Exploring $t_{\rm age}$ as a Free Parameter} 

Given the constraints discussed in Sections~\ref{j2226.sub1} and~\ref{j2226.sub2}, the true age of the PWN is not likely lower than 4000 years or much greater than 7000 years.  
To further explore this parameter, we performed a fit using TIDE with $t_{\rm age}$ left free, alongside the same set of free parameters described in Section~\ref{j2226.sub2}. 
The resulting best-fit age is $4880.3^{+1180.1}_{-1008.9}$ years, consistent with the previously inferred range. 
The model SED closely matches the one shown in Figure~\ref{fig3}, and the best-fit parameter values, aside from $t_{\rm age}$, show significant overlap within their 1$\sigma$ confidence intervals compared to the fixed $t_{\rm age} = 7000$ years case (see Table~\ref{tab1} and Section~\ref{j2226.sub2}). 
This indicates that the two solutions are statistically equivalent within the current model precision. 

Additionally, this fit yields a relatively large PWN radius of $R_{\rm PWN} = 8.26$ pc and a lower magnetic field of $B_{\rm PWN} = 1.87\,\mu$G, similar to the results obtained with the fixed $t_{\rm age}=7000$ years case. These implications are discussed further in the following section.

\subsection{Discussion and Concluding Remarks}
\label{j2226.discussion}

%
Let us summarize the main findings and discuss their implications. 

\begin{enumerate}
    \item \textbf{True age and braking index effects:} 
    Previous studies of this source typically fixed the age and braking index of the pulsar. 
    In contrast, our analysis explored a range of true age ($t_{age}$) between 4000 and 7000 years, finding this range most compatible with the observed MWL SED. 
    We found that varying the braking index within plausible ranges had a minimal impact on the overall SED.

    \item \textbf{MWL SED fitting:} 
    We obtained a best-fit model which can describe the MWL observations well.
    However, this model requires a very low magnetic field ($\sim 2 \, \mu$G) to reproduce the observed SED, comparable to the average Galactic magnetic field. 
    This is consistent with earlier studies \citep{joshi22, yu22} and other LHAASO-detected PeVatron candidates \citep{desarkar22, crestan21, li21, burgess22}. 
    Additionally, the current PWN radius was found to be quite large ($\sim 10$ pc), matching the extended source regions observed by VERITAS ($\sim 14$ pc) and LHAASO ($\sim 25.6$ pc). 
    This is significantly larger than the compact X-ray emission region, a known limitation of single-zone models. 

    \item \textbf{Impact of reverberation:} 
    We included the effects of reverberation, where the PWN interacts with the reverse shock from the SNR. 
    Our results indicate that this PWN is likely just entering the reverberation phase. 
    The fit quality declined as the assumed true age approached the characteristic age, suggesting that much older ages are unlikely if the gamma-ray emission originates from the PWN. 
    Consequently, including reverberation effect cannot resolve the large radius / low magnetic field problem. 

    \item \textbf{Distance uncertainty:} 
    The source distance remains a significant source of uncertainty. 
    We tested two extreme cases: a nearby distance of 800 pc and a more distant 7.5 kpc. 
    We obtained a worse fit under the case of 7.5 kpc, while the closer distance yielded a comparable fit. 
    But the overall large radius / low magnetic field problem persisted in both cases. 
    This suggests that the fundamental challenge lies in the physical interpretation of the PWN, rather than in distance uncertainties alone. 

    \item \textbf{Magnetic field constraints:} 
    The derived PWN magnetic field is close to the lower end of typical Galactic magnetic field estimates, which range from 2 to 11 $\mu$G \citep{bernardo13, desarkar21}. 
    This raises the question of whether the local environment around the PWN might have unusually low magnetic field strength, perhaps due to prior supernova explosions or other clearing effects. 
    Alternatively, this may point to a more complex PWN evolution than currently modeled, where the standard reverberation treatment (e.g., assuming constant ejecta pressure) is insufficient \citep{Bandiera:2020, Bandiera:2022}.

    \item \textbf{Future work and alternative scenarios:} 
    Given the unresolved large radius / low magnetic field problem, more sophisticated models, possibly including multi-zone or hybrid leptonic-hadronic components, may be necessary. 
    Further observational constraints, including better distance measurements and higher resolution imaging, will be critical in resolving this puzzle. 
\end{enumerate}

In conclusion, while the one-zone PWN model presented here represents a significant step forward in understanding LHAASO J2226+6057, the magnetic field issue, which couldn't be solved by age, braking index, and the effect of reverberation in present PWN models, cannot be removed and is a strong caveat of the interpretation; as it would need to be of the same order or lower than that of the ISM in order for the model to work, casting doubts on the scenario. 

\section{eHWC J2019+368}
\label{J2019}

\subsection{Introduction}
\label{J2019.intro}

%
The HAWC source eHWC J2019+368 is an extended very high energy (VHE) gamma-ray emitter ($>$ 100 TeV) with an integrated flux above 56 TeV of approximately \(1.6 \times 10^{-14}~\mathrm{ph~cm^{-2}~s^{-1}}\) \citep{Abeysekara2020}. 
Located in the Cygnus region, it was first detected by the Very Energetic Radiation Imaging Telescope Array System (VERITAS) in 2014 and named VER J2019+368 \citep{Aliu2014}. 
It lies within the extended MGRO J2019+37 region, which also hosts the star formation region Sh 2-104, the hard X-ray transient IGR J20188+3657, and the pulsar PSR J2021+3651 with its associated PWN G75.2+0.1 \citep{Aliu2014}.

%
PSR J2021+3651 is estimated to be at a distance of \(1.8_{-1.4}^{+1.7}~\mathrm{kpc}\) \citep{Kirichenko2015}, with a rotational period of \(P = 0.104~\mathrm{ms}\), a period derivative of \(\dot{P} = 9.57 \times 10^{-14}~\mathrm{s~s^{-1}}\) \citep{Roberts2002}, a characteristic age of 17.23 kyr, and a spin-down luminosity of \(3.4 \times 10^{36}~\mathrm{erg~s^{-1}}\). 
The true age is uncertain but likely below 10 kyr. 
Given its young and energetic nature, the PWN is considered a likely source of the VHE gamma-ray emission.

%
Hard X-ray observations of the Sh 2-104 region by NuSTAR revealed that the X-ray emission coincides with the TeV emission detected by VERITAS \citep{Gotthelf2016}. 
Further observations using Suzaku detected extended X-ray emission only around PSR J2021+3651 and its PWN \citep{Mizuno2017}. 
Combining these X-ray data with multi-band observations from the Very Large Array (VLA), \fermi, and VERITAS, they proposed a synchrotron plus IC) model, which accounts for about 80\% of the observed TeV flux, suggesting that the X-ray PWN is the primary source of VER J2019+368.

%
\citet{Fang2020} modeled the source using a time-dependent leptonic framework to evaluate whether PWN G75.2+0.1 could account for the observed TeV emission. 
Assuming a broken PL injection particles (\(\alpha_1 = 1.5\), \(\alpha_2 = 2.5\), \(\gamma_b = 2.0 \times 10^5\), \(\epsilon = 1/3\)), they examined ages of 8, 10, and 12 kyr, identifying the best fit at 10 kyr with a magnetic field of 2.7 \(\mu\mathrm{G}\).

To interpret the X-ray and TeV emission, \citet{Albert2021} applied both one-zone and two-zone models, considering synchrotron and IC processes. 
In the one-zone model, the cut-off energy, birth period, and conversion efficiency were free parameters, yielding values of 300 TeV, 80 ms (corresponding to an age of \(\approx 7\) kyr), and 5\%, respectively. 
The derived magnetic field was 2 \(\mu\mathrm{G}\), but the model predicted identical X-ray and TeV sizes, inconsistent with observations. 
To resolve this, they developed a two-zone model in which X-ray-emitting electrons form a subset of the \(\gamma\)-ray-emitting population. 
This model yielded a magnetic field of 3.5 \(\mu\mathrm{G}\) and an age of 2 kyr for the X-ray-emitting population after adjusting the SED to match observed X-ray fluxes.

\subsection{Results}
\label{J2019.results}

\begin{table}[H]
    \centering
    \caption{Summary of the physical magnitudes of eHWC J2019+368}. \label{tab:mag.J2019}
    \label{tab.pevatron.j2019}
    \begin{tabular}{@{}llll@{}}
    \toprule
		Parameters & Symbol &Values &Fitting Range\\
		\hline
         Measured or assumed parameters: \\
        		\hline
		Age (kyr) &$t_{age}$  &6\\
		Period (ms) &P  &104 \\
        Period derivative $(s~s^{-1})$ &$\dot{P}$ &$9.57\times10^{-14}$ &\\
		Characteristic age (kyr) &$\tau_c$ &17.23\\
		Spin-down luminosity now $(erg~s^{-1})$ &L &$3.4\times10^{36}$\\
		Braking index &n &3 &\\
		Initial spin-down luminosity $(erg~s^{-1})$ &$L_0$ &$8.004\times10^{36}$\\
		Initial spin-down age (kyr) &$\tau_0$ &11.23\\
		Distance (kpc) &d &1.8\\
		SN explosion energy (erg) &$E_{sn}$ &$10^{51}$ &\\
		ISM density ($cm^{-3}$) &$n_{ism}$ &0.1\\
		Minimum energy at injection &$\gamma_{min}$ &1 &\\
        Containment factor  &$\epsilon$ &0.5 & \\
        CMB temperature (K)  &$T_{cmb}$ &2.73 &\\
		CMB energy density $(eV~cm^{-3})$  &$U_{cmb}$ &0.25 &\\
        FIR temperature (K) &$T_{fir}$ &30\\
        NIR temperature (K) &$T_{nir}$ &2900\\
                \hline
        Fitted parameters: \\
        		\hline
		Break energy ($10^4$)&$\gamma_b$ &0.67 (0.19, 1.32) &0.1 - 1000\\
		Low energy index &$\alpha_1$ &1.48 (1, 1.75) &1 - 4\\
		High energy index &$\alpha_2$  &2.11 (2.03, 2.13) &1 - 4\\
		Ejected mass $(M_{\odot})$  &$M_{ej}$  &8.00 (8, 13.88) &8 - 15\\
		Magnetic energy fraction ($10^{-2}$)  &$\eta$ &1.16 (0.91, 3.47) &0.01 - 50\\
		FIR energy density $(eV~cm^{-3})$ &$U_{fir}$ &0.01 (0.01, 0.69) &0.01 - 5\\
		NIR energy density $(eV~cm^{-3})$ &$U_{nir}$ &0.01 (0.01, 5) &0.01 - 5\\
		PWN radius now (pc) &$R_{pwn}$ &5.83\\
		Magnetic field now ($\mu$G) &B &2.14\\
		Reduced $\chi^2$ &$\chi^2/D.O.F.$ &23.18/18 (1.29)\\
		Systematic uncertainty &$\sigma$ &0.36 &0.01-0.5\\
    \bottomrule
    \end{tabular}
\end{table}

\begin{figure}[H]
    \centering
  	\includegraphics[width=0.45\columnwidth]{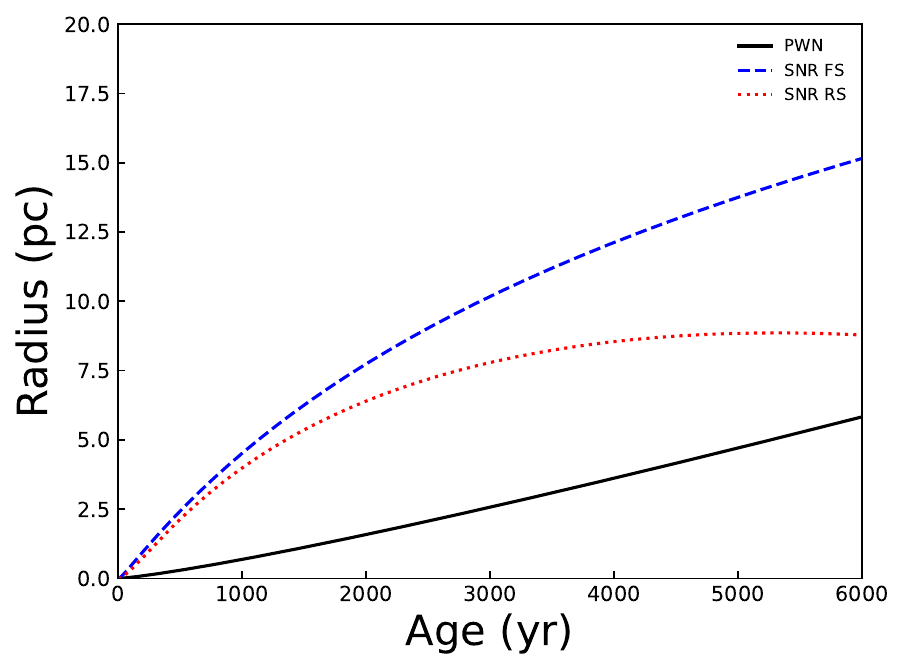}
        \includegraphics[width=0.75\columnwidth]{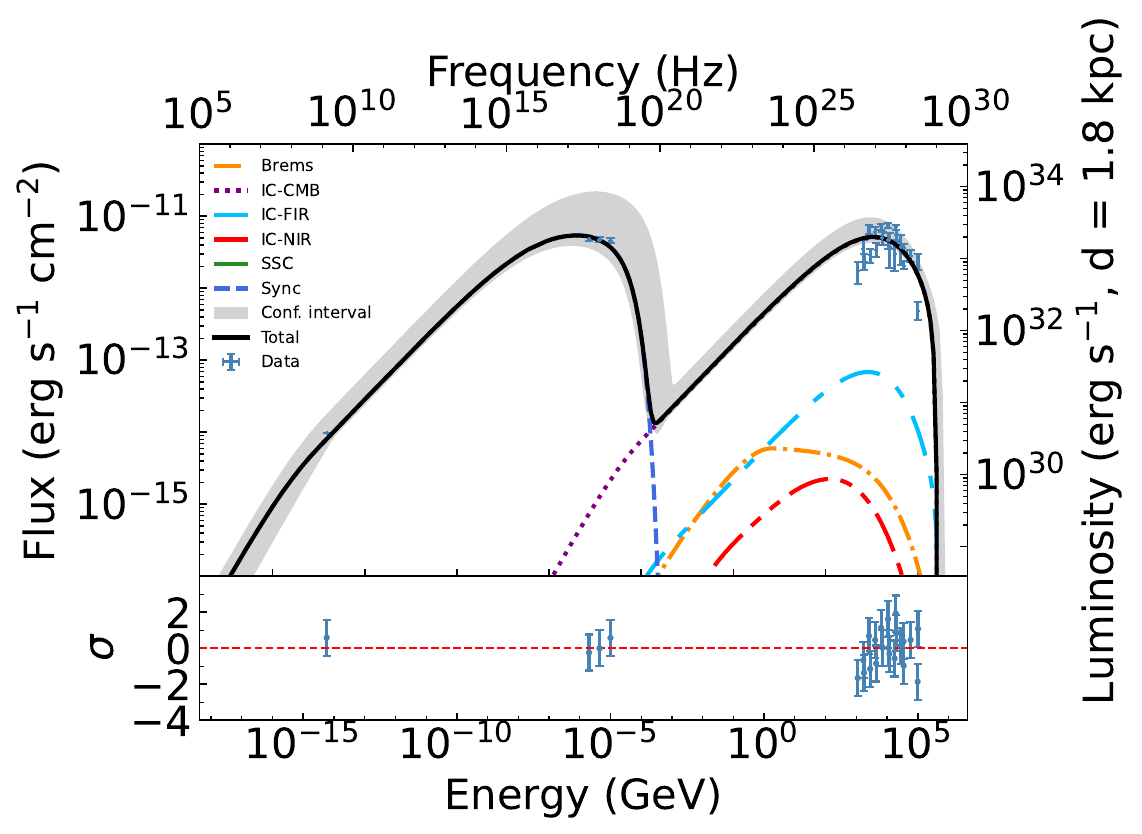}
    \caption{The resulting SED from our model for eHWC J2019+368.The multi-band data are detected by VLA \citep{Paredes2009}, Suzaku and XMM-Newton \citep{Mizuno2017}, VERITAS \citep{Aliu2014}, HAWC \citep{Abeysekara2020}, MILAGRO \citep{Abdo2007a,Abdo2007b,Abdo2009} and LHAASO \citep{Cao2021}.
    }
    \label{pic.pevatron.j2019}
\end{figure}


Unlike the models presented in \citet{Albert2021} and \citet{Fang2020}, our approach incorporates a more comprehensive set of radiation processes, including bremsstrahlung and synchrotron self-Compton (SSC) emission. 
However, these additions primarily enhance the physical completeness of the model, as both processes typically contribute negligibly to the overall emission in most PWNe (with the notable exception of the SSC component in the Crab Nebula). 
Moreover, our model explicitly includes the PWN radius as a constraint, allowing for a more consistent match to both the multi-band spectrum and the observed morphology. 
In the one-zone scenario, we achieved a reduced \(\chi^2\) of 1.288 and a systematic uncertainty of 0.36, indicating a significantly improved fit.

%
The physical parameters used in our model are detailed in Table~\ref{tab.pevatron.j2019}. 
The characteristic age \(\tau_c\) of PSR J2021+3651 is estimated to be approximately 17.2 kyr, based on its known period \(P\) and period derivative \(\dot{P}\). 
Assuming a typical braking index \(n = 3\) and a true age \(t_{\mathrm{age}} = 6\) kyr, the initial spin-down time \(\tau_0\) can be obtained as 
\(\tau_0 = \frac{2\tau_c}{n-1} - t_{\mathrm{age}} \approx 11.23~\text{kyr},\)
following \citet{Gaensler_Slane06a}. 
Given the current spin-down luminosity \(L = 3.4 \times 10^{36}~\text{erg~s}^{-1}\), the initial spin-down luminosity \(L_0\) is then calculated as approximately \(8.0 \times 10^{36}~\text{erg~s}^{-1}\). 
For the supernova explosion energy, ISM density, and containment factor, we adopt typical values of \(1 \times 10^{51}~\text{erg}\), 0.1~\(\text{cm}^{-3}\), and 0.5, respectively. 
The temperatures of the three background photon fields — CMB, FIR, and NIR — are derived from the \texttt{phfields} tool within the TIDE code.

Our model provides a consistent fit to the observed SED of eHWC J2019+368, spanning from radio to TeV bands, as illustrated in Figure~\ref{pic.pevatron.j2019}. 
Synchrotron radiation dominates from radio to X-ray bands, with a cutoff at several keV due to synchrotron cooling, while the TeV emission primarily arises from IC scattering off three ambient photon fields — CMB, FIR, and NIR — with the CMB being the most significant contributor. 
And our model suggests that eHWC J2019+368 is a potential PeVatron, predicting a maximum particle energy of approximately 0.4 PeV.

%
The derived physical parameters include a PWN radius \(R_{\mathrm{pwn}} = 5.83\) pc and a magnetic field \(B = 2.14~\mu G\) at an assumed age of 6 kyr. 
Furthermore, using the updated version of TIDE \citep{martin2022unique}, which now allows the age to be treated as a free parameter, we performed an additional fit with age unconstrained while keeping other parameters fixed. 
This resulted in a fitted age range of 5120 to 7719 years, in good agreement with the assumed 6 kyr.

As in the case of LHAASO~J2226+6057, the leptonic modeling of eHWC~J2019+368 also yields a low magnetic field, consistent with previous results by \citet{Fang2020} and \citet{Albert2021}. 
This again casts doubt on a purely PWN origin for the UHE source.
However, unlike LHAASO~J2226+6057, the predicted radius ($5.83$~pc) agrees well with the observed value ($5.652$~pc), within the uncertainties of the distance estimate. 
This suggests that the problem of accommodating both a large radius and a low magnetic field may be alleviated 
if having a distinct physical origins, at least in certain cases.

\section{HESS J1427-608}
\label{J1427}

\subsection{Introduction}
\label{J1427.intro}

%
HESS J1427-608 is an unidentified, slightly extended TeV source located at \(l=314.409^\circ\),  \(b=0.145^\circ\), first reported by \citet{Aharonian2008}. 
They measured an angular extension of \(0.063 \pm 0.01^\circ\), with a 1–10 TeV energy flux of \(F_{\mathrm{TeV}} = 4.0 \times 10^{-12}~\mathrm{erg~cm^{-2}~s^{-1}}\) and a spectral index of 2.16. 
However, subsequent observations by \citet{Abdalla2018} revised the extension to \(0.048 \pm 0.009^\circ\), which meets their criterion for a point-like source, leading them to model it as a Gaussian point source. 
\citet{Tibolla2011} proposed that HESS J1427-608 might be an ancient PWN, although no associated pulsar or supernova remnant (SNR) has been detected to date. 
Its potential X-ray counterpart, Suzaku J1427-051, was identified by \citet{Fujinaga2013}, but the diffuse nature of this X-ray source does not allow us to certify whether HESS J1427-608 originated from a PWN or a non-thermal SNR.

%
\citet{Vorster2013} employed a spatially independent model to support the hypothesis that HESS J1427-608 originates from an evolved PWN. 
Given the absence of a detected pulsar, they assumed a distance of 11 kpc and adopted an initial spin-down luminosity of \(L_0 = 5.5 \times 10^{38}~\mathrm{erg~s^{-1}}\), a characteristic age of 3 kyr, and a true age of 10 kyr. 
These parameters yield a present spin-down luminosity of \(L = 2.9 \times 10^{37}~\mathrm{erg~s^{-1}}\), along with a very weak present-day magnetic field of 0.4 \(\mu G\), consistent with an old PWN scenario characterized by low synchrotron luminosity. 
However, their model significantly underpredicts the observed X-ray flux, prompting them to test a younger scenario with an age of 6.4 kyr and a stronger magnetic field of 4 \(\mu G\). 
While this configuration could better reproduce the X-ray data, it failed to match the radio observations, possibly indicating that some of the data used may not be directly associated with HESS J1427-608.

%
Separately, \citet{Guo2017} identified a possible GeV counterpart to HESS J1427-608 using \lat data, revealing a power-law spectrum with an index of 2.0 and no significant high-energy cutoff, consistent with a potential PeVatron. 
Assuming a distance of 8 kpc \citep{Fujinaga2013}, they estimated a characteristic age of 11 kyr and a present spin-down power of \(6.5 \times 10^{36}~\mathrm{erg~s^{-1}}\), based on empirical flux ratios \citep{Mattana2009, Acero2013}. 
They explored both a PL and a broken PL (BPL) injection models, finding better agreement with the latter, yielding magnetic fields around 5 \(\mu G\) (PL) and 3.5 \(\mu G\). 
While the low field strengths and lack of shell-like X-ray morphology support a PWN origin, the flat \(\gamma\)-ray spectrum still allows for an SNR interpretation.

%
\citet{Devin2021} conducted a multi-band analysis of HESS J1427-608 using a generic code to constrain the origin of the TeV emission. 
Their model, which incorporates radio spectral index measurements and mean magnetic field estimates under the leptonic scenario, successfully reproduces the broadband non-thermal spectrum, suggesting a magnetic field below 10 \(\mu\)G. 
The \lat data further reveals a pulsar-like spectrum, with the estimated spin-down power and characteristic age consistent with typical TeV PWNe, supporting the evolved PWN origin for HESS J1427-608.

\subsection{Results}
\label{j1427.results}

\begin{table}[H]
    \centering
    \caption{Summary of the physical magnitudes of HESS J1427-608}. \label{tab:mag.J1427}
    \label{tab.j1427}
    \begin{tabular}{@{}llll@{}}
    \toprule
		Parameters & Symbol &Values &Fitting Range\\
		\hline
         Measured or assumed parameters: \\
        		\hline
		Age (kyr) &$t_{age}$  &8\\
		Characteristic age (kyr) &$\tau_c$ &11\\
		Spin-down luminosity now $(erg~s^{-1})$ &L &$1.0\times10^{37}$\\
		Braking index &n &3 &\\
		Initial spin-down luminosity $(erg~s^{-1})$ &$L_0$ &$1.344\times10^{38}$\\
		Initial spin-down age (kyr) &$\tau_0$ &3\\
		Distance (kpc) &d &8.0\\
		SN explosion energy (erg) &$E_{sn}$ &$10^{51}$ &\\
		ISM density ($cm^{-3}$) &$n_{ism}$ &0.1\\
		Minimum energy at injection &$\gamma_{min}$ &1 &\\
        Containment factor  &$\epsilon$ &0.5 & \\
        CMB temperature (K)  &$T_{cmb}$ &2.73 &\\
		CMB energy density $(eV~cm^{-3})$  &$U_{cmb}$ &0.25 &\\
        FIR temperature (K) &$T_{fir}$ &31\\
        NIR temperature (K) &$T_{nir}$ &2800\\
                \hline
        Fitted parameters: \\
        		\hline
		Break energy ($10^5$)&$\gamma_b$ &4.83 (3.69, 6.40) &0.1 - 100\\
		Low energy index &$\alpha_1$ &1.21 (1, 1.25) &1 - 4\\
		High energy index &$\alpha_2$  &2.64 (2.57, 2.72) &1 - 4\\
		Ejected mass $(M_{\odot})$  &$M_{ej}$  &8.29 (8.14, 8.46) &8 - 15\\
		Magnetic energy fraction ($10^{-2}$)  &$\eta$ &0.24 (0.19, 0.29) &0.01 - 50\\
		FIR energy density $(eV~cm^{-3})$ &$U_{fir}$ &0.01 (0.01, 0.23) &0.01 - 5\\
		NIR energy density $(eV~cm^{-3})$ &$U_{nir}$ &0.01 (0.01, 4.37) &0.01 - 5\\
		PWN radius now (pc) &$R_{pwn}$ &6.71\\
		Magnetic field now ($\mu$G) &B &2.66\\
		Reduced $\chi^2$ &$\chi^2/D.O.F.$ &7.56/10 (0.76)\\
		Systematic uncertainty &$\sigma$ &0.001 &0.001-0.5\\
    \bottomrule
    \end{tabular}
\end{table}

\begin{figure}[H]
    \centering
  	\includegraphics[width=0.45\columnwidth]{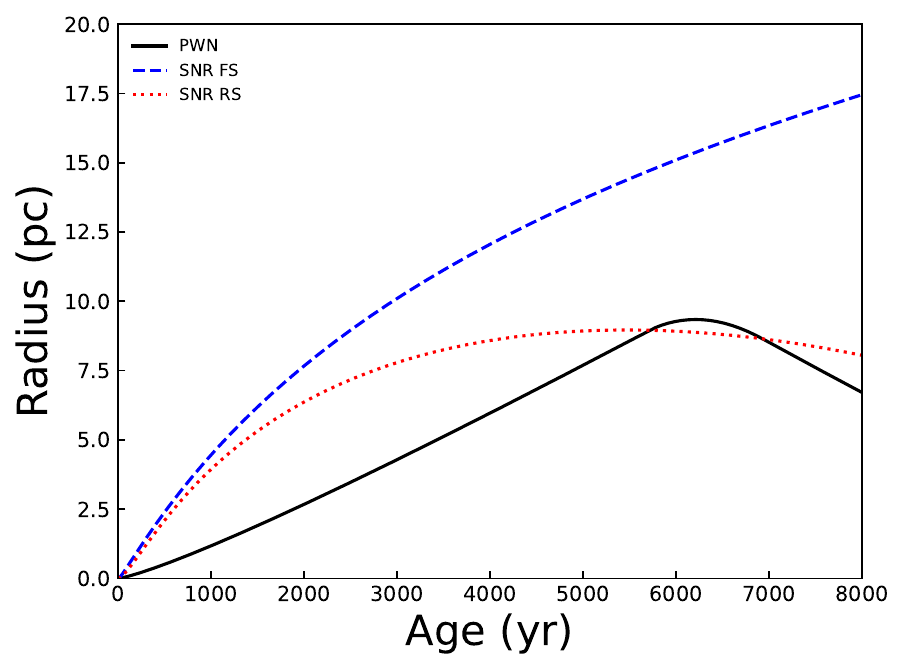}
        \includegraphics[width=0.45\columnwidth]{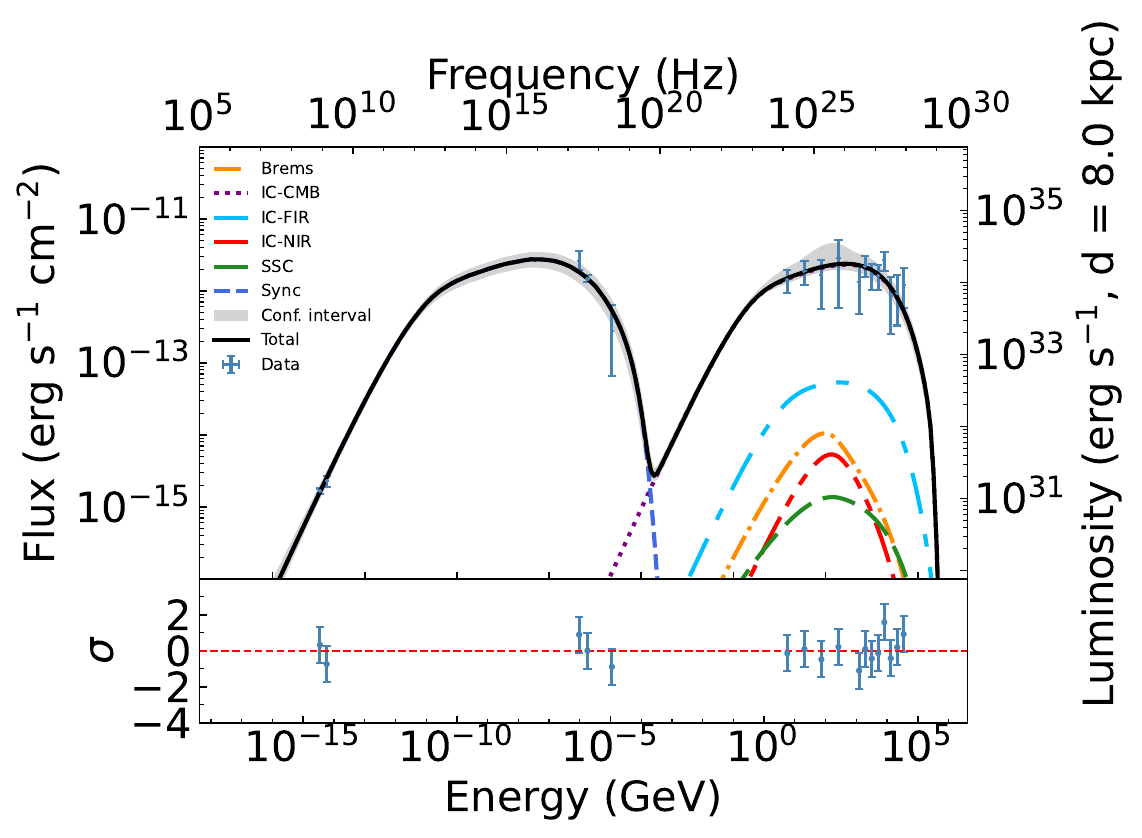}
    \caption{From left to right: the evolution of the radius and the resulting SED from our model for HESS J1427-608. The radio, X-ray (Suzaku), GeV (\lat) and TeV (HESS) data in the SED are from \citet{Devin2021}, \citet{Fujinaga2013}, \citet{Guo2017} and \citet{Aharonian2008}, respectively.}
    \label{pic.j1427}
\end{figure}

%
Similar to eHWC J2019+368, we applied our model to HESS J1427-608 using the parameters listed in Table~\ref{tab.j1427}. 
Given the absence of an identified pulsar, we adopted a characteristic age of 11 kyr as in \citet{Guo2017}, with the age, spin-down luminosity, braking index, and distance set to 8 kyr, \(10^{37}~\mathrm{erg~s^{-1}}\), 3, and 8 kpc, respectively, following \citet{Devin2021}.

%
As shown in Figure~\ref{pic.j1427}, the resulting SED achieves a good fit across a wide range from radio to TeV bands, with a \(\chi^2/\mathrm{D.O.F}\) of 7.56/10. 
Similar to LHAASO J2226+6057 and eHWC J2019+368, the synchrotron radiation dominates from radio to X-ray bands, while the TeV emission primarily arises from IC scattering of CMB photons. 
The model predicts a present-day PWN radius of 6.71~pc, consistent with the observed angular size of $0.048^\circ$ (corresponding to $\sim$6.70~pc at 8~kpc). 
The magnetic field is estimated to be 2.66~$\mu$G, in agreement with previous low-field estimates.
As shown in the left panel of Figure~\ref{pic.j1427}, the PWN exceeds the SNR reverse shock radius at $\sim$6~kyr, marking the onset of the reverberation phase.

In conclusion, a PWN scenario provides a good description of the multiwavelength emission and supports its classification as a potential PeVatron, similar to other UHE sources. 
And the pulsar-related parameters used in our model may provide useful guidance for the potential identification of the as-yet undetected pulsar associated with HESS J1427-608.   
However, the model again requires a low magnetic field, posing a persistent challenge to the leptonic interpretation. 
Interestingly, this case shares different features with eHWC~J2019+368 (low magnetic field but without a radius discrepancy) and LHAASO~J2226+6057 (entry into the reverberation phase).

\section{Conclusions}
\label{Pev.conclusion}

This chapter investigated the plausibility of a PWN origin for three UHE sources (LHAASO J2226+6057, eHWC~J2019+368, and HESS~J1427$-$608) and evaluated their potential as PeVatrons. 
While the leptonic model reproduces the multiwavelength emission of all three UHE sources, it consistently requires unexpected low magnetic fields - consistent with previous studies - which cannot be solved within reasonable parameter ranges. 
For LHAASO~J2226+6057, an additional discrepancy arises from the overestimated PWN radius. 
These results challenge a purely leptonic PWN origin for these sources. 
Given the agreement between the predicted and observed PWN radii for eHWC~J2019+368 and HESS~J1427$-$608, the low-$B_{\mathrm{pwn}}$ and large-$R_{\mathrm{pwn}}$ issues may have distinct physical origins. 
Possible explanations include a hybrid leptonic–hadronic scenario in which part of the $\gamma$-ray emission arises from hadronic processes, or improvements related to the limitations of current IC treatments in modeling VHE/UHE radiation \citep[see e.g.,][]{Xia2025}.  

\newpage
\thispagestyle{empty}
~
\newpage


\chapter{\texorpdfstring{Deep Search for Gamma-Ray Emission from the Accreting X-Ray Pulsar \\1A~0535+262}{Deep Search for Gamma-Ray Emission from the Accreting X-Ray Pulsar 1A~0535+262}}
\label{1A0535}
\textbf{\color{SectionBlue}\normalfont\Large\bfseries Contents of This Chapter\\\\}

This chapter presents a comprehensive investigation of gamma-ray emission from the high-mass X-ray binary pulsar 1A~0535+262, a candidate gamma-ray source that has been extensively studied but not yet confirmed. 
Building on previous efforts, we performed a binned likelihood analysis to the full dataset and X-ray outburst intervals, and used the weighted H-test to probe pulsations. 
Long-term flux variability and orbital phase-resolved analyses were also performed. 

No significant steady or pulsed gamma-ray emission was detected. 
We obtained the most stringent 95\% confidence upper limits on the gamma-ray luminosity, (2.3$-$4.7)$\times 10^{32}$ $\rm erg\, s^{-1}$, implying a gamma-to-X-ray luminosity ratio $L_{\gamma}/L_{\rm X}$ (2–150 keV) of (1.9$-$3.9)$\times10^{-6}$.
These results provide important constraints on gamma-ray production in 1A~0535+262 and serve as a reference for future studies of high-mass X-ray binaries. 

This work was published in the \apj\ \citep{Hou2023}. 
As the second author, I actively participated in the data analysis, cross-checking of results, and manuscript editing.

\newpage

\section{Introduction}
\label{1A0535.intro}

Binary systems are an important, albeit rare, class of gamma-ray emitters. 
Despite their small numbers, these systems display a remarkable diversity in emission mechanisms and physical configurations \citep{Dubus2015}. 
They include:

\begin{itemize}
    \item \textbf{Gamma-ray Binaries:} These are typically high-mass X-ray binaries (HMXBs) with massive O or B companions. 
    Their gamma-ray spectra often peak above 1 MeV and exhibit orbital modulation across a wide range of wavelengths, usually thought to be from the pulsar/stellar wind interaction. 
    Known examples include PSR B1259-63/LS~2883 \citep{HESS2020,Chernyakova2020a,Abdo2011B1259,Caliandro2015B1259,Chernyakova2015B1259}, PSR J2032+4127/MT91~213 \citep{Coe2019}, and LS~I~+61~303 \citep{Weng2022,Abdo2009LS,Papitto2012LS,Ackermann2013LS}, where the compact object has been confirmed as a pulsar.

    \item \textbf{Redbacks and Black Widows:} These are low-mass ($< 0.1 M_\odot$), tight-orbit ($P_{\mathrm{orb}} < 1 \mathrm{day}$) systems containing recycled MSPs and main sequence degenerate companions. 
    The pulsar wind ablates the companion, leading to radio eclipses, X-ray/radio anti-correlations, and pulsar nulling \citep{Bogdanov2018}. 
    Specially, the pulsars in these systems might switch between accretion-powered and rotation-powered states within weeks (could persist for years), known as transitional pulsars \citep{Archibald2009,Papitto2013,deMartino2015,Papitto2019}. 
    And this transition is usually accompanied by gamma-ray variability \citep{Stappers2014,Torres2017,Manca2025}. 

    \item \textbf{Microquasars:} Systems like Cyg~X-1 and Cyg~X-3 contain neutron stars or black holes with relativistic jets. 
    Their gamma-ray emission is thought to originate from these jets \citep{Albert2007,Zanin2016,Abdo2009c,Tavani2009}. 
    SS~433 is another notable example, as its jets produce gamma rays far from its jet \citep{HAWC2018,Rasul2019,Xing2019a,Fang2020,Li2020SS433}. 

    \item \textbf{Accreting Millisecond Pulsars:} Accreting millisecond pulsar SAX J1808.4-3658 is the only one candidate of such system detected in gamma rays, though current detection is not yet significant \citep{Emma2016}. 
\end{itemize}

1A~0535+262 is one of the best-studied HMXB accreting pulsars, discovered in 1975 with a 104 s pulsation period \citep{Rosenberg1975} and an orbital period of $\sim 111$ days \citep{Coe2006}. 
It is a highly magnetized pulsar accreting from the O9.7IIIe companion star \citep{Steele1998}, located at a distance of 1.8$\pm$0.1 kpc \citep{Bailer2018}. 
Its orbit is moderately eccentric ($e=0.47\pm0.02$) \citep{Finger1996}.

Since its discovery, 1A~0535+262 has exhibited several X-ray outbursts with peak fluxes ranging from 0.1 to 12.5 Crab. 
After the launch of \fermi satellite, three giant X-ray outbursts have been detected in 2009 \citep{Acciari2011}, 2011 \citep{Sartore2015}, and 2020 \citep{Kong2022,Mandal2022}, accompanied by non-thermal radio emission during the latest one \citep{Eijnden2020}. 
These episodes have motivated searches for correlated gamma-ray emission.

Past studies include a search for gamma rays during the 2009 outburst using \lat, which set a 99\% confidence upper limit of $F(>0.2{\,\rm GeV}) < 1.9 \times 10^{-8}$ photons cm$^{-2}$ s$^{-1}$ \citep{Acciari2011}. 
More recently, a marginal 3.5$\sigma$ persistent gamma-ray signal was reported, possibly originating from this HMXB \citep{Harvey2022}.

In this work, we present a comprehensive search for gamma-ray emission from 1A~0535+262 using over 13 years of \lat data, including the three giant outbursts and a earlier double-peaked outburst. 
Our analysis includes a detailed search for both steady and transient gamma-ray signals, as well as a pulsation search. 

\section{Analysis and Results}
\label{1A0535.analysis}

\subsection{Timing Solutions}
\label{1A0535.timing}

\begin{table}[h!]
	\begin{center}
		\caption{Timing solutions for 1A 0535+262 during different outbursts. Reproduced from \citet{Hou2023}. }
		\label{tab:spin}
		\begin{tabular}{c|c|c|c|c} 
			\hline
			Parameters                          & 2009 double outburst$^{a}$    & 2009 outburst$^{a}$        & 2011 outburst$^{b}$         & 2020 outburst$^{c}$ \\
		\hline
		Epoch (MJD)                             & 55040   & 55166.99     & 55616.202    & 59170         \\
		$T_{\rm start}$ (MJD)                   & 55040   & 55166.99    & 55608        & 59159.15      \\
		$T_{\rm stop}$ (MJD)                    & 55070   & 55201       & 55637        & 59207.92       \\
		$\nu_0$($\rm 10^{-3}\,Hz$)              & 9.66041(4)  & 9.6618(1)   & 9.6793(1)    & 9.66045(2)      \\
		$\nu_1$($\rm 10^{-12}\,Hz\,s^{-1}$)    &   0.67(3)    & -4.6(6)  & 6.43(5)    & 19.27(4)    \\
		$\nu_2$($\rm 10^{-17}\,Hz\,s^{-2}$)    &      & 3.3(0.3)     & 0.121(7)   & 1.17(4)    \\
		$\nu_3$($\rm 10^{-23}\,Hz\,s^{-3}$)    &      &    -4.8(6)      & -1.43(9)   & -5.8(2)    \\
		$\nu_4$($\rm 10^{-29}\,Hz\,s^{-4}$)    &      &     3(1)          & 1.67(5)   & -2.8(9)   \\
		$\nu_5$($\rm 10^{-36}\,Hz\,s^{-5}$)    &      &    -8(6)          &         & 737(7)     \\
		$\nu_6$($\rm 10^{-39}\,Hz\,s^{-6}$)    &      &                  &         & -1.6(2)      \\
		$\nu_7$($\rm 10^{-45}\,Hz\,s^{-6}$)    &      &                  &         & -5(2)      \\
		$\nu_8$($\rm 10^{-50}\,Hz\,s^{-6}$)    &      &                  &         & 3(1)      \\
		$\nu_9$($\rm 10^{-56}\,Hz\,s^{-6}$)    &      &                  &         & -8(2)      \\
		$\nu_{10}$($\rm 10^{-62}\,Hz\,s^{-6}$)    &      &                 &        & 7(1)      \\

		\hline
		\end{tabular}
	\begin{tablenotes}
     \scriptsize
     \item $^{a}$: Derived form the \fermi/GBM monitoring. 
     \item $^{b}$: Adopted from \cite{Sartore2015}.
     \item $^{c}$: Derived from \hxmt observations \citep[][see text]{Wang2022}.
    \end{tablenotes}
	\end{center}
\end{table}

To account for the spin and orbital evolution of 1A 0535+262 during different outbursts, we adopted the following timing solutions: 
\begin{itemize}
    \item For the 2009 double-peaked and giant outbursts, we used the spin and orbital ephemerides from the \fermi/GBM monitoring\footnote{\url{https://gammaray.nsstc.nasa.gov/gbm/science/pulsars/lightcurves/a0535.html}}. 
    \item For the 2011 outburst, we employed the \texttt{INTEGRAL} timing solution reported by \citet{Sartore2015}.
    \item For the 2020 outburst, we derived a high-precision spin evolution using the \hxmt data \citep{Wang2022} and a phase-connection technique \citep{Deeter1981}. 
    Practically, we extracted time-of-arrival (TOA) data by folding 1000 s background-subtracted light curves in the 25-80 keV range, where the pulse profile is relatively stable. 
    This interval matches the typical good time duration for \hxmt. 
    Each TOA was estimated by cross-correlating individual pulse profiles with an average template, and the resulting spin parameters were obtained using the software {\sc Tempo2} \citep{Hobbs2006}. 
\end{itemize}

 The final timing solutions for the further analysis are summarized in Table~\ref{tab:spin}.

\subsection{Data Set and Spectral Analysis}
\label{1A0535.spectral_analysis}

We analyzed approximate 14-years (from 2008 August 4 to 2022 June 9) of \lat Pass 8 data \citep{pass8Atwood} in the 0.1-300 GeV energy range. 
The data were selected using the SOURCE event class with a zenith angle cut of $<90^\circ$ to reduce Earth limb contamination. 
The GTIs were defined by the standard filtering criteria \texttt{``DATA\_QUAL>0 \&\& LAT\_CONFIG==1"} to ensure data quality. 

The ROI was a $10^\circ$ circle centered on 1A 0535+262: $\alpha=84.7274^\circ$, $\delta=26.3158^\circ$) (J2000 coordinates from SIMBAD\footnote{\url{http://simbad.u-strasbg.fr/simbad/}}. 
The analysis was performed using the P8R3\_SOURCE\_V3 IRFs and the Fermitools (v2.2.0).

To construct the source model, we included all 4FGL-DR3 sources within a $20^\circ$ radius of 1A 0535+262 \citep{4DFL-DR3}, along with the latest Galactic diffuse model (``gll\_iem\_v07.fits'') and isotropic background (``iso\_P8R3\_SOURCE\_V3\_v1.txt''). 
Sources within $5^\circ$ of 1A 0535+262 had their normalizations and spectral indices set free, except for 4FGL J0534.5+2201i, which was fixed to account for the inverse Compton component of the Crab Nebula. 
1A 0535+262 was manually added as a point source with a PL spectrum. 

We performed a binned likelihood fit in a $14^\circ \times 14^\circ$ ROI, using a $0.1^\circ \times 0.1^\circ$ spatial grid and 30 logarithmically spaced energy bins. 
The significance of 1A 0535+262 was assessed using the TS. 
Initial global binned likelihood fits were performed with fixed spectral indices (2.0, 2.3, 3.0) to evaluate potential emission scenarios. 
A harder spectrum would improve detection prospects at higher energies. 
We then repeated the fitting for individual X-ray outbursts and a combined stack of all outbursts to enhance the detection possibility and statistics. 

No significant gamma-ray signal from 1A~0535+262 was found in any of these cases. 
Accordingly, we computed 95\% confidence level flux upper limits, as summarized in Table~\ref{tab:fermiflux}. 

\begin{table}
\scriptsize
\begin{center}
\begin{threeparttable}
\caption{\lat spectral analysis results for 1A~0535+262. Reproduced from \citet{Hou2023}. } 
\label{tab:fermiflux}
\begin{tabular}{lcccc}
\toprule
Period$^{a}$ & Time Range (MJD) & Spectral Index & TS & Energy Flux Upper Limit$^{b}$ \\[0.8ex]
 &  &  &  & $(10^{-12}\, \rm erg \, cm^{-2}\,s^{-1})$ \\[0.8ex]
\midrule
\multicolumn{5}{c}{Whole Dataset} \\  
\midrule
Full & 54682-59739 & 2.0 & 0.0 & 0.6 \\ 
Full & 54682-59739 & 2.3 & 0.0 & 0.7 \\ 
Full & 54682-59739 & 3.0 & 0.0 & 1.2 \\ 
\midrule
\multicolumn{5}{c}{Stacked Outbursts} \\  
\midrule
Rising+Falling & 55040-59207 & 2.0 & 0.0 & 11.9 \\ 
Rising+Falling & 55040-59207 & 2.3 & 0.0 & 9.3 \\ 
Rising+Falling & 55040-59207 & 3.0 & 0.0 & 20.9 \\ 
\midrule
\multicolumn{5}{c}{2009 Double-Peaked Outburst} \\  
\midrule
Rising+Falling & 55040-55070 & 2.0 & 0.4 & 34.9 \\ 
Rising+Falling & 55040-55070 & 2.3 & 0.4 & 27.3 \\ 
Rising+Falling & 55040-55070 & 3.0 & 0.0 & 21.9 \\ 
\midrule
\multicolumn{5}{c}{2009 Giant Outburst} \\  
\midrule
ALL & 55165.9-55249.6 & 2.0 & 0.0 & 9.9 \\ 
ALL & 55165.9-55249.6 & 2.3 & 0.0 & 10.5 \\ 
ALL & 55165.9-55249.6 & 3.0 & 0.0 & 16.5 \\ 
Rising+Falling & 55165.9-55193.6 & 2.0 & 0.0 & 27.9 \\ 
Rising+Falling & 55165.9-55193.6 & 2.3 & 0.0 & 24.7 \\ 
Rising+Falling & 55165.9-55193.6 & 3.0 & 0.0 & 28.4 \\ 
Rising & 55165.9-55177.6 & 2.0 & 0.0 & 49.3 \\ 
Rising & 55165.9-55177.6 & 2.3 & 0.0 & 45.8 \\ 
Rising & 55165.9-55177.6 & 3.0 & 0.0 & 45.4 \\ 
Falling & 55178.4-55193.6 & 2.0 & 0.0 & 39.4 \\ 
Falling & 55178.4-55193.6 & 2.3 & 0.0 & 31.7 \\ 
Falling & 55178.4-55193.6 & 3.0 & 0.0 & 29.5 \\ 
Apastron & 55199.4-55216.6 & 2.0 & 0.0 & 23.6 \\ 
Apastron & 55199.4-55216.6 & 2.3 & 0.0 & 18.1 \\ 
Apastron & 55199.4-55216.6 & 3.0 & 0.0 & 18.6 \\ 
Periastron & 55230.4-55249.6 & 2.0 & 0.0 & 44.9 \\ 
Periastron & 55230.4-55249.6 & 2.3 & 0.0 & 39.3 \\ 
Periastron & 55230.4-55249.6 & 3.0 & 0.0 & 34.5 \\ 
\midrule
\multicolumn{5}{c}{2011 Giant Outburst} \\  
\midrule
Rising+Falling & 55600-55645 & 2.0 & 0.0 & 14.6 \\ 
Rising+Falling & 55600-55645 & 2.3 & 0.0 & 13.4 \\ 
Rising+Falling & 55600-55645 & 3.0 & 0.0 & 19.6 \\ 
Rising & 55600-55617 & 2.0 & 0.0 & 33.0 \\ 
Rising & 55600-55617 & 2.3 & 0.0 & 30.7 \\ 
Rising & 55600-55617 & 3.0 & 0.0 & 40.0 \\ 
Falling & 55618-55645 & 2.0 & 0.0 & 30.9 \\ 
Falling & 55618-55645 & 2.3 & 0.0 & 30.6 \\ 
Falling & 55618-55645 & 3.0 & 0.0 & 20.0 \\ 
\midrule
\multicolumn{5}{c}{2020 Giant Outburst} \\  
\midrule
Rising+Falling & 59159-59207 & 2.0 & 0.0 & 33.4 \\ 
Rising+Falling & 59159-59207 & 2.3 & 0.0 & 27.2 \\ 
Rising+Falling & 59159-59207 & 3.0 & 0.0 & 31.4 \\ 
Rising & 59159-59172.5 & 2.0 & 4.2 & 131.5 \\ 
Rising & 59159-59172.5 & 2.3 & 3.2 & 91.8 \\ 
Rising & 59159-59172.5 & 3.0 & 0.5 & 90.0 \\ 
Falling & 59173-59207 & 2.0 & 0.0 & 24.2 \\ 
Falling & 59173-59207 & 2.3 & 0.0 & 22.0 \\ 
Falling & 59173-59207 & 3.0 & 0.0 & 30.5 \\ 
\bottomrule
\end{tabular}
\begin{tablenotes}
\scriptsize
\item{$^{a}$ Full: entire dataset; ALL: includes rising, falling, apastron, and periastron phases.} 
\item{$^{b}$ 95\% confidence level upper limits in the 0.1$-$300 GeV range.}
\vspace{0.5cm}
\end{tablenotes}
\end{threeparttable}
\end{center}
\end{table}

\subsection{Long-term and Orbital Variability}
\label{1A0535.variability}

\begin{figure}[htbp]
    \centering
    \includegraphics[width=0.32\linewidth]{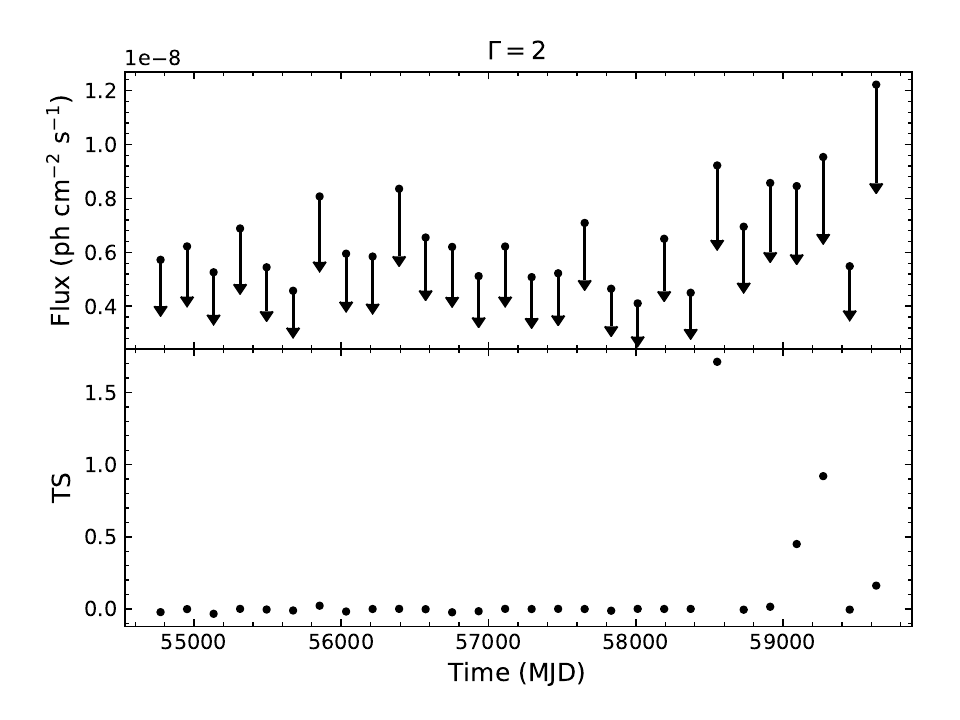}
    \includegraphics[width=0.32\linewidth]{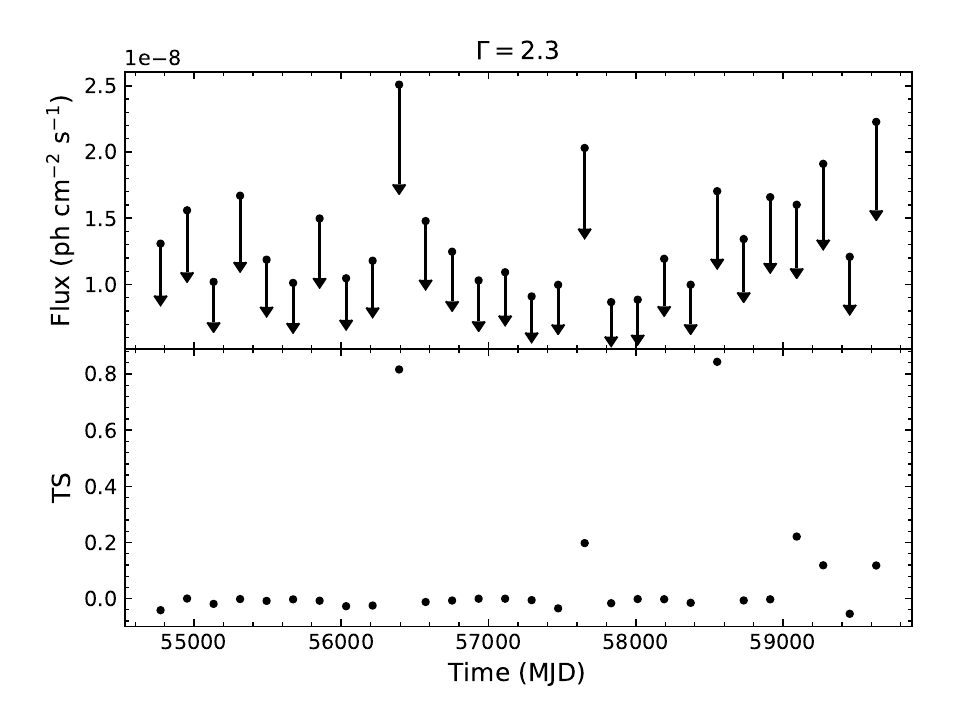} 
    \includegraphics[width=0.32\linewidth]{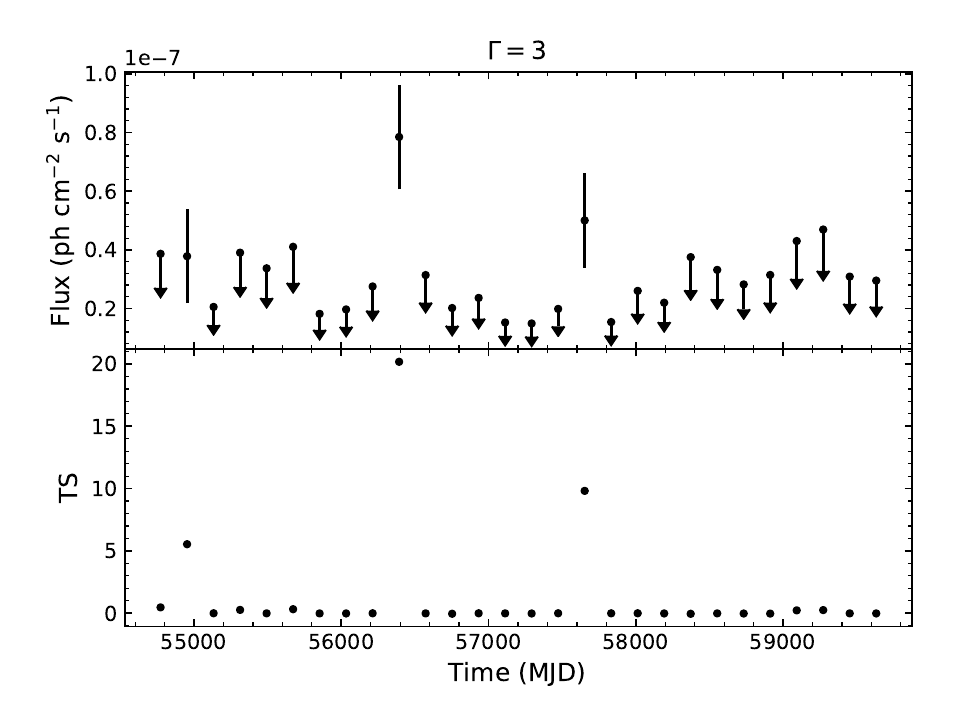} 
    \caption{Long-term gamma-ray light curves for 1A~0535+262 with different fixed spectral indices. Arrows indicate 95\% C.L. upper limits for bins with TS $<4$. Taken from \citet{Hou2023}.}
    \label{fig:lc}
\end{figure}

To assess long-term variability, we generated 180-day binned light curves (Figure~\ref{fig:lc}) in the 0.1-300 GeV energy range. 
In each time bin, the full model from the global fit was used, with normalizations free for sources within $3^\circ$ of 1A 0535+262. 
Bins with TS$<$4 were assigned 95\% confidence upper limits.

No significant variability was detected for PL indices of 2.0 and 2.3. 
For the index-3.0 case, two bins showed TS values of $\sim$10 and $\sim$20 (corresponding to about 3$\sigma$ and 4$\sigma$), but these periods did not coincide with major X-ray activities, indicating no clear correlation between its gamma and X rays. 
The overall TS$_{\rm var}$ for the light curve with a fixed index of 3 was only 1.7$\sigma$ (27 DOF), insufficient for a statistically significant detection. 

\begin{figure}[htbp]
    \centering
    \includegraphics[width=0.32\linewidth]{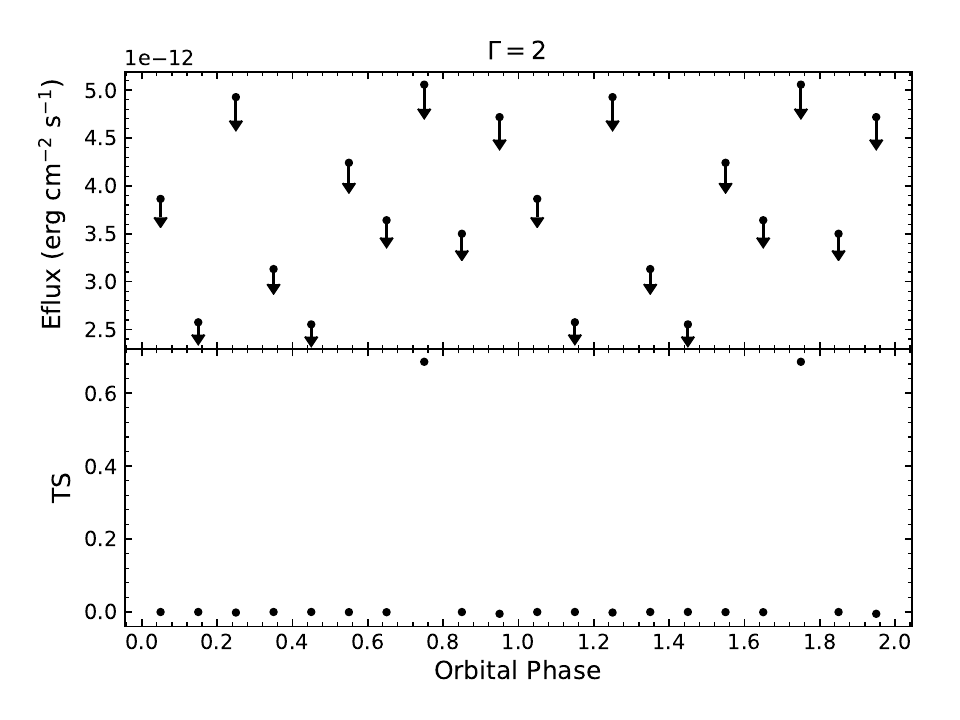}
    \includegraphics[width=0.32\linewidth]{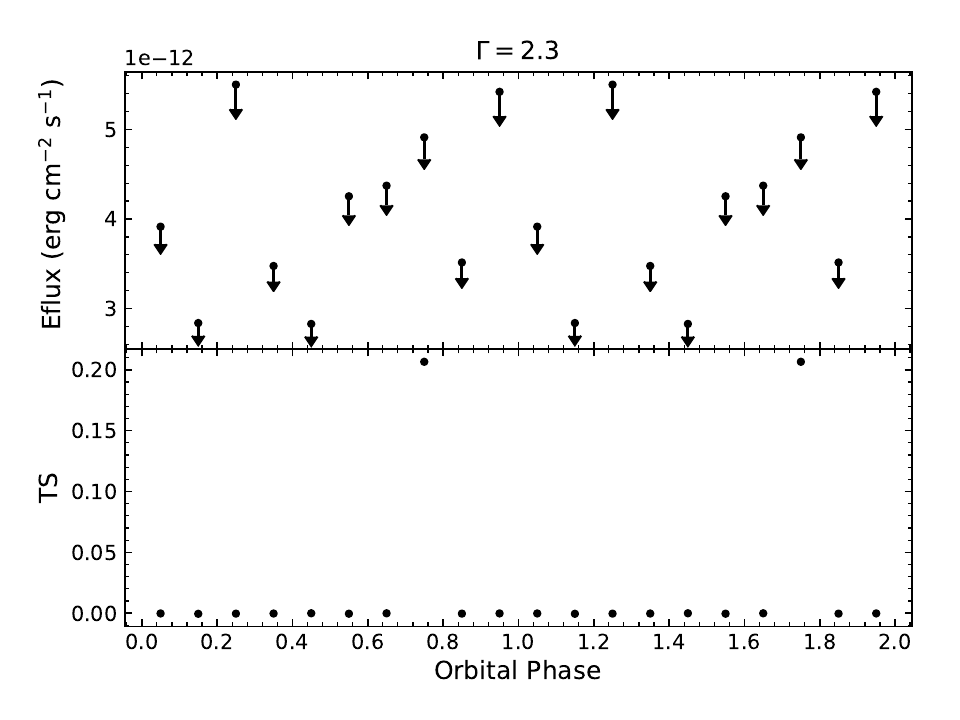} 
    \includegraphics[width=0.32\linewidth]{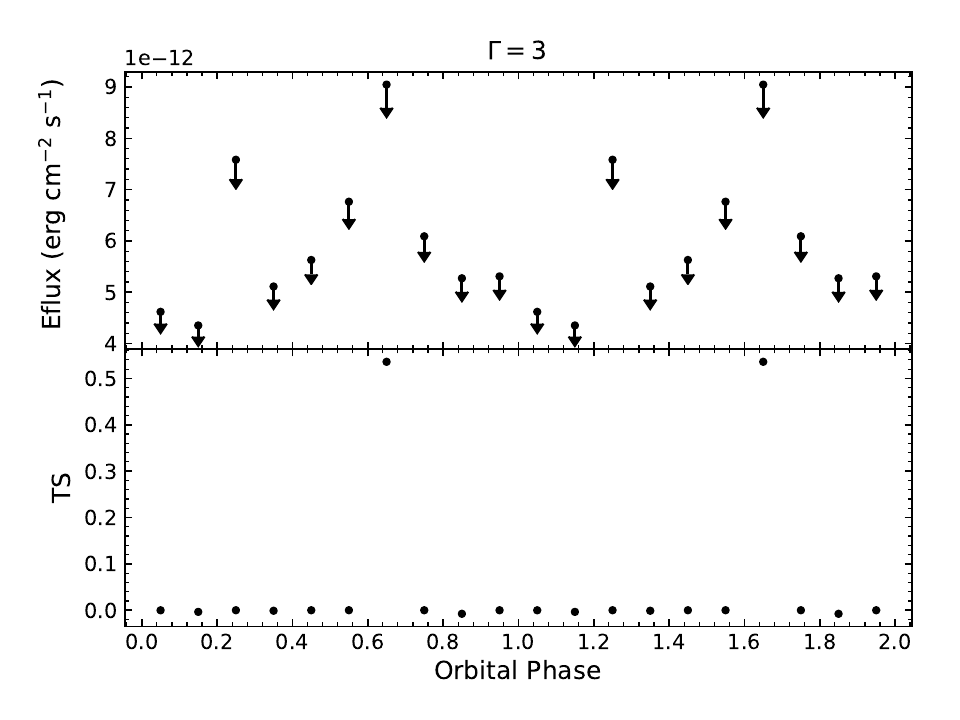} 
    \caption{Orbital light curves for 1A~0535+262 with different fixed spectral indices. Arrows indicate 95\% C.L. upper limits for bins with TS $<4$. Taken from \citet{Hou2023}.}
    \label{fig:orbit10bins}
\end{figure}

Orbital variability was also examined by folding the \lat data into 10 orbital bins, following the same fitting setup as for the long-term variability analysis. 
Again, no significant modulation was found, as shown in Figure~\ref{fig:orbit10bins}. 

\subsection{Pulsation Search}
\label{1A0535.pulsation}

We also searched for gamma-ray pulsations from 1A~0535+262 using \lat data within $1^\circ$ of the source in the 0.1-300 GeV range. 
Pulsation significance was assessed using the weighted H-test \citep{kerr2011}, which extends the original test by \citet{Jager1989} to include photon weights. 
The H-test statistic is defined as
\begin{equation}
    H_{mw} = \max \left[ Z^2_{iw} - c \times (i-1) \right], \,\,\, 1 \leq i \leq m,
\end{equation}
where 
\begin{equation}
    Z^2_{mw} = \frac{2}{\sum_{i=1}^N w_i^2} \sum_{k=1}^m \left(\alpha_{wk}^2 + \beta_{wk}^2 \right),
\end{equation}
and
\begin{equation}
    \alpha_{wk} = \sum_{i=1}^N w_i \cos(2\pi k \phi_i), \;\;\;\; \beta_{wk} = \sum_{i=1}^N w_i \sin(2\pi k \phi_i).
\end{equation}
$N$ is the number of photons, $\phi_i$ is the spin phase, $w_i$ is the weight, and $m$ is the maximum number of harmonics considered. 
For the offset for a harmonic (c), we adopted the standard offset $c=4$ and verified that varying the maximum harmonic up to $m=20$ did not change the results.

Photon weights, $w(E, \Delta \theta)$ (function of photon energy E and angular distance $\Delta \theta$ to the target), were computed using the Simple Weights method \citep{Bruel2019,Smith2019}, which provides a rough estimate of the probability that a given photon originates from the target source. 
Assuming a faint target compared to the background, the weights are given by
\begin{equation}
    w(E, \Delta \theta) = f(E) \times g(E, \Delta \theta),
\end{equation}
where
\begin{equation}
    f(E) = \exp \left(-2\log_{10}^2 \left( \frac{E}{E_{\rm ref}} \right) \right)
\end{equation}
is the weight at the target position, depending on the spectra of the target and background and also on the PSF of the \lat. 
\begin{equation}
    g(E, \Delta \theta) = \left(1 + \frac{9\Delta \theta^2}{4\sigma_{\rm psf}^2(E)}\right)^{-2}
\end{equation}
is the geometrical factor, with the PSF 68\% containment angle as
\begin{equation}
    \sigma_{\rm psf}(E) = \sqrt{p_0^2 (E/100)^{-2p_1} + p_2^2},
\end{equation}
where $p_0=5.445$, $p_1=0.848$, and $p_2=0.084$ for LAT P305 Pass 8 data \citep{pass8Atwood}. 
The reference energy is defined as $E_{\rm ref}=10^{\mu_w}$ \citep{Bruel2019}. 
Therefore, searching for pulsations is to identify the maximum H-test by scanning over $\mu_w$. 
Following \citet{Smith2019}, we used $\mu_w = 3.6$ as a reasonable choice, but also verified that scanning over a range of $\mu_w$ values did not significantly alter the results.

We folded \lat photons using the ephemerides listed in Table~\ref{tab:spin} and restricted the search to intervals where the ephemerides were valid. 
Despite this, no significant pulsations were detected in any of the observed outbursts.

To optimize the H-test further, we experimented with different energy, radius cuts and weighting methods (i.e. adding or not adding). 
While this approach produced signals approaching 4$\sigma$ in some cases, the large number of trials significantly reduced their significance, whore corrected value would decrease to 0 \citep{Hou2023}. 
In summary, no statistically significant pulsations were detected for 1A~0535+262.

\section{Discussion and Conclusions}
\label{1A0535.conclusion}

\subsection{Comparison with Previous Studies}

Our analysis yields significantly different results from those of \citet{Harvey2022}, where they reported correlated gamma-ray and X-ray outbursts from 1A~0535+262. 
In contrast, we did not find any significant persistent, transient, or pulsed gamma-ray emission from this source, despite analyzing over 13 years of \lat data. 
The two weak TS peaks ($\sim$20 and $\sim$10) in our long-term light curves do not align with known X-ray outburst periods, challenging their claim of a direct correlation.

Several methodological differences may explain this discrepancy. 
For instance, we utilized the newer 4FGL-DR3 catalog, P8R4\_SOURCE\_V3 IRFs, and updated diffuse and extended source templates, while they based their analysis on the older 4FGL-DR2 catalog, P8R3\_SOURCE\_V2 IRFs, and 8-year templates (see their Table 1). 
Our ROI was also larger, extending to $5^\circ$ for spectral fitting and orbital variability, and to $3^\circ$ for long-term variability studies, compared to their $1^\circ$. 
Additionally, we fixed the spectral index in certain cases to account for potential emission mechanisms, while they allowed it to vary freely. 
Finally, they did not account for energy dispersion, whereas we did. 

\subsection{Implications for Gamma-ray Emission Mechanisms}

Our non-detection places stringent constraints on gamma-ray production in this system. 
Gamma-ray binaries like LS I +61 303 are known to emit non-thermal radiation across multi-wavelengths, including radio and X-rays, often linked to particle acceleration in shocks or magnetospheric gaps. 
However, 1A~0535+262 lacks significant quiescent radio emission, suggesting a different physical emission mechanism. 
For instance, a hadronic model proposed by \citet{Cheng1991}, where protons accelerated in the magnetosphere impact a transient accretion disk, predict much higher gamma-ray fluxes than our upper limits. 
For 1A~0535+262, \citet{Orellana2007} estimated a flux of $3.8 \times 10^{-8} \, \rm ph \, cm^{-2} \, s^{-1}$ with this model, translating to a gamma-ray luminosity of $\sim 10^{33} \, \rm erg \, s^{-1}$ at 0.3 TeV , which should have been easily detected by the \lat. 
However, our limits, ranging from (2.3--4.7)$\times 10^{32} \, \rm erg \, s^{-1}$ depending on the assumed spectral index, are an order of magnitude lower. 

Additionally, using the X-ray luminosity in the 2-150 keV range reported for the largest 2020 outburst \citep{Kong2021}, estimated at $1.2 \times 10^{38} \, \rm erg \, s^{-1}$, the gamma-to-X-ray luminosity ratio, $L_\gamma / L_X$, is approximately (1.9$-$3.9)$\times 10^{-6}$.

\subsection{Alternative Explanations for the Non-Detection}

Several other factors might explain the lack of detectable gamma-ray emission. 
First, the shocks in this system may simply be too weak to accelerate particles to the required energies. 
Alternatively, the gamma rays might be significantly adsorbed by the dense surrounding material \citep[see e.g.,][]{Orellana2007}. 
Such absorption could vary over time, potentially masking even strong emission during certain phases. 
However, this would likely result in detectable flux variations, which we do not observe. 
Moreover, secondary electrons and positrons from pair production should also contribute to the MeV–GeV band, which remains undetected in our analysis.

Giant (Type II) outbursts, though rare and typically independent of the orbital period, can reach X-ray luminosities close to the Eddington limit, likely due to the formation of transient accretion disks. 
\citet{Eijnden2020} reported a radio counterpart of 1A~0535+262 during its 2020 outburst, marking the first observed coupling of increased X-ray and radio flux for this source. 
This indicates that the radio emission might relate to the accretion state, similar to the behaviour observed from transient Be X-ray binary Swift J0243.6+6124 \citep{Eijnden2018}.

Theoretical studies have proposed that HMXBs can emit gamma rays during accretion periods \citep{Bednarek2009,Bednarek2009b}. 
Observational hints include gamma-ray signals from the accreting millisecond pulsar SAX J1808.4-3658 \citep{Emma2016}, 
emissions during the sub-luminous states of transitional pulsars, potentially linked to propeller-mode outflows or mini pulsar wind nebulae \citep{Papitto2014,Papitto2015,Papitto2019,Veledina2019}, 
and pulsed optical and ultraviolet radiation from SAX J1808.4-3658 \citep{Ambrosino2021}, all suggesting possible particle acceleration in accreting systems.

In our case, the neutron star Eddington luminosity ($L_{\rm Edd} \sim 1.8 \times 10^{38}$ erg s$^{-1}$) far exceeds the observed gamma-ray flux, suggesting that if such emission exists, it is highly inefficient. 
Alternatively, gamma-ray emission might be confined to lower energies, such as below 100 MeV, as proposed for transients like the Be X-ray binary 4U 1036-56 (RX J1037.5-5647), potentially linked to \texttt{AGILE} detections \citep{Li2012}. 
%
Possible MeV missions like AMEGO \citep{McEnery2019}, e-ASTROGAM \citep{Angelis2018}, or the forthcoming COSI \citep{Beechert2022} can be critical for exploring this possibility.

\subsection{Conclusions}

In summary, we analyzed over 13 years of \lat data for 1A~0535+262 and found no significant persistent, transient, or pulsed gamma-ray emission, either over the entire dataset or during X-ray outbursts. 
The 95\% C.L. upper limits on the gamma-ray luminosity, assuming different spectral indices, range from (2.3$-$4.7)$\times 10^{32} \, \rm erg \, s^{-1}$, the most stringent constraints to date. 
Two time bins in the long-term light curve show excesses at roughly 3 and 4$\sigma$, but both occurred when the source was X-ray faint, suggesting no correlation between gamma-ray and X-ray activity. 
No significant orbital modulation was found, indicating that 1A~0535+262 is not a detectable gamma-ray emitter at current \lat sensitivity. 

\newpage
\thispagestyle{empty}
~
\newpage  
\chapter{Characterizing the Gamma-ray Emission Properties of the Globular Cluster M5 with the \lat}
\label{M5}
\textbf{\color{SectionBlue}\normalfont\Large\bfseries Contents of This Chapter\\\\}

This chapter presents the gamma-ray study of the globular cluster M5 (NGC 5904) using 15 years of \lat data, as published in \citet{Hou2024}, where I contributed as the second author, primarily responsible for part of the \fermi analysis and cross-checks.

We first searched for gamma-ray pulsations from the seven known MSPs in M5 using precise radio ephemerides from Arecibo and FAST but found no significant signals. 
We then investigated possible orbital modulation from the six binary MSPs, again with no significant detections.

The overall gamma-ray emission from M5 is well described by an exponentially cutoff power-law model and appears steady on monthly time scales. 
Using the observed luminosity and gamma-ray efficiencies of known MSPs, we estimate that M5 hosts 1 to 10 MSPs, consistent with the currently known sample, though a diffuse component cannot be excluded. 

\newpage
\section{Introduction}
\label{M5.intro}

Globular clusters (GCs) are among the oldest and most dense stellar systems, characterized by high stellar densities \citep[$>1000$ pc$^{-3}$,][]{Sollima2017} and frequent dynamical interactions. 
These conditions significantly enhance the formation rates of low-mass X-ray binaries (LMXBs) within GCs, making them orders of magnitude more common per unit mass than in the Galactic field \citep{Katz1975,Clark1975}. 
As a result, a robust correlation between the number of LMXBs in GCs and the stellar encounter rate ($\Gamma_{\rm c}$) has been established through numerous studies \citep{Pooley2003,Gendre2003,Menezes2023}.

Millisecond pulsars (MSPs) — defined as neutron stars with spin periods typically less than 30 ms — are widely believed to evolve from LMXBs via accretion-driven spin-up processes \citep{Alpar1982,Bhattacharya1991}. 
Currently, about 345 pulsars have been identified in 45 GCs within the Milky Way, with approximately 80\% classified as MSPs\footnote{\url{https://www3.mpifr-bonn.mpg.de/staff/pfreire/GCpsr.html}}. 
In contrast, only 10\% of known Galactic pulsars outside GCs are MSPs\footnote{\url{http://astro.phys.wvu.edu/GalacticMSPs/GalacticMSPs.txt}}. 
This stark difference underscores the unique dynamical environments within GCs that favor the formation of these compact objects. 
Indeed, a positive correlation between the MSP population in GCs and $\Gamma_{\rm c}$ has been observed \citep{Hui2010,Bahramian2013}, reinforcing the connection between LMXB formation and subsequent MSP evolution in dense stellar systems. 

GCs have also emerged as a distinct class of gamma-ray sources, primarily through observations by the \lat. 
To date, high-confidence gamma-ray counterparts have been identified for approximately 39 GCs, all of which are listed in the widely used \cite{Harris1996} catalog\footnote{2010 edition: \url{https://physics.mcmaster.ca/~harris/mwgc.dat}}. 
The prevailing interpretation is that the observed gamma-ray emission from GCs arises from their MSP populations, as first proposed by \cite{Chen1991} and also supported by spectral similarities between MSPs and GCs. 
Several studies have established a relationship between the gamma-ray luminosity ($L_{\gamma}$) of GCs and their encounter rates, further supporting the MSP origin of these $\gamma$-ray emissions \citep[see e.g.,][]{Abdo2010,Bahramian2013,Menezes2023,Feng2023}. 

Exceptionally, gamma-ray pulsations have been detected for several individual MSPs in GCs, including PSR J1823-3021A in NGC 6624 \citep{Freire2011}, PSR B1821-24 in NGC 6626 (M28) \citep{Wu2013,Johnson2013}, and PSR J1835-3259B in NGC 6652 \citep{Zhangp2022}. 
These energetic MSPs, characterized by high spin-down power, likely dominate the overall gamma-ray output of their respective host clusters.

\begin{table}[htbp]
\setlength\tabcolsep{1pt}
\scriptsize
\begin{center}
\caption{Basic properties of the seven MSPs in M5}
\label{tab:M5info}
\scalebox{0.8}{
\begin{tabular}{lccccccccc}
\toprule %
MSP & RA (deg) & Dec (deg) & $P$ (ms) & $P_{\rm b}$ (day) & $e$ & $\dot{E}$ ($10^{34}$ erg s$^{-1}$) & X-ray counterpart & Optical counterpart & Notes \\
\midrule %
M5A & 229.6388 & 2.0910 & 5.55 & ... & ... &$<1.56$ & No & No & isolated \\
M5B & 229.6311 & 2.0876 & 7.95 & 6.858 & 0.138 &$<0.25$ & No & No & likely heavy, not edge-on \\
M5C & 229.6366 & 2.0799 & 2.48 & 0.687 & 0 &$<12.36$ & Yes & Yes & eclipsing BW\\
M5D & 229.6268 & 2.0833 & 2.99 & 1.222 & lower than M5B &0.28-4.98 &Yes & Yes & He WD companion\\
M5E & 229.6388 & 2.0772 & 3.18 & 1.097 & lower than M5B &$<5.27$ &Yes & Yes & He WD companion\\
M5F & 229.6350 & 2.0867 & 2.65 & 1.610 & lower than M5B &$<8.30$ & No & Yes & He WD companion\\
M5G & 229.6197 & 2.0875 & 2.75 & 0.114 & 0 &0.04-3.34 & Yes & No & non-eclipsing BW\\
\bottomrule %
\multicolumn{10}{p{17cm}}{$^{a}$ The eccentricities of M5C and M5G are assumed to be zero due to tidal circularization typical in BW systems \citep[cf. Table 1 in ][]{Zhangl2023}.} \\
\multicolumn{10}{p{17cm}}{$^{b}$ Spin-down powers $\dot{E}$ are estimated based on the intrinsic spin-down rates corrected for gravitational acceleration within the cluster \citep[cf. Table 2 in ][]{Zhangl2023}.}
\end{tabular}}
\end{center}
\end{table}

M5 (NGC 5904) is a relatively bright GC, located at a distance of approximately 7.5 kpc. 
It has a half-mass radius of 5.6 pc, corresponding to an angular diameter of about 2.11$^{\prime}$ on the sky\footnote{\url{http://www.messier.seds.org/m/m005.html}}. 
M5 hosts at least seven MSPs, identified through radio observations using the Arecibo and FAST telescopes \citep{Anderson1997,Mott2003,Hessels2007,Pan2021,Zhangl2023}. 
The basic parameters of these seven MSPs are summarized in Table~\ref{tab:M5info}. 
Their spin periods range from 2.48 to 7.95 ms. 
M5A is an isolated MSP, while the others form binary systems with orbital periods ranging from 0.114 to 6.858 days. 
The companions include low-mass helium white dwarfs (He WDs) and black widow (BW) systems. 
The inferred spin-down powers ($\dot{E}$) of these MSPs also vary significantly, from as low as $4\times10^{32}$ erg s$^{-1}$ to as high as $1.2\times10^{35}$ erg s$^{-1}$, indicating a broad spectrum of potential gamma-ray contributions.

M5 was first suggested as a gamma-ray emitter by \citet{Zhou2015} based on a tentative detection using six years of \lat data, and this association was subsequently confirmed by \cite{Zhang2016} with a more significant detection using seven years of Pass 8 data. 
The cluster is now included as a gamma-ray source (4FGL J1518.8+0203) in the \lat Fourth Source Catalog (4FGL) \citep{4FGL}. 
Given the small angular separations (typically $0.01^{\circ} - 0.02^{\circ}$) between the known MSPs in M5, the \lat's angular resolution ($\sim 0.1^{\circ}$) is insufficient to resolve individual MSPs, making it challenging to distinguish their individual contributions to the overall gamma-ray emission. 

In this work, we present a comprehensive \lat analysis of M5, using the latest data and refined timing solutions for its known MSPs. 
Our primary objectives are to characterize the collective gamma-ray emission from this cluster, assess the contributions from individual MSPs, and evaluate the implications for MSP formation and evolution in dense stellar systems. 

\section{Data Set}
\label{M5.analysis}

We performed a detailed gamma-ray analysis of M5 using 15 years of Pass 8 \lat data (2008 August 4 – 2023 August 4) within a $10^\circ$ region centered on M5. 
The data selection included SOURCE class events in the energy range of 0.1–500 GeV with standard quality filters applied to remove contamination from Solar flares and Gamma-ray bursts (GRBs). 
To further reduce the Earth limb background, we adopted the energy-dependent zenith angle and point spread function (PSF) cuts similar to those in the LAT 4FGL catalog \citep{4FGL}, limiting the zenith angles to $<90^\circ$, $<100^\circ$ and $<105^\circ$ for the 0.1–0.3 GeV, 0.3–1 GeV, and 1–500 GeV bands, respectively. 

A spatial-spectral model was constructed for the ROI by including all 4FGL-DR4 sources within $20^\circ$ around M5, as well as the latest Galactic diffuse emission model (\texttt{gll\_iem\_v07.fits}) and the isotropic component (\texttt{iso\_P8R3\_SOURCE\_V3\_v1.txt}) to account for extragalactic contributions and residual instrumental backgrounds. 
Energy dispersion corrections were also applied, excluding the isotropic component.

We performed a summed likelihood analysis within a $14^\circ \times 14^\circ$ ROI using the \texttt{fermipy} package (v1.2.0) with Fermitools (v2.2.0). 

\section{Analyses and Results}
\label{M5.results}

\subsection{Spatial Analysis}  

\begin{table}[h]
\centering
\caption{Optimized positions with different spectral assumptions for M5 based on high-energy data (1–500 GeV).}
\label{tab:latfit}
\begin{tabular}{lcccc}
\hline
Model & R.A. (J2000) & Dec. (J2000) & 95\% containment radius & TS \\
\hline
LP & 229.6607$^\circ$ & 2.0517$^\circ$ & 0.07$^\circ$ & 73.3 \\
PL & 229.6592$^\circ$ & 2.0485$^\circ$  & 0.07$^\circ$ & 71.1 \\
PLEC4 & 229.6593$^\circ$ & 2.0524$^\circ$ & 0.07$^\circ$ & 73.8 \\
\hline
\end{tabular}
\end{table}

\begin{figure}[h]
\centering
\includegraphics[width=0.49\textwidth]{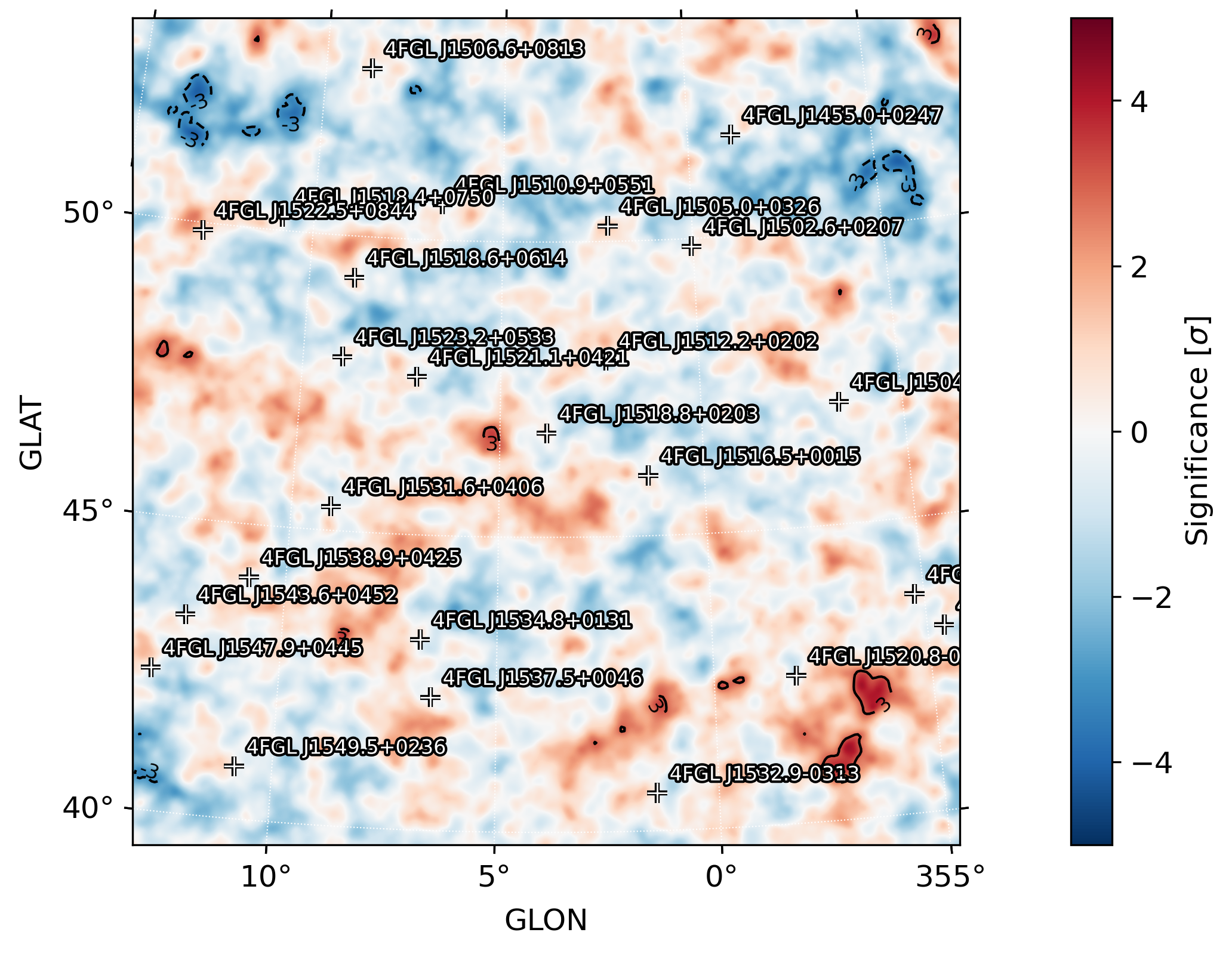}
\includegraphics[width=0.49\textwidth]{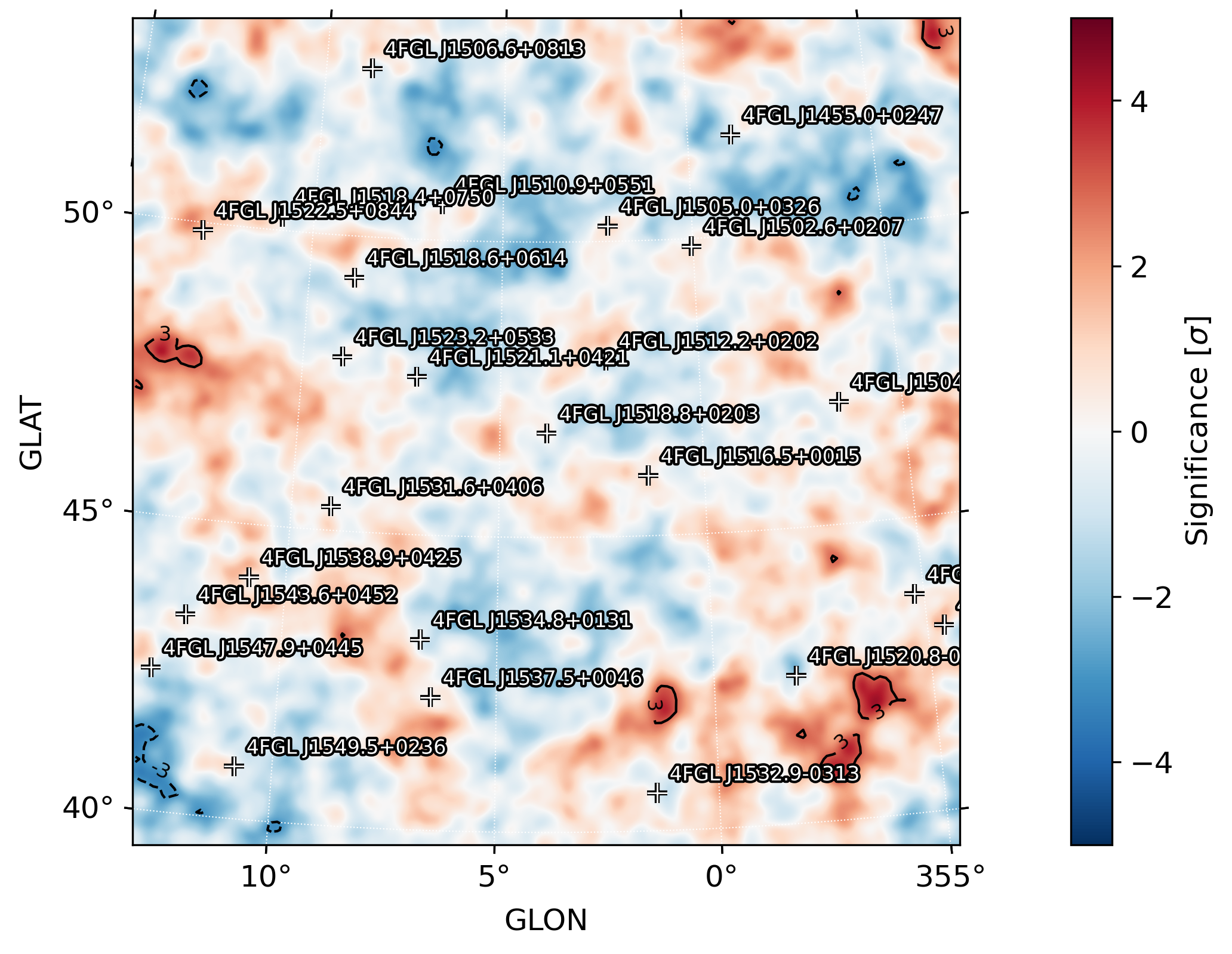}
\caption{Left: Residual map in the 1–500 GeV band for the source localization of M5. 
Right: Residual map in the 0.1–500 GeV band for the spectral fit of M5. 
Reproduced from \citet{Hou2024}.}
\label{fig:residualmap}
\end{figure}

\begin{figure}[htbp]
    \centering 
    \includegraphics[width=0.49\linewidth]{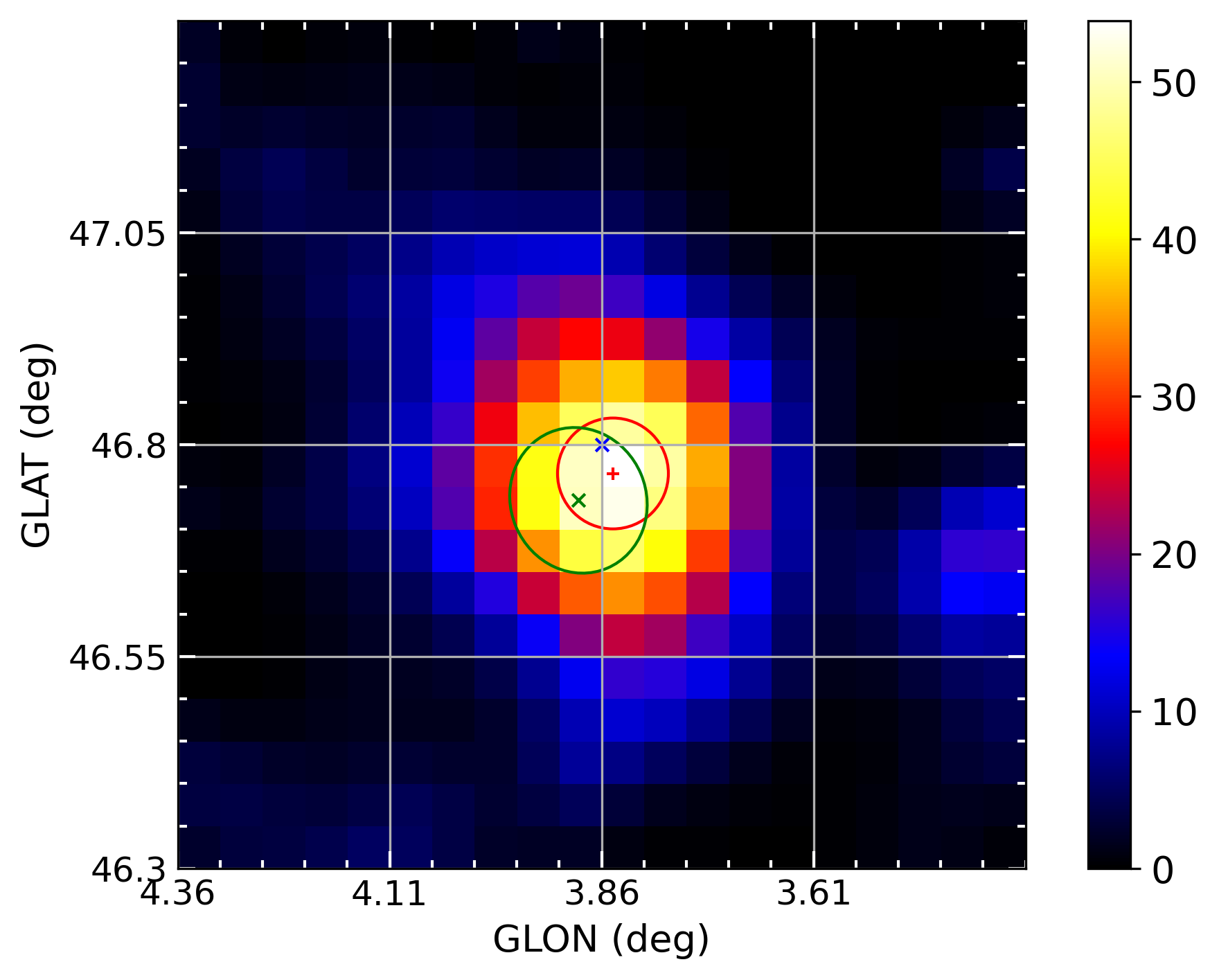}
    \includegraphics[width=0.49\linewidth]{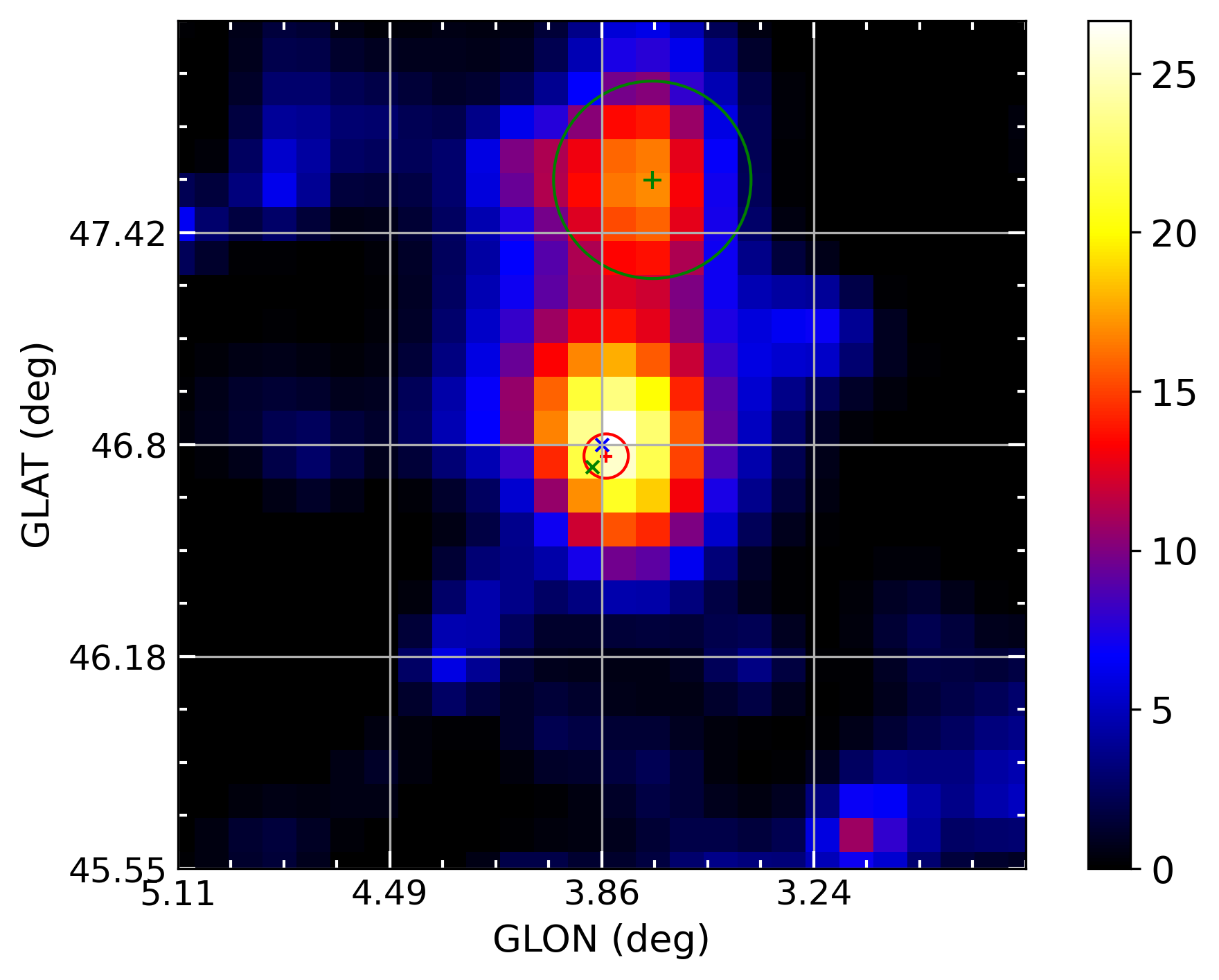}
    \caption{Left: TS excess map for M5 in the 1--500 GeV band, covering a $1^\circ \times 1^\circ$ region with $0.05^\circ$ bins. Shown are the best-fit position and 95\% containment radius derived from the source localization (red plus and circle), 4FGL-DR4 position and 95\% containment ellipse (green cross and ellipse), and the M5 center (blue cross). 
    Right: TS excess map in the 0.1--500 GeV band, covering a $2.5^\circ \times 2.5^\circ$ region with $0.1^\circ$ bins. The green plus and circle indicate the best-fit position and 95\% containment radius for the excess in the 90-day light curve bin 21 (see Section~\ref{lightcurve} and Figure~\ref{M5.fig:lc}). Reproduced from \citet{Hou2024}.}
    \label{fig:tsmap}
\end{figure}

To achieve a precise gamma-ray position for M5, a high-energy analysis (1–500 GeV) was conducted, exploiting the improved angular resolution in this energy band. 
The data were processed using a spatial binning of $0.05^{\circ} \times 0.05^{\circ}$ and ten logarithmic energy bins per decade. 
Maximum likelihood fitting was applied to assess the source location, with three common spectral models considered:  
\begin{itemize}
    \item LogParabola (LP, default model in 4FGL-DR4): 
         \begin{equation}
         \frac{dN}{dE} = N_0 \left(\frac{E}{E_0}\right)^{-(\alpha + \beta \ln(E/E_0))},
         \end{equation} 
    \item PL:
         \begin{equation}
         \frac{dN}{dE} = N_0 \left(\frac{E}{E_0}\right)^{-\Gamma},
         \end{equation} 
    \item PLSuperExpCutoff4 (PLEC4):
         \begin{equation}
         \frac{dN}{dE} = N_0 \left(\frac{E}{E_0}\right)^{-\Gamma_{0} + d/b} \exp\left[ \frac{d}{b^2} \left(1 - \left(\frac{E}{E_0}\right)^b \right) \right]. 
         \end{equation}
\end{itemize}

In which, $N_0$ is the normalization factor at the reference energy $E_0$, and the remaining parameters define the spectral curvature, particularly suited for pulsar-like spectra \citep{Smith2023}. 
The reference energy $E_0$ is fixed to the catalog value for the LP model and to 1 GeV for the PL and PLEC4 models, and index $b$ is fixed to $2/3$ for PLEC4 model. 

The localization results for these models are summarized in Table~\ref{tab:latfit}, revealing consistent source positions with minor differences. 
The results in the case of the PLEC4 model was ultimately selected for further analysis because it has a slightly higher TS value and PLEC4 is physically motivated as a superposition of curvature radiation spectra for a range of electron energies. 
The TS excess map (Figure~\ref{fig:tsmap}, left panel) also presents the results within a $1^\circ \times 1^\circ$ region centered on M5. 
The resulting residual TS map, displayed in the left panel of Figure~\ref{fig:residualmap}, indicates a well-fitted source model for the ROI, as no significant residuals ($>4\sigma$) are observed.

\subsection{Spectral Analysis}  

\begin{table}[h]
\setlength\tabcolsep{4pt}
\centering
\caption{Best-fit spectral parameters for M5 (0.1–500 GeV) using the PLEC4 model.}
\label{tab:spectralfit}
\scalebox{0.78}{
\begin{tabular}{lccccccc}
\hline
  &	 TS	 & $\alpha$  & $\beta$  & $\Gamma$ ($\Gamma_{0}$) & $d$ & Photon Flux  & Energy Flux\\
    &   &      &   &   &   &(10$^{-9}$~cm$^{-2}$~s$^{-1}$)   & (10$^{-12}$~erg cm$^{-2}$~s$^{-1}$)  \\\hline
LP  &90.9	   &$2.11\pm0.20$ & $0.44\pm0.18$	& ... & ...	&$1.41\pm0.60$ & $1.75\pm0.32$	\\
PL  &74.7	& ...  & ...  &$2.24\pm0.09$	& ... 	&$4.02\pm0.89$ & $2.90\pm0.41$	\\
PLEC4  &91.2	& ...   & ...  &$1.70\pm0.25$	& $0.63\pm0.24$	&$1.65\pm0.67$ & $1.78\pm0.32$	\\
\hline
\end{tabular}}
\end{table}

\begin{figure}[htbp]
    \centering 
        \includegraphics[width=0.5\linewidth]{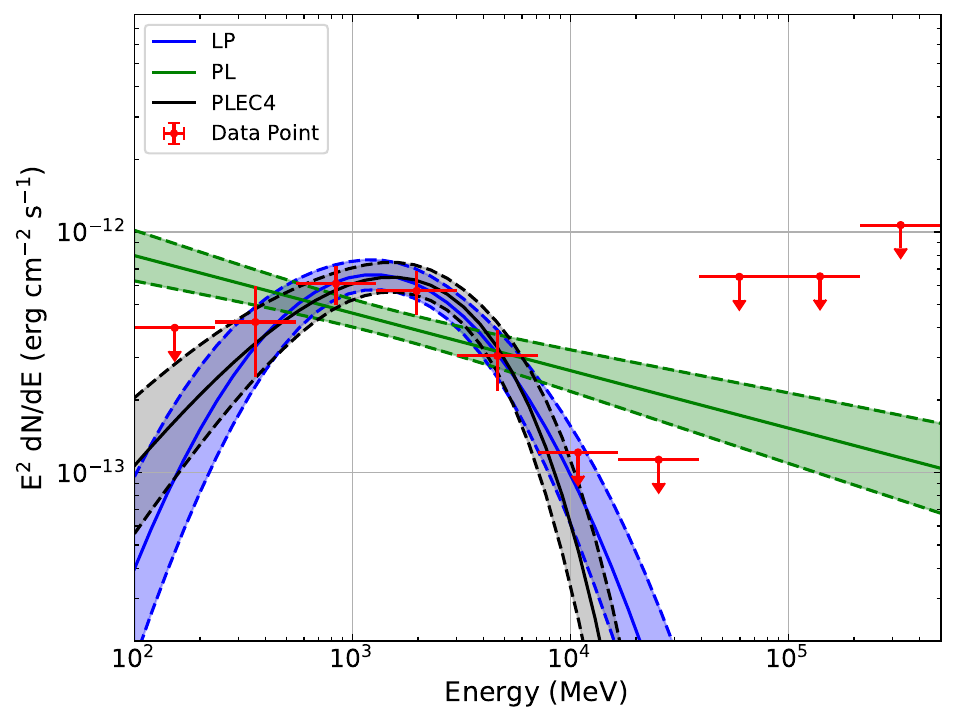}
        \caption{SED of M5. The measured flux points are shown as red dots, with 95\% confidence level upper limits (TS $<$ 4) indicated by red arrows. The best-fit models are overlaid: PLEC4 (black line and shaded region), LP (blue line and shaded region), and PL (green line and shaded region), with their respective uncertainties. 
        Reproduced from \citet{Hou2024}. }
    \label{fig:sed}
\end{figure}

Following localization, a broadband spectral analysis over the 0.1–500 GeV range was performed with three aforementioned spectral models. 
The parameters of sources within $5^\circ$ were allowed to vary freely, while those between $5^\circ$ and $8^\circ$ had their normalizations left free to accommodate potential variability. 
The resulting best-fit parameters are provided in Table~\ref{tab:spectralfit}. 
The PLEC4 and LP models show comparable performance, while the PL model fits poorly. Given its stronger physical foundation, PLEC4 was selected as the preferred model.

We computed the SED of M5 using the best-fit PLEC4 model over the 0.1--500 GeV range, dividing the data into 10 logarithmic bins. 
Background sources were treated as in the broad-band fit. 
For each bin, we fit the flux normalization with a PL model. 
The index was set to the local slope of the PLEC4 model for the first six bins and fixed at 4 for the higher energy bins to account for the steep spectrum and low normalization at those energies. 
For bins with TS$<$4, we calculated upper limits on the flux at the 95\% confidence level. 
Figure~\ref{fig:sed} presents the resulting SED along with the PLEC4, LP, and PL models.

\subsection{Temporal Analysis}  

\subsubsection{Pulsation Search}  
\label{pulsation}

We searched for gamma-ray pulsations from each pulsar within a $2^\circ$ radius. 
Spin phases were computed using the \fermi plugin \citep{Ray2011} for \texttt{TEMPO2}, with radio ephemerides from \citep{Zhangl2023}. 
Ephemerides for M5A, B, C, D, and E are valid from before \fermi's launch to November 2022, while those for M5F and G cover November 2020 to December 2022, as these MSPs were detected only by FAST. 
The weighted H-test \citep{kerr2011} was applied, for both the full \lat dataset and the relevant time range of ephemerides validity.

No significant pulsations were detected for any of the seven MSPs. 
The highest H-test value, 13.8 (corresponding to 2.9$\sigma$), was for M5A, which slightly decreased after accounting for the six trials.

\subsubsection{Long-term Light Curves}  
\label{lightcurve}

\begin{figure}[h]
\centering
\includegraphics[width=0.49\textwidth]{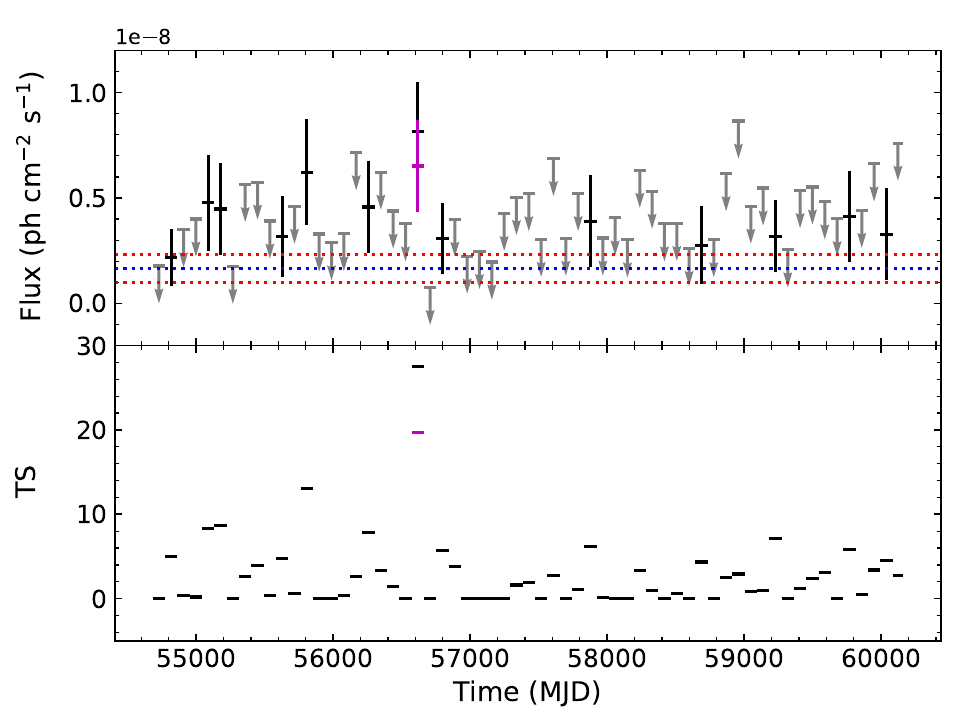}
\includegraphics[width=0.49\textwidth]{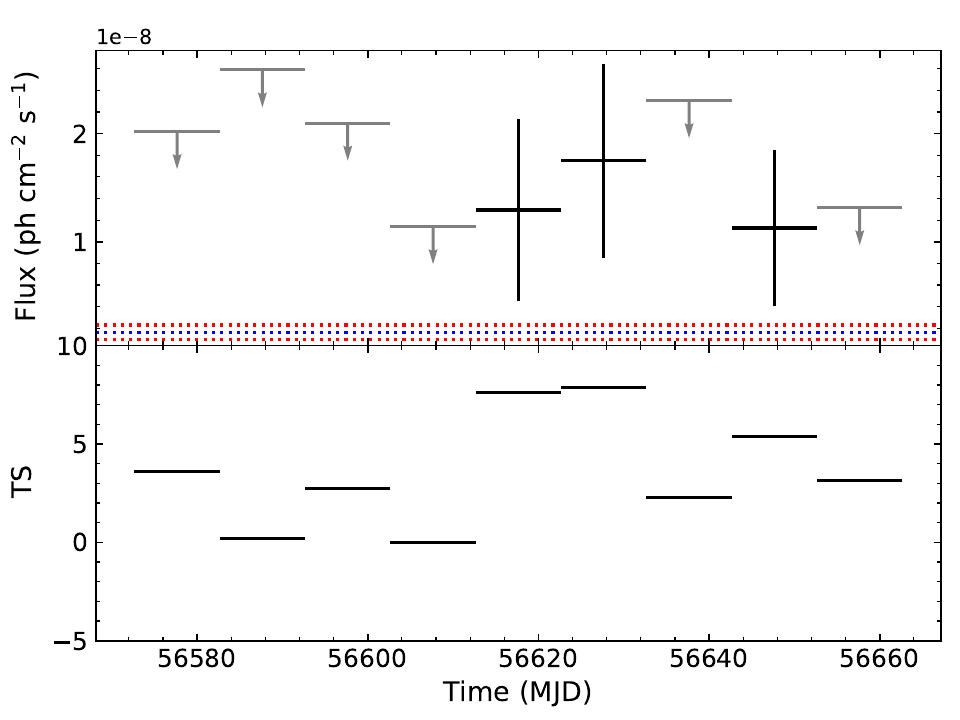}
\caption{The long-term light curve and TS evolution of M5 (0.1--500 GeV). 
Left: 90-day binning. Arrows indicate flux upper limits (95\% C.L.) for TS$<$4. The blue dashed line shows the average flux, with red dashed lines marking the 1$\sigma$ uncertainty. The magenta marker indicates the flux and TS in bin 21 after adjusting for the excess near M5.
Right: Zoom around bin 21 (from MJD 56572 to 56662) with 10-day bins. 
Reproduced from \citet{Hou2024}. }
\label{M5.fig:lc}
\end{figure}

We analyzed the long-term gamma-ray flux variability of M5 by constructing a 90-day binned light curve over the energy range 0.1--500 GeV (Figure~\ref{M5.fig:lc}, left panel). 
We used the PLEC4 model fitted to the broad-band data as a baseline for each bin, with only the normalizations allowed to vary. 
The variability was quantified using TS$_{\rm var}$ \citep{3FGL}, which yielded a value of 62.97. 
Considerably lower than the 99\% confidence threshold of 88.38 ($\chi^2$ distribution with 60 D.O.F.) indicated no significant variability on seasonally time scales.

However, a prominent TS peak of $\sim28$ appeared in bin 21 (MJD 56572-56662, 2013 Oct 7 – 2014 Jan 5). 
A TS map of this bin (Figure~\ref{fig:tsmap}, right panel) revealed an excess near M5, which we localized to $(\alpha, \delta) = (228.97^\circ, 2.49^\circ)$ with a 95\% containment radius of 0.29$^\circ$. 
Despite this proximity, the excess was not associated with M5 (0.78$^\circ$ offset). 
A likelihood fit including the excess gave a TS of about 20 for the excess and 10 for M5. 
Additionally, we didn't detect the excess before and after bin 21.

We searched multiple catalogs for counterparts within the 95\% containment radius of the excess, including the \lat Long-Term Transient Source Catalog \citep[1FLT,][]{Baldini2021}\footnote{\url{https://heasarc.gsfc.nasa.gov/W3Browse/all/fermiltrns.html}}, and blazar catalogs (FERMILBLAZ \citep{fermiblz2019}, BZCat \citep{Massaro2015}, CGRaBS \citep{Healey2008}, CRATES \citep{Healey2007}, WIBRaLs2 and KDEBLLACs \citep{Abrusco2019,Menezes2019}) through HEASARC\footnote{\url{https://heasarc.gsfc.nasa.gov/W3Browse/all/}}. 
No matches were found, and given M5's high Galactic latitude, we found no nova candidates reported during this period\footnote{\url{https://asd.gsfc.nasa.gov/Koji.Mukai/novae/novae.html}}.

After incorporating the excess, the TS$_{\rm var}$ for the 90-day light curve decreased to 57.9, but the TS in bin 21 remained an outlier. 
To assess the significance of this peak, we compared the likelihood of the best-fit model with the fit when fixing the flux of M5 to the average (the excess was included for both), resulting in a $\sqrt{2\Delta \log \mathcal{L}}$ of 2.8$\sigma$, suggesting the peak is likely a statistical fluctuation. 

Further analysis of bin 21 with a 10-day binned light curve (Figure~\ref{M5.fig:lc}, right panel) yielded TS$_{\rm var}$ = 15.2, below the 99\% confidence threshold of 20.1 for 8 DOF, confirming that the emission is consistent with steady flux and that the apparent peak is likely due to statistical noise.

\subsubsection{Orbital Modulation}  

\begin{figure}[htbp]
    \centering 
        \includegraphics[width=0.49\linewidth]{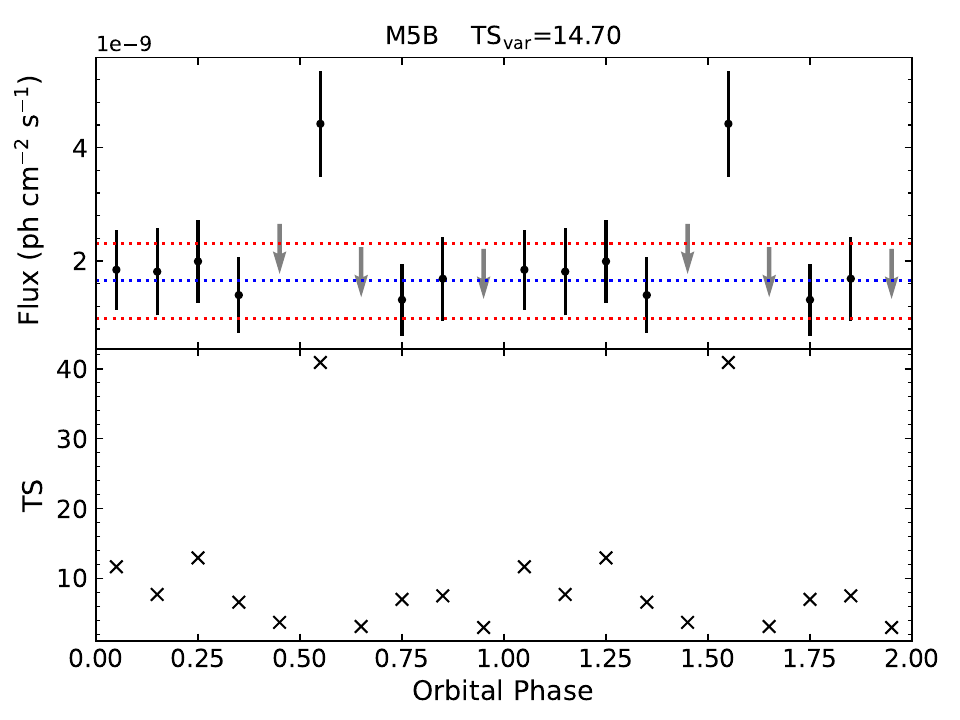}
        \includegraphics[width=0.49\linewidth]{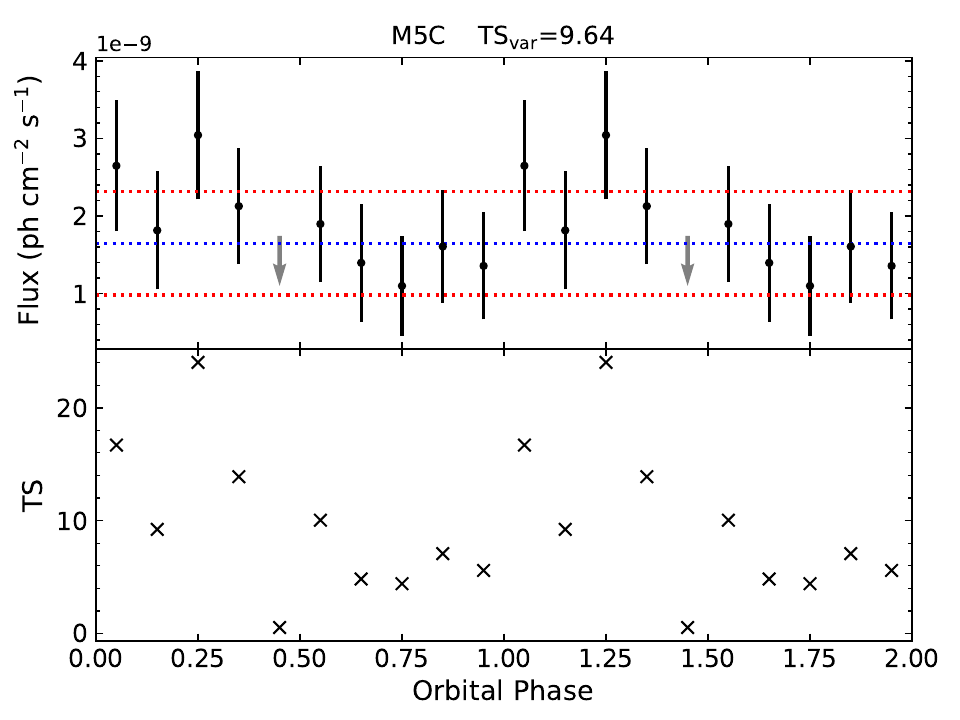}
        \includegraphics[width=0.49\linewidth]{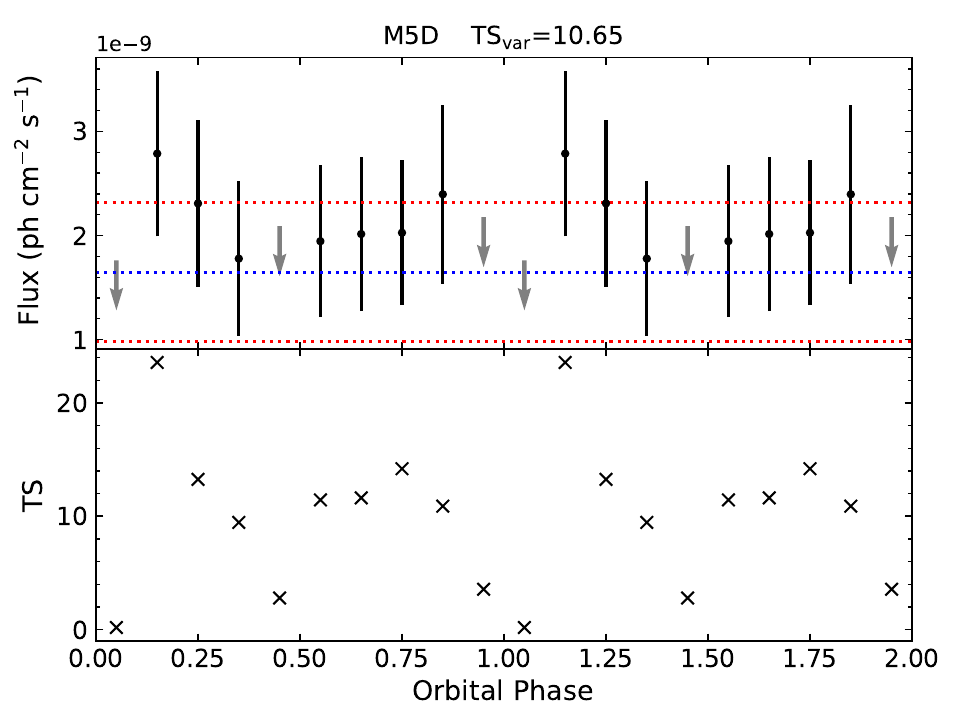}
        \includegraphics[width=0.49\linewidth]{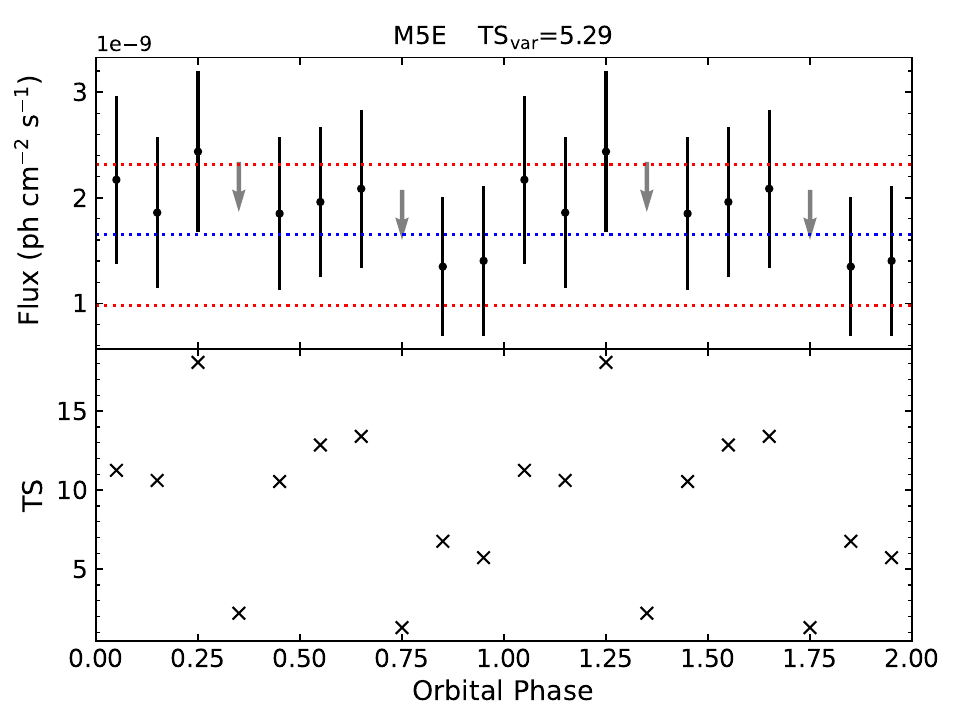}
        \includegraphics[width=0.49\linewidth]{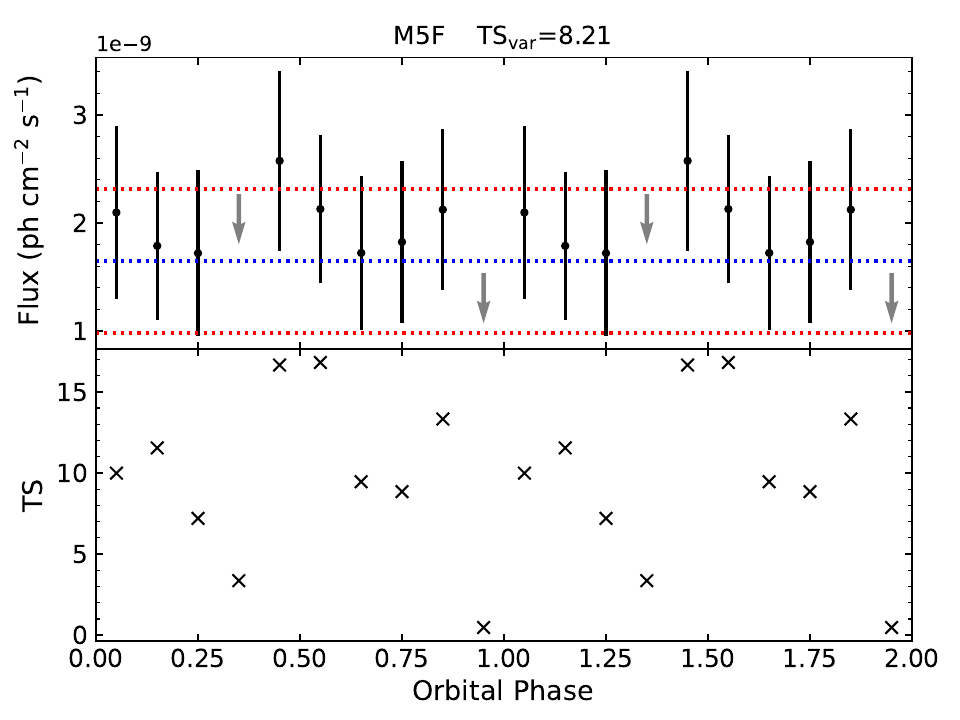}
        \includegraphics[width=0.49\linewidth]{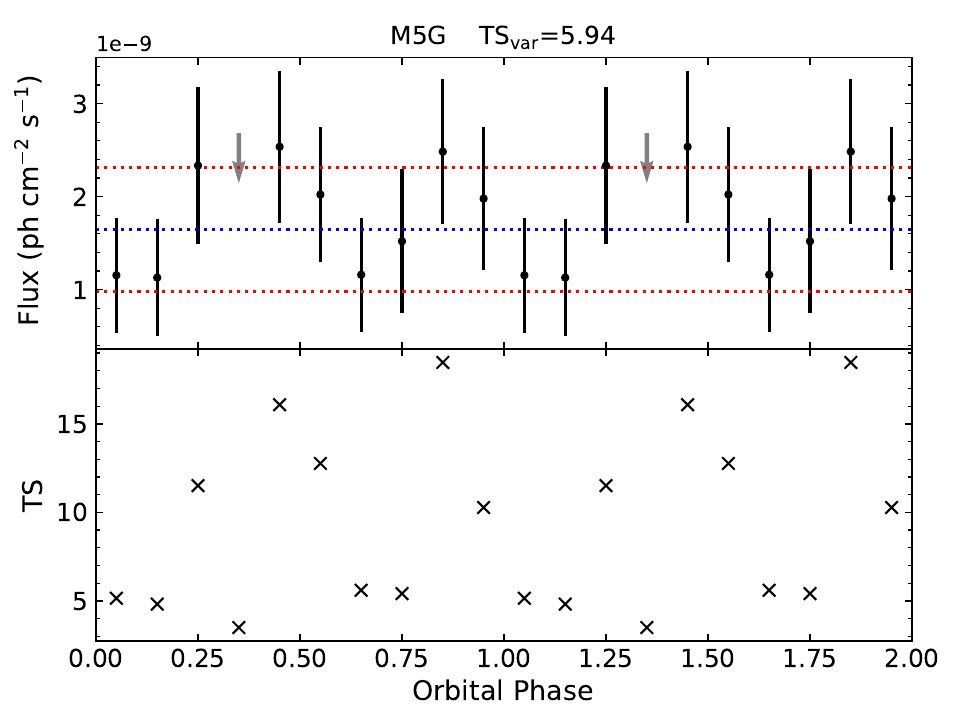}
        \caption{Orbital flux variations for the six binary MSPs in M5. Arrows indicate 95\% C.L. upper limits for bins with TS$<$4. The blue dotted line marks the average flux from the broad-band PLEC4 fit, with red dotted lines showing the corresponding 1$\sigma$ uncertainties. 
        Reproduced from \citet{Hou2024}. }
    \label{fig:orbi_lc}
\end{figure}

\begin{table}
\scriptsize
\begin{center}
\caption{Orbital-Phase Resolved Spectral Fits for M5B (copy from Table 3 in \citet{Hou2024}). }
\begin{tabular}{lccccc}
\toprule %
 Bin  & TS       & $\Gamma_{0}$     & $d$         &Photon Flux      & Energy Flux\\
     &    &  &  &  (10$^{-9}$~cm$^{-2}$~s$^{-1}$)   & (10$^{-12}$~erg cm$^{-2}$~s$^{-1}$)  \\
\midrule %
0.5-0.6 &42  & $2.08\pm0.33$	& $0.61\pm0.38$   & $7.70\pm4.67$  & $5.70\pm1.70$	\\
0.6-0.5 &61	 & $1.72\pm0.33$	& $0.56\pm0.28$   &$1.50\pm0.84$   & $1.58\pm0.39$	\\
\bottomrule %
\label{tab:orbifit_M5B}
\end{tabular}
\end{center}
\end{table}

Binary MSPs can exhibit orbitally modulated emission across multiple wavelengths, providing insights into their binary properties and emission mechanisms. 
We examined the orbital modulation of the six binary MSPs in M5 using two methods, following the approach in \citet{Johnson2015}.

We folded the gamma-ray events within $5^\circ$ of the M5 center using the orbital periods of each MSP, with orbital phases assigned via \texttt{TEMPO2} and radio ephemerides in \citet{Zhangl2023}. 
The time bin was set to 30 s. 
In the first method, to account for possible exposure variations, we generated a null distribution by binning the exposure into 1000 time bins and normalizing \citep{Ackermann2012}. 
We then corrected the exposure for each event within $2^\circ$ of M5, and applied the weighted H-test \citep{kerr2011} to evaluate modulation significance. 
No significant orbital signals were detected, with the largest H-test value being 10.3 (2.4$\sigma$).

The second method involved calculating phase-resolved orbital fluxes using 10 bins per orbit. 
We corrected for exposure variations by creating orbital phase-selected good time intervals (GTIs) from the 30 s counts light curve as done in the first method. 
Variability was quantified via TS$_{\rm var}$, similar to the long-term light curve analysis. 
As shown in Figure~\ref{fig:orbi_lc}, none of the MSPs exhibited significant orbital variability, with the highest TS$_{\rm var}$ of 14.7 for M5B, below the 99\% confidence threshold of 21.7 for the $\chi^2$ distribution with 9 DOF. 

M5B showed a moderately high TS and flux in the 0.5--0.6 orbital bin, suggesting possible phase-dependent variability. 
We quantified this by comparing the $\sqrt{2\Delta \log \mathcal{L}}$ relative to the phase-averaged flux (Table~\ref{tab:latfit}), obtaining a significance of 3.4$\sigma$. 
To test for spectral shape changes, we re-fitted the 0.5--0.6 and 0.6--0.5 orbital bins with all spectral parameters free, as listed in Table~\ref{tab:orbifit_M5B}. 
The spectral parameters remained consistent, indicating that the observed modulation is likely dominated by flux variation rather than shape changes.

\section{Discussion and Concluding Remarks}
\label{discuss}

\subsection{Spectral Characteristics of M5}

From the best-fit parameters, the SED peak energy, $E_{\rm p}$, and curvature, $d_{\rm p}$, can be estimated using the relations from 3PC \citep{Smith2023}:
\begin{equation}
E_{\rm p} = E_{0} \left[1 + \frac{b}{d} (2 - \Gamma_{0}) \right]^{\frac{1}{b}} \, ,
\end{equation}
\begin{equation}
d_{\rm p} = d + b (2 - \Gamma_{0}) \, .
\end{equation}
Here, the maximum theoretical $d_{\rm p} = 4/3$ corresponds to monoenergetic synchrotron or curvature radiation, while smaller values indicate a broader electron energy distribution. 
For M5, we obtained $E_{\rm p} = 1.5$ GeV and $d_{\rm p} = 0.83$, consistent with the dependence of the spectral curvature on the peak energy (Figure 20 in \citet{Smith2023}).

\subsection{Non-detection of Individual Pulsars}

Despite M5 hosting multiple MSPs, none were detected as individual gamma-ray pulsars. 
This is not unexpected, as gamma-ray detections generally require both high spin-down power ($\dot{E}$) and precise ephemerides, which many GC MSPs lack. 

From Table~\ref{tab:M5info}, several M5 MSPs have $\dot{E}$ upper limits comparable to known gamma-ray MSPs, such as M5C with $\dot{E} < 1.2 \times 10^{35}$ erg s$^{-1}$, which approaches the estimated $\dot{E} = 1.8 \times 10^{35}$ erg s$^{-1}$ for the gamma-ray MSP J1835-3259B. 
However, the non-detections likely indicate that the true $\dot{E}$ values are significantly lower than the upper limit, consistent with the generally low magnetic fields expected for recycled pulsars in low-density GCs like M5.

This pattern aligns with previous findings that gamma-ray MSPs are more commonly found in GCs with a very high encounter rate per formed binary ($\gamma_{\rm b}$), such as core-collapsed GCs, NGC 6624 and NGC 6652 \citep{Verbunt_Freire2014}. 
These clusters may host a larger fraction of isolated, slow pulsars due to frequent binary disruptions, leading to higher magnetic field strengths and, consequently, higher $\dot{E}$ values.

In contrast, in low-$\gamma_{\rm b}$ GCs like M5, LMXBs are unlikely to be disrupted, allowing their NSs to fully recycle into fast-spin, low-$B$ MSPs. 
This is supported by observations of 47 Tuc, where all MSPs in binaries exhibit small $\dot{P}$ values, consistent with typical Galactic MSPs \citep{Freire+2017}. 
The absence of individually detectable gamma-ray MSPs in 47 Tuc and $\omega$ Centauri further supports this view \citep{Dai2023}. 
Our endeavor to identify a high-significance gamma-ray MSP in M5 functions as a corroborating evaluation of this hypothesis.

\subsection{Gamma-ray Emission of M5 as a Whole}

Given the lack of individual detections, the observed gamma-ray emission from M5 likely arises from the collective contribution of its MSP population. 
Following the approach of \cite{Johnson2013}, the total number of MSPs can be estimated as
\begin{equation}
N_{\rm MSP} = \frac{L_\gamma}{\langle \dot{E} \rangle \langle \eta_\gamma \rangle} \, ,
\end{equation}
where $L_\gamma = (1.2 \pm 0.2 \pm 0.2) \times 10^{34}$ erg s$^{-1}$ is the gamma-ray luminosity from the fit (Table~\ref{tab:latfit}), $\langle \dot{E} \rangle = (1.8 \pm 0.7) \times 10^{34}$ erg s$^{-1}$ is the average spin-down power for MSPs in GCs \citep{Abdo2009}, and the gamma-ray efficiency $\eta_\gamma$ is estimated to lie in the range $0.07 - 0.4$ as Figure 24 in \citep{Smith2023}. 
This yields $N_{\rm MSP} \sim 1.7 - 9.5$, consistent with the seven detected MSPs in M5. 
However, these estimates should be taken with caution, as \citet{Smith2023} assumes a beaming factor of $f_\Omega = 1$, while recent studies suggest that in general $f_\Omega < 1$ \citep{Kalapotharakos2022}.

\subsection{Emission Mechanisms and Future Prospects}

Two main models have been proposed for the gamma-ray emission from GCs: direct magnetospheric emission and IC scattering. 
The former attributes the observed gamma-ray emission to curvature radiation within the magnetospheres of individual MSPs \citep[e.g.,][]{Venter2008,Venter2009}, while the latter involves IC scattering between ambient photon fields and relativistic electrons accelerated in the MSP wind \citep{Bednarek2007, Cheng2010}.

Both models can reproduce the GeV spectra observed from GCs \citep[e.g.,][]{Abdo2010}, though the IC model also predicts TeV emission, which remains largely undetected despite extensive searches with CANGAROO, VERITAS, H.E.S.S., and MAGIC \citep[see e.g.,][]{Kabuki2007,McCutcheon2009,Aharonian2009,Anderhub2009,HESS2013,MAGIC2019}, with Terzan 5 being a possible exception \citep{HESS2011}. 
However, IC model is also supported by diffuse radio and X-ray emissions observed in Terzan 5 and 47 Tuc \citep{Eger2010,Clapson2011,Wu2014}, potentially indicating non-thermal IC processes.

Future observations with more sensitive instruments like CTA, LHAASO, and SKA are expected to clarify the nature of GC gamma-ray emission and potentially confirm the existence of IC components, providing deeper insights into MSP populations and their environments. 

\newpage
\thispagestyle{empty}
~
\newpage

\chapter{Conclusion and Outlook}
\label{Conclusion}

This thesis presents an in-depth study of pulsar-powered systems, centered on the modeling and gamma-ray characterization of pulsar wind nebulae (PWNe). 
With the dedicated numerical framework of TIDE, the systematic analysis of \lat data, and the application of multi-wavelength constraints, this work deepen our understanding to gamma-ray features of pulsars and their environments.

\section{TIDE Model Development and Application}

A time-dependent, leptonic emission model (\texttt{TIDE}) was used to simulate the dynamic and spectral evolution of PWNe. 
The model includes synchrotron and inverse Compton cooling, magnetic field evolution, particle escape, and injection from pulsar spin-down. 
Its performance was validated on three representative sources: the Crab Nebula, 3C~58, and G11.2$-$0.3. 
These applications demonstrate the code’s capability to converge robustly, distinguish global versus local minima in multi-dimensional parameter space, and extract physically meaningful parameters even for sparsely observed systems.

Two Python3 scripts, \texttt{pwn\_sed\_prediction.py} and \texttt{tide\_cycle.py}, were developed by the author to streamline and improve modeling with the \texttt{TIDE} framework (v4.0). 
\texttt{pwn\_sed\_prediction.py} provides an automated interface for SED modeling by varying physical parameters over predefined ranges and systematically organizing the outputs. 
This facilitates efficient parameter space exploration and sensitivity analysis. 
Implementation details are presented in Appendix~\ref{script1}. 
To address convergence issues due to local optima - particularly in cases with sparse multiwavelength data - \texttt{tide\_cycle.py} performs repeated fits with randomized initial conditions. 
It automates the fitting cycle by invoking \texttt{tidefit.py} iteratively, sampling initial values of selected free parameters within specified bounds, and storing key physical and statistical results from each run. 
This approach ensures more robust parameter inference and enables statistical characterization of fitting outcomes (see Appendix~\ref{script2}).

\section{Systematic Search for MeV--GeV PWNe}

Building upon the modeling capability, a systematic search for GeV-bright PWNe without detected gamma-ray pulsars was conducted using \lat data. 
Among the 58 targeted ROIs, several new gamma-ray PWN candidates were identified. 
Five of them were modeled in detail using \texttt{TIDE}, and extended system checks were performed to ensure reliability. 
This work considerably expands the \fermi PWN catalog, provides a valuable reference for future investigations, and highlights promising targets for multiwavelength follow-up observations.

\section{TeV and UHE Studies: Detectability and Constraints}

To evaluate the detectability of possible TeV PWNe, four candidate systems were selected via a pulsar population clustering tool (the ``pulsar tree'') and modeled with \texttt{TIDE}. 
By comparing the predicted SEDs with the sensitivity curves of CTAO and LHAASO, this study evaluated their TeV detection prospects and demonstrated how intrinsic pulsar parameters influence the TeV emission from young PWNe.
By comparing predicted SEDs against the sensitivity of CTAO and LHAASO, the study evaluated their TeV detection prospects and demonstrated how intrinsic pulsar parameters influence the TeV emission from young PWNe.

For UHE sources such as eHWC~J2019+368, HESS~J1427$-$608, and LHAASO~J2226\allowbreak+6057, standard leptonic models require unrealistically low magnetic fields, comparable to the ISM background, to match the observed multiwavelength fluxes. 
This highlights the limitations of purely leptonic scenarios and points to the need for hybrid models or alternative interpretations.

\section{Additional Investigations: HMXB and GC Systems}

Beyond PWNe, two additional gamma-ray studies were performed. 
For the accreting X-ray pulsar 1A~0535+262, deep \lat analysis yielded the most stringent gamma-ray upper limits to date, placing tight constraints on potential emission mechanisms. 
In the case of the GC M5, a significant gamma-ray signal was detected and characterized, likely originating from the cumulative emission of its MSP population.

\section{Outlook}

Several natural extensions of this work are anticipated:
\begin{itemize}
  \item \textbf{Model Development:} Future development of \texttt{TIDE} will incorporate reverberation-phase dynamics, anisotropic expansion, and multidimensional geometries to enhance the physical fidelity of evolved and distorted PWNe. Inclusion of hadronic processes may also be necessary to account for sources exhibiting unusually low magnetic fields in purely leptonic scenario. 

  \item \textbf{Population Studies:} As the number of detected PWNe / PWN candidates increases, statistical analyses of their morphology, spectra, and evolutionary properties will become feasible. 
  These studies will support classification efforts and enable automated identification strategies, potentially integrated with machine learning frameworks, in the surveys such as \lat, CTAO, and LHAASO. 

  \item \textbf{Multi-wavelength Cross-validation:} Joint modeling across radio, X-ray, gamma-ray bands will be essential for breaking parameter degeneracies and refining estimates of key physical quantities, including magnetic field strength, particle injection spectra, and nebular age. 
\end{itemize}

This thesis establishes a comprehensive framework for studying pulsar environments, with a particular focus on PWNe, across multiple wavelengths, and offers critical insights into the mechanisms driving cosmic particle acceleration in our Galaxy. 

\newpage
\thispagestyle{empty}
~
\newpage

\appendix 



\chapter{Script and Description} 
\label{Appendix} 

This appendix presents a set of analysis and modeling scripts, all written by the author, which were used in support of the work described in some chapters of this thesis.

\section{Script 1: \texttt{pwn\_sed\_prediction.py}}  
\label{script1}

The \texttt{pwn\_sed\_prediction.py} script, written in Python 3, serves as an automated interface to perform SED modeling using the TIDE framework (v4.0).
By varying the model parameters over predefined ranges, it executes simulations and organizes the outputs of each iteration in a structured manner.
This approach facilitates a systematic exploration of the SED morphology in parameter space, as well as sensitivity studies of individual parameter effects.

\subsection{Workflow}
The overall workflow proceeds as follows:
\begin{enumerate}
\item \textbf{Initialization:} The user specifies the parameters to be explored, defines their variation ranges in \texttt{fitting\_par.txt}, and configures the output directory.
\item \textbf{Parameter Configuration:} Parameter sets are generated deterministically (by iterating through all user-defined values) or stochastically (by randomly sampling within the specified ranges). 
\item \textbf{Parameter Injection:} The script updates \texttt{fitting\_par.txt} with the current parameter set.
\item \textbf{Model Execution:} The modified parameter file is passed to \texttt{tidefit.py} for simulation.
\item \textbf{Output Archival:} Key outputs, including the resulting \texttt{fit\_results.txt} and \texttt{total.txt}, along with any additional user-specified files are saved with an iteration index. 
\item \textbf{Looping:} The process is repeated for each parameter combination or random sample.
\end{enumerate}

\subsection{How to Use}
\begin{enumerate}
\item Place \texttt{pwn\_sed\_prediction.py}, \texttt{fitting\_par.txt}, and \texttt{parameters.txt} in the working directory.
\item Ensure the following configuration:
\begin{itemize}
\item Except for \texttt{Systematic\_uncertainty\_correction\_factor}, all other free parameters in a standard \texttt{TIDE-fit} must be fixed in \texttt{fitting\_par.txt} to prevent an actual fit. Their values can be arbitrary, as they are later configured via \texttt{tide\_specify()} or \texttt{tide\_random()}. 
\item The \texttt{Spectrum\_file} entry in \texttt{fitting\_par.txt} should be set to 0 to indicate that no actual fit will be performed. 
\end{itemize}
\item Import the script and call the appropriate function for the fitting task:
\begin{itemize}
    \item Use \texttt{tide\_specify()} for user-defined parameter combinations.
    \item Use \texttt{tide\_random()} for random sampling within the specified parameter ranges.
\end{itemize}
\item An example usage is provided at the end of the full source code listing (see Sec.~\ref{A1.fullcode}) for reference.
\end{enumerate}

\subsection{Full Source Code}
\label{A1.fullcode}

The full source code of \texttt{pwn\_sed\_prediction.py} is provided below, where `$\sqcup$' represents a space character: 

\begin{lstlisting}[language=Python, breaklines=true, basicstyle=\ttfamily\tiny, caption={Full code: \texttt{pwn\_sed\_prediction.py}}, label={lst:fullcode1}] 

import numpy as np
import os
import shutil
import random
import matplotlib.pyplot as plt
import matplotlib.colors as colors

# List of physical constants
ElectronMass = 9.10938188e-28
ErgToGeV = 624.150934
ErgToTeV = 0.624150934
KpcToCm = 3.08568025e21
PlanckConstant = 6.62606876e-27
SpeedOfLight = 2.99792458e10

# Function to modify parameters specified by the user
def modifypar_specify(parameters, values):
    ## Open the fitting parameter file
    with open("fitting_par.txt", 'r') as fittingPar:
        file_data = ''
        lineIndex = 0

        ## Iterate through each line in the file
        for line in fittingPar:
            ## Check if the line starts with a comment character
            if line.startswith('#') or line.startswith('!'):
                ## If it does, keep the line unchanged and continue to the next line
                file_data += line
                continue
            else:
                ## If it doesn't, check if the line index matches any of the parameters to be modified
                if lineIndex in parameters:
                    newline = line.split()

                    ## Modify the parameter value according to the specified index
                    if lineIndex == 4:
                        index = np.where(parameters == lineIndex)[0][0]
                        newline[1] = str(values[index])
                    elif 5 <= lineIndex <= 6:
                        index = np.where(parameters == lineIndex)[0][0]
                        newline[2] = str(values[index])
                    else:
                        index = np.where(parameters == lineIndex)[0][0]
                        newline[1] = str(values[index])

                    ## Replace the original line with the modified line
                    line = line.replace(line, '      ' + '     '.join(newline) + '\n')
                ## Increment the line index counter
                lineIndex += 1
            ## Append the modified or unchanged line to the file data
            file_data += line

    ## Write the modified file data back to the fitting parameter file
    with open("fitting_par.txt", 'w', encoding="utf-8") as f:
        f.write(file_data)

# Function to randomly modify parameters within specified limits
def modifypar_random(pars, parameters, limits):
    with open("fitting_par.txt", 'r') as fittingPar:
        file_data = ''
        lineIndex1 = 0

        for line in fittingPar:
            if line.startswith('#') or line.startswith('!'):
                file_data += line
                continue
            else:
                if lineIndex1 in parameters:
                    newline = line.split()

                    if lineIndex1 == 9:
                        index = np.where(parameters == lineIndex1)[0][0]
                        min_val = np.float64(limits[index][0] * pars[0])
                        max_val = np.float64(limits[index][-1] * pars[0])
                        true_age = random.uniform(min_val, max_val)
                        newline[1] = '{:e}'.format(np.float64(true_age))

                        z = -1
                        braking_index = pars[2][z]
                        tau_0 = 2 * pars[0] / (braking_index - 1) - true_age
                        while tau_0 <= 0:
                            braking_index = pars[2][z - 1]
                            tau_0 = 2 * pars[0] / (braking_index - 1) - true_age
                        L0 = pars[1] * (1 + true_age / tau_0) ** ((braking_index + 1) / (braking_index - 1))

                    elif lineIndex1 == 11:
                        newline[1] = str(braking_index)

                    elif lineIndex1 == 12:
                        newline[1] = str(L0)

                    elif lineIndex1 == 13:
                        newline[1] = str(tau_0)

                    else:
                        index = np.where(parameters == lineIndex1)[0][0]
                        min_val = np.float64(limits[index][0])
                        max_val = np.float64(limits[index][-1])
                        newline[1] = '{:e}'.format(np.float64(random.uniform(min_val, max_val)))

                    line = line.replace(line, '      ' + '     '.join(newline) + '\n')

            lineIndex1 += 1
            file_data += line

    with open("fitting_par.txt", 'w', encoding="utf-8") as f:
        f.write(file_data)

# Function to specify parameter combinations and run model fits
def tide_specify(fix_pars, free_pars, n, save_dir, initial_index):
    parameters = np.array([4, 5, 6, 9, 11, 12, 13, 14, 18, 19, 20, 21, 22, 23, 24, 29, 34, 35, 36, 37])
    values = np.zeros((20))
    values[0] = fix_pars[0]
    values[1] = fix_pars[1]
    values[7] = fix_pars[2]
    values[11] = fix_pars[3]
    values[12] = fix_pars[4]
    values[16] = fix_pars[5]
    values[18] = fix_pars[6]
    values[2] = fix_pars[7]

    fit_number = initial_index

    for a in range(len(free_pars[0])):
        t_age = free_pars[0][a] * fix_pars[0]
        values[3] = t_age
        z = -1
        braking_index = n[z]
        tau_0 = ((2 / (braking_index - 1)) - free_pars[0])[a] * fix_pars[0]
        while tau_0 <= 0:
            braking_index = n[z - 1]
            tau_0 = ((2 / (braking_index - 1)) - free_pars[0])[a] * fix_pars[0]
        L0 = fix_pars[1] * (1 + t_age / tau_0) ** ((braking_index + 1) / (braking_index - 1))
        values[4] = braking_index
        values[6] = tau_0
        values[5] = L0

        for b in range(len(free_pars[1])):
            values[8] = free_pars[1][b]
            for c in range(len(free_pars[2])):
                values[9] = free_pars[2][c]
                for d in range(len(free_pars[3])):
                    values[10] = free_pars[3][d]
                    for g in range(len(free_pars[4])):
                        values[13] = free_pars[4][g]
                        for i in range(len(free_pars[5])):
                            values[15] = free_pars[5][i]
                            for j in range(len(free_pars[6])):
                                values[17] = free_pars[6][j]
                                for k in range(len(free_pars[7])):
                                    values[19] = free_pars[7][k]
                                    for f in range(len(free_pars[8])):
                                        values[14] = free_pars[8][f]

                                        modifypar_specify(parameters, values)
                                        os.system('python3 $TIDE_HOME/tidefit.py')

                                        file1 = 'fitting_par.txt'
                                        file4 = 'total.txt'

                                        newname = '_' + str(fit_number) + '.txt'
                                        newfile1 = file1.replace('.txt', newname)
                                        newfile4 = file4.replace('.txt', newname)
                                        shutil.copyfile(os.path.join(file1), os.path.join(save_dir + newfile1))
                                        shutil.copyfile(os.path.join(file4), os.path.join(save_dir + newfile4))
                                        fit_number += 1

# Function to randomly generate parameter combinations and run model fits
def tide_random(pars, pars_limits, save_dir, cycle_index, initial_index):
    parameters = np.array([9, 18, 19, 20, 23, 24, 29, 35, 37, 11, 12, 13])
    limits = np.array(pars_limits)

    for i in range(initial_index, cycle_index + initial_index):
        modifypar_random(pars, parameters, limits)
        os.system('python3 $TIDE_HOME/tidefit.py')

        file1 = 'fitting_par.txt'
        file4 = 'total.txt'

        newname = '_' + str(i) + '.txt'
        newfile1 = file1.replace('.txt', newname)
        newfile4 = file4.replace('.txt', newname)
        shutil.copyfile(os.path.join(file1), os.path.join(save_dir + newfile1))
        shutil.copyfile(os.path.join(file4), os.path.join(save_dir + newfile4))


# An example usage
'''
import numpy as np
import sys
sys.path.append('./') # the path to the pwn_sed_prediction.py script
import pwn_sed_prediction

# Define the parameters and their values
## Fixed parameters
distance = 6.2   # kpc
tau_c = 5600    # characteristic age, yr
L = 5.6E+37     # spin-down luminosity, erg/s
E_sn = 1.0E+51  # erg
containment_factor = 0.5
T_fir = 70      # Temperature of FIR, K
T_nir = 3000    # Temperature of NIR, K
R_pwn = np.array([6.06])  # radius of the PWN, pc

## Free parameters:
# format: [min, max] for use with tide_random(); [val1, val2, ...] for use with tide_specify()
U_nir = np.array([1.36, 4.08])      # energy density of NIR
U_fir = np.array([6.94, 20.82])     # energy density of FIR
n_ism = np.array([0.1, 1.0])        # ISM density in cm$^{-3}$
alpha1 = np.array([1.0, 1.6])       # Low-energy spectral index
alpha2 = np.array([2.2, 2.8])       # High-energy spectral index
energy_break = np.array([1.0E+5, 1.0E+6])  # Break energy in MeV
M_ej = np.array([8, 15])            # Ejecta mass in $M_\odot$
eta_B = np.array([0.02, 0.04])      # Magnetic energy fraction
k_age = np.array([0.7, 1.3])        # Age scaling factor

## Brake index
n = np.array([2, 3])

# Output folder
output_folder = './'

# Specify parameter combinations and run model fits, 
#Please do not change the order in which these parameters appear in the list
##initial_index: Which number does the fitting number start with
pwn_sed_prediction.tide_specify([tau_c, L, distance, containment_factor, E_sn, T_fir, T_nir, R_pwn], [k_age, energy_break, alpha1, alpha2, M_ej, eta_B, U_fir, U_nir, n_ism], n, output_folder, initial_index=1)

# Randomly generate parameter combinations and run model fits
#Please do not change the order in which these parameters appear in the list
##cycle_index: how many fits you want to run

pwn_sed_prediction.tide_random([tau_c, L, n], [k_age, energy_break, alpha1, alpha2, M_ej, n_ism, eta_B, U_fir, U_nir], output_folder, cycle_index=200, initial_index=1)
'''
\end{lstlisting}


\section{Script 2: \texttt{tide\_cycle.py}}  
\label{script2}

As demonstrated in Section~\ref{tide}, a single run of \texttt{TIDE} may converge to a local optimum, particularly when the MWL data are sparse. 
This issue can be effectively mitigated by performing multiple fits with randomized initial conditions for the free parameters. 

The \texttt{tide\_cycle.py} script, written in \texttt{Python 3}, automates this process. 
It conducts repeated fitting cycles with randomized parameter initialization, invoking \texttt{tidefit.py} at each iteration. 
For each run, the script samples the initial values of user-specified free parameters within predefined bounds, executes the fit, and records key physical and statistical outputs. 
This enables robust exploration of the parameter space and supports subsequent statistical or visual analysis of the fit distribution.

\subsection{Structure and Workflow}

The script consists of three major components:
\begin{enumerate}
    \item \textbf{Parameter Initialization}: The function \texttt{modifypar()} randomly initializes the free parameters within the ranges specified in \texttt{fitting\_par.txt}, optionally adjusting age-dependent quantities (e.g., initial spin-down luminosity $L_0$ and time scale $t_0$) if age is treated as a free parameter. 
    \item \textbf{Model Execution}: For each iteration, the script calls \texttt{tidefit.py} via the command line, executing the PWN radiation model fit based on the modified parameter file.
    \item \textbf{Result Recording and Archival}: The \texttt{saveresults()} function extracts key fitting results from \texttt{fit\_results.txt} and \texttt{pwn\_data.txt}, appending a summary to \texttt{cycle\_results.txt}. In parallel, a predefined set of output files is restored with a suffix denoting the iteration index.
\end{enumerate}

The entire loop is repeated a user-defined number of times (default: 10, configurable).

\subsection{How to Use}
To run the script:
\begin{enumerate}
    \item Place the script in the working directory containing the files \texttt{fitting\_par.txt}, \texttt{parameters.txt}, and any required input data for \texttt{TIDE}.
    \item Specify the number of loop iterations (default: 10) and the parameter line indices in the \texttt{modifypar()} function as needed for the intended use case. 
    \item Execute the script using:
    \begin{verbatim}
    python3 tide_cycle.py
    \end{verbatim}
\end{enumerate}

\subsection{Full Source Code}
\label{A2.fullcode}

The complete source code of \texttt{tide\_cycle.py} is provided below: 

\begin{lstlisting}[language=Python, breaklines=true, basicstyle=\ttfamily\tiny, caption={Full code: \texttt{tide\_cycle.py}}, label={lst:fullcode2}]

import numpy as np
import os, shutil, random


# Randomly selects initial values within fitting ranges for free parameters.
# Arguments p1-p7 correspond to line indices of free parameters in fitting_par.txt.
# Modify this function to change number of free parameters, value ranges, or other parameter adjustments.
def modifypar(p1, p2, p3, p4, p5, p6, p7, ageFree, brakingIndex):
    lineIndex = 0
    file_data = ''
    with open("fitting_par.txt", 'r') as fittingPar:
        for line in fittingPar:
            if line.startswith('#') or line.startswith('!'):
                file_data += line
                continue

            if lineIndex in {p1, p2, p3, p4, p5, p6, p7, 4, 5, 9, 12, 13}:
                parts = line.split()
                if lineIndex == 4:
                    tauc = np.float64(parts[1])
                elif lineIndex == 5:
                    Edot = np.float64(parts[2])
                elif lineIndex == 9:
                    if ageFree:
                        min_val, max_val = float(parts[2]), float(parts[3])
                        parts[1] = '{:e}'.format(random.uniform(min_val, max_val))
                        tage = float(parts[1])
                        t_initial = 2 * tauc / (brakingIndex - 1) - tage
                        while t_initial <= 0:
                            parts[1] = '{:e}'.format(random.uniform(min_val, max_val))
                            tage = float(parts[1])
                            t_initial = 2 * tauc / (brakingIndex - 1) - tage
                        L_initial = Edot * (1 + tage / t_initial) ** ((brakingIndex + 1) / (brakingIndex - 1))
                        line = '      ' + '     '.join(parts) + '\n'

                elif lineIndex == 12:
                    parts[1] = '{:e}'.format(L_initial)
                    line = '      ' + '     '.join(parts) + '\n'

                elif lineIndex == 13:
                    parts[1] = '{:e}'.format(t_initial)
                    line = '      ' + '     '.join(parts) + '\n'

                else:
                    min_val, max_val = float(parts[2]), float(parts[3])
                    parts[1] = '{:e}'.format(random.uniform(min_val, max_val))
                    line = '      ' + '     '.join(parts) + '\n'

            file_data += line
            lineIndex += 1

    with open("fitting_par.txt", 'w', encoding="utf-8") as f:
        f.write(file_data)


# Records the results of each fitting cycle to cycle_results.txt.
# Parameters:
#  i        - cycle index
#  ageFree  - boolean flag indicating if age parameter is free (True) or fixed (False)
def saveresults(i, ageFree):
    if ageFree:
        header = (
            '# Cycle results: each line corresponds to a cycle.\n'
            '# Columns: cycle_no, success_flag, dof, systematic_uncertainty, reduced_chi_sq, '
            'age_value, age_low_limit, age_upper_limit, energy_break_value, energy_break_low_limit, energy_break_upper_limit, '
            'low_energy_index_value, low_energy_index_low_limit, low_energy_index_upper_limit, '
            'high_energy_index_value, high_energy_index_low_limit, high_energy_index_upper_limit, '
            'ejected_mass_value, ejected_mass_low_limit, ejected_mass_upper_limit, '
            'magnetic_fraction_value, magnetic_fraction_low_limit, magnetic_fraction_upper_limit, '
            'FIR_density_value, FIR_density_low_limit, FIR_density_upper_limit, '
            'NIR_density_value, NIR_density_low_limit, NIR_density_upper_limit, '
            'PWN_radius_pc, PWN_magnetic_field_uG\n'
        )
        record_len = 31
    else:
        header = (
            '# Cycle results: each line corresponds to a cycle.\n'
            '# Columns: cycle_no, success_flag, dof, systematic_uncertainty, reduced_chi_sq, '
            'energy_break_value, energy_break_low_limit, energy_break_upper_limit, '
            'low_energy_index_value, low_energy_index_low_limit, low_energy_index_upper_limit, '
            'high_energy_index_value, high_energy_index_low_limit, high_energy_index_upper_limit, '
            'ejected_mass_value, ejected_mass_low_limit, ejected_mass_upper_limit, '
            'magnetic_fraction_value, magnetic_fraction_low_limit, magnetic_fraction_upper_limit, '
            'FIR_density_value, FIR_density_low_limit, FIR_density_upper_limit, '
            'NIR_density_value, NIR_density_low_limit, NIR_density_upper_limit, '
            'PWN_radius_pc, PWN_magnetic_field_uG\n'
        )
        record_len = 28

    ## Create results file with header if it doesn't exist
    if not os.path.exists('cycle_results.txt'):
        with open('cycle_results.txt', 'w', encoding='utf-8') as f:
            f.write(header)

    ## Read fitting results
    with open("fit_results.txt", 'r') as fitResults:
        lines = fitResults.readlines()

    record = [''] * record_len
    record[0] = str(i)
    record[1] = lines[9].split()[2]            # success flag
    record[2] = lines[5].split()[-1]           # degrees of freedom
    record[3] = lines[6].split()[-1]           # systematic uncertainty
    record[4] = lines[7].split()[-1]           # reduced chi-squared

    if ageFree:
        ## age and other parameters (columns 6-30)
        for idx, line_idx in enumerate(range(11, 19)):
            record[5 + idx * 3] = lines[line_idx].split()[-3]
            record[6 + idx * 3] = lines[line_idx].split()[-2]
            record[7 + idx * 3] = lines[line_idx].split()[-1]
        ## Read PWN radius and magnetic field from pwn_data.txt
        with open("pwn_data.txt", 'r') as pwndata:
            pwn_lines = pwndata.readlines()
        record[-2] = pwn_lines[-1].split()[2]  # PWN radius
        record[-1] = pwn_lines[-1].split()[3]  # PWN magnetic field
    else:
        ## parameters (columns 6-27)
        for idx, line_idx in enumerate(range(11, 18)):
            record[5 + idx * 3] = lines[line_idx].split()[-3]
            record[6 + idx * 3] = lines[line_idx].split()[-2]
            record[7 + idx * 3] = lines[line_idx].split()[-1]
        ## Read PWN radius and magnetic field from pwn_data.txt
        with open("pwn_data.txt", 'r') as pwndata:
            pwn_lines = pwndata.readlines()
        record[-2] = pwn_lines[-1].split()[2]
        record[-1] = pwn_lines[-1].split()[3]

    ## Format record and append to file
    formatted_record = '   '.join(record) + '\n'
    with open("cycle_results.txt", 'a', encoding="utf-8") as writeRecord:
        writeRecord.write(formatted_record)


# Main loop: perform fitting for a specified number of cycles.
# Modify 'range(10)' to change total fitting iterations.
for i in range(10):
    ageFree = 1  ## 1: age free; 0: age fixed
    braking_index = 2.16
    ## Indices of parameters (Eb, alpha1, alpha2, Mej, eta, Ufir, Unir) in fitting_par.txt to modify
    modifypar(18, 19, 20, 23, 29, 35, 37, ageFree, braking_index)
    os.system('python3 $TIDE_HOME/tidefit.py')  # Execute fitting script
    saveresults(i + 1, ageFree)

    ## Files to save per iteration
    file1, file2, file3, file4, file5 = 'fitting_par.txt', 'fit_results.txt', 'pwn_data.txt', 'sed.eps', 'total.txt'

    ## Rename and copy output files to preserve results for each cycle
    newname_txt = f'_{i+1}.txt'
    newname_eps = f'_{i+1}.eps'
    shutil.copyfile(file1, file1.replace('.txt', newname_txt))
    shutil.copyfile(file2, file2.replace('.txt', newname_txt))
    shutil.copyfile(file3, file3.replace('.txt', newname_txt))
    shutil.copyfile(file5, file5.replace('.txt', newname_txt))

    ## sed.eps may not be generated if fitting fails; check existence before copying
    if os.path.exists(file4):
        shutil.copyfile(file4, file4.replace('.eps', newname_eps))
        os.remove(file4)

\end{lstlisting}

\newpage
\thispagestyle{empty}
~
\newpage

\nocite{*}
\bibliographystyle{aasjournal}
\bibliography{main}

\begin{thebibliography}{}
\expandafter\ifx\csname natexlab\endcsname\relax\def\natexlab#1{#1}\fi
\providecommand{\url}[1]{\href{#1}{#1}}
\providecommand{\dodoi}[1]{doi:~\href{http://doi.org/#1}{\nolinkurl{#1}}}
\providecommand{\doeprint}[1]{\href{http://ascl.net/#1}{\nolinkurl{http://ascl.net/#1}}}
\providecommand{\doarXiv}[1]{\href{https://arxiv.org/abs/#1}{\nolinkurl{https://arxiv.org/abs/#1}}}

\bibitem[{{Abdalla} {et~al.}(2020){Abdalla}, {Adam}, {Aharonian}, {Ait
  Benkhali}, {Ang{\"u}ner}, {Arakawa}, {Arcaro}, {Armand}, {Ashkar}, {Backes},
  {Barbosa Martins}, {Barnard}, {Becherini}, {Berge}, {Bernl{\"o}hr},
  {Blackwell}, {B{\"o}ttcher}, {Boisson}, {Bolmont}, {Bonnefoy}, {Bregeon},
  {Breuhaus}, {Brun}, \& {Brun}}]{HESS2020}
{Abdalla}, H., {Adam}, R., {Aharonian}, F., {et~al.} 2020, \aap, 633, A102,
  \dodoi{10.1051/0004-6361/201936621}

\bibitem[{{Abdelmaguid} {et~al.}(2023){Abdelmaguid}, {Gelfand}, {Gotthelf}, \&
  {Straal}}]{Abdelmaguid2023}
{Abdelmaguid}, M., {Gelfand}, J.~D., {Gotthelf}, E., \& {Straal}, S. 2023,
  \apj, 946, 40, \dodoi{10.3847/1538-4357/acbd30}

\bibitem[{{Abdo} {et~al.}(2007{\natexlab{a}}){Abdo}, {Allen}, {Berley},
  {Blaufuss}, {Casanova}, {Chen}, {Coyne}, {Delay}, {Dingus}, {Ellsworth},
  {Fleysher}, {Fleysher}, {Gebauer}, {Gonzalez}, {Goodman}, {Hays}, {Hoffman},
  {Kolterman}, {Kelley}, {Lansdell}, {Linnemann}, {McEnery}, {Mincer},
  {Moskalenko}, {Nemethy}, {Noyes}, {Ryan}, {Samuelson}, {Saz Parkinson},
  {Schneider}, {Shoup}, {Sinnis}, {Smith}, {Strong}, {Sullivan}, {Vasileiou},
  {Walker}, {Williams}, {Xu}, \& {Yodh}}]{Abdo2007a}
{Abdo}, A.~A., {Allen}, B., {Berley}, D., {et~al.} 2007{\natexlab{a}}, \apjl,
  658, L33, \dodoi{10.1086/513696}

\bibitem[{{Abdo} {et~al.}(2007{\natexlab{b}}){Abdo}, {Allen}, {Berley},
  {Casanova}, {Chen}, {Coyne}, {Dingus}, {Ellsworth}, {Fleysher}, {Fleysher},
  {Gonzalez}, {Goodman}, {Hays}, {Hoffman}, {Hopper}, {H{\"u}ntemeyer},
  {Kolterman}, {Lansdell}, {Linnemann}, {McEnery}, {Mincer}, {Nemethy},
  {Noyes}, {Ryan}, {Saz Parkinson}, {Shoup}, {Sinnis}, {Smith}, {Sullivan},
  {Vasileiou}, {Walker}, {Williams}, {Xu}, \& {Yodh}}]{Abdo2007b}
---. 2007{\natexlab{b}}, \apjl, 664, L91, \dodoi{10.1086/520717}

\bibitem[{{Abdo} {et~al.}(2009{\natexlab{a}}){Abdo}, {Allen}, {Aune}, {Berley},
  {Chen}, {Christopher}, {DeYoung}, {Dingus}, {Ellsworth}, {Gonzalez},
  {Goodman}, {Hays}, {Hoffman}, {H{\"u}ntemeyer}, {Kolterman}, {Linnemann},
  {McEnery}, {Morgan}, {Mincer}, {Nemethy}, {Pretz}, {Ryan}, {Saz Parkinson},
  {Shoup}, {Sinnis}, {Smith}, {Vasileiou}, {Walker}, {Williams}, \&
  {Yodh}}]{milagro09}
{Abdo}, A.~A., {Allen}, B.~T., {Aune}, T., {et~al.} 2009{\natexlab{a}}, \apjl,
  700, L127, \dodoi{10.1088/0004-637X/700/2/L127}

\bibitem[{{Abdo} {et~al.}(2009{\natexlab{b}}){Abdo}, {Allen}, {Aune}, {Berley},
  {Chen}, {Christopher}, {De Young}, {Dingus}, {Ellsworth}, {Gonzalez},
  {Goodman}, {Hays}, {Hoffman}, {H{\"u}ntemeyer}, {Kolterman}, {Linnemann},
  {McEnery}, {Morgan}, {Mincer}, {Nemethy}, {Pretz}, {Ryan}, {Saz Parkinson},
  {Shoup}, {Sinnis}, {Smith}, {Vasileiou}, {Walker}, {Williams}, \&
  {Yodh}}]{Abdo2009}
---. 2009{\natexlab{b}}, \apjl, 703, L185, \dodoi{10.1088/0004-637X/703/2/L185}

\bibitem[{{Abdo} {et~al.}(2009{\natexlab{c}}){Abdo}, {Ackermann}, {Ajello},
  {Atwood}, {Axelsson}, {Baldini}, {Ballet}, {Barbiellini}, {Baring},
  {Bastieri}, {Baughman}, {Bechtol}, {Bellazzini}, {Berenji}, {Bloom},
  {Bonamente}, {Borgland}, {Bregeon}, {Brez}, {Brigida}, {Bruel}, {Caliandro},
  {Cameron}, {Camilo}, {Caraveo}, {Casandjian}, {Cecchi}, {Chekhtman},
  {Cheung}, {Chiang}, {Ciprini}, {Claus}, {Cognard}, {Cohen-Tanugi}, {Conrad},
  {de Angelis}, {de Palma}, {Dormody}, {Silva}, {Drell}, {Dubois}, {Dumora},
  {Farnier}, {Favuzzi}, {Frailis}, {Freire}, {Fukazawa}, {Funk}, {Fusco},
  {Gargano}, {Gehrels}, {Germani}, {Giebels}, {Giglietto}, {Giordano},
  {Glanzman}, {Godfrey}, {Grenier}, {Grondin}, {Grove}, {Guillemot}, {Guiriec},
  {Halpern}, {Hanabata}, {Harding}, {Hayashida}, {Hays}, {Hobbs}, {Hughes},
  {J{\'o}hannesson}, {Johnson}, {Johnson}, {Johnson}, {Johnson}, {Johnston},
  {Kamae}, {Katagiri}, {Kataoka}, {Kawai}, {Kerr}, {Kn{\"o}dlseder}, {Kocian},
  {Kramer}, {Kuehn}, {Kuss}, {Lande}, {Latronico}, {Lemoine-Goumard}, {Longo},
  {Loparco}, {Lott}, {Lovellette}, {Lubrano}, {Lyne}, {Makeev}, {Manchester},
  {Marelli}, {Mazziotta}, {McEnery}, {Meurer}, {Michelson}, {Mitthumsiri},
  {Mizuno}, {Moiseev}, {Monte}, {Monzani}, {Morselli}, {Moskalenko}, {Murgia},
  {Nolan}, {Norris}, {Noutsos}, {Nuss}, {Ohsugi}, {Omodei}, {Orlando}, {Ormes},
  {Ozaki}, {Paneque}, {Panetta}, {Parent}, {Pepe}, {Pesce-Rollins}, {Piron},
  {Porter}, {Rain{\`o}}, {Rando}, {Ransom}, {Razzano}, {Reimer}, {Reimer},
  {Reposeur}, {Rochester}, {Rodriguez}, {Romani}, {Roth}, {Ryde},
  {Sadrozinski}, {Sanchez}, {Sander}, {Saz Parkinson}, {Scargle}, {Sgr{\`o}},
  {Siskind}, {Smith}, {Smith}, {Spandre}, {Spinelli}, {Stappers}, {Strickman},
  {Suson}, {Tajima}, {Takahashi}, {Tanaka}, {Thayer}, {Thayer}, {Theureau},
  {Thompson}, {Thorsett}, {Tibaldo}, {Torres}, {Tosti}, {Uchiyama}, {Usher},
  {Van Etten}, {Vilchez}, {Vitale}, {Waite}, {Wang}, {Wang}, {Watters},
  {Weltevrede}, {Winer}, {Wood}, {Ylinen}, \& {Ziegler}}]{abdo09}
{Abdo}, A.~A., {Ackermann}, M., {Ajello}, M., {et~al.} 2009{\natexlab{c}},
  \apj, 706, 1331, \dodoi{10.1088/0004-637X/706/2/1331}

\bibitem[{{Abdo} {et~al.}(2009{\natexlab{d}}){Abdo}, {Ackermann}, {Ajello},
  {Atwood}, {Axelsson}, {Baldini}, {Ballet}, {Barbiellini}, {Bastieri},
  {Baughman}, {Bechtol}, {Bellazzini}, {Berenji}, {Blandford}, {Bloom},
  {Bonamente}, {Borgland}, {Bregeon}, {Brez}, {Brigida}, {Bruel}, {Burnett},
  {Caliandro}, {Cameron}, {Caraveo}, {Casandjian}, {Cavazzuti}, {Cecchi},
  {{\c{C}}elik}, {Charles}, {Chaty}, {Chekhtman}, {Cheung}, {Chiang},
  {Ciprini}, {Claus}, {Cohen-Tanugi}, {Cominsky}, {Conrad}, {Corbel}, {Corbet},
  {Cutini}, {Dermer}, {de Angelis}, {de Luca}, {de Palma}, {Digel}, {Dormody},
  {do Couto e Silva}, {Drell}, {Dubois}, {Dubus}, {Dumora}, {Farnier},
  {Favuzzi}, {Fegan}, {Focke}, {Frailis}, {Fukazawa}, {Funk}, {Fusco},
  {Gargano}, {Gasparrini}, {Gehrels}, {Germani}, {Giebels}, {Giglietto},
  {Giordano}, {Glanzman}, {Godfrey}, {Grenier}, {Grondin}, {Grove},
  {Guillemot}, {Guiriec}, {Hanabata}, {Harding}, {Hayashida}, {Hays}, {Hill},
  {Hughes}, {J{\'o}hannesson}, {Johnson}, {Johnson}, {Johnson}, {Johnson},
  {Kamae}, {Katagiri}, {Kataoka}, {Kawai}, {Kerr}, {Kn{\"o}dlseder}, {Kocian},
  {Kuehn}, {Kuss}, {Lande}, {Larsson}, {Latronico}, {Longo}, {Loparco}, {Lott},
  {Lovellette}, {Lubrano}, {Madejski}, {Makeev}, {Marelli}, {Mazziotta},
  {McEnery}, {Meurer}, {Michelson}, {Mitthumsiri}, {Mizuno}, {Monte},
  {Monzani}, {Morselli}, {Moskalenko}, {Murgia}, {Nolan}, {Nuss}, {Ohsugi},
  {Okumura}, {Omodei}, {Orlando}, {Ormes}, {Paneque}, {Panetta}, {Parent},
  {Pelassa}, {Pepe}, {Pesce-Rollins}, {Piron}, {Porter}, {Rain{\`o}}, {Rando},
  {Ray}, {Razzano}, {Rea}, {Reimer}, {Reimer}, {Reposeur}, {Ritz}, {Rochester},
  {Rodriguez}, {Romani}, {Ryde}, {Sadrozinski}, {Sanchez}, {Sander}, {Saz
  Parkinson}, {Scargle}, {Sgr{\`o}}, {Shaw}, {Sierpowska-Bartosik}, {Siskind},
  {Smith}, {Smith}, {Spandre}, {Spinelli}, {Striani}, {Strickman}, {Suson},
  {Tajima}, {Takahashi}, {Takahashi}, {Tanaka}, {Thayer}, {Thayer}, {Thompson},
  {Tibaldo}, {Torres}, {Tosti}, {Tramacere}, {Uchiyama}, {Usher}, {Vasileiou},
  {Vilchez}, {Vitale}, {Waite}, {Wang}, {Winer}, {Wood}, {Ylinen}, \&
  {Ziegler}}]{Abdo2009LS}
---. 2009{\natexlab{d}}, \apjl, 701, L123, \dodoi{10.1088/0004-637X/701/2/L123}

\bibitem[{{Abdo} {et~al.}(2009{\natexlab{e}}){Abdo}, {Ackermann}, {Ajello},
  {Axelsson}, {Baldini}, {Ballet}, {Barbiellini}, {Bastieri}, {Baughman},
  {Bechtol}, {Bellazzini}, {Berenji}, {Blandford}, {Bloom}, {Bonamente},
  {Borgland}, {Brez}, {Brigida}, {Bruel}, {Burnett}, {Buson}, {Caliandro},
  {Cameron}, {Caraveo}, {Casandjian}, {Cecchi}, {{\c{C}}elik}, {Chaty},
  {Cheung}, {Chiang}, {Ciprini}, {Claus}, {Cohen-Tanugi}, {Cominsky}, {Conrad},
  {Corbel}, {Corbet}, \& {Dermer}}]{Abdo2009c}
---. 2009{\natexlab{e}}, Science, 326, 1512, \dodoi{10.1126/science.1182174}

\bibitem[{{Abdo} {et~al.}(2010){Abdo}, {Ackermann}, {Ajello}, {Baldini},
  {Ballet}, {Barbiellini}, {Bastieri}, {Bellazzini}, {Blandford}, {Bloom},
  {Bonamente}, {Borgland}, {Bouvier}, {Brandt}, {Bregeon}, {Brigida}, {Bruel},
  {Buehler}, {Buson}, {Caliandro}, {Cameron}, {Caraveo}, {Carrigan},
  {Casandjian}, {Charles}, {Chaty}, {Chekhtman}, {Cheung}, {Chiang}, {Ciprini},
  {Claus}, {Cohen-Tanugi}, {Conrad}, {Decesar}, {Dermer}, \& {de
  Palma}}]{Abdo2010}
---. 2010, \aap, 524, A75, \dodoi{10.1051/0004-6361/201014458}

\bibitem[{{Abdo} {et~al.}(2011){Abdo}, {Ackermann}, {Ajello}, {Allafort},
  {Ballet}, {Barbiellini}, {Bastieri}, {Bechtol}, {Bellazzini}, {Berenji},
  {Blandford}, {Bonamente}, {Borgland}, {Bregeon}, {Brigida}, {Bruel},
  {Buehler}, {Buson}, {Caliandro}, {Cameron}, {Camilo}, {Caraveo}, {Cecchi},
  {Charles}, {Chaty}, {Chekhtman}, {Chernyakova}, {Cheung}, {Chiang},
  {Ciprini}, {Claus}, {Cohen-Tanugi}, {Cominsky}, {Corbel}, {Cutini},
  {D'Ammando}, {de Angelis}, {den Hartog}, {de Palma}, {Dermer}, {Digel},
  {Silva}, {Dormody}, {Drell}, {Drlica-Wagner}, {Dubois}, {Dubus}, {Dumora},
  {Enoto}, {Espinoza}, {Favuzzi}, {Fegan}, {Ferrara}, {Focke}, {Fortin},
  {Fukazawa}, {Funk}, {Fusco}, {Gargano}, {Gasparrini}, {Gehrels}, {Germani},
  {Giglietto}, {Giommi}, {Giordano}, {Giroletti}, {Glanzman}, {Godfrey},
  {Grenier}, {Grondin}, {Grove}, {Grundstrom}, {Guiriec}, {Gwon}, {Hadasch},
  {Harding}, {Hayashida}, {Hays}, {J{\'o}hannesson}, {Johnson}, {Johnson},
  {Johnston}, {Kamae}, {Katagiri}, {Kataoka}, {Keith}, {Kerr},
  {Kn{\"o}dlseder}, {Kramer}, {Kuss}, {Lande}, {Lee}, {Lemoine-Goumard},
  {Longo}, {Loparco}, {Lovellette}, {Lubrano}, {Manchester}, {Marelli},
  {Mazziotta}, {Michelson}, {Mitthumsiri}, {Mizuno}, {Moiseev}, {Monte},
  {Monzani}, {Morselli}, {Moskalenko}, {Murgia}, {Nakamori}, {Naumann-Godo},
  {Neronov}, {Nolan}, {Norris}, {Noutsos}, {Nuss}, {Ohsugi}, {Okumura},
  {Omodei}, {Orlando}, {Paneque}, {Parent}, {Pesce-Rollins}, {Pierbattista},
  {Piron}, {Porter}, {Possenti}, {Rain{\`o}}, {Rando}, {Ray}, {Razzano},
  {Razzaque}, {Reimer}, {Reimer}, {Reposeur}, {Ritz}, {Sadrozinski}, {Scargle},
  {Sgr{\`o}}, {Shannon}, {Siskind}, {Smith}, {Spandre}, {Spinelli},
  {Strickman}, {Suson}, {Takahashi}, {Tanaka}, {Thayer}, {Thayer}, {Thompson},
  {Thorsett}, {Tibaldo}, {Tibolla}, {Torres}, {Tosti}, {Troja}, {Uchiyama},
  {Usher}, {Vandenbroucke}, {Vasileiou}, {Vianello}, {Vitale}, {Waite}, {Wang},
  {Winer}, {Wolff}, {Wood}, {Wood}, {Yang}, {Ziegler}, \&
  {Zimmer}}]{Abdo2011B1259}
---. 2011, \apjl, 736, L11, \dodoi{10.1088/2041-8205/736/1/L11}

\bibitem[{{Abdo} {et~al.}(2013{\natexlab{a}}){Abdo}, {Ajello}, {Allafort},
  {Baldini}, {Ballet}, {Barbiellini}, {Baring}, {Bastieri}, {Belfiore},
  {Bellazzini}, {Bhattacharyya}, {Bissaldi}, {Bloom}, {Bonamente}, {Bottacini},
  {Brandt}, {Bregeon}, {Brigida}, {Bruel}, {Buehler}, {Burgay}, {Burnett},
  {Busetto}, {Buson}, {Caliandro}, {Cameron}, {Camilo}, {Caraveo},
  {Casandjian}, {Cecchi}, {{\c{C}}elik}, {Charles}, {Chaty}, {Chaves},
  {Chekhtman}, {Chen}, {Chiang}, {Chiaro}, {Ciprini}, {Claus}, {Cognard},
  {Cohen-Tanugi}, {Cominsky}, {Conrad}, {Cutini}, {D'Ammando}, {de Angelis},
  {DeCesar}, {De Luca}, {den Hartog}, {de Palma}, {Dermer}, {Desvignes},
  {Digel}, {Di Venere}, {Drell}, {Drlica-Wagner}, {Dubois}, {Dumora},
  {Espinoza}, {Falletti}, {Favuzzi}, {Ferrara}, {Focke}, {Franckowiak},
  {Freire}, {Funk}, {Fusco}, {Gargano}, {Gasparrini}, {Germani}, {Giglietto},
  {Giommi}, {Giordano}, {Giroletti}, {Glanzman}, {Godfrey}, {Gotthelf},
  {Grenier}, {Grondin}, {Grove}, {Guillemot}, {Guiriec}, {Hadasch}, {Hanabata},
  {Harding}, {Hayashida}, {Hays}, {Hessels}, {Hewitt}, {Hill}, {Horan}, {Hou},
  {Hughes}, {Jackson}, {Janssen}, {Jogler}, {J{\'o}hannesson}, {Johnson},
  {Johnson}, {Johnson}, {Johnson}, {Johnston}, {Kamae}, {Kataoka}, {Keith},
  {Kerr}, {Kn{\"o}dlseder}, {Kramer}, {Kuss}, {Lande}, {Larsson}, {Latronico},
  {Lemoine-Goumard}, {Longo}, {Loparco}, {Lovellette}, {Lubrano}, {Lyne},
  {Manchester}, {Marelli}, {Massaro}, {Mayer}, {Mazziotta}, {McEnery},
  {McLaughlin}, {Mehault}, {Michelson}, {Mignani}, {Mitthumsiri}, {Mizuno},
  {Moiseev}, {Monzani}, {Morselli}, {Moskalenko}, {Murgia}, {Nakamori},
  {Nemmen}, {Nuss}, {Ohno}, {Ohsugi}, {Orienti}, {Orlando}, {Ormes}, {Paneque},
  {Panetta}, {Parent}, {Perkins}, {Pesce-Rollins}, {Pierbattista}, {Piron},
  {Pivato}, {Pletsch}, {Porter}, {Possenti}, {Rain{\`o}}, {Rando}, {Ransom},
  {Ray}, {Razzano}, {Rea}, {Reimer}, {Reimer}, {Renault}, {Reposeur}, {Ritz},
  {Romani}, {Roth}, {Rousseau}, {Roy}, {Ruan}, {Sartori}, {Saz Parkinson},
  {Scargle}, {Schulz}, {Sgr{\`o}}, {Shannon}, {Siskind}, {Smith}, {Spandre},
  {Spinelli}, {Stappers}, {Strong}, {Suson}, {Takahashi}, {Thayer}, {Thayer},
  {Theureau}, {Thompson}, {Thorsett}, {Tibaldo}, {Tibolla}, {Tinivella},
  {Torres}, {Tosti}, {Troja}, {Uchiyama}, {Usher}, {Vandenbroucke}, \&
  {Vasileiou}}]{lat2013}
{Abdo}, A.~A., {Ajello}, M., {Allafort}, A., {et~al.} 2013{\natexlab{a}},
  \apjs, 208, 17, \dodoi{10.1088/0067-0049/208/2/17}

\bibitem[{{Abdo} {et~al.}(2013{\natexlab{b}}){Abdo}, {Ajello}, {Allafort},
  {Baldini}, {Ballet}, {Barbiellini}, {Baring}, {Bastieri}, {Belfiore},
  {Bellazzini}, {Bhattacharyya}, {Bissaldi}, {Bloom}, {Bonamente}, {Bottacini},
  {Brandt}, {Bregeon}, {Brigida}, {Bruel}, {Buehler}, {Burgay}, {Burnett},
  {Busetto}, {Buson}, {Caliandro}, {Cameron}, {Camilo}, {Caraveo},
  {Casandjian}, {Cecchi}, {{\c{C}}elik}, {Charles}, {Chaty}, {Chaves},
  {Chekhtman}, {Chen}, {Chiang}, {Chiaro}, {Ciprini}, {Claus}, {Cognard},
  {Cohen-Tanugi}, {Cominsky}, {Conrad}, {Cutini}, {D'Ammando}, {de Angelis},
  {DeCesar}, {De Luca}, {den Hartog}, {de Palma}, {Dermer}, {Desvignes},
  {Digel}, {Di Venere}, {Drell}, {Drlica-Wagner}, {Dubois}, {Dumora},
  {Espinoza}, {Falletti}, {Favuzzi}, {Ferrara}, {Focke}, {Franckowiak},
  {Freire}, {Funk}, {Fusco}, {Gargano}, {Gasparrini}, {Germani}, {Giglietto},
  {Giommi}, {Giordano}, {Giroletti}, {Glanzman}, {Godfrey}, {Gotthelf},
  {Grenier}, {Grondin}, {Grove}, {Guillemot}, {Guiriec}, {Hadasch}, {Hanabata},
  {Harding}, {Hayashida}, {Hays}, {Hessels}, {Hewitt}, {Hill}, {Horan}, {Hou},
  {Hughes}, {Jackson}, {Janssen}, {Jogler}, {J{\'o}hannesson}, {Johnson},
  {Johnson}, {Johnson}, {Johnson}, {Johnston}, {Kamae}, {Kataoka}, {Keith},
  {Kerr}, {Kn{\"o}dlseder}, {Kramer}, {Kuss}, {Lande}, {Larsson}, {Latronico},
  {Lemoine-Goumard}, {Longo}, {Loparco}, {Lovellette}, {Lubrano}, {Lyne},
  {Manchester}, {Marelli}, {Massaro}, {Mayer}, {Mazziotta}, {McEnery},
  {McLaughlin}, {Mehault}, {Michelson}, {Mignani}, {Mitthumsiri}, {Mizuno},
  {Moiseev}, {Monzani}, {Morselli}, {Moskalenko}, {Murgia}, {Nakamori},
  {Nemmen}, {Nuss}, {Ohno}, {Ohsugi}, {Orienti}, {Orlando}, {Ormes}, {Paneque},
  {Panetta}, {Parent}, {Perkins}, {Pesce-Rollins}, {Pierbattista}, {Piron},
  {Pivato}, {Pletsch}, {Porter}, {Possenti}, {Rain{\`o}}, {Rando}, {Ransom},
  {Ray}, {Razzano}, {Rea}, {Reimer}, {Reimer}, {Renault}, {Reposeur}, {Ritz},
  {Romani}, {Roth}, {Rousseau}, {Roy}, {Ruan}, {Sartori}, {Saz Parkinson},
  {Scargle}, {Schulz}, {Sgr{\`o}}, {Shannon}, {Siskind}, {Smith}, {Spandre},
  {Spinelli}, {Stappers}, {Strong}, {Suson}, {Takahashi}, {Thayer}, {Thayer},
  {Theureau}, {Thompson}, {Thorsett}, {Tibaldo}, {Tibolla}, {Tinivella},
  {Torres}, {Tosti}, {Troja}, {Uchiyama}, {Usher}, {Vandenbroucke},
  {Vasileiou}, {Venter}, {Vianello}, {Vitale}, {Wang}, {Weltevrede}, {Winer},
  {Wolff}, {Wood}, {Wood}, {Wood}, \& {Yang}}]{abdo2013second}
---. 2013{\natexlab{b}}, \apjs, 208, 17, \dodoi{10.1088/0067-0049/208/2/17}

\bibitem[{{Abdollahi} {et~al.}(2020){Abdollahi}, {Acero}, {Ackermann},
  {Ajello}, {Atwood}, {Axelsson}, {Baldini}, {Ballet}, {Barbiellini},
  {Bastieri}, {Becerra Gonzalez}, {Bellazzini}, {Berretta}, {Bissaldi},
  {Blandford}, \& {Bloom}}]{4FGL}
{Abdollahi}, S., {Acero}, F., {Ackermann}, M., {et~al.} 2020, \apjs, 247, 33,
  \dodoi{10.3847/1538-4365/ab6bcb}

\bibitem[{{Abdollahi} {et~al.}(2022){Abdollahi}, {Acero}, {Baldini}, {Ballet},
  {Bastieri}, {Bellazzini}, {Berenji}, {Berretta}, {Bissaldi}, {Blandford},
  {Bloom}, {Bonino}, {Brill}, {Britto}, {Bruel}, {Burnett}, {Buson}, {Cameron},
  {Caputo}, {Caraveo}, {Castro}, {Chaty}, {Cheung}, {Chiaro}, {Cibrario},
  {Ciprini}, {Coronado-Bl{\'a}zquez}, {Crnogorcevic}, {Cutini}, {D'Ammando},
  {De Gaetano}, {Digel}, {Di Lalla}, {Dirirsa}, {Di Venere}, {Dom{\'\i}nguez},
  {Fallah Ramazani}, {Fegan}, {Ferrara}, {Fiori}, {Fleischhack}, {Franckowiak},
  {Fukazawa}, {Funk}, {Fusco}, {Galanti}, {Gammaldi}, {Gargano}, {Garrappa},
  {Gasparrini}, {Giacchino}, {Giglietto}, {Giordano}, {Giroletti}, {Glanzman},
  {Green}, {Grenier}, {Grondin}, {Guillemot}, {Guiriec}, {Gustafsson},
  {Harding}, {Hays}, {Hewitt}, {Horan}, {Hou}, {J{\'o}hannesson}, {Karwin},
  {Kayanoki}, {Kerr}, {Kuss}, {Landriu}, {Larsson}, {Latronico},
  {Lemoine-Goumard}, {Li}, {Liodakis}, {Longo}, {Loparco}, {Lott}, {Lubrano},
  {Maldera}, {Malyshev}, {Manfreda}, {Mart{\'\i}-Devesa}, {Mazziotta}, {Mereu},
  {Meyer}, {Michelson}, {Mirabal}, {Mitthumsiri}, {Mizuno}, {Moiseev},
  {Monzani}, {Morselli}, {Moskalenko}, {Negro}, {Nuss}, {Omodei}, {Orienti},
  {Orlando}, {Paneque}, {Pei}, {Perkins}, {Persic}, {Pesce-Rollins},
  {Petrosian}, {Pillera}, {Poon}, {Porter}, {Principe}, {Rain{\`o}}, {Rando},
  {Rani}, {Razzano}, {Razzaque}, {Reimer}, {Reimer}, {Reposeur},
  {S{\'a}nchez-Conde}, {Saz Parkinson}, {Scotton}, {Serini}, {Sgr{\`o}},
  {Siskind}, {Smith}, {Spandre}, {Spinelli}, {Sueoka}, {Suson}, {Tajima},
  {Tak}, {Thayer}, {Thompson}, {Torres}, {Troja}, {Valverde}, {Wood}, \&
  {Zaharijas}}]{Abdollahi2022}
{Abdollahi}, S., {Acero}, F., {Baldini}, L., {et~al.} 2022, \apjs, 260, 53,
  \dodoi{10.3847/1538-4365/ac6751}

\bibitem[{{Abeysekara} {et~al.}(2017){Abeysekara}, {Albert}, {Alfaro},
  {Alvarez}, {{\'A}lvarez}, {Arceo}, {Arteaga-Vel{\'a}zquez}, {Ayala Solares},
  {Barber}, {Baughman}, {Bautista-Elivar}, {Becerra Gonzalez}, {Becerril},
  {Belmont-Moreno}, {BenZvi}, {Berley}, {Bernal}, {Braun}, {Brisbois},
  {Caballero-Mora}, {Capistr{\'a}n}, {Carrami{\~n}ana}, {Casanova}, {Castillo},
  {Cotti}, {Cotzomi}, {Couti{\~n}o de Le{\'o}n}, {de la Fuente}, {De Le{\'o}n},
  {Diaz Hernandez}, {Dingus}, {DuVernois}, {D{\'\i}az-V{\'e}lez}, {Ellsworth},
  {Engel}, {Fiorino}, {Fraija}, {Garc{\'\i}a-Gonz{\'a}lez}, {Garfias},
  {Gerhardt}, {Gonz{\'a}lez Mu{\~n}oz}, {Gonz{\'a}lez}, {Goodman},
  {Hampel-Arias}, {Harding}, {Hernandez}, {Hernandez-Almada}, {Hinton}, {Hui},
  {H{\"u}ntemeyer}, {Iriarte}, {Jardin-Blicq}, {Joshi}, {Kaufmann}, {Kieda},
  {Lara}, {Lauer}, {Lee}, {Lennarz}, {Le{\'o}n Vargas}, {Linnemann},
  {Longinotti}, {Raya}, {Luna-Garc{\'\i}a}, {L{\'o}pez-Coto}, {Malone},
  {Marinelli}, {Martinez}, {Martinez-Castellanos}, {Mart{\'\i}nez-Castro},
  {Mart{\'\i}nez-Huerta}, {Matthews}, {Miranda-Romagnoli}, {Moreno},
  {Mostaf{\'a}}, {Nellen}, {Newbold}, {Nisa}, {Noriega-Papaqui}, {Pelayo},
  {Pretz}, {P{\'e}rez-P{\'e}rez}, {Ren}, {Rho}, {Rivi{\`e}re},
  {Rosa-Gonz{\'a}lez}, {Rosenberg}, {Ruiz-Velasco}, {Salazar}, {Salesa Greus},
  {Sandoval}, {Schneider}, {Schoorlemmer}, {Sinnis}, {Smith}, {Springer},
  {Surajbali}, {Taboada}, {Tibolla}, {Tollefson}, {Torres}, {Ukwatta},
  {Vianello}, {Villase{\~n}or}, {Weisgarber}, {Westerhoff}, {Wisher}, {Wood},
  {Yapici}, {Younk}, {Zepeda}, \& {Zhou}}]{Abeysekara2017}
{Abeysekara}, A.~U., {Albert}, A., {Alfaro}, R., {et~al.} 2017, \apj, 843, 40,
  \dodoi{10.3847/1538-4357/aa7556}

\bibitem[{{Abeysekara} {et~al.}(2018{\natexlab{a}}){Abeysekara}, {Albert},
  {Alfaro}, {Alvarez}, {{\'A}lvarez}, {Arceo}, {Arteaga-Vel{\'a}zquez}, {Avila
  Rojas}, {Ayala Solares}, {Belmont-Moreno}, {BenZvi}, {Brisbois},
  {Caballero-Mora}, {Capistr{\'a}n}, {Carrami{\~n}ana}, {Casanova}, {Castillo},
  {Cotti}, {Cotzomi}, {Couti{\~n}o de Le{\'o}n}, {De Le{\'o}n}, {De la Fuente},
  {D{\'\i}az-V{\'e}lez}, {Dichiara}, {Dingus}, \& {DuVernois}}]{HAWC2018}
---. 2018{\natexlab{a}}, \nat, 562, 82, \dodoi{10.1038/s41586-018-0565-5}

\bibitem[{{Abeysekara} {et~al.}(2018{\natexlab{b}}){Abeysekara}, {Archer},
  {Benbow}, {Bird}, {Brose}, {Buchovecky}, {Buckley}, {Bugaev}, {Chromey},
  {Connolly}, {Cui}, {Daniel}, {Falcone}, {Feng}, {Finley}, {Fortson},
  {Furniss}, {H{\"u}tten}, {Hanna}, {Hervet}, {Holder}, {Hughes}, {Humensky},
  {Johnson}, {Kaaret}, {Kar}, {Kertzman}, {Kieda}, {Krause}, {Krennrich},
  {Kumar}, {Lang}, {Lin}, {McArthur}, {Moriarty}, {Mukherjee}, {O'Brien},
  {Ong}, {Otte}, {Park}, {Petrashyk}, {Pohl}, {Pueschel}, {Quinn}, {Ragan},
  {Reynolds}, {Richards}, {Roache}, {Rulten}, {Sadeh}, {Santander},
  {Sembroski}, {Shahinyan}, {Sushch}, {Tyler}, {Wakely}, {Weinstein}, {Wells},
  {Wilcox}, {Wilhelm}, {Williams}, {Williamson}, {Zitzer}, {VERITAS
  Collaboration}, {Abdollahi}, {Ajello}, {Baldini}, {Barbiellini}, {Bastieri},
  {Bellazzini}, {Berenji}, {Bissaldi}, {Blandford}, {Bonino}, {Bottacini},
  {Brandt}, {Bruel}, {Buehler}, {Cameron}, {Caputo}, {Caraveo}, {Castro},
  {Cavazzuti}, {Charles}, {Chiaro}, {Ciprini}, {Cohen-Tanugi}, {Costantin},
  {Cutini}, {D'Ammando}, {de Palma}, {Di Lalla}, {Di Mauro}, {Di Venere},
  {Dom{\'\i}nguez}, {Favuzzi}, {Fegan}, {Franckowiak}, {Fukazawa}, {Funk},
  {Fusco}, {Gargano}, {Gasparrini}, {Giglietto}, {Giordano}, {Giroletti},
  {Green}, {Grenier}, {Guillemot}, {Guiriec}, {Hays}, {Hewitt}, {Horan},
  {J{\'o}hannesson}, {Kensei}, {Kuss}, {Larsson}, {Latronico},
  {Lemoine-Goumard}, {Li}, {Longo}, {Loparco}, {Lovellette}, {Lubrano},
  {Magill}, {Maldera}, {Mazziotta}, {McEnery}, {Michelson}, {Mitthumsiri},
  {Mizuno}, {Monzani}, {Morselli}, {Moskalenko}, {Negro}, {Nuss}, {Ojha},
  {Omodei}, {Orienti}, {Orlando}, {Palatiello}, {Paliya}, {Paneque}, {Perkins},
  {Persic}, {Pesce-Rollins}, {Petrosian}, {Piron}, {Porter}, {Principe},
  {Rain{\`o}}, {Rando}, {Rani}, {Razzano}, {Razzaque}, {Reimer}, {Reimer},
  {Reposeur}, {Sgr{\`o}}, {Siskind}, {Spandre}, {Spinelli}, {Suson}, {Tajima},
  {Thayer}, {Thompson}, {Torres}, {Tosti}, {Troja}, {Valverde}, {Vianello},
  {Vogel}, {Wood}, {Yassine}, {Fermi-LAT Collaboration}, {Alfaro},
  {{\'A}lvarez}, {{\'A}lvarez}, {Arceo}, {Arteaga-Vel{\'a}zquez}, {Avila
  Rojas}, {Ayala Solares}, {Becerril}, {Belmont-Moreno}, {BenZvi}, {Bernal},
  {Braun}, {Brisbois}, {Caballero-Mora}, {Capistr{\'a}n}, {Carrami{\~n}ana},
  {Casanova}, {Castillo}, {Cotti}, {Cotzomi}, {Couti{\~n}o de Le{\'o}n}, {De
  Le{\'o}n}, {De la Fuente}, {Dichiara}, {Dingus}, {DuVernois},
  {D{\'\i}az-V{\'e}lez}, {Engel}, {Enriquez-Rivera}, {Fiorino}, {Fleischhack},
  {Fraija}, {Garc{\'\i}a-Gonz{\'a}lez}, {Garfias}, {Gonz{\'a}lez Mu{\~n}oz},
  {Gonz{\'a}lez}, {Goodman}, {Hampel-Arias}, {Harding}, {Hernandez},
  {Hernandez-Almada}, {Hona}, {Hueyotl-Zahuantitla}, {Hui}, {H{\"u}ntemeyer},
  {Iriarte}, {Jardin-Blicq}, {Joshi}, {Kaufmann}, {Lara}, {Lauer}, {Lee},
  {Lennarz}, {Le{\'o}n Vargas}, {Linnemann}, {Longinotti}, {Luis-Raya},
  {Luna-Garc{\'\i}a}, {L{\'o}pez-Coto}, {Malone}, {Marinelli}, {Martinez},
  {Martinez-Castellanos}, {Mart{\'\i}nez-Castro}, {Mart{\'\i}nez-Huerta},
  {Matthews}, {Miranda-Romagnoli}, {Moreno}, {Mostaf{\'a}}, {Nayerhoda},
  {Nellen}, {Newbold}, {Nisa}, {Noriega-Papaqui}, {Pelayo}, {Pretz},
  {P{\'e}rez-P{\'e}rez}, {Ren}, {Rho}, {Rivi{\`e}re}, {Rosa-Gonz{\'a}lez},
  {Rosenberg}, {Ruiz-Velasco}, {Salazar}, {Salesa Greus}, {Sandoval},
  {Schneider}, {Seglar Arroyo}, {Sinnis}, {Smith}, {Springer}, {Surajbali},
  {Taboada}, {Tibolla}, {Tollefson}, {Torres}, {Ukwatta}, {Villase{\~n}or},
  {Weisgarber}, {Westerhoff}, {Wisher}, {Wood}, {Yapici}, {Yodh}, {Zepeda},
  {Zhou}, \& {HAWC Collaboration}}]{abeysekara2018}
{Abeysekara}, A.~U., {Archer}, A., {Benbow}, W., {et~al.} 2018{\natexlab{b}},
  \apj, 866, 24, \dodoi{10.3847/1538-4357/aade4e}

\bibitem[{{Abeysekara} {et~al.}(2020){Abeysekara}, {Albert}, {Alfaro}, {Angeles
  Camacho}, {Arteaga-Vel{\'a}zquez}, {Arunbabu}, {Avila Rojas}, {Ayala
  Solares}, {Baghmanyan}, {Belmont-Moreno}, {BenZvi}, {Brisbois},
  {Caballero-Mora}, {Capistr{\'a}n}, {Carrami{\~n}ana}, {Casanova}, {Cotti},
  {Cotzomi}, {Couti{\~n}o de Le{\'o}n}, {De la Fuente}, {de Le{\'o}n},
  {Dichiara}, {Dingus}, {DuVernois}, {D{\'\i}az-V{\'e}lez}, {Ellsworth},
  {Engel}, {Espinoza}, {Fleischhack}, {Fraija}, {Galv{\'a}n-G{\'a}mez},
  {Garcia}, {Garc{\'\i}a-Gonz{\'a}lez}, {Garfias}, {Gonz{\'a}lez}, {Goodman},
  {Harding}, {Hernandez}, {Hinton}, {Hona}, {Huang}, {Hueyotl-Zahuantitla},
  {H{\"u}ntemeyer}, {Iriarte}, {Jardin-Blicq}, {Joshi}, {Kaufmann}, {Kieda},
  {Lara}, {Lee}, {Le{\'o}n Vargas}, {Linnemann}, {Longinotti}, {Luis-Raya},
  {Lundeen}, {L{\'o}pez-Coto}, {Malone}, {Marinelli}, {Martinez},
  {Martinez-Castellanos}, {Mart{\'\i}nez-Castro}, {Mart{\'\i}nez-Huerta},
  {Matthews}, {Miranda-Romagnoli}, {Morales-Soto}, {Moreno}, {Mostaf{\'a}},
  {Nayerhoda}, {Nellen}, {Newbold}, {Nisa}, {Noriega-Papaqui}, {Peisker},
  {P{\'e}rez-P{\'e}rez}, {Pretz}, {Ren}, {Rho}, {Rivi{\`e}re},
  {Rosa-Gonz{\'a}lez}, {Rosenberg}, {Ruiz-Velasco}, {Salesa Greus}, {Sandoval},
  {Schneider}, {Schoorlemmer}, {Sinnis}, {Smith}, {Springer}, {Surajbali},
  {Tabachnick}, {Tanner}, {Tibolla}, {Tollefson}, {Torres}, {Torres-Escobedo},
  {Villase{\~n}or}, {Weisgarber}, {Wood}, {Yapici}, {Zhang}, {Zhou}, \& {HAWC
  Collaboration}}]{Abeysekara2020}
{Abeysekara}, A.~U., {Albert}, A., {Alfaro}, R., {et~al.} 2020, \prl, 124,
  021102, \dodoi{10.1103/PhysRevLett.124.021102}

\bibitem[{{Acciari} {et~al.}(2009){Acciari}, {Aliu}, {Arlen}, {Aune},
  {Bautista}, {Beilicke}, {Benbow}, {Boltuch}, {Bradbury}, {Buckley}, {Bugaev},
  {Butt}, {Byrum}, {Cannon}, {Cesarini}, {Chow}, {Ciupik}, {Cogan}, {Cui},
  {Dickherber}, {Ergin}, {Fegan}, {Finley}, {Fortin}, {Fortson}, {Furniss},
  {Gall}, {Gillanders}, {Gotthelf}, {Grube}, {Guenette}, {Gyuk}, {Hanna},
  {Holder}, {Horan}, {Hui}, {Humensky}, {Kaaret}, {Karlsson}, {Kertzman},
  {Kieda}, {Konopelko}, {Krawczynski}, {Krennrich}, {Lang}, {LeBohec}, {Maier},
  {McCann}, {McCutcheon}, {Millis}, {Moriarty}, {Mukherjee}, {Ong}, {Otte},
  {Pandel}, {Perkins}, {Pohl}, {Quinn}, {Ragan}, {Reyes}, {Reynolds}, {Roache},
  {Rose}, {Schroedter}, {Sembroski}, {Smith}, {Steele}, {Swordy}, {Theiling},
  {Toner}, {Vassiliev}, {Vincent}, {Wagner}, {Wakely}, {Ward}, {Weekes},
  {Weinstein}, {Weisgarber}, {Williams}, {Wissel}, {Wood}, \&
  {Zitzer}}]{acciari09}
{Acciari}, V.~A., {Aliu}, E., {Arlen}, T., {et~al.} 2009, \apjl, 703, L6,
  \dodoi{10.1088/0004-637X/703/1/L6}

\bibitem[{{Acciari} {et~al.}(2010){Acciari}, {Aliu}, {Arlen}, {Aune},
  {Bautista}, {Beilicke}, {Benbow}, {Boltuch}, {Bradbury}, {Buckley}, {Bugaev},
  {Butt}, {Byrum}, {Cesarini}, {Ciupik}, {Cui}, {Dickherber}, {Duke}, {Finley},
  {Finnegan}, {Fortson}, {Furniss}, {Galante}, {Gall}, {Gillanders}, {Godambe},
  {Gotthelf}, {Grube}, {Guenette}, {Gyuk}, {Hanna}, {Holder}, {Hui},
  {Humensky}, {Imran}, {Kaaret}, {Karlsson}, {Kertzman}, {Kieda}, {Konopelko},
  {Krawczynski}, {Krennrich}, {Lang}, {LeBohec}, {Maier}, {McArthur}, {McCann},
  {McCutcheon}, {Moriarty}, {Muhkerjee}, {Ong}, {Otte}, {Pandel}, {Perkins},
  {Pohl}, {Quinn}, {Ragan}, {Reyes}, {Reynolds}, {Roache}, {Rose},
  {Schroedter}, {Sembroski}, {Senturk}, {Slane}, {Smith}, {Steele}, {Swordy},
  {T{\v{e}}si{\'c}}, {Theiling}, {Thibadeau}, {Vassiliev}, {Vincent}, {Wakely},
  {Ward}, {Weekes}, {Weinstein}, {Weisgarber}, {Williams}, {Wissel}, {Wood}, \&
  {Zitzer}}]{acciari2010}
---. 2010, \apjl, 719, L69, \dodoi{10.1088/2041-8205/719/1/L69}

\bibitem[{{Acciari} {et~al.}(2011){Acciari}, {Aliu}, {Araya}, {Arlen}, {Aune},
  {Beilicke}, {Benbow}, {Bradbury}, {Buckley}, {Bugaev}, {Byrum}, {Cannon},
  {Cesarini}, {Ciupik}, {Collins-Hughes}, {Cui}, {Dickherber}, {Duke},
  {Falcone}, {Finley}, {Fortson}, {Furniss}, {Galante}, {Gall}, {Godambe},
  {Griffin}, {Guenette}, {Gyuk}, {Hanna}, {Holder}, {Hughes}, {Hui},
  {Humensky}, {Imran}, {Kaaret}, {Kertzman}, {Krawczynski}, {Krennrich},
  {Madhavan}, {Maier}, {Majumdar}, {McArthur}, {Moriarty}, {Ong}, {Otte},
  {Pandel}, {Park}, {Perkins}, {Pohl}, {Prokoph}, {Quinn}, {Ragan}, {Reyes},
  {Reynolds}, {Roache}, {Rose}, {Saxon}, {Sembroski}, {{\c{S}}ent{\"u}rk},
  {Smith}, {Te{\v{s}}i{\'c}}, {Theiling}, {Thibadeau}, {Varlotta}, {Vincent},
  {Vivier}, {Wakely}, {Ward}, {Weekes}, {Weinstein}, {Weisgarber}, {Weng},
  {Williams}, {Wood}, \& {Zitzer}}]{Acciari2011}
{Acciari}, V.~A., {Aliu}, E., {Araya}, M., {et~al.} 2011, \apj, 733, 96,
  \dodoi{10.1088/0004-637X/733/2/96}

\bibitem[{{Acero} {et~al.}(2013){Acero}, {Ackermann}, {Ajello}, {Allafort},
  {Baldini}, {Ballet}, {Barbiellini}, {Bastieri}, {Bechtol}, {Bellazzini},
  {Blandford}, {Bloom}, {Bonamente}, {Bottacini}, {Brandt}, {Bregeon},
  {Brigida}, {Bruel}, {Buehler}, {Buson}, {Caliandro}, {Cameron}, {Caraveo},
  {Cecchi}, {Charles}, {Chaves}, {Chekhtman}, {Chiang}, {Chiaro}, {Ciprini},
  {Claus}, {Cohen-Tanugi}, {Conrad}, {Cutini}, {Dalton}, {D'Ammando}, {de
  Palma}, {Dermer}, {Di Venere}, {Silva}, {Drell}, {Drlica-Wagner}, {Falletti},
  {Favuzzi}, {Fegan}, {Ferrara}, {Focke}, {Franckowiak}, {Fukazawa}, {Funk},
  {Fusco}, {Gargano}, {Gasparrini}, {Giglietto}, {Giordano}, {Giroletti},
  {Glanzman}, {Godfrey}, {Gr{\'e}goire}, {Grenier}, {Grondin}, {Grove},
  {Guiriec}, {Hadasch}, {Hanabata}, {Harding}, {Hayashida}, {Hayashi}, {Hays},
  {Hewitt}, {Hill}, {Horan}, {Hou}, {Hughes}, {Inoue}, {Jackson}, {Jogler},
  {J{\'o}hannesson}, {Johnson}, {Kamae}, {Kawano}, {Kerr}, {Kn{\"o}dlseder},
  {Kuss}, {Lande}, {Larsson}, {Latronico}, {Lemoine-Goumard}, {Longo},
  {Loparco}, {Lovellette}, {Lubrano}, {Marelli}, {Massaro}, {Mayer},
  {Mazziotta}, {McEnery}, {Mehault}, {Michelson}, {Mitthumsiri}, {Mizuno},
  {Monte}, {Monzani}, {Morselli}, {Moskalenko}, {Murgia}, {Nakamori}, {Nemmen},
  {Nuss}, {Ohsugi}, {Okumura}, {Orienti}, {Orlando}, {Ormes}, {Paneque},
  {Panetta}, {Perkins}, {Pesce-Rollins}, {Piron}, {Pivato}, {Porter},
  {Rain{\`o}}, {Rando}, {Razzano}, {Reimer}, {Reimer}, {Reposeur}, {Ritz},
  {Roth}, {Rousseau}, {Saz Parkinson}, {Schulz}, {Sgr{\`o}}, {Siskind},
  {Smith}, {Spandre}, {Spinelli}, {Suson}, {Takahashi}, {Takeuchi}, {Thayer},
  {Thayer}, {Thompson}, {Tibaldo}, {Tibolla}, {Tinivella}, {Torres}, {Tosti},
  {Troja}, {Uchiyama}, {Vandenbroucke}, {Vasileiou}, {Vianello}, {Vitale},
  {Werner}, {Winer}, {Wood}, \& {Yang}}]{Acero2013}
{Acero}, F., {Ackermann}, M., {Ajello}, M., {et~al.} 2013, \apj, 773, 77,
  \dodoi{10.1088/0004-637X/773/1/77}

\bibitem[{{Acero} {et~al.}(2015){Acero}, {Ackermann}, {Ajello}, {Albert},
  {Atwood}, {Axelsson}, {Baldini}, {Ballet}, {Barbiellini}, \&
  {Bastieri}}]{3FGL}
---. 2015, \apjs, 218, 23, \dodoi{10.1088/0067-0049/218/2/23}

\bibitem[{{Ackermann} {et~al.}(2012){Ackermann}, {Ajello}, {Ballet},
  {Barbiellini}, {Bastieri}, {Belfiore}, {Bellazzini}, {Berenji}, {Blandford},
  \& {Bloom}}]{Ackermann2012}
{Ackermann}, M., {Ajello}, M., {Ballet}, J., {et~al.} 2012, Science, 335, 189,
  \dodoi{10.1126/science.1213974}

\bibitem[{{Ackermann} {et~al.}(2013{\natexlab{a}}){Ackermann}, {Ajello},
  {Allafort}, {Atwood}, {Baldini}, {Ballet}, {Barbiellini}, {Bastieri},
  {Bechtol}, {Belfiore}, {Bellazzini}, {Bernieri}, {Bissaldi}, {Bloom},
  {Bonamente}, {Brandt}, {Bregeon}, {Brigida}, {Bruel}, {Buehler}, {Burnett},
  {Buson}, {Caliandro}, {Cameron}, {Campana}, {Caraveo}, {Casandjian},
  {Cavazzuti}, {Cecchi}, {Charles}, {Chaves}, {Chekhtman}, {Cheung}, {Chiang},
  {Chiaro}, {Ciprini}, {Claus}, {Cohen-Tanugi}, {Cominsky}, {Conrad}, {Cutini},
  {D'Ammando}, {de Angelis}, {de Palma}, {Dermer}, {Desiante}, {Digel}, {Di
  Venere}, {Drell}, {Drlica-Wagner}, {Favuzzi}, {Fegan}, {Ferrara}, {Focke},
  {Fortin}, {Franckowiak}, {Funk}, {Fusco}, {Gargano}, {Gasparrini}, {Gehrels},
  {Germani}, {Giglietto}, {Giommi}, {Giordano}, {Giroletti}, {Godfrey},
  {Gomez-Vargas}, {Grenier}, {Guiriec}, {Hadasch}, {Hanabata}, {Harding},
  {Hayashida}, {Hays}, {Hewitt}, {Hill}, {Horan}, {Hughes}, {Jogler},
  {J{\'o}hannesson}, {Johnson}, {Johnson}, {Johnson}, {Kamae}, {Kataoka},
  {Kawano}, {Kn{\"o}dlseder}, {Kuss}, {Lande}, {Larsson}, {Latronico},
  {Lemoine-Goumard}, {Longo}, {Loparco}, {Lott}, {Lovellette}, {Lubrano},
  {Massaro}, {Mayer}, {Mazziotta}, {McEnery}, {Mehault}, {Michelson}, {Mizuno},
  {Moiseev}, {Monzani}, {Morselli}, {Moskalenko}, {Murgia}, {Nemmen}, {Nuss},
  {Ohsugi}, {Okumura}, {Orienti}, {Ormes}, {Paneque}, {Perkins},
  {Pesce-Rollins}, {Piron}, {Pivato}, {Porter}, {Rain{\`o}}, {Razzano},
  {Reimer}, {Reimer}, {Reposeur}, {Ritz}, {Romani}, {Roth}, {Saz Parkinson},
  {Schulz}, {Sgr{\`o}}, {Siskind}, {Smith}, {Spandre}, {Spinelli}, {Stawarz},
  {Strong}, {Suson}, {Takahashi}, {Thayer}, {Thayer}, {Thompson}, {Tibaldo},
  {Tinivella}, {Torres}, {Tosti}, {Troja}, {Uchiyama}, {Usher},
  {Vandenbroucke}, {Vasileiou}, {Vianello}, {Vitale}, {Werner}, {Winer},
  {Wood}, \& {Wood}}]{Ackermann2013}
{Ackermann}, M., {Ajello}, M., {Allafort}, A., {et~al.} 2013{\natexlab{a}},
  \apjs, 209, 34, \dodoi{10.1088/0067-0049/209/2/34}

\bibitem[{{Ackermann} {et~al.}(2013{\natexlab{b}}){Ackermann}, {Ajello},
  {Ballet}, {Barbiellini}, {Bastieri}, {Bellazzini}, {Bonamente}, {Brandt},
  {Bregeon}, {Brigida}, {Bruel}, {Buehler}, {Buson}, {Caliandro}, {Cameron},
  {Caraveo}, {Casandjian}, {Cavazzuti}, {Cecchi}, {Chekhtman}, {Chiang},
  {Chiaro}, {Ciprini}, {Claus}, {Cohen-Tanugi}, {Cominsky}, {Conrad}, {Cutini},
  {Dalton}, {D'Ammando}, {de Angelis}, {den Hartog}, {de Palma}, {Dermer},
  {Digel}, {Di Venere}, {Drell}, {Dubois}, {Favuzzi}, {Fegan}, {Ferrara},
  {Focke}, {Franckowiak}, {Funk}, {Fusco}, {Gargano}, {Gasparrini}, {Germani},
  {Giglietto}, {Giordano}, {Giroletti}, {Glanzman}, {Godfrey}, {Grenier},
  {Guiriec}, {Hadasch}, {Hanabata}, {Harding}, {Hayashida}, {Hays}, {Hill},
  {Horan}, {Hughes}, {Jogler}, {J{\'o}hannesson}, {Johnson}, {Johnson},
  {Kawano}, {Kerr}, {Kn{\"o}dlseder}, {Kuss}, {Lande}, {Larsson}, {Latronico},
  {Lemoine-Goumard}, {Li}, {Longo}, {Lovellette}, {Lubrano}, {Mayer},
  {Mazziotta}, {McEnery}, {Michelson}, {Mizuno}, {Monzani}, {Morselli},
  {Moskalenko}, {Murgia}, {Nemmen}, {Nuss}, {Ohsugi}, {Okumura}, {Orienti},
  {Orlando}, {Ormes}, {Paneque}, {Papitto}, {Perkins}, {Pesce-Rollins},
  {Piron}, {Pivato}, {Rain{\`o}}, {Rando}, {Razzano}, {Rea}, {Reimer},
  {Reimer}, {Scargle}, {Schulz}, {Sgr{\`o}}, {Siskind}, {Spandre}, {Spinelli},
  {Takahashi}, {Thayer}, {Thayer}, {Tinivella}, {Torres}, {Tosti}, {Troja},
  {Uchiyama}, {Usher}, {Vandenbroucke}, {Vasileiou}, {Vianello}, {Vitale},
  {Werner}, {Winer}, \& {Wood}}]{Ackermann2013LS}
{Ackermann}, M., {Ajello}, M., {Ballet}, J., {et~al.} 2013{\natexlab{b}},
  \apjl, 773, L35, \dodoi{10.1088/2041-8205/773/2/L35}

\bibitem[{{Ackermann} {et~al.}(2016){Ackermann}, {Ajello}, {Atwood}, {Baldini},
  {Ballet}, {Barbiellini}, {Bastieri}, {Becerra Gonzalez}, {Bellazzini},
  {Bissaldi}, {Blandford}, {Bloom}, {Bonino}, {Bottacini}, {Brandt}, {Bregeon},
  {Bruel}, {Buehler}, {Buson}, {Caliandro}, {Cameron}, {Caputo}, {Caragiulo},
  {Caraveo}, {Cavazzuti}, {Cecchi}, {Charles}, {Chekhtman}, {Cheung}, {Chiang},
  {Chiaro}, {Ciprini}, {Cohen}, {Cohen-Tanugi}, {Cominsky}, {Conrad}, {Cuoco},
  {Cutini}, {D'Ammando}, {de Angelis}, {de Palma}, {Desiante}, {Di Mauro}, {Di
  Venere}, {Dom{\'\i}nguez}, {Drell}, {Favuzzi}, {Fegan}, {Ferrara}, {Focke},
  {Fortin}, {Franckowiak}, {Fukazawa}, {Funk}, {Furniss}, {Fusco}, {Gargano},
  {Gasparrini}, {Giglietto}, {Giommi}, {Giordano}, {Giroletti}, {Glanzman},
  {Godfrey}, {Grenier}, {Grondin}, {Guillemot}, {Guiriec}, {Harding}, {Hays},
  {Hewitt}, {Hill}, {Horan}, {Iafrate}, {Hartmann}, {Jogler},
  {J{\'o}hannesson}, {Johnson}, {Kamae}, {Kataoka}, {Kn{\"o}dlseder}, {Kuss},
  {La Mura}, {Larsson}, {Latronico}, {Lemoine-Goumard}, {Li}, {Li}, {Longo},
  {Loparco}, {Lott}, {Lovellette}, {Lubrano}, {Madejski}, {Maldera},
  {Manfreda}, {Mayer}, {Mazziotta}, {Michelson}, {Mirabal}, {Mitthumsiri},
  {Mizuno}, {Moiseev}, {Monzani}, {Morselli}, {Moskalenko}, {Murgia}, {Nuss},
  {Ohsugi}, {Omodei}, {Orienti}, {Orlando}, {Ormes}, {Paneque}, {Perkins},
  {Pesce-Rollins}, {Petrosian}, {Piron}, {Pivato}, {Porter}, {Rain{\`o}},
  {Rando}, {Razzano}, {Razzaque}, {Reimer}, {Reimer}, {Reposeur}, {Romani},
  {S{\'a}nchez-Conde}, {Saz Parkinson}, {Schmid}, {Schulz}, {Sgr{\`o}},
  {Siskind}, {Spada}, {Spandre}, {Spinelli}, {Suson}, {Tajima}, {Takahashi},
  {Takahashi}, {Takahashi}, {Thayer}, {Thompson}, {Tibaldo}, {Torres}, {Tosti},
  {Troja}, {Vianello}, {Wood}, {Wood}, {Yassine}, {Zaharijas}, \&
  {Zimmer}}]{ackermann2016}
{Ackermann}, M., {Ajello}, M., {Atwood}, W.~B., {et~al.} 2016, \apjs, 222, 5,
  \dodoi{10.3847/0067-0049/222/1/5}

\bibitem[{{Aharonian} {et~al.}(2006){Aharonian}, {Akhperjanian}, {Bazer-Bachi},
  {Beilicke}, {Benbow}, {Berge}, {Bernl{\"o}hr}, {Boisson}, {Bolz}, {Borrel},
  {Braun}, {Breitling}, {Brown}, {Chadwick}, {Chounet}, {Cornils},
  {Costamante}, {Degrange}, {Dickinson}, {Djannati-Ata{\"\i}}, {Drury},
  {Dubus}, {Emmanoulopoulos}, {Espigat}, {Feinstein}, {Fontaine}, {Fuchs},
  {Funk}, {Gallant}, {Giebels}, {Gillessen}, {Glicenstein}, {Goret},
  {Hadjichristidis}, {Hauser}, {Heinzelmann}, {Henri}, {Hermann}, {Hinton},
  {Hofmann}, {Holleran}, {Horns}, {Jacholkowska}, {de Jager}, {Kh{\'e}lifi},
  {Komin}, {Konopelko}, {Latham}, {Le Gallou}, {Lemi{\`e}re},
  {Lemoine-Goumard}, {Leroy}, {Lohse}, {Martin}, {Martineau-Huynh},
  {Marcowith}, {Masterson}, {McComb}, {de Naurois}, {Nolan}, {Noutsos},
  {Orford}, {Osborne}, {Ouchrif}, {Panter}, {Pelletier}, {Pita},
  {P{\"u}hlhofer}, {Punch}, {Raubenheimer}, {Raue}, {Raux}, {Rayner}, {Reimer},
  {Reimer}, {Ripken}, {Rob}, {Rolland}, {Rowell}, {Sahakian}, {Saug{\'e}},
  {Schlenker}, {Schlickeiser}, {Schuster}, {Schwanke}, {Siewert}, {Sol},
  {Spangler}, {Steenkamp}, {Stegmann}, {Tavernet}, {Terrier}, {Th{\'e}oret},
  {Tluczykont}, {Vasileiadis}, {Venter}, {Vincent}, {V{\"o}lk}, \&
  {Wagner}}]{aharonian2006}
{Aharonian}, F., {Akhperjanian}, A.~G., {Bazer-Bachi}, A.~R., {et~al.} 2006,
  \apj, 636, 777, \dodoi{10.1086/498013}

\bibitem[{{Aharonian} {et~al.}(2007){Aharonian}, {Akhperjanian}, {Bazer-Bachi},
  {Behera}, {Beilicke}, {Benbow}, {Berge}, {Bernl{\"o}hr}, {Boisson}, {Bolz},
  {Borrel}, {Braun}, {Brion}, {Brown}, {B{\"u}hler}, {B{\"u}sching},
  {Boutelier}, {Carrigan}, {Chadwick}, {Chounet}, {Coignet}, {Cornils},
  {Costamante}, {Degrange}, {Dickinson}, {Djannati-Ata{\"\i}}, {Domainko},
  {Drury}, {Dubus}, {Egberts}, {Emmanoulopoulos}, {Espigat}, {Farnier},
  {Feinstein}, {Fiasson}, {F{\"o}rster}, {Fontaine}, {Funk}, {Funk},
  {F{\"u}{\ss}ling}, {Gallant}, {Giebels}, {Glicenstein}, {Gl{\"u}ck}, {Goret},
  {Hadjichristidis}, {Hauser}, {Hauser}, {Heinzelmann}, {Henri}, {Hermann},
  {Hinton}, {Hoffmann}, {Hofmann}, {Holleran}, {Hoppe}, {Horns},
  {Jacholkowska}, {de Jager}, {Kendziorra}, {Kerschhaggl}, {Kh{\'e}lifi},
  {Komin}, {Kosack}, {Lamanna}, {Latham}, {Le Gallou}, {Lemi{\`e}re},
  {Lemoine-Goumard}, {Lohse}, {Martin}, {Martineau-Huynh}, {Marcowith},
  {Masterson}, {Maurin}, {McComb}, {Moulin}, {de Naurois}, {Nedbal}, {Nolan},
  {Noutsos}, {Olive}, {Orford}, {Osborne}, {Panter}, {Pedaletti}, {Pelletier},
  {Petrucci}, {Pita}, {P{\"u}hlhofer}, {Punch}, {Ranchon}, {Raubenheimer},
  {Raue}, {Rayner}, {Ripken}, {Rob}, {Rolland}, {Rosier-Lees}, {Rowell},
  {Ruppel}, {Sahakian}, {Santangelo}, {Saug{\'e}}, {Schlenker}, {Schlickeiser},
  {Schr{\"o}der}, {Schwanke}, {Schwarzburg}, {Schwemmer}, {Shalchi}, {Sol},
  {Spangler}, {Steenkamp}, {Stegmann}, {Superina}, {Tam}, {Tavernet},
  {Terrier}, {Tluczykont}, {van Eldik}, {Vasileiadis}, {Venter}, {Vialle},
  {Vincent}, {V{\"o}lk}, {Wagner}, \& {Ward}}]{Aharonian2007}
---. 2007, \aap, 472, 489, \dodoi{10.1051/0004-6361:20077280}

\bibitem[{{Aharonian} {et~al.}(2008){Aharonian}, {Akhperjanian}, {Barres de
  Almeida}, {Bazer-Bachi}, {Behera}, {Beilicke}, {Benbow}, {Bernl{\"o}hr},
  {Boisson}, {Bolz}, {Borrel}, {Braun}, {Brion}, {Brown}, {B{\"u}hler},
  {Bulik}, {B{\"u}sching}, {Boutelier}, {Carrigan}, {Chadwick}, {Chounet},
  {Clapson}, {Coignet}, {Cornils}, {Costamante}, {Dalton}, {Degrange},
  {Dickinson}, {Djannati-Ata{\"\i}}, {Domainko}, {Drury}, {Dubois}, {Dubus},
  {Dyks}, {Egberts}, {Emmanoulopoulos}, {Espigat}, {Farnier}, {Feinstein},
  {Fiasson}, {F{\"o}rster}, {Fontaine}, {Funk}, {F{\"u}{\ss}ling}, {Gallant},
  {Giebels}, {Glicenstein}, {Gl{\"u}ck}, {Goret}, {Hadjichristidis}, {Hauser},
  {Hauser}, {Heinzelmann}, {Henri}, {Hermann}, {Hinton}, {Hoffmann}, {Hofmann},
  {Holleran}, {Hoppe}, {Horns}, {Jacholkowska}, {de Jager}, {Jung},
  {Katarzy{\'n}ski}, {Kendziorra}, {Kerschhaggl}, {Kh{\'e}lifi}, {Keogh},
  {Komin}, {Kosack}, {Lamanna}, {Latham}, {Lemi{\`e}re}, {Lemoine-Goumard},
  {Lenain}, {Lohse}, {Martin}, {Martineau-Huynh}, {Marcowith}, {Masterson},
  {Maurin}, {Maurin}, {McComb}, {Moderski}, {Moulin}, {de Naurois}, {Nedbal},
  {Nolan}, {Ohm}, {Olive}, {de O{\~n}a Wilhelmi}, {Orford}, {Osborne},
  {Ostrowski}, {Panter}, {Pedaletti}, {Pelletier}, {Petrucci}, {Pita},
  {P{\"u}hlhofer}, {Punch}, {Ranchon}, {Raubenheimer}, {Raue}, {Rayner},
  {Renaud}, {Ripken}, {Rob}, {Rolland}, {Rosier-Lees}, {Rowell}, {Rudak},
  {Ruppel}, {Sahakian}, {Santangelo}, {Schlickeiser}, {Sch{\"o}ck},
  {Schr{\"o}der}, {Schwanke}, {Schwarzburg}, {Schwemmer}, {Shalchi}, {Sol},
  {Spangler}, {Stawarz}, {Steenkamp}, {Stegmann}, {Superina}, {Tam},
  {Tavernet}, {Terrier}, {van Eldik}, {Vasileiadis}, {Venter}, {Vialle},
  {Vincent}, {Vivier}, {V{\"o}lk}, {Volpe}, {Wagner}, {Ward}, {Zdziarski}, \&
  {Zech}}]{Aharonian2008}
{Aharonian}, F., {Akhperjanian}, A.~G., {Barres de Almeida}, U., {et~al.} 2008,
  \aap, 477, 353, \dodoi{10.1051/0004-6361:20078516}

\bibitem[{{Aharonian} {et~al.}(2009){Aharonian}, {Akhperjanian}, {Anton},
  {Barres de Almeida}, {Bazer-Bachi}, {Becherini}, {Behera}, {Bernl{\"o}hr},
  {Boisson}, \& {Bochow}}]{Aharonian2009}
{Aharonian}, F., {Akhperjanian}, A.~G., {Anton}, G., {et~al.} 2009, \aap, 499,
  273, \dodoi{10.1051/0004-6361/200811564}

\bibitem[{{Aharonian} {et~al.}(1997){Aharonian}, {Atoyan}, \&
  {Kifune}}]{Aharonian1997}
{Aharonian}, F.~A., {Atoyan}, A.~M., \& {Kifune}, T. 1997, \mnras, 291, 162,
  \dodoi{10.1093/mnras/291.1.162}

\bibitem[{{Ajello} {et~al.}(2017){Ajello}, {Atwood}, {Baldini}, {Ballet},
  {Barbiellini}, {Bastieri}, {Bellazzini}, {Bissaldi}, {Blandford}, {Bloom},
  {Bonino}, {Bregeon}, {Britto}, {Bruel}, {Buehler}, {Buson}, {Cameron},
  {Caputo}, {Caragiulo}, {Caraveo}, {Cavazzuti}, {Cecchi}, {Charles},
  {Chekhtman}, {Cheung}, {Chiaro}, {Ciprini}, {Cohen}, {Costantin}, {Costanza},
  {Cuoco}, {Cutini}, {D'Ammando}, {de Palma}, {Desiante}, {Digel}, {Di Lalla},
  {Di Mauro}, {Di Venere}, {Dom{\'\i}nguez}, {Drell}, {Dumora}, {Favuzzi},
  {Fegan}, {Ferrara}, {Fortin}, {Franckowiak}, {Fukazawa}, {Funk}, {Fusco},
  {Gargano}, {Gasparrini}, {Giglietto}, {Giommi}, {Giordano}, {Giroletti},
  {Glanzman}, {Green}, {Grenier}, {Grondin}, {Grove}, {Guillemot}, {Guiriec},
  {Harding}, {Hays}, {Hewitt}, {Horan}, {J{\'o}hannesson}, {Kensei}, {Kuss},
  {La Mura}, {Larsson}, {Latronico}, {Lemoine-Goumard}, {Li}, {Longo},
  {Loparco}, {Lott}, {Lubrano}, {Magill}, {Maldera}, {Manfreda}, {Mazziotta},
  {McEnery}, {Meyer}, {Michelson}, {Mirabal}, {Mitthumsiri}, {Mizuno},
  {Moiseev}, {Monzani}, {Morselli}, {Moskalenko}, {Negro}, {Nuss}, {Ohsugi},
  {Omodei}, {Orienti}, {Orlando}, {Palatiello}, {Paliya}, {Paneque}, {Perkins},
  {Persic}, {Pesce-Rollins}, {Piron}, {Porter}, {Principe}, {Rain{\`o}},
  {Rando}, {Razzano}, {Razzaque}, {Reimer}, {Reimer}, {Reposeur}, {Saz
  Parkinson}, {Sgr{\`o}}, {Simone}, {Siskind}, {Spada}, {Spandre}, {Spinelli},
  {Stawarz}, {Suson}, {Takahashi}, {Tak}, {Thayer}, {Thayer}, {Thompson},
  {Torres}, {Torresi}, {Troja}, {Vianello}, {Wood}, \& {Wood}}]{ajello2017}
{Ajello}, M., {Atwood}, W.~B., {Baldini}, L., {et~al.} 2017, \apjs, 232, 18,
  \dodoi{10.3847/1538-4365/aa8221}

\bibitem[{Albert {et~al.}(2021)Albert, Alfaro, Alvarez, {\'{A}}lvarez, Camacho,
  Arteaga-Vel{\'{a}}zquez, Arunbabu, Rojas, Solares, Baghmanyan,
  Belmont-Moreno, BenZvi, Brisbois, Caballero-Mora, Capistr{\'{a}}n,
  Carrami{\~{n}}ana, Casanova, Cotti, Cotzomi, de~Le{\'{o}}n, la~Fuente,
  de~Le{\'{o}}n, Hernandez, Dingus, DuVernois, Durocher,
  D{\'{\i}}az-V{\'{e}}lez, Ellsworth, Engel, Espinoza, Fan, Alonso, Fraija,
  Galv{\'{a}}n-G{\'{a}}mez, Garc{\'{\i}}a-Gonz{\'{a}}lez, Garfias, Giacinti,
  Gonz{\'{a}}lez, Goodman, Harding, Hernandez, Hona, Huang,
  Hueyotl-Zahuantitla, Hüntemeyer, Iriarte, Jardin-Blicq, Joshi, Kieda, Lara,
  Lee, Lee, Vargas, Linnemann, Longinotti, Luis-Raya, Lundeen, Malone,
  Marandon, Martinez, Mart{\'{\i}}nez-Castro, Matthews, Miranda-Romagnoli,
  Morales-Soto, Moreno, Mostaf{\'{a}}, Nayerhoda, Nellen, Newbold, Nisa,
  Noriega-Papaqui, Olivera-Nieto, Omodei, Peisker, Araujo,
  P{\'{e}}rez-P{\'{e}}rez, Rho, Roh, Rosa-Gonz{\'{a}}lez, Ruiz-Velasco,
  Salazar, Greus, Sandoval, Schneider, Schoorlemmer, Serna-Franco, Smith,
  Springer, Surajbali, Tanner, Tollefson, Torres, Torres-Escobedo, Turner,
  Ure{\~{n}}a-Mena, Villase{\~{n}}or, Weisgarber, Willox, \& Zhou}]{albert21}
Albert, A., Alfaro, R., Alvarez, C., {et~al.} 2021, \apjl, 911, L27,
  \dodoi{10.3847/2041-8213/abf4dc}

\bibitem[{{Albert} {et~al.}(2021){Albert}, {Alfaro}, {Alvarez},
  {Arteaga-Vel{\'a}zquez}, {Arunbabu}, {Avila Rojas}, {Ayala Solares},
  {Baghmanyan}, {Belmont-Moreno}, {Brisbois}, {Caballero-Mora},
  {Capistr{\'a}n}, {Carrami{\~n}ana}, {Casanova}, {Cotzomi}, {Couti{\~n}o de
  Le{\'o}n}, {De la Fuente}, {Diaz Hernandez}, {Dingus}, {DuVernois},
  {Durocher}, {Engel}, {Espinoza}, {Fraija}, {Garcia},
  {Garc{\'\i}a-Gonz{\'a}lez}, {Giacinti}, {Gonz{\'a}lez}, {Goodman}, {Harding},
  {Hinton}, {Hona}, {Huang}, {Hueyotl-Zahuantitla}, {Huentemeyer},
  {Jardin-Blicq}, {Joshi}, {Lee}, {Le{\'o}n Vargas}, {Linnemann}, {Longinotti},
  {Luis-Raya}, {Lundeen}, {L{\'o}pez-Coto}, {Malone}, {Martinez},
  {Mart{\'\i}nez-Castro}, {Matthews}, {Miranda-Romagnoli}, {Morales-Soto},
  {Moreno}, {Mostaf{\'a}}, {Nayerhoda}, {Nellen}, {Newbold}, {Nisa},
  {Noriega-Papaqui}, {Olivera-Nieto}, {Omodei}, {Peisker}, {P{\'e}rez Araujo},
  {P{\'e}rez-P{\'e}rez}, {Rho}, {Rosa-Gonz{\'a}lez}, {Ruiz-Velasco}, {Salazar},
  {Salesa Greus}, {Sandoval}, {Schneider}, {Schoorlemmer}, {Serna-Franco},
  {Smith}, {Springer}, {Surajbali}, {Tollefson}, {Torres}, {Turner},
  {Ure{\~n}a-Mena}, {Weisgarber}, {Willox}, {Zhou}, {de Le{\'o}n}, \& {HAWC
  Collaboration}}]{Albert2021}
{Albert}, A., {Alfaro}, R., {Alvarez}, C., {et~al.} 2021, \apj, 911, 143,
  \dodoi{10.3847/1538-4357/abecda}

\bibitem[{{Albert} {et~al.}(2007){Albert}, {Aliu}, {Anderhub}, {Antoranz},
  {Armada}, {Baixeras}, {Barrio}, {Bartko}, {Bastieri}, {Becker}, {Bednarek},
  {Berger}, {Bigongiari}, {Biland}, {Bock}, {Bordas}, {Bosch-Ramon}, {Bretz},
  {Britvitch}, {Camara}, {Carmona}, {Chilingarian}, {Coarasa}, {Commichau},
  {Contreras}, {Cortina}, {Costado}, {Curtef}, {Danielyan}, {Dazzi}, \& {De
  Angelis}}]{Albert2007}
{Albert}, J., {Aliu}, E., {Anderhub}, H., {et~al.} 2007, \apjl, 665, L51,
  \dodoi{10.1086/521145}

\bibitem[{{Aleksi{\'c}} {et~al.}(2014){Aleksi{\'c}}, {Ansoldi}, {Antonelli},
  {Antoranz}, {Babic}, {Bangale}, {Barrio}, {Becerra Gonz{\'a}lez}, {Bednarek},
  {Bernardini}, {Biasuzzi}, {Biland}, {Blanch}, {Bonnefoy}, {Bonnoli},
  {Borracci}, {Bretz}, {Carmona}, {Carosi}, {Colin}, {Colombo}, {Contreras},
  {Cortina}, {Covino}, {Da Vela}, {Dazzi}, {De Angelis}, {De Caneva}, {De
  Lotto}, {de O{\~n}a Wilhelmi}, {Delgado Mendez}, {Dominis Prester}, {Dorner},
  {Doro}, {Einecke}, {Eisenacher}, {Elsaesser}, {Fonseca}, {Font}, {Frantzen},
  {Fruck}, {Galindo}, {Garc{\'\i}a L{\'o}pez}, {Garczarczyk}, {Garrido
  Terrats}, {Gaug}, {Godinovi{\'c}}, {Gonz{\'a}lez Mu{\~n}oz}, {Gozzini},
  {Hadasch}, {Hanabata}, {Hayashida}, {Herrera}, {Hildebrand}, {Hose},
  {Hrupec}, {Idec}, {Kadenius}, {Kellermann}, {Kodani}, {Konno}, {Krause},
  {Kubo}, {Kushida}, {La Barbera}, {Lelas}, {Lewandowska}, {Lindfors},
  {Lombardi}, {L{\'o}pez}, {L{\'o}pez-Coto}, {L{\'o}pez-Oramas}, {Lorenz},
  {Lozano}, {Makariev}, {Mallot}, {Maneva}, {Mankuzhiyil}, {Mannheim},
  {Maraschi}, {Marcote}, {Mariotti}, {Mart{\'\i}nez}, {Mazin}, {Menzel},
  {Miranda}, {Mirzoyan}, {Moralejo}, {Munar-Adrover}, {Nakajima},
  {Niedzwiecki}, {Nilsson}, {Nishijima}, {Noda}, {Orito}, {Overkemping},
  {Paiano}, {Palatiello}, {Paneque}, {Paoletti}, {Paredes}, {Paredes-Fortuny},
  {Persic}, {Prada Moroni}, {Prandini}, {Puljak}, {Reinthal}, {Rhode},
  {Rib{\'o}}, {Rico}, {Rodriguez Garcia}, {R{\"u}gamer}, {Saito}, {Saito},
  {Satalecka}, {Scalzotto}, {Scapin}, {Schultz}, {Schweizer}, {Shore},
  {Sillanp{\"a}{\"a}}, {Sitarek}, {Snidaric}, {Sobczynska}, {Spanier},
  {Stamatescu}, {Stamerra}, {Steinbring}, {Storz}, {Strzys}, {Takalo},
  {Takami}, {Tavecchio}, {Temnikov}, {Terzi{\'c}}, {Tescaro}, {Teshima},
  {Thaele}, {Tibolla}, {Torres}, {Toyama}, {Treves}, {Uellenbeck}, {Vogler}, \&
  {Zanin}}]{Aleksic2014}
{Aleksi{\'c}}, J., {Ansoldi}, S., {Antonelli}, L.~A., {et~al.} 2014, \aap, 567,
  L8, \dodoi{10.1051/0004-6361/201424261}

\bibitem[{{Aliu} {et~al.}(2014){Aliu}, {Aune}, {Behera}, {Beilicke}, {Benbow},
  {Berger}, {Bird}, {Bouvier}, {Buckley}, {Bugaev}, {Cerruti}, {Chen},
  {Ciupik}, {Connolly}, {Cui}, {Dumm}, {Dwarkadas}, {Errando}, {Falcone},
  {Federici}, {Feng}, {Finley}, {Fleischhack}, {Fortin}, {Fortson}, {Furniss},
  {Galante}, {Gillanders}, {Gotthelf}, {Griffin}, {Griffiths}, {Grube}, {Gyuk},
  {Hanna}, {Holder}, {Hughes}, {Humensky}, {Johnson}, {Kaaret}, {Kargaltsev},
  {Kertzman}, {Khassen}, {Kieda}, {Krennrich}, {Lang}, {Madhavan}, {Maier},
  {McArthur}, {McCann}, {Millis}, {Moriarty}, {Mukherjee}, {Nieto},
  {O'Faol{\'a}in de Bhr{\'o}ithe}, {Ong}, {Otte}, {Pandel}, {Park}, {Pohl},
  {Popkow}, {Prokoph}, {Quinn}, {Ragan}, {Rajotte}, {Reyes}, {Reynolds},
  {Richards}, {Roache}, {Roberts}, {Sembroski}, {Shahinyan}, {Smith},
  {Staszak}, {Telezhinsky}, {Tucci}, {Tyler}, {Vincent}, {Wakely}, {Weinstein},
  {Welsing}, {Wilhelm}, {Williams}, \& {Zitzer}}]{Aliu2014}
{Aliu}, E., {Aune}, T., {Behera}, B., {et~al.} 2014, \apj, 788, 78,
  \dodoi{10.1088/0004-637X/788/1/78}

\bibitem[{{Alpar} {et~al.}(1982){Alpar}, {Cheng}, {Ruderman}, \&
  {Shaham}}]{Alpar1982}
{Alpar}, M.~A., {Cheng}, A.~F., {Ruderman}, M.~A., \& {Shaham}, J. 1982, \nat,
  300, 728, \dodoi{10.1038/300728a0}

\bibitem[{{Ambrosino} {et~al.}(2021){Ambrosino}, {Miraval Zanon}, {Papitto},
  {Coti Zelati}, {Campana}, {D'Avanzo}, {Stella}, {Di Salvo}, {Burderi},
  {Casella}, {Sanna}, {de Martino}, {Cadelano}, {Ghedina}, {Leone}, {Meddi},
  {Cretaro}, {Baglio}, {Poretti}, {Mignani}, {Torres}, {Israel}, {Cecconi},
  {Russell}, {Gonzalez Gomez}, {Riverol Rodriguez}, {Perez Ventura}, {Hernandez
  Diaz}, {San Juan}, {Bramich}, \& {Lewis}}]{Ambrosino2021}
{Ambrosino}, F., {Miraval Zanon}, A., {Papitto}, A., {et~al.} 2021, Nature
  Astronomy, 5, 552, \dodoi{10.1038/s41550-021-01308-0}

\bibitem[{{Amenomori} {et~al.}(2019){Amenomori}, {Bao}, {Bi}, {Chen}, {Chen},
  {Chen}, {Chen}, {Chen}, {Cirennima}, {Cui}, {Danzengluobu}, {Ding}, {Fang},
  {Fang}, {Feng}, {Feng}, {Feng}, {Gao}, {Gou}, {Guo}, {He}, {He}, {Hibino},
  {Hotta}, {Hu}, {Hu}, {Huang}, {Jia}, {Jiang}, {Jin}, {Kajino}, {Kasahara},
  {Katayose}, {Kato}, {Kato}, {Kawata}, {Kozai}, {Labaciren}, {Le}, {Li}, {Li},
  {Li}, {Lin}, {Liu}, {Liu}, {Liu}, {Liu}, {Lou}, {Lu}, {Meng}, {Mitsui},
  {Munakata}, {Nakamura}, {Nanjo}, {Nishizawa}, {Ohnishi}, {Ohta}, {Ozawa},
  {Qian}, {Qu}, {Saito}, {Sakata}, {Sako}, {Sengoku}, {Shao}, {Shibata},
  {Shiomi}, {Sugimoto}, {Takita}, {Tan}, {Tateyama}, {Torii}, {Tsuchiya},
  {Udo}, {Wang}, {Wu}, {Xue}, {Yagisawa}, {Yamamoto}, {Yang}, {Yuan}, {Zhai},
  {Zhang}, {Zhang}, {Zhang}, {Zhang}, {Zhang}, {Zhang}, {Zhang},
  {Zhaxisangzhu}, {Zhou}, \& {Tibet AS {\ensuremath{\gamma}}
  Collaboration}}]{amenomori19}
{Amenomori}, M., {Bao}, Y.~W., {Bi}, X.~J., {et~al.} 2019, \prl, 123, 051101,
  \dodoi{10.1103/PhysRevLett.123.051101}

\bibitem[{Amenomori {et~al.}(2021)Amenomori, Bao, Bi, Chen, Chen, Chen, Chen,
  Chen, Cirennima, Cui, Danzengluobu, Ding, Fang, Fang, Feng, Feng, Feng, Gao,
  Gomi, Gou, Guo, Guo, He, He, Hibino, Hotta, Hu, Hu, Huang, Jia, Jiang, Jiang,
  Jin, Kasahara, Katayose, Kato, Kato, Kawata, Kozai, Kurashige, Labaciren, Le,
  Li, Li, Li, Li, Lin, Liu, Liu, Liu, Liu, Liu, Liu, Liu, Lou, Lu, Meng,
  Munakata, Nakada, Nakamura, Nakazawa, Nanjo, Ning, Nishizawa, Ohnishi, Ohura,
  Okukawa, Ozawa, Qian, Qian, Qian, Qu, Saito, Sakata, Sako, Sako, Shao,
  Shibata, Shiomi, Sugimoto, Takano, Takita, Tan, Tateyama, Torii, Tsuchiya,
  Udo, Wang, Wang, Wangdui, Wu, Wu, Xu, Xue, Yamamoto, Yang, Yao, Yin, Yokoe,
  Yu, Yuan, Zhai, Zhang, Zhang, Zhang, Zhang, Zhang, Zhang, Zhang, Zhang, Zhao,
  Zhaxisangzhu, \& Zhou}]{amenomori21}
Amenomori, M., Bao, Y.~W., Bi, X.~J., {et~al.} 2021, Phys. Rev. Lett., 127,
  031102, \dodoi{10.1103/PhysRevLett.127.031102}

\bibitem[{{Amenomori} {et~al.}(2022){Amenomori}, {Asano}, {Bao}, {Bi}, {Chen},
  {Chen}, {Chen}, {Chen}, {Chen}, {Cirennima}, {Cui}, {Danzengluobu}, {Ding},
  {Fang}, {Fang}, {Feng}, {Feng}, {Feng}, {Gao}, {Gomi}, {Gou}, {Guo}, {Guo},
  {He}, {He}, {Hibino}, {Hotta}, {Hu}, {Hu}, {Hu}, {Huang}, {Jia}, {Jiang},
  {Jiang}, {Jin}, {Kasahara}, {Katayose}, {Kato}, {Kato}, {Kawashima},
  {Kawata}, {Kozai}, {Kurashige}, {Labaciren}, {Le}, {Li}, {Li}, {Li}, {Li},
  {Lin}, {Liu}, {Liu}, {Liu}, {Liu}, {Liu}, {Liu}, {Liu}, {Lou}, {Lu}, {Meng},
  {Meng}, {Munakata}, {Nagaya}, {Nakamura}, {Nakazawa}, {Nanjo}, {Ning},
  {Nishizawa}, {Ohnishi}, {Okukawa}, {Ozawa}, {Qian}, {Qian}, {Qian}, {Qu},
  {Saito}, {Sakakibara}, {Sakata}, {Sako}, {Sako}, {Shao}, {Shibata}, {Shiomi},
  {Sugimoto}, {Takano}, {Takita}, {Tan}, {Tateyama}, {Torii}, {Tsuchiya},
  {Udo}, {Wang}, {Wang}, {Wangdui}, {Wu}, {Wu}, {Xu}, {Xue}, {Yang}, {Yao},
  {Yin}, {Yokoe}, {Yu}, {Yuan}, {Zhai}, {Zhang}, {Zhang}, {Zhang}, {Zhang},
  {Zhang}, {Zhang}, {Zhang}, {Zhang}, {Zhao}, {Zhaxisangzhu}, \&
  {Zhou}}]{amenomori2022measurement}
{Amenomori}, M., {Asano}, S., {Bao}, Y.~W., {et~al.} 2022, \apj, 932, 120,
  \dodoi{10.3847/1538-4357/ac6ef4}

\bibitem[{{Anderhub} {et~al.}(2009){Anderhub}, {Antonelli}, {Antoranz},
  {Backes}, {Baixeras}, {Balestra}, {Barrio}, {Bastieri}, {Becerra
  Gonz{\'a}lez}, {Becker}, {Bednarek}, {Berger}, \&
  {Bernardini}}]{Anderhub2009}
{Anderhub}, H., {Antonelli}, L.~A., {Antoranz}, P., {et~al.} 2009, \apj, 702,
  266, \dodoi{10.1088/0004-637X/702/1/266}

\bibitem[{{Anderson} {et~al.}(1997){Anderson}, {Wolszczan}, {Kulkarni}, \&
  {Prince}}]{Anderson1997}
{Anderson}, S.~B., {Wolszczan}, A., {Kulkarni}, S.~R., \& {Prince}, T.~A. 1997,
  \apj, 482, 870, \dodoi{10.1086/304162}

\bibitem[{{Arad} {et~al.}(2021){Arad}, {Lavi}, \& {Keshet}}]{arad2021maximally}
{Arad}, O., {Lavi}, A., \& {Keshet}, U. 2021, \mnras, 504, 4952,
  \dodoi{10.1093/mnras/stab1044}

\bibitem[{{Araya}(2018)}]{Araya2018}
{Araya}, M. 2018, \apj, 859, 69, \dodoi{10.3847/1538-4357/aabd7e}

\bibitem[{{Archibald} {et~al.}(2009){Archibald}, {Stairs}, {Ransom}, {Kaspi},
  {Kondratiev}, {Lorimer}, {McLaughlin}, {Boyles}, {Hessels}, {Lynch}, {van
  Leeuwen}, {Roberts}, {Jenet}, {Champion}, {Rosen}, {Barlow}, {Dunlap}, \&
  {Remillard}}]{Archibald2009}
{Archibald}, A.~M., {Stairs}, I.~H., {Ransom}, S.~M., {et~al.} 2009, Science,
  324, 1411, \dodoi{10.1126/science.1172740}

\bibitem[{{Archibald} {et~al.}(2016){Archibald}, {Gotthelf}, {Ferdman},
  {Kaspi}, {Guillot}, {Harrison}, {Keane}, {Pivovaroff}, {Stern}, {Tendulkar},
  \& {Tomsick}}]{archibald16}
{Archibald}, R.~F., {Gotthelf}, E.~V., {Ferdman}, R.~D., {et~al.} 2016, \apjl,
  819, L16, \dodoi{10.3847/2041-8205/819/1/L16}

\bibitem[{{Arzoumanian} {et~al.}(2011){Arzoumanian}, {Gotthelf}, {Ransom},
  {Safi-Harb}, {Kothes}, \& {Landecker}}]{arzoumanian2011}
{Arzoumanian}, Z., {Gotthelf}, E.~V., {Ransom}, S.~M., {et~al.} 2011, \apj,
  739, 39, \dodoi{10.1088/0004-637X/739/1/39}

\bibitem[{{Atoyan} \& {Aharonian}(1996)}]{Atoyan1996}
{Atoyan}, A.~M., \& {Aharonian}, F.~A. 1996, \mnras, 278, 525,
  \dodoi{10.1093/mnras/278.2.525}

\bibitem[{{Atwood} {et~al.}(2013){Atwood}, {Albert}, {Baldini}, {Tinivella},
  {Bregeon}, {Pesce-Rollins}, {Sgr{\`o}}, {Bruel}, {Charles}, {Drlica-Wagner},
  {Franckowiak}, {Jogler}, {Rochester}, {Usher}, {Wood}, {Cohen-Tanugi}, \&
  {S.~Zimmer for the Fermi-LAT Collaboration}}]{pass8Atwood}
{Atwood}, W., {Albert}, A., {Baldini}, L., {et~al.} 2013, ArXiv e-prints.
\newblock \doarXiv{1303.3514}

\bibitem[{{Atwood} {et~al.}(2009){Atwood}, {Abdo}, {Ackermann}, {Althouse},
  {Anderson}, {Axelsson}, {Baldini}, {Ballet}, {Band}, {Barbiellini},
  {Bartelt}, {Bastieri}, {Baughman}, {Bechtol}, {B{\'e}d{\'e}r{\`e}de},
  {Bellardi}, {Bellazzini}, {Berenji}, {Bignami}, {Bisello}, {Bissaldi},
  {Blandford}, {Bloom}, {Bogart}, {Bonamente}, {Bonnell}, {Borgland},
  {Bouvier}, {Bregeon}, {Brez}, {Brigida}, {Bruel}, {Burnett}, {Busetto},
  {Caliandro}, {Cameron}, {Caraveo}, {Carius}, {Carlson}, {Casandjian},
  {Cavazzuti}, {Ceccanti}, {Cecchi}, {Charles}, {Chekhtman}, {Cheung},
  {Chiang}, {Chipaux}, {Cillis}, {Ciprini}, {Claus}, {Cohen-Tanugi},
  {Condamoor}, {Conrad}, {Corbet}, {Corucci}, {Costamante}, {Cutini}, {Davis},
  {Decotigny}, {DeKlotz}, {Dermer}, {de Angelis}, {Digel}, {do Couto e Silva},
  {Drell}, {Dubois}, {Dumora}, {Edmonds}, {Fabiani}, {Farnier}, {Favuzzi},
  {Flath}, {Fleury}, {Focke}, {Funk}, {Fusco}, {Gargano}, {Gasparrini},
  {Gehrels}, {Gentit}, {Germani}, {Giebels}, {Giglietto}, {Giommi}, {Giordano},
  {Glanzman}, {Godfrey}, {Grenier}, {Grondin}, {Grove}, {Guillemot}, {Guiriec},
  {Haller}, {Harding}, {Hart}, {Hays}, {Healey}, {Hirayama}, {Hjalmarsdotter},
  {Horn}, {Hughes}, {J{\'o}hannesson}, {Johansson}, {Johnson}, {Johnson},
  {Johnson}, {Johnson}, {Kamae}, {Katagiri}, {Kataoka}, {Kavelaars}, {Kawai},
  {Kelly}, {Kerr}, {Klamra}, {Kn{\"o}dlseder}, {Kocian}, {Komin}, {Kuehn},
  {Kuss}, {Landriu}, {Latronico}, {Lee}, {Lee}, {Lemoine-Goumard}, {Lionetto},
  {Longo}, {Loparco}, {Lott}, {Lovellette}, {Lubrano}, {Madejski}, {Makeev},
  {Marangelli}, {Massai}, {Mazziotta}, {McEnery}, {Menon}, {Meurer},
  {Michelson}, {Minuti}, {Mirizzi}, {Mitthumsiri}, {Mizuno}, {Moiseev},
  {Monte}, {Monzani}, {Moretti}, {Morselli}, {Moskalenko}, {Murgia},
  {Nakamori}, {Nishino}, {Nolan}, {Norris}, {Nuss}, {Ohno}, {Ohsugi}, {Omodei},
  {Orlando}, {Ormes}, {Paccagnella}, {Paneque}, {Panetta}, {Parent}, {Pearce},
  {Pepe}, {Perazzo}, {Pesce-Rollins}, {Picozza}, {Pieri}, {Pinchera}, {Piron},
  {Porter}, {Poupard}, {Rain{\`o}}, {Rando}, {Rapposelli}, {Razzano}, {Reimer},
  {Reimer}, {Reposeur}, {Reyes}, {Ritz}, {Rochester}, {Rodriguez}, {Romani},
  {Roth}, {Russell}, {Ryde}, {Sabatini}, {Sadrozinski}, {Sanchez}, {Sander},
  {Sapozhnikov}, {Parkinson}, {Scargle}, {Schalk}, \&
  {Scolieri}}]{Fermi2009ApJ}
{Atwood}, W.~B., {Abdo}, A.~A., {Ackermann}, M., {et~al.} 2009, \apj, 697,
  1071, \dodoi{10.1088/0004-637X/697/2/1071}

\bibitem[{{Bahramian} {et~al.}(2013){Bahramian}, {Heinke}, {Sivakoff}, \&
  {Gladstone}}]{Bahramian2013}
{Bahramian}, A., {Heinke}, C.~O., {Sivakoff}, G.~R., \& {Gladstone}, J.~C.
  2013, \apj, 766, 136, \dodoi{10.1088/0004-637X/766/2/136}

\bibitem[{{Bailer-Jones} {et~al.}(2018){Bailer-Jones}, {Rybizki}, {Fouesneau},
  {Mantelet}, \& {Andrae}}]{Bailer2018}
{Bailer-Jones}, C.~A.~L., {Rybizki}, J., {Fouesneau}, M., {Mantelet}, G., \&
  {Andrae}, R. 2018, \aj, 156, 58, \dodoi{10.3847/1538-3881/aacb21}

\bibitem[{{Baldini} {et~al.}(2021){Baldini}, {Ballet}, {Bastieri}, {Becerra
  Gonzalez}, {Bellazzini}, {Berretta}, {Bissaldi}, {Blandford}, {Bloom}, \&
  {Bonino}}]{Baldini2021}
{Baldini}, L., {Ballet}, J., {Bastieri}, D., {et~al.} 2021, \apjs, 256, 13,
  \dodoi{10.3847/1538-4365/ac072a}

\bibitem[{{Baldwin}(1971)}]{Baldwin1971}
{Baldwin}, J.~E. 1971, in IAU Symposium, Vol.~46, The Crab Nebula, ed. R.~D.
  {Davies} \& F.~{Graham-Smith}, 22

\bibitem[{{Ballet} {et~al.}(2023){Ballet}, {Bruel}, {Burnett}, {Lott}, \& {The
  Fermi-LAT collaboration}}]{Ballet2023}
{Ballet}, J., {Bruel}, P., {Burnett}, T.~H., {Lott}, B., \& {The Fermi-LAT
  collaboration}. 2023, arXiv e-prints, arXiv:2307.12546,
  \dodoi{10.48550/arXiv.2307.12546}

\bibitem[{{Bamba} {et~al.}(2022){Bamba}, {Shibata}, {Tanaka}, {Mori}, {Uchida},
  {Terada}, \& {Ishizaki}}]{bamba2022spectral}
{Bamba}, A., {Shibata}, S., {Tanaka}, S.~J., {et~al.} 2022, \pasj, 74, 1186,
  \dodoi{10.1093/pasj/psac062}

\bibitem[{{Bamba} {et~al.}(2001){Bamba}, {Ueno}, {Koyama}, \&
  {Yamauchi}}]{bamba2001diffuse}
{Bamba}, A., {Ueno}, M., {Koyama}, K., \& {Yamauchi}, S. 2001, \pasj, 53, L21,
  \dodoi{10.1093/pasj/53.4.L21}

\bibitem[{{Bamba} {et~al.}(2020){Bamba}, {Watanabe}, {Mori}, {Shibata},
  {Terada}, {Sano}, \& {Filipovi{\'c}}}]{bamba2020low}
{Bamba}, A., {Watanabe}, E., {Mori}, K., {et~al.} 2020, \apss, 365, 178,
  \dodoi{10.1007/s10509-020-03891-6}

\bibitem[{{Bandiera} {et~al.}(2020){Bandiera}, {Bucciantini}, {Mart{\'\i}n},
  {Olmi}, \& {Torres}}]{Bandiera:2020}
{Bandiera}, R., {Bucciantini}, N., {Mart{\'\i}n}, J., {Olmi}, B., \& {Torres},
  D.~F. 2020, \mnras, 499, 2051, \dodoi{10.1093/mnras/staa2956}

\bibitem[{{Bandiera} {et~al.}(2021){Bandiera}, {Bucciantini}, {Mart{\'\i}n},
  {Olmi}, \& {Torres}}]{Bandiera2021}
---. 2021, \mnras, 508, 3194, \dodoi{10.1093/mnras/stab2600}

\bibitem[{{Bandiera} {et~al.}(2023{\natexlab{a}}){Bandiera}, {Bucciantini},
  {Mart{\'\i}n}, {Olmi}, \& {Torres}}]{Bandiera:2022}
---. 2023{\natexlab{a}}, \mnras, 520, 2451, \dodoi{10.1093/mnras/stad134}

\bibitem[{{Bandiera} {et~al.}(2023{\natexlab{b}}){Bandiera}, {Bucciantini},
  {Olmi}, \& {Torres}}]{bandiera2023reverberation}
{Bandiera}, R., {Bucciantini}, N., {Olmi}, B., \& {Torres}, D.~F.
  2023{\natexlab{b}}, \mnras, 525, 2839, \dodoi{10.1093/mnras/stad2387}

\bibitem[{Becker {et~al.}(1985)Becker, Markert, \& Donahue}]{Becker1985}
Becker, R., Markert, T., \& Donahue, M. 1985, The Astrophysical Journal, 296,
  461

\bibitem[{{Becker} \& {Huang}(2007)}]{Becker2007}
{Becker}, W., \& {Huang}, H.~H., eds. 2007, {Proceedings of the 363. WE-Heraeus
  Seminar on Neutron Stars and Pulsars 40 years after the discovery.}

\bibitem[{{Bednarek}(2009{\natexlab{a}})}]{Bednarek2009}
{Bednarek}, W. 2009{\natexlab{a}}, \mnras, 397, 1420,
  \dodoi{10.1111/j.1365-2966.2009.14893.x}

\bibitem[{{Bednarek}(2009{\natexlab{b}})}]{Bednarek2009b}
---. 2009{\natexlab{b}}, \prd, 79, 123010, \dodoi{10.1103/PhysRevD.79.123010}

\bibitem[{{Bednarek} \& {Sitarek}(2007)}]{Bednarek2007}
{Bednarek}, W., \& {Sitarek}, J. 2007, \mnras, 377, 920,
  \dodoi{10.1111/j.1365-2966.2007.11664.x}

\bibitem[{{Beechert} {et~al.}(2022){Beechert}, {Lazar}, {Boggs}, {Brandt},
  {Chang}, {Chu}, {Gulick}, {Kierans}, {Lowell}, {Pellegrini}, {Roberts},
  {Siegert}, {Sleator}, {Tomsick}, \& {Zoglauer}}]{Beechert2022}
{Beechert}, J., {Lazar}, H., {Boggs}, S.~E., {et~al.} 2022, Nuclear Instruments
  and Methods in Physics Research A, 1031, 166510,
  \dodoi{10.1016/j.nima.2022.166510}

\bibitem[{{Benli} {et~al.}(2021){Benli}, {P{\'e}tri}, \& {Mitra}}]{Benli2021}
{Benli}, O., {P{\'e}tri}, J., \& {Mitra}, D. 2021, \aap, 647, A101,
  \dodoi{10.1051/0004-6361/202039853}

\bibitem[{{Bhattacharya} \& {van den Heuvel}(1991)}]{Bhattacharya1991}
{Bhattacharya}, D., \& {van den Heuvel}, E.~P.~J. 1991, \physrep, 203, 1,
  \dodoi{10.1016/0370-1573(91)90064-S}

\bibitem[{{Blondin} {et~al.}(2001){Blondin}, {Chevalier}, \&
  {Frierson}}]{Blondin2001}
{Blondin}, J.~M., {Chevalier}, R.~A., \& {Frierson}, D.~M. 2001, \apj, 563,
  806, \dodoi{10.1086/324042}

\bibitem[{{Bogdanov} {et~al.}(2018){Bogdanov}, {Deller}, {Miller-Jones},
  {Archibald}, {Hessels}, {Jaodand}, {Patruno}, {Bassa}, \&
  {D'Angelo}}]{Bogdanov2018}
{Bogdanov}, S., {Deller}, A.~T., {Miller-Jones}, J. C.~A., {et~al.} 2018, \apj,
  856, 54, \dodoi{10.3847/1538-4357/aaaeb9}

\bibitem[{{Bordas} \& {Zhang}(2020)}]{bordas2020}
{Bordas}, P., \& {Zhang}, X. 2020, \aap, 644, L4,
  \dodoi{10.1051/0004-6361/202039327}

\bibitem[{{Breuhaus} {et~al.}(2022){Breuhaus}, {Reville}, \&
  {Hinton}}]{breuhaus2022}
{Breuhaus}, M., {Reville}, B., \& {Hinton}, J.~A. 2022, \aap, 660, A8,
  \dodoi{10.1051/0004-6361/202142097}

\bibitem[{{Brogan} {et~al.}(2004){Brogan}, {Devine}, {Lazio}, {Kassim}, {Tam},
  {Brisken}, {Dyer}, \& {Roberts}}]{Brogan2004}
{Brogan}, C.~L., {Devine}, K.~E., {Lazio}, T.~J., {et~al.} 2004, \aj, 127, 355,
  \dodoi{10.1086/379856}

\bibitem[{{Bruel}(2019)}]{Bruel2019}
{Bruel}, P. 2019, \aap, 622, A108, \dodoi{10.1051/0004-6361/201834555}

\bibitem[{{Bucciantini} {et~al.}(2011){Bucciantini}, {Arons}, \&
  {Amato}}]{Bucciantini2011}
{Bucciantini}, N., {Arons}, J., \& {Amato}, E. 2011, \mnras, 410, 381,
  \dodoi{10.1111/j.1365-2966.2010.17449.x}

\bibitem[{Buehler \& Blandford(2013)}]{Buehler2013}
Buehler, R., \& Blandford, R. 2013, The Crab pulsar wind nebula: Our laboratory
  of the non-thermal Universe, Tech. rep.

\bibitem[{{Burgess} {et~al.}(2022){Burgess}, {Mori}, {Gelfand}, {Hailey},
  {Tokayer}, {Woo}, {An}, {Malone}, {Reynolds}, {Safi-Harb}, \&
  {Temim}}]{burgess22}
{Burgess}, D.~A., {Mori}, K., {Gelfand}, J.~D., {et~al.} 2022, \apj, 930, 148,
  \dodoi{10.3847/1538-4357/ac650a}

\bibitem[{{Caliandro} {et~al.}(2015){Caliandro}, {Cheung}, {Li}, {Scargle},
  {Torres}, {Wood}, \& {Chernyakova}}]{Caliandro2015B1259}
{Caliandro}, G.~A., {Cheung}, C.~C., {Li}, J., {et~al.} 2015, \apj, 811, 68,
  \dodoi{10.1088/0004-637X/811/1/68}

\bibitem[{Cao {et~al.}(2024)Cao, Yang, \& Zhang}]{Cao2024psr}
Cao, G., Yang, X., \& Zhang, L. 2024, Universe, 10, 130

\bibitem[{{Caswell} {et~al.}(1992){Caswell}, {Kesteven}, {Stewart}, {Milne}, \&
  {Haynes}}]{caswell1992g308}
{Caswell}, J.~L., {Kesteven}, M.~J., {Stewart}, R.~T., {Milne}, D.~K., \&
  {Haynes}, R.~F. 1992, \apjl, 399, L151, \dodoi{10.1086/186629}

\bibitem[{{Cerutti} {et~al.}(2025){Cerutti}, {Figueiredo}, \&
  {Dubus}}]{Cerutti2025}
{Cerutti}, B., {Figueiredo}, E., \& {Dubus}, G. 2025, \aap, 695, A93,
  \dodoi{10.1051/0004-6361/202451948}

\bibitem[{{Chen}(1991)}]{Chen1991}
{Chen}, K. 1991, \nat, 352, 695, \dodoi{10.1038/352695a0}

\bibitem[{{Cheng} {et~al.}(2010){Cheng}, {Chernyshov}, {Dogiel}, {Hui}, \&
  {Kong}}]{Cheng2010}
{Cheng}, K.~S., {Chernyshov}, D.~O., {Dogiel}, V.~A., {Hui}, C.~Y., \& {Kong},
  A.~K.~H. 2010, \apj, 723, 1219, \dodoi{10.1088/0004-637X/723/2/1219}

\bibitem[{{Cheng} {et~al.}(1986){Cheng}, {Ho}, \& {Ruderman}}]{Cheng1986}
{Cheng}, K.~S., {Ho}, C., \& {Ruderman}, M. 1986, \apj, 300, 500,
  \dodoi{10.1086/163829}

\bibitem[{{Cheng} \& {Ruderman}(1991)}]{Cheng1991}
{Cheng}, K.~S., \& {Ruderman}, M. 1991, \apj, 373, 187, \dodoi{10.1086/170036}

\bibitem[{{Cheng} {et~al.}(2000){Cheng}, {Ruderman}, \& {Zhang}}]{Cheng2000}
{Cheng}, K.~S., {Ruderman}, M., \& {Zhang}, L. 2000, \apj, 537, 964,
  \dodoi{10.1086/309051}

\bibitem[{{Chernyakova} {et~al.}(2020){Chernyakova}, {Malyshev}, {Mc Keague},
  {van Soelen}, {Marais}, {Martin-Carrillo}, \& {Murphy}}]{Chernyakova2020a}
{Chernyakova}, M., {Malyshev}, D., {Mc Keague}, S., {et~al.} 2020, \mnras, 497,
  648, \dodoi{10.1093/mnras/staa1876}

\bibitem[{{Chernyakova} {et~al.}(2015){Chernyakova}, {Neronov}, {van Soelen},
  {Callanan}, {O'Shaughnessy}, {Babyk}, {Tsygankov}, {Vovk}, {Krivonos},
  {Tomsick}, {Malyshev}, {Li}, {Wood}, {Torres}, {Zhang}, {Kretschmar},
  {McSwain}, {Buckley}, \& {Koen}}]{Chernyakova2015B1259}
{Chernyakova}, M., {Neronov}, A., {van Soelen}, B., {et~al.} 2015, \mnras, 454,
  1358, \dodoi{10.1093/mnras/stv1988}

\bibitem[{{Chevalier}(1976)}]{Chevalier1976}
{Chevalier}, R.~A. 1976, \apj, 207, 872, \dodoi{10.1086/154557}

\bibitem[{{Chevalier}(2004)}]{Chevalier2004}
---. 2004, Advances in Space Research, 33, 456,
  \dodoi{10.1016/j.asr.2003.04.018}

\bibitem[{Chevalier(2005)}]{Chevalier2005}
Chevalier, R.~A. 2005, The Astrophysical Journal, 619, 839

\bibitem[{{Chevalier} \& {Reynolds}(2011)}]{Chevalier2011}
{Chevalier}, R.~A., \& {Reynolds}, S.~P. 2011, \apjl, 740, L26,
  \dodoi{10.1088/2041-8205/740/1/L26}

\bibitem[{{Clapson} {et~al.}(2011){Clapson}, {Domainko}, {Jamrozy}, {Dyrda}, \&
  {Eger}}]{Clapson2011}
{Clapson}, A.~C., {Domainko}, W., {Jamrozy}, M., {Dyrda}, M., \& {Eger}, P.
  2011, \aap, 532, A47, \dodoi{10.1051/0004-6361/201015559}

\bibitem[{{Clark} {et~al.}(2016){Clark}, {Pletsch}, {Wu}, {Guillemot},
  {Camilo}, {Johnson}, {Kerr}, {Allen}, {Aulbert}, {Beer}, {Bock},
  {Cu{\'e}llar}, {Eggenstein}, {Fehrmann}, {Kramer}, {Machenschalk}, \&
  {Nieder}}]{clark2016braking}
{Clark}, C.~J., {Pletsch}, H.~J., {Wu}, J., {et~al.} 2016, \apjl, 832, L15,
  \dodoi{10.3847/2041-8205/832/1/L15}

\bibitem[{{Clark} {et~al.}(2017){Clark}, {Wu}, {Pletsch}, {Guillemot}, {Allen},
  {Aulbert}, {Beer}, {Bock}, {Cu{\'e}llar}, {Eggenstein}, {Fehrmann}, {Kramer},
  {Machenschalk}, \& {Nieder}}]{clark2017einstein}
{Clark}, C.~J., {Wu}, J., {Pletsch}, H.~J., {et~al.} 2017, \apj, 834, 106,
  \dodoi{10.3847/1538-4357/834/2/106}

\bibitem[{{Clark} {et~al.}(1977){Clark}, {McCrea}, \& {Stephenson}}]{Clark1977}
{Clark}, D.~H., {McCrea}, W.~H., \& {Stephenson}, F.~R. 1977, \nat, 265, 318,
  \dodoi{10.1038/265318a0}

\bibitem[{{Clark}(1975)}]{Clark1975}
{Clark}, G.~W. 1975, \apjl, 199, L143, \dodoi{10.1086/181869}

\bibitem[{{Coe} {et~al.}(2006){Coe}, {Reig}, {McBride}, {Galache}, \&
  {Fabregat}}]{Coe2006}
{Coe}, M.~J., {Reig}, P., {McBride}, V.~A., {Galache}, J.~L., \& {Fabregat}, J.
  2006, \mnras, 368, 447, \dodoi{10.1111/j.1365-2966.2006.10127.x}

\bibitem[{{Coe} {et~al.}(2019){Coe}, {Okazaki}, {Steele}, {Ng}, {Ho}, {Lyne},
  {Stappers}, {Johnson}, {Ray}, \& {Kerr}}]{Coe2019}
{Coe}, M.~J., {Okazaki}, A.~T., {Steele}, I.~A., {et~al.} 2019, \mnras, 485,
  1864, \dodoi{10.1093/mnras/stz515}

\bibitem[{{Cordes} \& {Lazio}(2002)}]{Cordes2002}
{Cordes}, J.~M., \& {Lazio}, T.~J.~W. 2002, arXiv e-prints, astro,
  \dodoi{10.48550/arXiv.astro-ph/0207156}

\bibitem[{{Coti Zelati} {et~al.}(2020){Coti Zelati}, {Torres}, {Li}, \&
  {Vigan{\`o}}}]{Zelati2020}
{Coti Zelati}, F., {Torres}, D.~F., {Li}, J., \& {Vigan{\`o}}, D. 2020, \mnras,
  492, 1025, \dodoi{10.1093/mnras/stz3485}

\bibitem[{{Crestan} {et~al.}(2021){Crestan}, {Giuliani}, {Mereghetti},
  {Sidoli}, {Pintore}, \& {La Palombara}}]{crestan21}
{Crestan}, S., {Giuliani}, A., {Mereghetti}, S., {et~al.} 2021, \mnras, 505,
  2309, \dodoi{10.1093/mnras/stab1422}

\bibitem[{{D'Abrusco} {et~al.}(2019){D'Abrusco}, {{\'A}lvarez Crespo},
  {Massaro}, {Campana}, {Chavushyan}, {Landoni}, {La Franca}, {Masetti},
  {Milisavljevic}, {Paggi}, {Ricci}, \& {Smith}}]{Abrusco2019}
{D'Abrusco}, R., {{\'A}lvarez Crespo}, N., {Massaro}, F., {et~al.} 2019, \apjs,
  242, 4, \dodoi{10.3847/1538-4365/ab16f4}

\bibitem[{{Dai} {et~al.}(2023){Dai}, {Johnston}, {Kerr}, {Berteaud},
  {Bhattacharyya}, {Camilo}, \& {Keane}}]{Dai2023}
{Dai}, S., {Johnston}, S., {Kerr}, M., {et~al.} 2023, \mnras, 521, 2616,
  \dodoi{10.1093/mnras/stad704}

\bibitem[{{Daugherty} \& {Harding}(1982)}]{Daugherty1982}
{Daugherty}, J.~K., \& {Harding}, A.~K. 1982, \apj, 252, 337,
  \dodoi{10.1086/159561}

\bibitem[{{de Angelis} {et~al.}(2018){de Angelis}, {Tatischeff}, {Grenier},
  {McEnery}, {Mallamaci}, {Tavani}, {Oberlack}, {Hanlon}, {Walter}, {Argan},
  {von Ballmoos}, {Bulgarelli}, {Bykov}, {Hernanz}, {Kanbach}, {Kuvvetli},
  {Pearce}, {Zdziarski}, {Conrad}, \& {Ghisellini}}]{Angelis2018}
{de Angelis}, A., {Tatischeff}, V., {Grenier}, I.~A., {et~al.} 2018, Journal of
  High Energy Astrophysics, 19, 1, \dodoi{10.1016/j.jheap.2018.07.001}

\bibitem[{{de Jager} \& {Djannati-Ata{\"\i}}(2009{\natexlab{a}})}]{Jager2009}
{de Jager}, O.~C., \& {Djannati-Ata{\"\i}}, A. 2009{\natexlab{a}}, in
  Astrophysics and Space Science Library, Vol. 357, Astrophysics and Space
  Science Library, ed. W.~{Becker}, 451, \dodoi{10.1007/978-3-540-76965-1_17}

\bibitem[{{de Jager} \& {Djannati-Ata{\"\i}}(2009{\natexlab{b}})}]{dejager2009}
{de Jager}, O.~C., \& {Djannati-Ata{\"\i}}, A. 2009{\natexlab{b}}, in
  Astrophysics and Space Science Library, Vol. 357, Astrophysics and Space
  Science Library, ed. W.~{Becker}, 451, \dodoi{10.1007/978-3-540-76965-1_17}

\bibitem[{{de Jager} {et~al.}(1989){de Jager}, {Raubenheimer}, \&
  {Swanepoel}}]{Jager1989}
{de Jager}, O.~C., {Raubenheimer}, B.~C., \& {Swanepoel}, J.~W.~H. 1989, \aap,
  221, 180

\bibitem[{{de Martino} {et~al.}(2015){de Martino}, {Papitto}, {Belloni},
  {Burgay}, {De Ona Wilhelmi}, {Li}, {Pellizzoni}, {Possenti}, {Rea}, \&
  {Torres}}]{deMartino2015}
{de Martino}, D., {Papitto}, A., {Belloni}, T., {et~al.} 2015, \mnras, 454,
  2190, \dodoi{10.1093/mnras/stv2109}

\bibitem[{{de Menezes} {et~al.}(2019){de Menezes}, {Cafardo}, \&
  {Nemmen}}]{Menezes2019}
{de Menezes}, R., {Cafardo}, F., \& {Nemmen}, R. 2019, \mnras, 486, 851,
  \dodoi{10.1093/mnras/stz898}

\bibitem[{{de Menezes} {et~al.}(2023){de Menezes}, {Di Pierro}, \&
  {Chiavassa}}]{Menezes2023}
{de Menezes}, R., {Di Pierro}, F., \& {Chiavassa}, A. 2023, \mnras, 523, 4455,
  \dodoi{10.1093/mnras/stad1694}

\bibitem[{{de O{\~n}a Wilhelmi} {et~al.}(2022){de O{\~n}a Wilhelmi},
  {L{\'o}pez-Coto}, {Amato}, \& {Aharonian}}]{wilhelmi22}
{de O{\~n}a Wilhelmi}, E., {L{\'o}pez-Coto}, R., {Amato}, E., \& {Aharonian},
  F. 2022, \apjl, 930, L2, \dodoi{10.3847/2041-8213/ac66cf}

\bibitem[{{de O{\~n}a Wilhelmi} {et~al.}(2016){de O{\~n}a Wilhelmi}, {Papitto},
  {Li}, {Rea}, {Torres}, {Burderi}, {Di Salvo}, {Iaria}, {Riggio}, \&
  {Sanna}}]{Emma2016}
{de O{\~n}a Wilhelmi}, E., {Papitto}, A., {Li}, J., {et~al.} 2016, \mnras, 456,
  2647, \dodoi{10.1093/mnras/stv2695}

\bibitem[{{De Sarkar} {et~al.}(2021){De Sarkar}, {Biswas}, \&
  {Gupta}}]{desarkar21}
{De Sarkar}, A., {Biswas}, S., \& {Gupta}, N. 2021, Journal of High Energy
  Astrophysics, 29, 1, \dodoi{10.1016/j.jheap.2020.11.001}

\bibitem[{{De Sarkar} \& {Gupta}(2022)}]{desarkar22}
{De Sarkar}, A., \& {Gupta}, N. 2022, \apj, 934, 118,
  \dodoi{10.3847/1538-4357/ac6ce5}

\bibitem[{{De Sarkar} {et~al.}(2022){De Sarkar}, {Zhang}, {Mart{\'\i}n},
  {Torres}, {Li}, \& {Hou}}]{Sarkar2022}
{De Sarkar}, A., {Zhang}, W., {Mart{\'\i}n}, J., {et~al.} 2022, \aap, 668, A23,
  \dodoi{10.1051/0004-6361/202244841}

\bibitem[{{Dean} {et~al.}(2008){Dean}, {de Rosa}, {McBride}, {Landi}, {Hill},
  {Bassani}, {Bazzano}, {Bird}, \& {Ubertini}}]{Dean2008}
{Dean}, A.~J., {de Rosa}, A., {McBride}, V.~A., {et~al.} 2008, \mnras, 384,
  L29, \dodoi{10.1111/j.1745-3933.2007.00415.x}

\bibitem[{{Deeter} {et~al.}(1981){Deeter}, {Boynton}, \& {Pravdo}}]{Deeter1981}
{Deeter}, J.~E., {Boynton}, P.~E., \& {Pravdo}, S.~H. 1981, \apj, 247, 1003,
  \dodoi{10.1086/159110}

\bibitem[{{Deutsch}(1955)}]{Deutsch1955}
{Deutsch}, A.~J. 1955, Annales d'Astrophysique, 18, 1

\bibitem[{{Devin} {et~al.}(2021){Devin}, {Renaud}, {Lemoine-Goumard}, \&
  {Vasileiadis}}]{Devin2021}
{Devin}, J., {Renaud}, M., {Lemoine-Goumard}, M., \& {Vasileiadis}, G. 2021,
  \aap, 647, A68, \dodoi{10.1051/0004-6361/202039563}

\bibitem[{{Di Bernardo} {et~al.}(2013){Di Bernardo}, {Evoli}, {Gaggero},
  {Grasso}, \& {Maccione}}]{bernardo13}
{Di Bernardo}, G., {Evoli}, C., {Gaggero}, D., {Grasso}, D., \& {Maccione}, L.
  2013, \jcap, 2013, 036, \dodoi{10.1088/1475-7516/2013/03/036}

\bibitem[{{Di Mauro} {et~al.}(2020){Di Mauro}, {Manconi}, \&
  {Donato}}]{di2020evidences}
{Di Mauro}, M., {Manconi}, S., \& {Donato}, F. 2020, \prd, 101, 103035,
  \dodoi{10.1103/PhysRevD.101.103035}

\bibitem[{{Doherty} {et~al.}(2003){Doherty}, {Johnston}, {Green}, {Roberts},
  {Romani}, {Gaensler}, \& {Crawford}}]{doherty2003}
{Doherty}, M., {Johnston}, S., {Green}, A.~J., {et~al.} 2003, \mnras, 339,
  1048, \dodoi{10.1046/j.1365-8711.2003.06265.x}

\bibitem[{{Dubus}(2015)}]{Dubus2015}
{Dubus}, G. 2015, Comptes Rendus Physique, 16, 661,
  \dodoi{10.1016/j.crhy.2015.08.014}

\bibitem[{{Duvidovich} {et~al.}(2019){Duvidovich}, {Giacani}, {Castelletti},
  {Petriella}, \& {Sup{\'a}n}}]{duvidovich2019}
{Duvidovich}, L., {Giacani}, E., {Castelletti}, G., {Petriella}, A., \&
  {Sup{\'a}n}, L. 2019, \aap, 623, A115, \dodoi{10.1051/0004-6361/201834590}

\bibitem[{{Dyks} \& {Rudak}(2003)}]{Dyks2003}
{Dyks}, J., \& {Rudak}, B. 2003, \apj, 598, 1201, \dodoi{10.1086/379052}

\bibitem[{{Eagle} {et~al.}(2025){Eagle}, {Castro}, {Zhang}, {Torres}, {Ballet},
  \& {The Fermi-LAT Collaboration}}]{Eagle2025}
{Eagle}, J., {Castro}, D., {Zhang}, W., {et~al.} 2025, arXiv e-prints,
  arXiv:2506.18599.
\newblock \doarXiv{2506.18599}

\bibitem[{{Eagle}(2022)}]{Eagle2022}
{Eagle}, J.~L. 2022, PhD thesis, Clemson University, South Carolina

\bibitem[{{Eger} {et~al.}(2010){Eger}, {Domainko}, \& {Clapson}}]{Eger2010}
{Eger}, P., {Domainko}, W., \& {Clapson}, A.~C. 2010, \aap, 513, A66,
  \dodoi{10.1051/0004-6361/200913732}

\bibitem[{{Espinoza} {et~al.}(2011){Espinoza}, {Lyne}, {Kramer}, {Manchester},
  \& {Kaspi}}]{espinoza11b}
{Espinoza}, C.~M., {Lyne}, A.~G., {Kramer}, M., {Manchester}, R.~N., \&
  {Kaspi}, V.~M. 2011, \apjl, 741, L13, \dodoi{10.1088/2041-8205/741/1/L13}

\bibitem[{Espinoza {et~al.}(2011)Espinoza, Lyne, Stappers, \&
  Kramer}]{espinoza11}
Espinoza, C.~M., Lyne, A.~G., Stappers, B.~W., \& Kramer, M. 2011, \mnras, 414,
  1679, \dodoi{10.1111/j.1365-2966.2011.18503.x}

\bibitem[{{Fang} {et~al.}(2020){Fang}, {Wen}, {Yu}, \& {Chen}}]{Fang2020}
{Fang}, J., {Wen}, L., {Yu}, H., \& {Chen}, S. 2020, \mnras, 498, 4901,
  \dodoi{10.1093/mnras/staa2703}

\bibitem[{{Fang} \& {Zhang}(2010)}]{fang2010}
{Fang}, J., \& {Zhang}, L. 2010, \aap, 515, A20,
  \dodoi{10.1051/0004-6361/200913615}

\bibitem[{{Faucher-Gigu{\`e}re} \& {Kaspi}(2006)}]{FG2006}
{Faucher-Gigu{\`e}re}, C.-A., \& {Kaspi}, V.~M. 2006, \apj, 643, 332,
  \dodoi{10.1086/501516}

\bibitem[{{Feng} {et~al.}(2023){Feng}, {Cheng}, {Wang}, {Li}, \&
  {Chen}}]{Feng2023}
{Feng}, L., {Cheng}, Z., {Wang}, W., {Li}, Z., \& {Chen}, Y. 2023, arXiv
  e-prints, arXiv:2310.15859, \dodoi{10.48550/arXiv.2310.15859}

\bibitem[{{Fermi-LAT collaboration}(2022)}]{4DFL-DR3}
{Fermi-LAT collaboration}. 2022, arXiv e-prints, arXiv:2201.11184.
\newblock \doarXiv{2201.11184}

\bibitem[{{Ferrand} \& {Safi-Harb}(2012)}]{ferrand2012census}
{Ferrand}, G., \& {Safi-Harb}, S. 2012, Advances in Space Research, 49, 1313,
  \dodoi{10.1016/j.asr.2012.02.004}

\bibitem[{{Finger} {et~al.}(1996){Finger}, {Wilson}, \& {Harmon}}]{Finger1996}
{Finger}, M.~H., {Wilson}, R.~B., \& {Harmon}, B.~A. 1996, \apj, 459, 288,
  \dodoi{10.1086/176892}

\bibitem[{{Freire} {et~al.}(2011){Freire}, {Abdo}, {Ajello}, {Allafort},
  {Ballet}, {Barbiellini}, {Bastieri}, {Bechtol}, {Bellazzini}, {Blandford},
  {Bloom}, {Bonamente}, {Borgland}, {Brigida}, {Bruel}, {Buehler}, \&
  {Buson}}]{Freire2011}
{Freire}, P.~C.~C., {Abdo}, A.~A., {Ajello}, M., {et~al.} 2011, Science, 334,
  1107, \dodoi{10.1126/science.1207141}

\bibitem[{{Freire} {et~al.}(2017){Freire}, {Ridolfi}, {Kramer}, {Jordan},
  {Manchester}, {Torne}, {Sarkissian}, {Heinke}, {D'Amico}, {Camilo},
  {Lorimer}, \& {Lyne}}]{Freire+2017}
{Freire}, P.~C.~C., {Ridolfi}, A., {Kramer}, M., {et~al.} 2017, \mnras, 471,
  857, \dodoi{10.1093/mnras/stx1533}

\bibitem[{{Fujinaga} {et~al.}(2013){Fujinaga}, {Mori}, {Bamba}, {Kimura},
  {Dotani}, {Ozaki}, {Matsuta}, {P{\"u}lhofer}, {Uchiyama}, {Hiraga},
  {Matsumoto}, \& {Terada}}]{Fujinaga2013}
{Fujinaga}, T., {Mori}, K., {Bamba}, A., {et~al.} 2013, \pasj, 65, 61,
  \dodoi{10.1093/pasj/65.3.61}

\bibitem[{Fujita {et~al.}(2021)Fujita, Bamba, Nobukawa, \&
  Matsumoto}]{fujita21}
Fujita, Y., Bamba, A., Nobukawa, K.~K., \& Matsumoto, H. 2021, \apj, 912, 133,
  \dodoi{10.3847/1538-4357/abf14a}

\bibitem[{{Gaensler} \& {Frail}(2000)}]{gaensler2000}
{Gaensler}, B.~M., \& {Frail}, D.~A. 2000, \nat, 406, 158,
  \dodoi{10.48550/arXiv.astro-ph/0005526}

\bibitem[{{Gaensler} \& {Slane}(2006)}]{Gaensler_Slane06a}
{Gaensler}, B.~M., \& {Slane}, P.~O. 2006, \araa, 44, 17,
  \dodoi{10.1146/annurev.astro.44.051905.092528}

\bibitem[{{Gaensler} \& {Wallace}(2003)}]{gaensler2003}
{Gaensler}, B.~M., \& {Wallace}, B.~J. 2003, \apj, 594, 326,
  \dodoi{10.1086/376861}

\bibitem[{Gao \& Han(2012)}]{Gao2012}
Gao, F., \& Han, L. 2012, Computational Optimization and Applications, 51, 259

\bibitem[{{Garc{\'\i}a} {et~al.}(2024{\natexlab{a}}){Garc{\'\i}a}, {Illiano},
  {Torres}, {Papitto}, {Coti Zelati}, {de Martino}, \&
  {Patruno}}]{Garcia2024msp}
{Garc{\'\i}a}, C.~R., {Illiano}, G., {Torres}, D.~F., {et~al.}
  2024{\natexlab{a}}, \aap, 692, A187, \dodoi{10.1051/0004-6361/202450758}

\bibitem[{{Garc{\'\i}a} \& {Torres}(2023)}]{MST-II}
{Garc{\'\i}a}, C.~R., \& {Torres}, D.~F. 2023, \mnras, 520, 599,
  \dodoi{10.1093/mnras/stad183}

\bibitem[{{Garc{\'\i}a} {et~al.}(2022){Garc{\'\i}a}, {Torres}, \&
  {Patruno}}]{MST-I}
{Garc{\'\i}a}, C.~R., {Torres}, D.~F., \& {Patruno}, A. 2022, \mnras, 515,
  3883, \dodoi{10.1093/mnras/stac1997}

\bibitem[{{Garc{\'\i}a} {et~al.}(2024{\natexlab{b}}){Garc{\'\i}a}, {Torres},
  {Zhu-Ge}, \& {Zhang}}]{Garcia2024frb}
{Garc{\'\i}a}, C.~R., {Torres}, D.~F., {Zhu-Ge}, J.-M., \& {Zhang}, B.
  2024{\natexlab{b}}, \apj, 977, 273, \dodoi{10.3847/1538-4357/ad9020}

\bibitem[{{Gelfand} {et~al.}(2009){Gelfand}, {Slane}, \& {Zhang}}]{gelfand2009}
{Gelfand}, J.~D., {Slane}, P.~O., \& {Zhang}, W. 2009, \apj, 703, 2051,
  \dodoi{10.1088/0004-637X/703/2/2051}

\bibitem[{{Gendre} {et~al.}(2003){Gendre}, {Barret}, \& {Webb}}]{Gendre2003}
{Gendre}, B., {Barret}, D., \& {Webb}, N. 2003, \aap, 403, L11,
  \dodoi{10.1051/0004-6361:20030423}

\bibitem[{{Giacinti} {et~al.}(2020){Giacinti}, {Mitchell}, {L{\'o}pez-Coto},
  {Joshi}, {Parsons}, \& {Hinton}}]{Giacinti2020}
{Giacinti}, G., {Mitchell}, A.~M.~W., {L{\'o}pez-Coto}, R., {et~al.} 2020,
  \aap, 636, A113, \dodoi{10.1051/0004-6361/201936505}

\bibitem[{{Ginzburg} \& {Ptuskin}(1985)}]{ginzburg1985}
{Ginzburg}, V.~L., \& {Ptuskin}, V.~S. 1985, 4, 161

\bibitem[{{Ginzburg} \& {Syrovatskii}(1964)}]{ginzburg1964}
{Ginzburg}, V.~L., \& {Syrovatskii}, S.~I. 1964, {The Origin of Cosmic Rays}

\bibitem[{{Goldreich} \& {Julian}(1969)}]{Goldreich1969}
{Goldreich}, P., \& {Julian}, W.~H. 1969, \apj, 157, 869,
  \dodoi{10.1086/150119}

\bibitem[{G{\'o}mez de~la G{\'a}ndara~P{\'e}rez(2020)}]{gomez2020predicting}
G{\'o}mez de~la G{\'a}ndara~P{\'e}rez, J. 2020, PhD thesis, Universidad de
  Cantabria.
\newblock
  \url{https://repositorio.unican.es/xmlui/bitstream/handle/10902/21329/Gomez%20de%20la%20Gandara%20Perez%20Javier.pdf}

\bibitem[{{Gonzalez} {et~al.}(2006){Gonzalez}, {Kaspi}, {Pivovaroff}, \&
  {Gaensler}}]{gonzalez2006}
{Gonzalez}, M.~E., {Kaspi}, V.~M., {Pivovaroff}, M.~J., \& {Gaensler}, B.~M.
  2006, \apj, 652, 569, \dodoi{10.1086/507125}

\bibitem[{{Gotthelf} {et~al.}(2016){Gotthelf}, {Mori}, {Aliu}, {Paredes},
  {Tomsick}, {Boggs}, {Christensen}, {Craig}, {Hailey}, {Harrison}, {Hong},
  {Rahoui}, {Stern}, \& {Zhang}}]{Gotthelf2016}
{Gotthelf}, E.~V., {Mori}, K., {Aliu}, E., {et~al.} 2016, \apj, 826, 25,
  \dodoi{10.3847/0004-637X/826/1/25}

\bibitem[{{Green} {et~al.}(1997){Green}, {Frail}, {Goss}, \&
  {Otrupcek}}]{green1997continuation}
{Green}, A.~J., {Frail}, D.~A., {Goss}, W.~M., \& {Otrupcek}, R. 1997, \aj,
  114, 2058, \dodoi{10.1086/118626}

\bibitem[{Green {et~al.}(1988)Green, Gull, Tan, \& Simon}]{Green1988}
Green, D., Gull, S., Tan, S., \& Simon, A. 1988, Monthly Notices of the Royal
  Astronomical Society, 231, 735

\bibitem[{{Green}(1986)}]{Green1986}
{Green}, D.~A. 1986, \mnras, 218, 533, \dodoi{10.1093/mnras/218.3.533}

\bibitem[{{Green}(1994)}]{Green1994}
---. 1994, \apjs, 90, 817, \dodoi{10.1086/191908}

\bibitem[{{Green}(2004)}]{Green2004}
---. 2004, Bulletin of the Astronomical Society of India, 32, 335,
  \dodoi{10.48550/arXiv.astro-ph/0411083}

\bibitem[{{Guo} {et~al.}(2017){Guo}, {Xin}, {Liao}, {Yuan}, {Gao}, {He}, {Fan},
  \& {Liu}}]{Guo2017}
{Guo}, X.-L., {Xin}, Y.-L., {Liao}, N.-H., {et~al.} 2017, \apj, 835, 42,
  \dodoi{10.3847/1538-4357/835/1/42}

\bibitem[{{H.~E.~S.~S. Collaboration}(2011)}]{HESS2011}
{H.~E.~S.~S. Collaboration}. 2011, \aap, 531, L18,
  \dodoi{10.1051/0004-6361/201117171}

\bibitem[{{H.~E.~S.~S. Collaboration}(2013)}]{HESS2013}
---. 2013, \aap, 551, A26, \dodoi{10.1051/0004-6361/201220719}

\bibitem[{{H.~E.~S.~S. Collaboration} {et~al.}(2014){H.~E.~S.~S.
  Collaboration}, {Abramowski}, {Aharonian}, {Ait Benkhali}, {Akhperjanian},
  {Ang{\"u}ner}, {Anton}, {Balenderan}, {Balzer}, {Barnacka}, {Becherini},
  {Becker Tjus}, {Bernl{\"o}hr}, {Birsin}, {Bissaldi}, {Biteau},
  {B{\"o}ttcher}, {Boisson}, {Bolmont}, {Bordas}, {Brucker}, {Brun}, {Brun},
  {Bulik}, {Carrigan}, {Casanova}, {Cerruti}, {Chadwick}, {Chalme-Calvet},
  {Chaves}, {Cheesebrough}, {Chr{\'e}tien}, {Colafrancesco}, {Cologna},
  {Conrad}, {Couturier}, {Cui}, {Dalton}, {Daniel}, {Davids}, {Degrange},
  {Deil}, {deWilt}, {Dickinson}, {Djannati-Ata{\"\i}}, {Domainko}, {Drury},
  {Dubus}, {Dutson}, {Dyks}, {Dyrda}, {Edwards}, {Egberts}, {Eger}, {Espigat},
  {Farnier}, {Fegan}, {Feinstein}, {Fernandes}, {Fernandez}, {Fiasson},
  {Fontaine}, {F{\"o}rster}, {F{\"u}{\ss}ling}, {Gajdus}, {Gallant},
  {Garrigoux}, {Giavitto}, {Giebels}, {Glicenstein}, {Grondin},
  {Grudzi{\'n}ska}, {H{\"a}ffner}, {Hahn}, {Harris}, {Heinzelmann}, {Henri},
  {Hermann}, {Hervet}, {Hillert}, {Hinton}, {Hofmann}, {Hofverberg}, {Holler},
  {Horns}, {Jacholkowska}, {Jahn}, {Jamrozy}, {Janiak}, {Jankowsky}, {Jung},
  {Kastendieck}, {Katarzy{\'n}ski}, {Katz}, {Kaufmann}, {Kh{\'e}lifi},
  {Kieffer}, {Klepser}, {Klochkov}, {Klu{\'z}niak}, {Kneiske}, {Kolitzus},
  {Komin}, {Kosack}, {Krakau}, {Krayzel}, {Kr{\"u}ger}, {Laffon}, {Lamanna},
  {Lefaucheur}, {Lemi{\`e}re}, {Lemoine-Goumard}, {Lenain}, {Lennarz}, {Lohse},
  {Lopatin}, {Lu}, {Marandon}, {Marcowith}, {Marx}, {Maurin}, {Maxted},
  {Mayer}, {McComb}, {M{\'e}hault}, {Meintjes}, {Menzler}, {Meyer}, {Moderski},
  {Mohamed}, {Moulin}, {Murach}, {Naumann}, {de Naurois}, {Niemiec}, {Nolan},
  {Oakes}, {Ohm}, {de O{\~n}a Wilhelmi}, {Opitz}, {Ostrowski}, {Oya}, {Panter},
  {Parsons}, {Paz Arribas}, {Pekeur}, {Pelletier}, {Perez}, {Petrucci},
  {Peyaud}, {Pita}, {Poon}, {P{\"u}hlhofer}, {Punch}, {Quirrenbach}, {Raab},
  {Raue}, {Reimer}, {Reimer}, {Renaud}, {de los Reyes}, {Rieger}, {Rob},
  {Romoli}, {Rosier-Lees}, {Rowell}, {Rudak}, {Rulten}, {Sahakian}, {Sanchez},
  {Santangelo}, {Schlickeiser}, {Sch{\"u}ssler}, {Schulz}, {Schwanke},
  {Schwarzburg}, {Schwemmer}, {Sol}, {Spengler}, {Spies}, {Stawarz},
  {Steenkamp}, {Stegmann}, {Stinzing}, {Stycz}, {Sushch}, {Szostek},
  {Tavernet}, {Tavernier}, {Taylor}, {Terrier}, {Tluczykont}, {Trichard},
  {Valerius}, {van Eldik}, {van Soelen}, {Vasileiadis}, {Venter}, {Viana}, \&
  {Vincent}}]{HESS2014}
{H.~E.~S.~S. Collaboration}, {Abramowski}, A., {Aharonian}, F., {et~al.} 2014,
  \aap, 562, A40, \dodoi{10.1051/0004-6361/201322914}

\bibitem[{{H.~E.~S.~S. Collaboration} {et~al.}(2018{\natexlab{a}}){H.~E.~S.~S.
  Collaboration}, {Abdalla}, {Abramowski}, {Aharonian}, {Ait Benkhali},
  {Akhperjanian}, {Andersson}, {Ang{\"u}ner}, {Arrieta}, {Aubert}, {Backes},
  {Balzer}, {Barnard}, {Becherini}, {Becker Tjus}, {Berge}, {Bernhard},
  {Bernl{\"o}hr}, {Blackwell}, {B{\"o}ttcher}, {Boisson}, {Bolmont}, {Bordas},
  {Bregeon}, {Brun}, {Brun}, {Bryan}, {Bulik}, {Capasso}, {Carr}, {Carrigan},
  {Casanova}, {Cerruti}, {Chakraborty}, {Chalme-Calvet}, {Chaves}, {Chen},
  {Chevalier}, {Chr{\'e}tien}, {Colafrancesco}, {Cologna}, {Condon}, {Conrad},
  {Couturier}, {Cui}, {Davids}, {Degrange}, {Deil}, {Devin}, {deWilt},
  {Dirson}, {Djannati-Ata{\"\i}}, {Domainko}, {Donath}, {Drury}, {Dubus},
  {Dutson}, {Dyks}, {Edwards}, {Egberts}, {Eger}, {Ernenwein}, {Eschbach},
  {Farnier}, {Fegan}, {Fernandes}, {Fiasson}, {Fontaine}, {F{\"o}rster},
  {Funk}, {F{\"u}{\ss}ling}, {Gabici}, {Gajdus}, {Gallant}, {Garrigoux},
  {Giavitto}, {Giebels}, {Glicenstein}, {Gottschall}, {Goyal}, {Grondin},
  {Hadasch}, {Hahn}, {Haupt}, {Hawkes}, {Heinzelmann}, {Henri}, {Hermann},
  {Hervet}, {Hillert}, {Hinton}, {Hofmann}, {Hoischen}, {Holler}, {Horns},
  {Ivascenko}, {Jacholkowska}, {Jamrozy}, {Janiak}, {Jankowsky}, {Jankowsky},
  {Jingo}, {Jogler}, {Jouvin}, {Jung-Richardt}, {Kastendieck},
  {Katarzy{\'n}ski}, {Katz}, {Kerszberg}, {Kh{\'e}lifi}, {Kieffer}, {King},
  {Klepser}, {Klochkov}, {Klu{\'z}niak}, {Kolitzus}, {Komin}, {Kosack},
  {Krakau}, {Kraus}, {Krayzel}, {Kr{\"u}ger}, {Laffon}, {Lamanna}, {Lau},
  {Lees}, {Lefaucheur}, {Lefranc}, {Lemi{\`e}re}, {Lemoine-Goumard}, {Lenain},
  {Leser}, {Lohse}, {Lorentz}, {Liu}, {L{\'o}pez-Coto}, {Lypova}, {Marandon},
  {Marcowith}, {Mariaud}, {Marx}, {Maurin}, {Maxted}, {Mayer}, {Meintjes},
  {Meyer}, {Mitchell}, {Moderski}, {Mohamed}, {Mohrmann}, {Mor{\r{a}}},
  {Moulin}, {Murach}, {de Naurois}, {Niederwanger}, {Niemiec}, {Oakes},
  {O'Brien}, {Odaka}, {{\"O}ttl}, {Ohm}, {de O{\~n}a Wilhelmi}, {Ostrowski},
  {Oya}, {Padovani}, {Panter}, {Parsons}, {Paz Arribas}, {Pekeur}, {Pelletier},
  {Perennes}, {Petrucci}, {Peyaud}, {Pita}, {Poon}, {Prokhorov}, {Prokoph},
  {P{\"u}hlhofer}, {Punch}, {Quirrenbach}, {Raab}, {Reimer}, {Reimer},
  {Renaud}, {de los Reyes}, {Rieger}, {Romoli}, {Rosier-Lees}, {Rowell},
  {Rudak}, {Rulten}, {Sahakian}, {Salek}, {Sanchez}, {Santangelo}, {Sasaki},
  {Schlickeiser}, {Sch{\"u}ssler}, {Schulz}, {Schwanke}, {Schwemmer},
  {Settimo}, {Seyffert}, {Shafi}, {Shilon}, {Simoni}, {Sol}, {Spanier},
  {Spengler}, {Spies}, {Stawarz}, {Steenkamp}, {Stegmann}, {Stinzing}, {Stycz},
  {Sushch}, {Tavernet}, {Tavernier}, {Taylor}, {Terrier}, {Tibaldo}, {Tiziani},
  {Tluczykont}, {Trichard}, {Tuffs}, {Uchiyama}, {Valerius}, {van der Walt},
  {van Eldik}, {van Soelen}, {Vasileiadis}, {Veh}, {Venter}, {Viana},
  {Vincent}, {Vink}, {Voisin}, {V{\"o}lk}, {Vuillaume}, {Wadiasingh}, {Wagner},
  {Wagner}, {Wagner}, {White}, {Wierzcholska}, {Willmann}, {W{\"o}rnlein},
  {Wouters}, {Yang}, {Zabalza}, {Zaborov}, {Zacharias}, {Zdziarski}, {Zech},
  {Zefi}, {Ziegler}, \& {{\.Z}ywucka}}]{hess2018population}
{H.~E.~S.~S. Collaboration}, {Abdalla}, H., {Abramowski}, A., {et~al.}
  2018{\natexlab{a}}, \aap, 612, A2, \dodoi{10.1051/0004-6361/201629377}

\bibitem[{{H.~E.~S.~S. Collaboration} {et~al.}(2018{\natexlab{b}}){H.~E.~S.~S.
  Collaboration}, {Abdalla}, {Abramowski}, {Aharonian}, {Ait Benkhali},
  {Ang{\"u}ner}, {Arakawa}, {Arrieta}, {Aubert}, {Backes}, {Balzer}, {Barnard},
  {Becherini}, {Becker Tjus}, {Berge}, {Bernhard}, {Bernl{\"o}hr}, {Blackwell},
  {B{\"o}ttcher}, {Boisson}, {Bolmont}, {Bonnefoy}, {Bordas}, {Bregeon},
  {Brun}, {Brun}, {Bryan}, {B{\"u}chele}, {Bulik}, {Capasso}, {Carrigan},
  {Caroff}, {Carosi}, {Casanova}, {Cerruti}, {Chakraborty}, {Chaves}, {Chen},
  {Chevalier}, {Colafrancesco}, {Condon}, {Conrad}, {Davids}, {Decock}, {Deil},
  {Devin}, {deWilt}, {Dirson}, {Djannati-Ata{\"\i}}, {Domainko}, {Donath},
  {Drury}, {Dutson}, {Dyks}, {Edwards}, {Egberts}, {Eger}, {Emery},
  {Ernenwein}, {Eschbach}, {Farnier}, {Fegan}, {Fernandes}, {Fiasson},
  {Fontaine}, {F{\"o}rster}, {Funk}, {F{\"u}{\ss}ling}, {Gabici}, {Gallant},
  {Garrigoux}, {Gast}, {Gat{\'e}}, {Giavitto}, {Giebels}, {Glawion},
  {Glicenstein}, {Gottschall}, {Grondin}, {Hahn}, {Haupt}, {Hawkes},
  {Heinzelmann}, {Henri}, {Hermann}, {Hinton}, {Hofmann}, {Hoischen}, {Holch},
  {Holler}, {Horns}, {Ivascenko}, {Iwasaki}, {Jacholkowska}, {Jamrozy},
  {Jankowsky}, {Jankowsky}, {Jingo}, {Jouvin}, {Jung-Richardt}, {Kastendieck},
  {Katarzy{\'n}ski}, {Katsuragawa}, {Katz}, {Kerszberg}, {Khangulyan},
  {Kh{\'e}lifi}, {King}, {Klepser}, {Klochkov}, {Klu{\'z}niak}, {Komin},
  {Kosack}, {Krakau}, {Kraus}, {Kr{\"u}ger}, {Laffon}, {Lamanna}, {Lau},
  {Lees}, {Lefaucheur}, {Lemi{\`e}re}, {Lemoine-Goumard}, {Lenain}, {Leser},
  {Lohse}, {Lorentz}, {Liu}, {L{\'o}pez-Coto}, {Lypova}, {Marandon},
  {Malyshev}, {Marcowith}, {Mariaud}, {Marx}, {Maurin}, {Maxted}, {Mayer},
  {Meintjes}, {Meyer}, {Mitchell}, {Moderski}, {Mohamed}, {Mohrmann},
  {Mor{\r{a}}}, {Moulin}, {Murach}, {Nakashima}, {de Naurois}, {Ndiyavala},
  {Niederwanger}, {Niemiec}, {Oakes}, {O'Brien}, {Odaka}, {Ohm}, {Ostrowski},
  {Oya}, {Padovani}, {Panter}, {Parsons}, {Paz Arribas}, {Pekeur}, {Pelletier},
  {Perennes}, {Petrucci}, {Peyaud}, {Piel}, {Pita}, {Poireau}, {Poon},
  {Prokhorov}, {Prokoph}, {P{\"u}hlhofer}, {Punch}, {Quirrenbach}, {Raab},
  {Rauth}, {Reimer}, {Reimer}, {Renaud}, {de los Reyes}, {Rieger}, {Rinchiuso},
  {Romoli}, {Rowell}, {Rudak}, {Rulten}, {Safi-Harb}, {Sahakian}, {Saito},
  {Sanchez}, {Santangelo}, {Sasaki}, {Schandri}, {Schlickeiser},
  {Sch{\"u}ssler}, {Schulz}, {Schwanke}, \& {Schwemmer}}]{Abdalla2018}
---. 2018{\natexlab{b}}, \aap, 612, A1, \dodoi{10.1051/0004-6361/201732098}

\bibitem[{Halpern {et~al.}(2001)Halpern, {Camilo}, {Gotthelf}, {Helfand},
  {Kramer}, {Lyne}, {Leighly}, \& {Eracleous}}]{halpern01}
Halpern, J.~P., {Camilo}, F., {Gotthelf}, E.~V., {et~al.} 2001, \apjl, 552,
  L125, \dodoi{10.1086/320347}

\bibitem[{{Halpern} {et~al.}(2001){Halpern}, {Gotthelf}, {Leighly}, \&
  {Helfand}}]{Halpern_2001b}
{Halpern}, J.~P., {Gotthelf}, E.~V., {Leighly}, K.~M., \& {Helfand}, D.~J.
  2001, \apj, 547, 323, \dodoi{10.1086/318361}

\bibitem[{{Halpern} {et~al.}(2014){Halpern}, {Tomsick}, {Gotthelf}, {Camilo},
  {Ng}, {Bodaghee}, {Rodriguez}, {Chaty}, \& {Rahoui}}]{Halpern2014}
{Halpern}, J.~P., {Tomsick}, J.~A., {Gotthelf}, E.~V., {et~al.} 2014, \apjl,
  795, L27, \dodoi{10.1088/2041-8205/795/2/L27}

\bibitem[{{Harding} \& {Kalapotharakos}(2015)}]{Harding2015}
{Harding}, A.~K., \& {Kalapotharakos}, C. 2015, \apj, 811, 63,
  \dodoi{10.1088/0004-637X/811/1/63}

\bibitem[{{Harding} \& {Kalapotharakos}(2017)}]{Harding2017}
---. 2017, \apj, 840, 73, \dodoi{10.3847/1538-4357/aa6ead}

\bibitem[{{Harding} {et~al.}(2021){Harding}, {Venter}, \&
  {Kalapotharakos}}]{Harding2021}
{Harding}, A.~K., {Venter}, C., \& {Kalapotharakos}, C. 2021, \apj, 923, 194,
  \dodoi{10.3847/1538-4357/ac3084}

\bibitem[{{Harris}(1996)}]{Harris1996}
{Harris}, W.~E. 1996, \aj, 112, 1487, \dodoi{10.1086/118116}

\bibitem[{{Harvey} {et~al.}(2022){Harvey}, {Rulten}, \&
  {Chadwick}}]{Harvey2022}
{Harvey}, M., {Rulten}, C.~B., \& {Chadwick}, P.~M. 2022, \mnras, 512, 1141,
  \dodoi{10.1093/mnras/stac375}

\bibitem[{{Healey} {et~al.}(2007){Healey}, {Romani}, {Taylor}, {Sadler},
  {Ricci}, {Murphy}, {Ulvestad}, \& {Winn}}]{Healey2007}
{Healey}, S.~E., {Romani}, R.~W., {Taylor}, G.~B., {et~al.} 2007, \apjs, 171,
  61, \dodoi{10.1086/513742}

\bibitem[{{Healey} {et~al.}(2008){Healey}, {Romani}, {Cotter}, {Michelson},
  {Schlafly}, {Readhead}, {Giommi}, {Chaty}, {Grenier}, \&
  {Weintraub}}]{Healey2008}
{Healey}, S.~E., {Romani}, R.~W., {Cotter}, G., {et~al.} 2008, \apjs, 175, 97,
  \dodoi{10.1086/523302}

\bibitem[{{Hessels} {et~al.}(2007){Hessels}, {Ransom}, {Stairs}, {Kaspi}, \&
  {Freire}}]{Hessels2007}
{Hessels}, J.~W.~T., {Ransom}, S.~M., {Stairs}, I.~H., {Kaspi}, V.~M., \&
  {Freire}, P.~C.~C. 2007, \apj, 670, 363, \dodoi{10.1086/521780}

\bibitem[{{Hewish} {et~al.}(1969){Hewish}, {Bell}, {Pilkington}, {Scott}, \&
  {Collins}}]{Hewish1969}
{Hewish}, A., {Bell}, S.~J., {Pilkington}, J.~D.~H., {Scott}, P.~F., \&
  {Collins}, R.~A. 1969, \nat, 224, 472, \dodoi{10.1038/224472b0}

\bibitem[{{Hobbs} {et~al.}(2006){Hobbs}, {Edwards}, \&
  {Manchester}}]{Hobbs2006}
{Hobbs}, G.~B., {Edwards}, R.~T., \& {Manchester}, R.~N. 2006, \mnras, 369,
  655, \dodoi{10.1111/j.1365-2966.2006.10302.x}

\bibitem[{{Holler} {et~al.}(2012){Holler}, {Sch{\"o}ck}, {Eger},
  {Kie{\ss}ling}, {Valerius}, \& {Stegmann}}]{Holler2012}
{Holler}, M., {Sch{\"o}ck}, F.~M., {Eger}, P., {et~al.} 2012, \aap, 539, A24,
  \dodoi{10.1051/0004-6361/201118121}

\bibitem[{{Hou} {et~al.}(2023){Hou}, {Zhang}, {Torres}, {Ji}, \&
  {Li}}]{Hou2023}
{Hou}, X., {Zhang}, W., {Torres}, D.~F., {Ji}, L., \& {Li}, J. 2023, \apj, 944,
  57, \dodoi{10.3847/1538-4357/acaec7}

\bibitem[{{Hou} {et~al.}(2024){Hou}, {Zhang}, {Freire}, {Torres}, {Ballet},
  {Smith}, {Johnson}, {Kerr}, {Cheung}, {Guillemot}, {Li}, {Zhang}, {Ridolfi},
  {Wang}, {Li}, {Yuan}, \& {Wang}}]{Hou2024}
{Hou}, X., {Zhang}, W., {Freire}, P.~C.~C., {et~al.} 2024, \apj, 964, 118,
  \dodoi{10.3847/1538-4357/ad3210}

\bibitem[{{Hui} {et~al.}(2010){Hui}, {Cheng}, \& {Taam}}]{Hui2010}
{Hui}, C.~Y., {Cheng}, K.~S., \& {Taam}, R.~E. 2010, \apj, 714, 1149,
  \dodoi{10.1088/0004-637X/714/2/1149}

\bibitem[{{{\'I}{\~n}iguez-Pascual} {et~al.}(2024){{\'I}{\~n}iguez-Pascual},
  {Torres}, \& {Vigan{\`o}}}]{2024Iniguez}
{{\'I}{\~n}iguez-Pascual}, D., {Torres}, D.~F., \& {Vigan{\`o}}, D. 2024,
  \mnras, 530, 1550, \dodoi{10.1093/mnras/stae933}

\bibitem[{{{\'I}{\~n}iguez-Pascual} {et~al.}(2025){{\'I}{\~n}iguez-Pascual},
  {Torres}, \& {Vigan{\`o}}}]{2025Iniguez}
---. 2025, arXiv e-prints, arXiv:2504.01892, \dodoi{10.48550/arXiv.2504.01892}

\bibitem[{{Johnson} {et~al.}(2013){Johnson}, {Guillemot}, {Kerr}, {Cognard},
  {Ray}, {Wolff}, {B{\'e}gin}, {Janssen}, {Romani}, {Venter}, {Grove},
  {Freire}, {Wood}, {Cheung}, {Casandjian}, {Stairs}, {Camilo}, {Espinoza},
  {Ferrara}, {Harding}, {Johnston}, {Kramer}, {Lyne}, {Michelson}, {Ransom},
  {Shannon}, {Smith}, {Stappers}, {Theureau}, \& {Thorsett}}]{Johnson2013}
{Johnson}, T.~J., {Guillemot}, L., {Kerr}, M., {et~al.} 2013, \apj, 778, 106,
  \dodoi{10.1088/0004-637X/778/2/106}

\bibitem[{{Johnson} {et~al.}(2015){Johnson}, {Ray}, {Roy}, {Cheung}, {Harding},
  {Pletsch}, {Fort}, {Camilo}, {Deneva}, {Bhattacharyya}, {Stappers}, \&
  {Kerr}}]{Johnson2015}
{Johnson}, T.~J., {Ray}, P.~S., {Roy}, J., {et~al.} 2015, \apj, 806, 91,
  \dodoi{10.1088/0004-637X/806/1/91}

\bibitem[{{Joshi} {et~al.}(2022){Joshi}, {Tanaka}, {Miranda}, \&
  {Razzaque}}]{joshi22}
{Joshi}, J.~C., {Tanaka}, S.~J., {Miranda}, L.~S., \& {Razzaque}, S. 2022,
  arXiv e-prints, arXiv:2205.00521.
\newblock \doarXiv{2205.00521}

\bibitem[{{Joshi} {et~al.}(2023){Joshi}, {Maurya}, {John}, {Panchal}, {Joshi},
  \& {Kumar}}]{Joshi2023}
{Joshi}, Y.~C., {Maurya}, J., {John}, A.~A., {et~al.} 2023, {VizieR Online Data
  Catalog: Study of the young open cluster NGC 1960 (Joshi+, 2020)}, VizieR
  On-line Data Catalog: J/MNRAS/492/3602. Originally published in:
  2020MNRAS.492.3602J

\bibitem[{{Kabuki} {et~al.}(2007){Kabuki}, {Enomoto}, {Bicknell}, {Clay},
  {Edwards}, {Gunji}, {Hara}, {Hattori}, {Hayashi}, {Higashi}, {Inoue}, \&
  {Itoh}}]{Kabuki2007}
{Kabuki}, S., {Enomoto}, R., {Bicknell}, G.~V., {et~al.} 2007, \apj, 668, 968,
  \dodoi{10.1086/520767}

\bibitem[{{Kalapotharakos} {et~al.}(2012){Kalapotharakos}, {Harding},
  {Kazanas}, \& {Contopoulos}}]{Kalapotharakos2012}
{Kalapotharakos}, C., {Harding}, A.~K., {Kazanas}, D., \& {Contopoulos}, I.
  2012, \apjl, 754, L1, \dodoi{10.1088/2041-8205/754/1/L1}

\bibitem[{{Kalapotharakos} {et~al.}(2022){Kalapotharakos}, {Wadiasingh},
  {Harding}, \& {Kazanas}}]{Kalapotharakos2022}
{Kalapotharakos}, C., {Wadiasingh}, Z., {Harding}, A.~K., \& {Kazanas}, D.
  2022, \apj, 934, 65, \dodoi{10.3847/1538-4357/ac78e3}

\bibitem[{{Kaplan} {et~al.}(2004){Kaplan}, {Frail}, {Gaensler}, {Gotthelf},
  {Kulkarni}, {Slane}, \& {Nechita}}]{Kaplan2004}
{Kaplan}, D.~L., {Frail}, D.~A., {Gaensler}, B.~M., {et~al.} 2004, \apjs, 153,
  269, \dodoi{10.1086/421065}

\bibitem[{{Kargaltsev} \& {Pavlov}(2007)}]{Kargaltsev2007}
{Kargaltsev}, O., \& {Pavlov}, G.~G. 2007, \apj, 670, 655,
  \dodoi{10.1086/521814}

\bibitem[{{Karpova} {et~al.}(2019){Karpova}, {Zyuzin}, \&
  {Shibanov}}]{karpova2019}
{Karpova}, A.~V., {Zyuzin}, D.~A., \& {Shibanov}, Y.~A. 2019, \mnras, 487,
  1964, \dodoi{10.1093/mnras/stz1387}

\bibitem[{{Kaspi} {et~al.}(2001){Kaspi}, {Gotthelf}, {Gaensler}, \&
  {Lyutikov}}]{kaspi2001}
{Kaspi}, V.~M., {Gotthelf}, E.~V., {Gaensler}, B.~M., \& {Lyutikov}, M. 2001,
  \apjl, 562, L163, \dodoi{10.1086/324757}

\bibitem[{{Kaspi} {et~al.}(1992){Kaspi}, {Manchester}, {Johnston}, {Lyne}, \&
  {D'Amico}}]{kaspi1992psr}
{Kaspi}, V.~M., {Manchester}, R.~N., {Johnston}, S., {Lyne}, A.~G., \&
  {D'Amico}, N. 1992, \apjl, 399, L155, \dodoi{10.1086/186630}

\bibitem[{{Katz}(1975)}]{Katz1975}
{Katz}, J.~I. 1975, \nat, 253, 698, \dodoi{10.1038/253698a0}

\bibitem[{{Kerr}(2011)}]{kerr2011}
{Kerr}, M. 2011, \apj, 732, 38, \dodoi{10.1088/0004-637X/732/1/38}

\bibitem[{{Kilpatrick} {et~al.}(2016){Kilpatrick}, {Bieging}, \&
  {Rieke}}]{Kilpatrick2015}
{Kilpatrick}, C.~D., {Bieging}, J.~H., \& {Rieke}, G.~H. 2016, \apj, 816, 1,
  \dodoi{10.3847/0004-637X/816/1/1}

\bibitem[{{Kirichenko} {et~al.}(2015){Kirichenko}, {Danilenko}, {Shternin},
  {Shibanov}, {Ryspaeva}, {Zyuzin}, {Durant}, {Kargaltsev}, {Pavlov}, \&
  {Cabrera-Lavers}}]{Kirichenko2015}
{Kirichenko}, A., {Danilenko}, A., {Shternin}, P., {et~al.} 2015, \apj, 802,
  17, \dodoi{10.1088/0004-637X/802/1/17}

\bibitem[{{Klingler} {et~al.}(2018){Klingler}, {Kargaltsev}, {Pavlov}, {Ng},
  {Beniamini}, \& {Volkov}}]{klingler2018}
{Klingler}, N., {Kargaltsev}, O., {Pavlov}, G.~G., {et~al.} 2018, \apj, 861, 5,
  \dodoi{10.3847/1538-4357/aac6e0}

\bibitem[{{Kohler}(2020)}]{Kohler2020}
{Kohler}, S. 2020, {An Infant Pulsar Defies Categorization}, AAS Nova
  Highlight, 27 Jul 2020, id.6836

\bibitem[{{Kong} {et~al.}(2021){Kong}, {Zhang}, {Ji}, {Reig}, {Doroshenko},
  {Santangelo}, {Staubert}, {Zhang}, {Soria}, {Chang}, {Chen}, {Wang}, {Tao},
  \& {Qu}}]{Kong2021}
{Kong}, L.~D., {Zhang}, S., {Ji}, L., {et~al.} 2021, \apjl, 917, L38,
  \dodoi{10.3847/2041-8213/ac1ad3}

\bibitem[{{Kong} {et~al.}(2022){Kong}, {Zhang}, {Ji}, {Doroshenko},
  {Santangelo}, {Orlandini}, {Frontera}, {Li}, {Chen}, {Wang}, {Chang}, {Qu},
  \& {Zhang}}]{Kong2022}
{Kong}, L.-D., {Zhang}, S., {Ji}, L., {et~al.} 2022, \apj, 932, 106,
  \dodoi{10.3847/1538-4357/ac6e66}

\bibitem[{Kothes(2010)}]{Kothes2010}
Kothes, R. 2010, arXiv preprint arXiv:1010.4586

\bibitem[{Kothes {et~al.}(2001)Kothes, Uyaniker, \& Pineault}]{kothes01}
Kothes, R., Uyaniker, B., \& Pineault, S. 2001, \apj, 560, 236,
  \dodoi{10.1086/322511}

\bibitem[{{Kova{\v{c}}evi{\'c}} {et~al.}(2019){Kova{\v{c}}evi{\'c}}, {},
  {Chiaro}, {Cutini}, \& {Tosti}}]{fermiblz2019}
{Kova{\v{c}}evi{\'c}}, {}, M., {Chiaro}, G., {Cutini}, S., \& {Tosti}, G. 2019,
  \mnras, 490, 4770, \dodoi{10.1093/mnras/stz2920}

\bibitem[{{Kuiper} \& {Hermsen}(2015)}]{kuiper2015soft}
{Kuiper}, L., \& {Hermsen}, W. 2015, \mnras, 449, 3827,
  \dodoi{10.1093/mnras/stv426}

\bibitem[{{Kundu} {et~al.}(2024){Kundu}, {Joshi}, {Venter}, {Engelbrecht},
  {Zhang}, {Torres}, {Sushch}, \& {Tanaka}}]{Kundu2024}
{Kundu}, A., {Joshi}, J.~C., {Venter}, C., {et~al.} 2024, \mnras, 535, 2415,
  \dodoi{10.1093/mnras/stae2435}

\bibitem[{{Lemiere} {et~al.}(2009){Lemiere}, {Slane}, {Gaensler}, \&
  {Murray}}]{lemiere2009}
{Lemiere}, A., {Slane}, P., {Gaensler}, B.~M., \& {Murray}, S. 2009, \apj, 706,
  1269, \dodoi{10.1088/0004-637X/706/2/1269}

\bibitem[{{Lhaaso Collaboration} {et~al.}(2021{\natexlab{a}}){Lhaaso
  Collaboration}, {Cao}, {Aharonian}, {An}, {Axikegu}, {Bai}, {Bai}, {Bao},
  {Bastieri}, {Bi}, {Bi}, {Cai}, {Cai}, {Cao}, {Chang}, {Chang}, {Chen},
  {Chen}, {Chen}, {Chen}, {Chen}, {Chen}, {Chen}, {Chen}, {Chen}, {Chen},
  {Chen}, {Chen}, {Chen}, {Chen}, {Cheng}, {Cheng}, {Cui}, {Cui}, {Cui},
  {D'Ettorre Piazzoli}, {Dai}, {Dai}, {Dai}, {Danzengluobu}, {Della Volpe},
  {Dong}, {Duan}, {Fan}, {Fan}, {Fan}, {Fang}, {Fang}, {Feng}, {Feng}, {Feng},
  {Feng}, {Gao}, {Gao}, {Gao}, {Gao}, {Gao}, {Ge}, {Geng}, {Gong}, {Gou}, {Gu},
  {Guo}, {Guo}, {Guo}, {Guo}, {Guo}, {Han}, {He}, {He}, {He}, {He}, {He}, {He},
  {Heller}, {Hor}, {Hou}, {Hou}, {Hu}, {Hu}, {Hu}, {Hu}, {Huang}, {Huang},
  {Huang}, {Huang}, {Huang}, {Huang}, {Ji}, {Ji}, {Jia}, {Jiang}, {Jiang},
  {Jin}, {Ke}, {Kuleshov}, {Levochkin}, {Li}, {Li}, {Li}, {Li}, {Li}, {Li},
  {Li}, {Li}, {Li}, {Li}, {Li}, {Li}, {Li}, {Li}, {Li}, {Li}, {Li}, {Li},
  {Liang}, {Liang}, {Lin}, {Liu}, {Liu}, {Liu}, {Liu}, {Liu}, {Liu}, {Liu},
  {Liu}, {Liu}, {Liu}, {Liu}, {Liu}, {Liu}, {Liu}, {Liu}, {Liu}, {Long}, {Lu},
  {Lv}, {Ma}, {Ma}, {Ma}, {Mao}, {Masood}, {Min}, {Mitthumsiri}, {Montaruli},
  {Nan}, {Pang}, {Pattarakijwanich}, {Pei}, {Qi}, {Qi}, {Qiao}, {Qin},
  {Ruffolo}, {Rulev}, {Saiz}, {Shao}, {Shchegolev}, {Sheng}, {Shi}, {Song},
  {Stenkin}, {Stepanov}, {Su}, {Sun}, {Sun}, {Sun}, {Tam}, {Tang}, {Tian},
  {Wang}, {Wang}, {Wang}, {Wang}, {Wang}, {Wang}, {Wang}, {Wang}, {Wang},
  {Wang}, {Wang}, {Wang}, {Wang}, {Wang}, {Wang}, {Wang}, {Wang}, {Wang},
  {Wang}, {Wang}, {Wang}, {Wang}, {Wei}, {Wei}, {Wei}, {Wen}, {Wu}, {Wu}, {Wu},
  \& {Wu}}]{lhaaso2021}
{Lhaaso Collaboration}, {Cao}, Z., {Aharonian}, F., {et~al.}
  2021{\natexlab{a}}, Science, 373, 425, \dodoi{10.1126/science.abg5137}

\bibitem[{{Lhaaso Collaboration} {et~al.}(2021{\natexlab{b}}){Lhaaso
  Collaboration}, {Cao}, {Aharonian}, {An}, {Axikegu}, {Bai}, {Bai}, {Bao},
  {Bastieri}, {Bi}, {Bi}, {Cai}, {Cai}, {Cao}, {Chang}, {Chang}, {Chen},
  {Chen}, {Chen}, {Chen}, {Chen}, {Chen}, {Chen}, {Chen}, {Chen}, {Chen},
  {Chen}, {Chen}, {Chen}, {Chen}, {Cheng}, {Cheng}, {Cui}, {Cui}, {Cui},
  {D'Ettorre Piazzoli}, {Dai}, {Dai}, {Dai}, {Danzengluobu}, {Della Volpe},
  {Dong}, {Duan}, {Fan}, {Fan}, {Fan}, {Fang}, {Fang}, {Feng}, {Feng}, {Feng},
  {Feng}, {Gao}, {Gao}, {Gao}, {Gao}, {Gao}, {Ge}, {Geng}, {Gong}, {Gou}, {Gu},
  {Guo}, {Guo}, {Guo}, {Guo}, {Guo}, {Han}, {He}, {He}, {He}, {He}, {He}, {He},
  {Heller}, {Hor}, {Hou}, {Hou}, {Hu}, {Hu}, {Hu}, {Hu}, {Huang}, {Huang},
  {Huang}, {Huang}, {Huang}, {Huang}, {Ji}, {Ji}, {Jia}, {Jiang}, {Jiang},
  {Jin}, {Ke}, {Kuleshov}, {Levochkin}, {Li}, {Li}, {Li}, {Li}, {Li}, {Li},
  {Li}, {Li}, {Li}, {Li}, {Li}, {Li}, {Li}, {Li}, {Li}, {Li}, {Li}, {Li},
  {Liang}, {Liang}, {Lin}, {Liu}, {Liu}, {Liu}, {Liu}, {Liu}, {Liu}, {Liu},
  {Liu}, {Liu}, {Liu}, {Liu}, {Liu}, {Liu}, {Liu}, {Liu}, {Liu}, {Long}, {Lu},
  {Lv}, {Ma}, {Ma}, {Ma}, {Mao}, {Masood}, {Min}, {Mitthumsiri}, {Montaruli},
  {Nan}, {Pang}, {Pattarakijwanich}, {Pei}, {Qi}, {Qi}, {Qiao}, {Qin},
  {Ruffolo}, {Rulev}, {Saiz}, {Shao}, {Shchegolev}, {Sheng}, {Shi}, {Song},
  {Stenkin}, {Stepanov}, {Su}, {Sun}, {Sun}, {Sun}, {Tam}, {Tang}, {Tian},
  {Wang}, {Wang}, {Wang}, {Wang}, {Wang}, {Wang}, {Wang}, {Wang}, {Wang},
  {Wang}, {Wang}, {Wang}, {Wang}, {Wang}, {Wang}, {Wang}, {Wang}, {Wang},
  {Wang}, {Wang}, {Wang}, {Wang}, {Wei}, {Wei}, {Wei}, {Wen}, {Wu}, {Wu}, {Wu},
  \& {Wu}}]{Cao2021}
---. 2021{\natexlab{b}}, Science, 373, 425, \dodoi{10.1126/science.abg5137}

\bibitem[{{Li} {et~al.}(2018){Li}, {Torres}, {Lin}, {Grondin}, {Kerr},
  {Lemoine-Goumard}, \& {de O{\~n}a Wilhelmi}}]{Li2018}
{Li}, J., {Torres}, D.~F., {Lin}, T.~T., {et~al.} 2018, \apj, 858, 84,
  \dodoi{10.3847/1538-4357/aabac9}

\bibitem[{{Li} {et~al.}(2020){Li}, {Torres}, {Liu}, {Kerr}, {de O{\~n}a
  Wilhelmi}, \& {Su}}]{Li2020SS433}
{Li}, J., {Torres}, D.~F., {Liu}, R.-Y., {et~al.} 2020, Nature Astronomy, 4,
  1177, \dodoi{10.1038/s41550-020-1164-6}

\bibitem[{{Li} {et~al.}(2012){Li}, {Torres}, {Zhang}, {Papitto}, {Chen}, \&
  {Wang}}]{Li2012}
{Li}, J., {Torres}, D.~F., {Zhang}, S., {et~al.} 2012, \apj, 761, 49,
  \dodoi{10.1088/0004-637X/761/1/49}

\bibitem[{{Li} {et~al.}(2021){Li}, {Liu}, {de O{\~n}a Wilhelmi}, {Torres},
  {Liu}, {Kerr}, {B{\"u}hler}, {Su}, {He}, \& {Xiao}}]{li21}
{Li}, J., {Liu}, R.-Y., {de O{\~n}a Wilhelmi}, E., {et~al.} 2021, \apjl, 913,
  L33, \dodoi{10.3847/2041-8213/abf925}

\bibitem[{{Liu} {et~al.}(2019){Liu}, {Yang}, {Sun}, {Aharonian}, \&
  {Chen}}]{Liu2019G8}
{Liu}, B., {Yang}, R.-z., {Sun}, X.-n., {Aharonian}, F., \& {Chen}, Y. 2019,
  \apj, 881, 94, \dodoi{10.3847/1538-4357/ab2df8}

\bibitem[{{Livingstone} {et~al.}(2011){Livingstone}, {Ng}, {Kaspi}, {Gavriil},
  \& {Gotthelf}}]{Livingston2011}
{Livingstone}, M.~A., {Ng}, C.~Y., {Kaspi}, V.~M., {Gavriil}, F.~P., \&
  {Gotthelf}, E.~V. 2011, \apj, 730, 66, \dodoi{10.1088/0004-637X/730/2/66}

\bibitem[{Lorimer \& Kramer(2005)}]{lorimer2005handbook}
Lorimer, D.~R., \& Kramer, M. 2005, Handbook of pulsar astronomy, Vol.~4
  (Cambridge university press)

\bibitem[{{Lower} {et~al.}(2021){Lower}, {Johnston}, {Dunn}, {Shannon},
  {Bailes}, {Dai}, {Kerr}, {Manchester}, {Melatos}, {Oswald}, {Parthasarathy},
  {Sobey}, \& {Weltevrede}}]{lower2021impact}
{Lower}, M.~E., {Johnston}, S., {Dunn}, L., {et~al.} 2021, \mnras, 508, 3251,
  \dodoi{10.1093/mnras/stab2678}

\bibitem[{Lu(2014)}]{Lu2014}
Lu, T. 2014, Modern Astrophysics (BEIJING BOOK CO. INC.)

\bibitem[{{Mac{\'\i}as-P{\'e}rez} {et~al.}(2010){Mac{\'\i}as-P{\'e}rez},
  {Mayet}, {Aumont}, \& {D{\'e}sert}}]{Macias2010}
{Mac{\'\i}as-P{\'e}rez}, J.~F., {Mayet}, F., {Aumont}, J., \& {D{\'e}sert},
  F.~X. 2010, \apj, 711, 417, \dodoi{10.1088/0004-637X/711/1/417}

\bibitem[{{Madsen} {et~al.}(2020){Madsen}, {Fryer}, {Grefenstette}, {Lopez},
  {Reynolds}, \& {Zoglauer}}]{Madsen2020}
{Madsen}, K.~K., {Fryer}, C.~L., {Grefenstette}, B.~W., {et~al.} 2020, \apj,
  889, 23, \dodoi{10.3847/1538-4357/ab54ca}

\bibitem[{{MAGIC Collaboration}(2019)}]{MAGIC2019}
{MAGIC Collaboration}. 2019, \mnras, 484, 2876, \dodoi{10.1093/mnras/stz179}

\bibitem[{{MAGIC Collaboration} {et~al.}(2020){MAGIC Collaboration}, {Acciari},
  {Ansoldi}, {Antonelli}, {Arbet Engels}, {Asano}, {Baack}, {Babi{\'c}},
  {Banerjee}, {Baquero}, {Barres de Almeida}, {Barrio}, {Becerra Gonz{\'a}lez},
  {Bednarek}, {Bellizzi}, {Bernardini}, {Bernardos}, {Berti}, {Besenrieder},
  {Bhattacharyya}, {Bigongiari}, {Biland}, {Blanch}, {Bonnoli},
  {Bo{\v{s}}njak}, {Busetto}, {Carosi}, {Ceribella}, {Cerruti}, {Chai},
  {Chilingarian}, {Cikota}, {Colak}, {Colombo}, {Contreras}, {Cortina},
  {Covino}, {D'Amico}, {D'Elia}, {da Vela}, {Dazzi}, {de Angelis}, {de Lotto},
  {Delfino}, {Delgado}, {Delgado Mendez}, {Depaoli}, {di Girolamo}, {di
  Pierro}, {di Venere}, {Do Souto Espi{\~n}eira}, {Dominis Prester}, {Donini},
  {Dorner}, {Doro}, {Elsaesser}, {Fallah Ramazani}, {Fattorini}, {Ferrara},
  {Foffano}, {Fonseca}, {Font}, {Fruck}, {Fukami}, {Garc{\'\i}a L{\'o}pez},
  {Garczarczyk}, {Gasparyan}, {Gaug}, {Giglietto}, {Giordano}, {Gliwny},
  {Godinovi{\'c}}, {Green}, {Hadasch}, {Hahn}, {Heckmann}, {Herrera}, {Hoang},
  {Hrupec}, {H{\"u}tten}, {Inada}, {Inoue}, {Ishio}, {Iwamura}, {Jouvin},
  {Kajiwara}, {Karjalainen}, {Kerszberg}, {Kobayashi}, {Kubo}, {Kushida},
  {Lamastra}, {Lelas}, {Leone}, {Lindfors}, {Lombardi}, {Longo}, {L{\'o}pez},
  {L{\'o}pez-Coto}, {L{\'o}pez-Oramas}, {Loporchio}, {Machado de Oliveira
  Fraga}, {Maggio}, {Majumdar}, {Makariev}, {Mallamaci}, {Maneva}, {Manganaro},
  {Mannheim}, {Maraschi}, {Mariotti}, {Mart{\'\i}nez}, {Mazin}, {Mender},
  {Mi{\'c}anovi{\'c}}, {Miceli}, {Miener}, {Minev}, {Miranda}, {Mirzoyan},
  {Molina}, {Moralejo}, {Morcuende}, {Moreno}, {Moretti}, {Munar-Adrover},
  {Neustroev}, {Nigro}, {Nilsson}, {Ninci}, {Nishijima}, {Noda}, {Nozaki},
  {Ohtani}, {Oka}, {Otero-Santos}, {Palatiello}, {Paneque}, {Paoletti},
  {Paredes}, {Pavleti{\'c}}, {Pe{\~n}il}, {Perennes}, {Persic}, {Prada Moroni},
  {Prandini}, {Priyadarshi}, {Puljak}, {Rhode}, {Rib{\'o}}, {Rico}, {Righi},
  {Rugliancich}, {Saha}, {Sahakyan}, {Saito}, {Sakurai}, {Satalecka},
  {Schleicher}, {Schmidt}, {Schweizer}, {Sitarek}, {{\v{S}}nidari{\'c}},
  {Sobczynska}, {Spolon}, {Stamerra}, {Strom}, {Strzys}, {Suda}, {Suri{\'c}},
  {Takahashi}, {Tavecchio}, {Temnikov}, {Terzi{\'c}}, {Teshima},
  {Torres-Alb{\`a}}, {Tosti}, {Truzzi}, {van Scherpenberg}, {Vanzo}, {Vazquez
  Acosta}, {Ventura}, {Verguilov}, {Vigorito}, {Vitale}, {Vovk}, {Will}, \&
  {Zari{\'c}}}]{magic2020studying}
{MAGIC Collaboration}, {Acciari}, V.~A., {Ansoldi}, S., {et~al.} 2020, \mnras,
  497, 3734, \dodoi{10.1093/mnras/staa2135}

\bibitem[{{MAGIC Collaboration} {et~al.}(2023){MAGIC Collaboration}, {Abe},
  {Abe}, {Acciari}, {Agudo}, {Aniello}, {Ansoldi}, {Antonelli}, {Arbet Engels},
  {Arcaro}, {Artero}, {Asano}, {Baack}, {Babi{\'c}}, {Baquero}, {Barres de
  Almeida}, {Barrio}, {Batkovi{\'c}}, {Baxter}, {Becerra Gonz{\'a}lez},
  {Bednarek}, {Bernardini}, {Bernardos}, {Berti}, {Besenrieder},
  {Bhattacharyya}, {Bigongiari}, {Biland}, {Blanch}, {Bonnoli},
  {Bo{\v{s}}njak}, {Burelli}, {Busetto}, {Carosi}, {Carretero-Castrillo},
  {Castro-Tirado}, {Ceribella}, {Chai}, {Chilingarian}, {Cikota}, {Colombo},
  {Contreras}, {Cortina}, {Covino}, {D'Amico}, {D'Elia}, {da Vela}, {Dazzi},
  {de Angelis}, {de Lotto}, {Del Popolo}, {Delfino}, {Delgado}, {Delgado
  Mendez}, {Depaoli}, {di Pierro}, {di Venere}, {Do Souto Espi{\~n}eira},
  {Dominis Prester}, {Donini}, {Dorner}, {Doro}, {Elsaesser}, {Emery},
  {Escudero}, {Fallah Ramazani}, {Fari{\~n}a}, {Fattorini}, {Font}, {Fruck},
  {Fukami}, {Fukazawa}, {Garc{\'\i}a L{\'o}pez}, {Garczarczyk}, {Gasparyan},
  {Gaug}, {Giesbrecht Paiva}, {Giglietto}, {Giordano}, {Gliwny},
  {Godinovi{\'c}}, {Grau}, {Green}, {Green}, {Hadasch}, {Hahn}, {Hassan},
  {Heckmann}, {Herrera}, {Hrupec}, {H{\"u}tten}, {Imazawa}, {Inada}, {Iotov},
  {Ishio}, {Jim{\'e}nez Mart{\'\i}nez}, {Jormanainen}, {Kerszberg},
  {Kobayashi}, {Kubo}, {Kushida}, {Lamastra}, {Lelas}, {Leone}, {Lindfors},
  {Linhoff}, {Lombardi}, {Longo}, {L{\'o}pez-Coto}, {L{\'o}pez-Moya},
  {L{\'o}pez-Oramas}, {Loporchio}, {Lorini}, {Lyard}, {Machado de Oliveira
  Fraga}, {Majumdar}, {Makariev}, {Maneva}, {Mang}, {Manganaro}, {Mangano},
  {Mannheim}, {Mariotti}, {Mart{\'\i}nez}, {Mas Aguilar}, {Mazin}, {Menchiari},
  {Mender}, {Mi{\'c}anovi{\'c}}, {Miceli}, {Miener}, {Miranda}, {Mirzoyan},
  {Molina}, {Mondal}, {Moralejo}, {Morcuende}, {Moreno}, {Nakamori}, {Nanci},
  {Nava}, {Neustroev}, {Nievas Rosillo}, {Nigro}, {Nilsson}, {Nishijima}, {Njoh
  Ekoume}, {Noda}, {Nozaki}, {Ohtani}, {Oka}, {Okumura}, {Otero-Santos},
  {Paiano}, {Palatiello}, {Paneque}, {Paoletti}, {Paredes}, {Pavleti{\'c}},
  {Persic}, {Pihet}, {Pirola}, {Podobnik}, {Prada Moroni}, {Prandini},
  {Principe}, {Priyadarshi}, {Rhode}, {Rib{\'o}}, {Rico}, {Righi},
  {Rugliancich}, {Sahakyan}, {Saito}, {Sakurai}, {Satalecka}, {Saturni},
  {Schleicher}, {Schmidt}, {Schmuckermaier}, {Schubert}, {Schweizer},
  {Sitarek}, {Sliusar}, {Sobczynska}, {Spolon}, {Stamerra},
  {Stri{\v{s}}kovi{\'c}}, {Strom}, {Strzys}, {Suda}, {Suri{\'c}}, {Tajima},
  {Takahashi}, {Takeishi}, {Tavecchio}, {Temnikov}, {Terauchi}, {Terzi{\'c}},
  {Teshima}, {Tosti}, {Truzzi}, {Tutone}, {Ubach}, {van Scherpenberg}, {Vazquez
  Acosta}, {Ventura}, {Verguilov}, {Viale}, {Vigorito}, {Vitale}, {Vovk},
  {Walter}, {Will}, {Wunderlich}, {Yamamoto}, \& {Zari{\'c}}}]{magic2023}
{MAGIC Collaboration}, {Abe}, H., {Abe}, S., {et~al.} 2023, \aap, 671, A12,
  \dodoi{10.1051/0004-6361/202244931}

\bibitem[{{Malizia} {et~al.}(2021){Malizia}, {Fiocchi}, {Natalucci}, {Sguera},
  {Stephen}, {Bassani}, {Bazzano}, {Ubertini}, {Pian}, \&
  {Bird}}]{malizia2021integral}
{Malizia}, A., {Fiocchi}, M., {Natalucci}, L., {et~al.} 2021, Universe, 7, 135,
  \dodoi{10.3390/universe7050135}

\bibitem[{{Manca} {et~al.}(2025){Manca}, {Coti Zelati}, {Li}, {Torres},
  {Ballet}, {Marino}, {Sanna}, {Rea}, {Di Salvo}, {Riggio}, {Burderi}, \&
  {Iaria}}]{Manca2025}
{Manca}, A., {Coti Zelati}, F., {Li}, J., {et~al.} 2025, \aap, 695, A187,
  \dodoi{10.1051/0004-6361/202453010}

\bibitem[{{Manchester} {et~al.}(1985{\natexlab{a}}){Manchester}, {Damico}, \&
  {Tuohy}}]{manchester1985search}
{Manchester}, R.~N., {Damico}, N., \& {Tuohy}, I.~R. 1985{\natexlab{a}},
  \mnras, 212, 975, \dodoi{10.1093/mnras/212.4.975}

\bibitem[{{Manchester} {et~al.}(1985{\natexlab{b}}){Manchester}, {Durdin}, \&
  {Newton}}]{manchester85}
{Manchester}, R.~N., {Durdin}, J.~M., \& {Newton}, L.~M. 1985{\natexlab{b}},
  \nat, 313, 374, \dodoi{10.1038/313374a0}

\bibitem[{{Manchester} {et~al.}(2005){Manchester}, {Hobbs}, {Teoh}, \&
  {Hobbs}}]{manchester2005australia}
{Manchester}, R.~N., {Hobbs}, G.~B., {Teoh}, A., \& {Hobbs}, M. 2005, \aj, 129,
  1993, \dodoi{10.1086/428488}

\bibitem[{{Manchester} {et~al.}(1993){Manchester}, {Staveley-Smith}, \&
  {Kesteven}}]{manchester1993}
{Manchester}, R.~N., {Staveley-Smith}, L., \& {Kesteven}, M.~J. 1993, \apj,
  411, 756, \dodoi{10.1086/172877}

\bibitem[{Manchester \& Taylor(1977)}]{manchester77}
Manchester, R.~N., \& Taylor, J.~H. 1977.
\newblock \url{https://www.osti.gov/biblio/6581440}

\bibitem[{{Mandal} \& {Pal}(2022)}]{Mandal2022}
{Mandal}, M., \& {Pal}, S. 2022, \mnras, 511, 1121,
  \dodoi{10.1093/mnras/stac111}

\bibitem[{{Marelli} {et~al.}(2014){Marelli}, {Harding}, {Pizzocaro}, {De Luca},
  {Wood}, {Caraveo}, {Salvetti}, {Saz Parkinson}, \&
  {Acero}}]{marelli2014puzzling}
{Marelli}, M., {Harding}, A., {Pizzocaro}, D., {et~al.} 2014, \apj, 795, 168,
  \dodoi{10.1088/0004-637X/795/2/168}

\bibitem[{{Marthi} {et~al.}(2011){Marthi}, {Chengalur}, {Gupta}, {Dewangan}, \&
  {Bhattacharya}}]{marthi2011}
{Marthi}, V.~R., {Chengalur}, J.~N., {Gupta}, Y., {Dewangan}, G.~C., \&
  {Bhattacharya}, D. 2011, \mnras, 416, 2560,
  \dodoi{10.1111/j.1365-2966.2011.19155.x}

\bibitem[{{Martin} \& {Torres}(2022)}]{martin2022unique}
{Martin}, J., \& {Torres}, D.~F. 2022, Journal of High Energy Astrophysics, 36,
  128, \dodoi{10.1016/j.jheap.2022.09.003}

\bibitem[{{Mart{\'\i}n} {et~al.}(2016){Mart{\'\i}n}, {Torres}, \&
  {Pedaletti}}]{Martin2016}
{Mart{\'\i}n}, J., {Torres}, D.~F., \& {Pedaletti}, G. 2016, \mnras, 459, 3868,
  \dodoi{10.1093/mnras/stw684}

\bibitem[{{Mart{\'\i}n} {et~al.}(2012){Mart{\'\i}n}, {Torres}, \&
  {Rea}}]{Martin2012}
{Mart{\'\i}n}, J., {Torres}, D.~F., \& {Rea}, N. 2012, \mnras, 427, 415,
  \dodoi{10.1111/j.1365-2966.2012.22014.x}

\bibitem[{{Massaro} {et~al.}(2015){Massaro}, {Maselli}, {Leto}, {Marchegiani},
  {Perri}, {Giommi}, \& {Piranomonte}}]{Massaro2015}
{Massaro}, E., {Maselli}, A., {Leto}, C., {et~al.} 2015, \apss, 357, 75,
  \dodoi{10.1007/s10509-015-2254-2}

\bibitem[{{Mattana} {et~al.}(2009){Mattana}, {Falanga}, {G{\"o}tz}, {Terrier},
  {Esposito}, {Pellizzoni}, {De Luca}, {Marandon}, {Goldwurm}, \&
  {Caraveo}}]{Mattana2009}
{Mattana}, F., {Falanga}, M., {G{\"o}tz}, D., {et~al.} 2009, \apj, 694, 12,
  \dodoi{10.1088/0004-637X/694/1/12}

\bibitem[{{Mattox} {et~al.}(1996){Mattox}, {Bertsch}, {Chiang}, {Dingus},
  {Digel}, {Esposito}, {Fierro}, {Hartman}, {Hunter}, {Kanbach}, {Kniffen},
  {Lin}, {Macomb}, {Mayer-Hasselwander}, {Michelson}, {von Montigny},
  {Mukherjee}, {Nolan}, {Ramanamurthy}, {Schneid}, {Sreekumar}, {Thompson}, \&
  {Willis}}]{Mattox1996}
{Mattox}, J.~R., {Bertsch}, D.~L., {Chiang}, J., {et~al.} 1996, \apj, 461, 396,
  \dodoi{10.1086/177068}

\bibitem[{{McCutcheon}(2009)}]{McCutcheon2009}
{McCutcheon}, M. 2009, arXiv e-prints, arXiv:0907.4974,
  \dodoi{10.48550/arXiv.0907.4974}

\bibitem[{{McEnery} {et~al.}(2019){McEnery}, {van der Horst}, {Dominguez},
  {Moiseev}, {Marcowith}, {Harding}, {Lien}, {Giuliani}, {Inglis}, {Ansoldi},
  {Stamerra}, {Manousakis}, {Strong}, {Bambi}, {Patricelli}, {Baring},
  {Barrio}, {Bastieri}, {Fields}, {Beacom}, {Beckmann}, \&
  {Bednarek}}]{McEnery2019}
{McEnery}, J., {van der Horst}, A., {Dominguez}, A., {et~al.} 2019, in Bulletin
  of the American Astronomical Society, Vol.~51, 245.
\newblock \doarXiv{1907.07558}

\bibitem[{{Meegan} {et~al.}(2009){Meegan}, {Lichti}, {Bhat}, {Bissaldi},
  {Briggs}, {Connaughton}, {Diehl}, {Fishman}, {Greiner}, {Hoover}, {van der
  Horst}, {von Kienlin}, {Kippen}, {Kouveliotou}, {McBreen}, {Paciesas},
  {Preece}, {Steinle}, {Wallace}, {Wilson}, \& {Wilson-Hodge}}]{GBM2009ApJ}
{Meegan}, C., {Lichti}, G., {Bhat}, P.~N., {et~al.} 2009, \apj, 702, 791,
  \dodoi{10.1088/0004-637X/702/1/791}

\bibitem[{{Michel}(1969)}]{michel69}
{Michel}, F.~C. 1969, \apj, 158, 727, \dodoi{10.1086/150233}

\bibitem[{{Mignani} {et~al.}(2012){Mignani}, {De Luca}, {Hummel}, {Zajczyk},
  {Rudak}, {Kanbach}, \& {S{\l}owikowska}}]{mignani2012}
{Mignani}, R.~P., {De Luca}, A., {Hummel}, W., {et~al.} 2012, \aap, 544, A100,
  \dodoi{10.1051/0004-6361/201219177}

\bibitem[{{Minter} {et~al.}(2008){Minter}, {Camilo}, {Ransom}, {Halpern}, \&
  {Zimmerman}}]{Minter2008}
{Minter}, A.~H., {Camilo}, F., {Ransom}, S.~M., {Halpern}, J.~P., \&
  {Zimmerman}, N. 2008, \apj, 676, 1189, \dodoi{10.1086/529005}

\bibitem[{{Mizuno} {et~al.}(2017){Mizuno}, {Tanaka}, {Takahashi}, {Katsuta},
  {Hayashi}, \& {Yamazaki}}]{Mizuno2017}
{Mizuno}, T., {Tanaka}, N., {Takahashi}, H., {et~al.} 2017, \apj, 841, 104,
  \dodoi{10.3847/1538-4357/aa7201}

\bibitem[{{Morris} {et~al.}(2002){Morris}, {Hobbs}, {Lyne}, {Stairs}, {Camilo},
  {Manchester}, {Possenti}, {Bell}, {Kaspi}, {Amico}, {McKay}, {Crawford}, \&
  {Kramer}}]{Morris2002}
{Morris}, D.~J., {Hobbs}, G., {Lyne}, A.~G., {et~al.} 2002, \mnras, 335, 275,
  \dodoi{10.1046/j.1365-8711.2002.05551.x}

\bibitem[{{Morsi} \& {Reich}(1987)}]{Morsi1987}
{Morsi}, H.~W., \& {Reich}, W. 1987, \aaps, 71, 189

\bibitem[{{Mott} \& {Freire}(2003)}]{Mott2003}
{Mott}, A.~J., \& {Freire}, P.~C. 2003, in American Astronomical Society
  Meeting Abstracts, Vol. 203, American Astronomical Society Meeting Abstracts,
  53.07

\bibitem[{{Murray} {et~al.}(2002){Murray}, {Slane}, {Seward}, {Ransom}, \&
  {Gaensler}}]{Murray2002}
{Murray}, S.~S., {Slane}, P.~O., {Seward}, F.~D., {Ransom}, S.~M., \&
  {Gaensler}, B.~M. 2002, \apj, 568, 226, \dodoi{10.1086/338766}

\bibitem[{{Muslimov} \& {Harding}(2004)}]{Muslimov2004}
{Muslimov}, A.~G., \& {Harding}, A.~K. 2004, \apj, 606, 1143,
  \dodoi{10.1086/383079}

\bibitem[{{Ng} {et~al.}(2017){Ng}, {Bandiera}, {Hunstead}, \&
  {Johnston}}]{ng2017discovery}
{Ng}, C.~Y., {Bandiera}, R., {Hunstead}, R.~W., \& {Johnston}, S. 2017, \apj,
  842, 100, \dodoi{10.3847/1538-4357/aa762e}

\bibitem[{{O'Brien} {et~al.}(2008){O'Brien}, {Johnston}, {Kramer}, {Lyne},
  {Bailes}, {Possenti}, {Burgay}, {Lorimer}, {McLaughlin}, {Hobbs}, {Parent},
  \& {Guillemot}}]{o2008psr}
{O'Brien}, J.~T., {Johnston}, S., {Kramer}, M., {et~al.} 2008, \mnras, 388, L1,
  \dodoi{10.1111/j.1745-3933.2008.00481.x}

\bibitem[{{Olmi}(2024)}]{Olmi2024}
{Olmi}, B. 2024, in High Energy Phenomena in Relativistic Outflows VIII, 16

\bibitem[{{Olmi} \& {Bucciantini}(2023{\natexlab{a}})}]{Olmi2023}
{Olmi}, B., \& {Bucciantini}, N. 2023{\natexlab{a}}, \pasa, 40, e042,
  \dodoi{10.1017/pasa.2023.41}

\bibitem[{{Olmi} \& {Bucciantini}(2023{\natexlab{b}})}]{Olmi_Bucciantini:2023}
---. 2023{\natexlab{b}}, \pasa, 40, e007, \dodoi{10.1017/pasa.2023.5}

\bibitem[{{Olmi} \& {Torres}(2020)}]{Olmi2020}
{Olmi}, B., \& {Torres}, D.~F. 2020, \mnras, 494, 4357,
  \dodoi{10.1093/mnras/staa1052}

\bibitem[{{Orellana} {et~al.}(2007){Orellana}, {Romero}, {Pellizza}, \&
  {Vidrih}}]{Orellana2007}
{Orellana}, M., {Romero}, G.~E., {Pellizza}, L.~J., \& {Vidrih}, S. 2007, \aap,
  465, 703, \dodoi{10.1051/0004-6361:20066238}

\bibitem[{{Pan} {et~al.}(2021){Pan}, {Qian}, {Ma}, {Liu}, {Wang}, {Luo}, {Yan},
  {Ransom}, {Lorimer}, {Li}, \& {Jiang}}]{Pan2021}
{Pan}, Z., {Qian}, L., {Ma}, X., {et~al.} 2021, \apjl, 915, L28,
  \dodoi{10.3847/2041-8213/ac0bbd}

\bibitem[{{Papitto} \& {Torres}(2015)}]{Papitto2015}
{Papitto}, A., \& {Torres}, D.~F. 2015, \apj, 807, 33,
  \dodoi{10.1088/0004-637X/807/1/33}

\bibitem[{{Papitto} {et~al.}(2014){Papitto}, {Torres}, \& {Li}}]{Papitto2014}
{Papitto}, A., {Torres}, D.~F., \& {Li}, J. 2014, \mnras, 438, 2105,
  \dodoi{10.1093/mnras/stt2336}

\bibitem[{{Papitto} {et~al.}(2012){Papitto}, {Torres}, \&
  {Rea}}]{Papitto2012LS}
{Papitto}, A., {Torres}, D.~F., \& {Rea}, N. 2012, \apj, 756, 188,
  \dodoi{10.1088/0004-637X/756/2/188}

\bibitem[{{Papitto} {et~al.}(2013){Papitto}, {Ferrigno}, {Bozzo}, {Rea},
  {Pavan}, {Burderi}, {Burgay}, {Campana}, {di Salvo}, {Falanga},
  {Filipovi{\'c}}, {Freire}, {Hessels}, {Possenti}, {Ransom}, {Riggio},
  {Romano}, {Sarkissian}, {Stairs}, {Stella}, {Torres}, {Wieringa}, \&
  {Wong}}]{Papitto2013}
{Papitto}, A., {Ferrigno}, C., {Bozzo}, E., {et~al.} 2013, \nat, 501, 517,
  \dodoi{10.1038/nature12470}

\bibitem[{{Papitto} {et~al.}(2019){Papitto}, {Ambrosino}, {Stella}, {Torres},
  {Coti Zelati}, {Ghedina}, {Meddi}, {Sanna}, {Casella}, {Dallilar},
  {Eikenberry}, {Israel}, {Onori}, {Piranomonte}, {Bozzo}, {Burderi},
  {Campana}, {de Martino}, {Di Salvo}, {Ferrigno}, {Rea}, {Riggio}, {Serrano},
  {Veledina}, \& {Zampieri}}]{Papitto2019}
{Papitto}, A., {Ambrosino}, F., {Stella}, L., {et~al.} 2019, \apj, 882, 104,
  \dodoi{10.3847/1538-4357/ab2fdf}

\bibitem[{{Paredes} {et~al.}(2009){Paredes}, {Mart{\'\i}}, {Ishwara-Chandra},
  {S{\'a}nchez-Sutil}, {Mu{\~n}oz-Arjonilla}, {Mold{\'o}n}, {Peracaula},
  {Luque-Escamilla}, {Zabalza}, {Bosch-Ramon}, {Bordas}, {Romero}, \&
  {Rib{\'o}}}]{Paredes2009}
{Paredes}, J.~M., {Mart{\'\i}}, J., {Ishwara-Chandra}, C.~H., {et~al.} 2009,
  \aap, 507, 241, \dodoi{10.1051/0004-6361/200912448}

\bibitem[{{Park} {et~al.}(2007){Park}, {Hughes}, {Slane}, {Burrows},
  {Gaensler}, \& {Ghavamian}}]{park2007}
{Park}, S., {Hughes}, J.~P., {Slane}, P.~O., {et~al.} 2007, \apjl, 670, L121,
  \dodoi{10.1086/524406}

\bibitem[{{Pavlovic} {et~al.}(2014){Pavlovic}, {Dobardzic}, {Vukotic}, \&
  {Urosevic}}]{Pavlovic2014}
{Pavlovic}, M.~Z., {Dobardzic}, A., {Vukotic}, B., \& {Urosevic}, D. 2014,
  Serbian Astronomical Journal, 189, 25, \dodoi{10.2298/SAJ1489025P}

\bibitem[{{P{\'e}tri} \& {Mitra}(2021)}]{Petri2021}
{P{\'e}tri}, J., \& {Mitra}, D. 2021, \aap, 654, A106,
  \dodoi{10.1051/0004-6361/202141272}

\bibitem[{Philippov \& Kramer(2022)}]{Philippov2022psr}
Philippov, A., \& Kramer, M. 2022, Annual Review of Astronomy and Astrophysics,
  60, 495

\bibitem[{{Pineault} \& {Joncas}(2000)}]{pineault2000}
{Pineault}, S., \& {Joncas}, G. 2000, \aj, 120, 3218, \dodoi{10.1086/316863}

\bibitem[{{Pletsch} {et~al.}(2012){Pletsch}, {Guillemot}, {Allen}, {Kramer},
  {Aulbert}, {Fehrmann}, {Baring}, {Camilo}, {Caraveo}, {Grove}, {Kerr},
  {Marelli}, {Ransom}, {Ray}, \& {Saz Parkinson}}]{pletsch2012psr}
{Pletsch}, H.~J., {Guillemot}, L., {Allen}, B., {et~al.} 2012, \apjl, 755, L20,
  \dodoi{10.1088/2041-8205/755/1/L20}

\bibitem[{{Pons} {et~al.}(2012){Pons}, {Vigan{\`o}}, \& {Geppert}}]{pons12}
{Pons}, J.~A., {Vigan{\`o}}, D., \& {Geppert}, U. 2012, \aap, 547, A9,
  \dodoi{10.1051/0004-6361/201220091}

\bibitem[{{Pooley} {et~al.}(2003){Pooley}, {Lewin}, {Anderson}, {Baumgardt},
  {Filippenko}, {Gaensler}, {Homer}, {Hut}, {Kaspi}, {Makino}, {Margon},
  {McMillan}, {Portegies Zwart}, {van der Klis}, \& {Verbunt}}]{Pooley2003}
{Pooley}, D., {Lewin}, W. H.~G., {Anderson}, S.~F., {et~al.} 2003, \apjl, 591,
  L131, \dodoi{10.1086/377074}

\bibitem[{{Pope} {et~al.}(2024){Pope}, {Mori}, {Abdelmaguid}, {Gelfand},
  {Reynolds}, {Safi-Harb}, {Hailey}, {An}, {Bangale}, {Batista}, {Benbow},
  {Buckley}, {Capasso}, {Christiansen}, {Chromey}, {Falcone}, {Feng}, {Finley},
  {Foote}, {Gallagher}, {Hanlon}, {Hanna}, {Hervet}, {Holder}, {Humensky},
  {Jin}, {Kaaret}, {Kertzman}, {Kieda}, {Kleiner}, {Korzoun}, {Krennrich},
  {Kumar}, {Lang}, {Maier}, {McGrath}, {Mooney}, {Moriarty}, {Mukherjee},
  {O'Brien}, {Ong}, {Park}, {Patel}, {Pfrang}, {Pohl}, {Pueschel}, {Quinn},
  {Ragan}, {Reynolds}, {Roache}, {Sadeh}, {Saha}, {Sembroski}, {Tak}, {Tucci},
  {Weinstein}, {Williams}, {Woo}, \& {VERITAS Collaboration}}]{pope2024}
{Pope}, I., {Mori}, K., {Abdelmaguid}, M., {et~al.} 2024, \apj, 960, 75,
  \dodoi{10.3847/1538-4357/ad0120}

\bibitem[{{Porter} {et~al.}(2022){Porter}, {J{\'o}hannesson}, \&
  {Moskalenko}}]{porter2022galprop}
{Porter}, T.~A., {J{\'o}hannesson}, G., \& {Moskalenko}, I.~V. 2022, \apjs,
  262, 30, \dodoi{10.3847/1538-4365/ac80f6}

\bibitem[{Porter {et~al.}(2006)Porter, Moskalenko, \& Strong}]{porter06}
Porter, T.~A., Moskalenko, I.~V., \& Strong, A.~W. 2006, \apj, 648, L29,
  \dodoi{10.1086/507770}

\bibitem[{Press {et~al.}(1992)Press, Teukolsky, Vetterling, \&
  Flannery}]{Press1992}
Press, W.~H., Teukolsky, S.~A., Vetterling, W.~T., \& Flannery, B.~P. 1992, The
  art of scientific computing, 1

\bibitem[{{Priestley} {et~al.}(2022){Priestley}, {Chawner}, {Barlow}, {De
  Looze}, {Gomez}, \& {Matsuura}}]{Priestley2022}
{Priestley}, F.~D., {Chawner}, H., {Barlow}, M.~J., {et~al.} 2022, \mnras, 516,
  2314, \dodoi{10.1093/mnras/stac2408}

\bibitem[{{Prinz} \& {Becker}(2015)}]{Prinz2015}
{Prinz}, T., \& {Becker}, W. 2015, arXiv e-prints, arXiv:1511.07713,
  \dodoi{10.48550/arXiv.1511.07713}

\bibitem[{{Radhakrishnan} {et~al.}(1972){Radhakrishnan}, {Goss}, {Murray}, \&
  {Brooks}}]{Radhakrishnan1972}
{Radhakrishnan}, V., {Goss}, W.~M., {Murray}, J.~D., \& {Brooks}, J.~W. 1972,
  \apjs, 24, 49, \dodoi{10.1086/190249}

\bibitem[{{Ranasinghe} \& {Leahy}(2022)}]{Ranasinghe2022}
{Ranasinghe}, S., \& {Leahy}, D. 2022, \apj, 940, 63,
  \dodoi{10.3847/1538-4357/ac940a}

\bibitem[{{Rasul} {et~al.}(2019){Rasul}, {Chadwick}, {Graham}, \&
  {Brown}}]{Rasul2019}
{Rasul}, K., {Chadwick}, P.~M., {Graham}, J.~A., \& {Brown}, A.~M. 2019,
  \mnras, 485, 2970, \dodoi{10.1093/mnras/stz559}

\bibitem[{{Ray} {et~al.}(2011){Ray}, {Kerr}, {Parent}, {Abdo}, {Guillemot},
  {Ransom}, {Rea}, {Wolff}, {Makeev}, {Roberts}, {Camilo}, {Dormody}, {Freire},
  {Grove}, {Gwon}, {Harding}, {Johnston}, {Keith}, {Kramer}, {Michelson},
  {Romani}, {Saz Parkinson}, {Thompson}, {Weltevrede}, {Wood}, \&
  {Ziegler}}]{Ray2011}
{Ray}, P.~S., {Kerr}, M., {Parent}, D., {et~al.} 2011, \apjs, 194, 17,
  \dodoi{10.1088/0067-0049/194/2/17}

\bibitem[{{Remy} {et~al.}(2020){Remy}, {Gallant}, \&
  {Renaud}}]{remy2020prospects}
{Remy}, Q., {Gallant}, Y.~A., \& {Renaud}, M. 2020, Astroparticle Physics, 122,
  102462, \dodoi{10.1016/j.astropartphys.2020.102462}

\bibitem[{{Renaud} {et~al.}(2010){Renaud}, {Marandon}, {Gotthelf}, {Rodriguez},
  {Terrier}, {Mattana}, {Lebrun}, {Tomsick}, \& {Manchester}}]{renaud2010}
{Renaud}, M., {Marandon}, V., {Gotthelf}, E.~V., {et~al.} 2010, \apj, 716, 663,
  \dodoi{10.1088/0004-637X/716/1/663}

\bibitem[{Renault-Tinacci(2014)}]{Nicolas2014}
Renault-Tinacci, N. 2014, PhD thesis, Universit{\'e} Paris 7-Denis Diderot

\bibitem[{{Roberts} {et~al.}(2007){Roberts}, {Gotthelf}, {Halpern}, {Brogan},
  \& {Ransom}}]{roberts2007}
{Roberts}, M. S.~E., {Gotthelf}, E.~V., {Halpern}, J.~P., {Brogan}, C.~L., \&
  {Ransom}, S.~M. 2007, in WE-Heraeus Seminar on Neutron Stars and Pulsars 40
  years after the Discovery, ed. W.~{Becker} \& H.~H. {Huang}, 24,
  \dodoi{10.48550/arXiv.astro-ph/0612631}

\bibitem[{{Roberts} {et~al.}(2002){Roberts}, {Hessels}, {Ransom}, {Kaspi},
  {Freire}, {Crawford}, \& {Lorimer}}]{Roberts2002}
{Roberts}, M. S.~E., {Hessels}, J. W.~T., {Ransom}, S.~M., {et~al.} 2002,
  \apjl, 577, L19, \dodoi{10.1086/344082}

\bibitem[{{Roberts} {et~al.}(2003){Roberts}, {Tam}, {Kaspi}, {Lyutikov},
  {Vasisht}, {Pivovaroff}, {Gotthelf}, \& {Kawai}}]{Roberts2003}
{Roberts}, M. S.~E., {Tam}, C.~R., {Kaspi}, V.~M., {et~al.} 2003, \apj, 588,
  992, \dodoi{10.1086/374266}

\bibitem[{{Roberts} {et~al.}(2008){Roberts}, {Brogan}, {Ransom}, {Lyutikov},
  {de O{\~n}a Wilhelmi}, {Djannati-Atai}, {Terrier}, {Dougherty}, {Grundstrom},
  {Hessels}, {Johnston}, {McSwain}, {Ray}, {Wood}, {Pooley}, \&
  {Weinstein}}]{roberts2008}
{Roberts}, M. S.~E., {Brogan}, C., {Ransom}, S., {et~al.} 2008, in American
  Institute of Physics Conference Series, Vol. 1085, American Institute of
  Physics Conference Series, ed. F.~A. {Aharonian}, W.~{Hofmann}, \&
  F.~{Rieger}, 328--331, \dodoi{10.1063/1.3076673}

\bibitem[{{Rosenberg} {et~al.}(1975){Rosenberg}, {Eyles}, {Skinner}, \&
  {Willmore}}]{Rosenberg1975}
{Rosenberg}, F.~D., {Eyles}, C.~J., {Skinner}, G.~K., \& {Willmore}, A.~P.
  1975, \nat, 256, 628, \dodoi{10.1038/256628a0}

\bibitem[{{Ruderman} \& {Sutherland}(1975)}]{Ruderman1975}
{Ruderman}, M.~A., \& {Sutherland}, P.~G. 1975, \apj, 196, 51,
  \dodoi{10.1086/153393}

\bibitem[{{Salter} {et~al.}(1989){Salter}, {Reynolds}, {Hogg}, {Payne}, \&
  {Rhodes}}]{Salter1989}
{Salter}, C.~J., {Reynolds}, S.~P., {Hogg}, D.~E., {Payne}, J.~M., \& {Rhodes},
  P.~J. 1989, \apj, 338, 171, \dodoi{10.1086/167191}

\bibitem[{{Sartore} {et~al.}(2015){Sartore}, {Jourdain}, \&
  {Roques}}]{Sartore2015}
{Sartore}, N., {Jourdain}, E., \& {Roques}, J.~P. 2015, \apj, 806, 193,
  \dodoi{10.1088/0004-637X/806/2/193}

\bibitem[{{Saz Parkinson} {et~al.}(2010){Saz Parkinson}, {Dormody}, {Ziegler},
  {Ray}, {Abdo}, {Ballet}, {Baring}, {Belfiore}, {Burnett}, {Caliandro},
  {Camilo}, {Caraveo}, {de Luca}, {Ferrara}, {Freire}, {Grove}, {Gwon},
  {Harding}, {Johnson}, {Johnson}, {Johnston}, {Keith}, {Kerr},
  {Kn{\"o}dlseder}, {Makeev}, {Marelli}, {Michelson}, {Parent}, {Ransom},
  {Reimer}, {Romani}, {Smith}, {Thompson}, {Watters}, {Weltevrede}, {Wolff}, \&
  {Wood}}]{parkinson2010eight}
{Saz Parkinson}, P.~M., {Dormody}, M., {Ziegler}, M., {et~al.} 2010, \apj, 725,
  571, \dodoi{10.1088/0004-637X/725/1/571}

\bibitem[{{Sezer} {et~al.}(2011){Sezer}, {G{\"o}k}, {Hudaverdi}, {Kimura}, \&
  {Ercan}}]{Sezer2011}
{Sezer}, A., {G{\"o}k}, F., {Hudaverdi}, M., {Kimura}, M., \& {Ercan}, E.~N.
  2011, \mnras, 415, 301, \dodoi{10.1111/j.1365-2966.2011.18710.x}

\bibitem[{{Slane}(2017)}]{Slane:2017}
{Slane}, P. 2017, in Handbook of Supernovae, ed. A.~W. {Alsabti} \&
  P.~{Murdin}, 2159, \dodoi{10.1007/978-3-319-21846-5_95}

\bibitem[{{Slane} {et~al.}(2010){Slane}, {Castro}, {Funk}, {Uchiyama},
  {Lemiere}, {Gelfand}, \& {Lemoine-Goumard}}]{slane2010fermi}
{Slane}, P., {Castro}, D., {Funk}, S., {et~al.} 2010, \apj, 720, 266,
  \dodoi{10.1088/0004-637X/720/1/266}

\bibitem[{{Slane} {et~al.}(2008){Slane}, {Helfand}, {Reynolds}, {Gaensler},
  {Lemiere}, \& {Wang}}]{Slane2008}
{Slane}, P., {Helfand}, D.~J., {Reynolds}, S.~P., {et~al.} 2008, \apjl, 676,
  L33, \dodoi{10.1086/587031}

\bibitem[{{Slane} {et~al.}(2012){Slane}, {Hughes}, {Temim}, {Rousseau},
  {Castro}, {Foight}, {Gaensler}, {Funk}, {Lemoine-Goumard}, {Gelfand},
  {Moffett}, {Dodson}, \& {Bernstein}}]{slane2012broadband}
{Slane}, P., {Hughes}, J.~P., {Temim}, T., {et~al.} 2012, \apj, 749, 131,
  \dodoi{10.1088/0004-637X/749/2/131}

\bibitem[{{Smith} {et~al.}(2008){Smith}, {Guillemot}, {Camilo}, {Cognard},
  {Dumora}, {Espinoza}, {Freire}, {Gotthelf}, {Harding}, {Hobbs}, {Johnston},
  {Kaspi}, {Kramer}, {Livingstone}, {Lyne}, {Manchester}, {Marshall},
  {McLaughlin}, {Noutsos}, {Ransom}, {Roberts}, {Romani}, {Stappers},
  {Theureau}, {Thompson}, {Thorsett}, {Wang}, \& {Weltevrede}}]{Smith2008}
{Smith}, D.~A., {Guillemot}, L., {Camilo}, F., {et~al.} 2008, \aap, 492, 923,
  \dodoi{10.1051/0004-6361:200810285}

\bibitem[{{Smith} {et~al.}(2019){Smith}, {Bruel}, {Cognard}, {Cameron},
  {Camilo}, {Dai}, {Guillemot}, {Johnson}, {Johnston}, {Keith}, {Kerr},
  {Kramer}, {Lyne}, {Manchester}, {Shannon}, {Sobey}, {Stappers}, \&
  {Weltevrede}}]{Smith2019}
{Smith}, D.~A., {Bruel}, P., {Cognard}, I., {et~al.} 2019, \apj, 871, 78,
  \dodoi{10.3847/1538-4357/aaf57d}

\bibitem[{{Smith} {et~al.}(2023){Smith}, {Bruel}, {Clark}, {Guillemot}, {Kerr},
  {Ray}, {Abdollahi}, {Ajello}, {Baldini}, {Ballet}, {Baring}, \&
  {Bassa}}]{Smith2023}
{Smith}, D.~A., {Bruel}, P., {Clark}, C.~J., {et~al.} 2023, arXiv e-prints,
  arXiv:2307.11132, \dodoi{10.48550/arXiv.2307.11132}

\bibitem[{{Sollima} \& {Baumgardt}(2017)}]{Sollima2017}
{Sollima}, A., \& {Baumgardt}, H. 2017, \mnras, 471, 3668,
  \dodoi{10.1093/mnras/stx1856}

\bibitem[{{Stappers} {et~al.}(2014){Stappers}, {Archibald}, {Hessels}, {Bassa},
  {Bogdanov}, {Janssen}, {Kaspi}, {Lyne}, {Patruno}, {Tendulkar}, {Hill}, \&
  {Glanzman}}]{Stappers2014}
{Stappers}, B.~W., {Archibald}, A.~M., {Hessels}, J.~W.~T., {et~al.} 2014,
  \apj, 790, 39, \dodoi{10.1088/0004-637X/790/1/39}

\bibitem[{{Steele} {et~al.}(1998){Steele}, {Negueruela}, {Coe}, \&
  {Roche}}]{Steele1998}
{Steele}, I.~A., {Negueruela}, I., {Coe}, M.~J., \& {Roche}, P. 1998, \mnras,
  297, L5, \dodoi{10.1046/j.1365-8711.1998.01593.x}

\bibitem[{{Straal} {et~al.}(2023){Straal}, {Gelfand}, \& {Eagle}}]{Straal2023}
{Straal}, S.~M., {Gelfand}, J.~D., \& {Eagle}, J.~L. 2023, \apj, 942, 103,
  \dodoi{10.3847/1538-4357/aca1a9}

\bibitem[{{Su} {et~al.}(2017){Su}, {Zhang}, {Zhu}, \& {Wu}}]{Su2017}
{Su}, H.-Q., {Zhang}, M.-F., {Zhu}, H., \& {Wu}, D. 2017, Research in Astronomy
  and Astrophysics, 17, 109, \dodoi{10.1088/1674-4527/17/10/109}

\bibitem[{{Sudoh} {et~al.}(2021){Sudoh}, {Linden}, \&
  {Hooper}}]{sudoh2021highest}
{Sudoh}, T., {Linden}, T., \& {Hooper}, D. 2021, \jcap, 2021, 010,
  \dodoi{10.1088/1475-7516/2021/08/010}

\bibitem[{{Tam} {et~al.}(2002){Tam}, {Roberts}, \& {Kaspi}}]{Tam2002}
{Tam}, C., {Roberts}, M. S.~E., \& {Kaspi}, V.~M. 2002, \apj, 572, 202,
  \dodoi{10.1086/340229}

\bibitem[{{Tanaka} \& {Takahara}(2010)}]{Tanaka2010}
{Tanaka}, S.~J., \& {Takahara}, F. 2010, \apj, 715, 1248,
  \dodoi{10.1088/0004-637X/715/2/1248}

\bibitem[{{Tanaka} \& {Takahara}(2011)}]{tanaka2011}
---. 2011, \apj, 741, 40, \dodoi{10.1088/0004-637X/741/1/40}

\bibitem[{{Tanaka} \& {Takahara}(2013)}]{Tanaka2013}
---. 2013, \mnras, 429, 2945, \dodoi{10.1093/mnras/sts528}

\bibitem[{{Tavani} {et~al.}(2009){Tavani}, {Bulgarelli}, {Piano}, {Sabatini},
  {Striani}, {Evangelista}, {Trois}, {Pooley}, {Trushkin}, {Nizhelskij},
  {McCollough}, {Koljonen}, {Pucella}, {Giuliani}, {Chen}, {Costa},
  {Vittorini}, {Trifoglio}, {Gianotti}, {Argan}, {Barbiellini}, {Caraveo},
  {Cattaneo}, {Cocco}, {Contessi}, {D'Ammando}, {Del Monte}, {de Paris}, {Di
  Cocco}, {di Persio}, {Donnarumma}, {Feroci}, {Ferrari}, {Fuschino}, {Galli},
  {Labanti}, {Lapshov}, {Lazzarotto}, {Lipari}, {Longo}, {Mattaini},
  {Marisaldi}, {Mastropietro}, {Mauri}, {Mereghetti}, {Morelli}, {Morselli},
  {Pacciani}, {Pellizzoni}, {Perotti}, {Picozza}, {Pilia}, {Prest},
  {Rapisarda}, {Rappoldi}, {Rossi}, {Rubini}, {Scalise}, {Soffitta},
  {Vallazza}, {Vercellone}, {Zambra}, {Zanello}, {Pittori}, {Verrecchia},
  {Giommi}, {Colafrancesco}, {Santolamazza}, {Antonelli}, \&
  {Salotti}}]{Tavani2009}
{Tavani}, M., {Bulgarelli}, A., {Piano}, G., {et~al.} 2009, \nat, 462, 620,
  \dodoi{10.1038/nature08578}

\bibitem[{{Temim} {et~al.}(2015){Temim}, {Slane}, {Kolb}, {Blondin}, {Hughes},
  \& {Bucciantini}}]{Temim2015}
{Temim}, T., {Slane}, P., {Kolb}, C., {et~al.} 2015, \apj, 808, 100,
  \dodoi{10.1088/0004-637X/808/1/100}

\bibitem[{{Temim} {et~al.}(2010){Temim}, {Slane}, {Reynolds}, {Raymond}, \&
  {Borkowski}}]{temim2010}
{Temim}, T., {Slane}, P., {Reynolds}, S.~P., {Raymond}, J.~C., \& {Borkowski},
  K.~J. 2010, \apj, 710, 309, \dodoi{10.1088/0004-637X/710/1/309}

\bibitem[{{Tibet AS{\ensuremath{\gamma}} Collaboration} {et~al.}(2021){Tibet
  AS{\ensuremath{\gamma}} Collaboration}, {Amenomori}, {Bao}, {Bi}, {Chen},
  {Chen}, {Chen}, {Chen}, {Chen}, {Cirennima}, {Danzengluobu}, {Fang}, {Fang},
  {Feng}, {Feng}, {Feng}, {Gao}, {Gou}, {Guo}, {Guo}, {He}, {He}, {Hibino},
  {Hotta}, {Hu}, {Hu}, {Huang}, {Jia}, {Jiang}, {Jin}, {Kasahara}, {Katayose},
  {Kato}, {Kato}, {Kawata}, {Kihara}, {Ko}, {Kozai}, {Labaciren}, {Li}, {Li},
  {Li}, {Lin}, {Liu}, {Liu}, {Liu}, {Liu}, {Liu}, {Lou}, {Lu}, {Meng},
  {Munakata}, {Nakada}, {Nakamura}, {Nanjo}, {Nishizawa}, {Ohnishi}, {Ohura},
  {Ozawa}, {Qian}, {Qu}, {Saito}, {Sakata}, {Sako}, {Shao}, {Shibata},
  {Shiomi}, {Sugimoto}, {Takano}, {Takita}, {Tan}, {Tateyama}, {Torii},
  {Tsuchiya}, {Udo}, {Wang}, {Wu}, {Xue}, {Yamamoto}, {Yang}, {Yokoe}, {Yuan},
  {Zhai}, {Zhang}, {Zhang}, {Zhang}, {Zhang}, {Zhang}, {Zhang}, {Zhang},
  {Zhao}, \& {Zhaxisangzhu}}]{tibet21}
{Tibet AS{\ensuremath{\gamma}} Collaboration}, {Amenomori}, M., {Bao}, Y.~W.,
  {et~al.} 2021, Nature Astronomy, 5, 460, \dodoi{10.1038/s41550-020-01294-9}

\bibitem[{{Tibolla}(2011)}]{Tibolla2011}
{Tibolla}, O. 2011, in International Cosmic Ray Conference, Vol.~6,
  International Cosmic Ray Conference, 202, \dodoi{10.7529/ICRC2011/V06/1233}

\bibitem[{Tibolla {et~al.}(2022)Tibolla, Kaufmann, \&
  Chadwick}]{tibolla2022pulsar}
Tibolla, O., Kaufmann, S., \& Chadwick, P. 2022, Multidisciplinary Scientific
  Journal, 5, 318, \dodoi{10.3390/j5030022}

\bibitem[{{Torii} {et~al.}(1997){Torii}, {Tsunemi}, {Dotani}, \&
  {Mitsuda}}]{Torii1997}
{Torii}, K., {Tsunemi}, H., {Dotani}, T., \& {Mitsuda}, K. 1997, \apjl, 489,
  L145, \dodoi{10.1086/316798}

\bibitem[{{Torii} {et~al.}(1999){Torii}, {Tsunemi}, {Dotani}, {Mitsuda},
  {Kawai}, {Kinugasa}, {Saito}, \& {Shibata}}]{Torii1999}
{Torii}, K., {Tsunemi}, H., {Dotani}, T., {et~al.} 1999, \apjl, 523, L69,
  \dodoi{10.1086/312251}

\bibitem[{{Torres}(2018)}]{Torres2018p}
{Torres}, D.~F. 2018, Nature Astronomy, 2, 247,
  \dodoi{10.1038/s41550-018-0384-5}

\bibitem[{{Torres} {et~al.}(2014){Torres}, {Cillis}, {Mart{\'\i}n}, \& {de
  O{\~n}a Wilhelmi}}]{Torres2014}
{Torres}, D.~F., {Cillis}, A., {Mart{\'\i}n}, J., \& {de O{\~n}a Wilhelmi}, E.
  2014, Journal of High Energy Astrophysics, 1, 31,
  \dodoi{10.1016/j.jheap.2014.02.001}

\bibitem[{{Torres} {et~al.}(2017){Torres}, {Ji}, {Li}, {Papitto}, {Rea}, {de
  O{\~n}a Wilhelmi}, \& {Zhang}}]{Torres2017}
{Torres}, D.~F., {Ji}, L., {Li}, J., {et~al.} 2017, \apj, 836, 68,
  \dodoi{10.3847/1538-4357/836/1/68}

\bibitem[{{Torres} \& {Lin}(2018)}]{torres2018}
{Torres}, D.~F., \& {Lin}, T. 2018, \apjl, 864, L2,
  \dodoi{10.3847/2041-8213/aad6e1}

\bibitem[{{Torres} {et~al.}(2013){Torres}, {Mart{\'\i}n}, {de O{\~n}a
  Wilhelmi}, \& {Cillis}}]{Torres2013}
{Torres}, D.~F., {Mart{\'\i}n}, J., {de O{\~n}a Wilhelmi}, E., \& {Cillis}, A.
  2013, \mnras, 436, 3112, \dodoi{10.1093/mnras/stt1793}

\bibitem[{{Torres} {et~al.}(2019){Torres}, {Vigan{\`o}}, {Coti Zelati}, \&
  {Li}}]{Torres2019p}
{Torres}, D.~F., {Vigan{\`o}}, D., {Coti Zelati}, F., \& {Li}, J. 2019, \mnras,
  489, 5494, \dodoi{10.1093/mnras/stz2403}

\bibitem[{{Townsley} {et~al.}(2011){Townsley}, {Broos}, {Chu}, {Gruendl},
  {Oey}, \& {Pittard}}]{townsley2011integrated}
{Townsley}, L.~K., {Broos}, P.~S., {Chu}, Y.-H., {et~al.} 2011, \apjs, 194, 16,
  \dodoi{10.1088/0067-0049/194/1/16}

\bibitem[{{Ueno} {et~al.}(2003){Ueno}, {Bamba}, {Koyama}, \&
  {Ebisawa}}]{ueno2003chandra}
{Ueno}, M., {Bamba}, A., {Koyama}, K., \& {Ebisawa}, K. 2003, \apj, 588, 338,
  \dodoi{10.1086/368355}

\bibitem[{{van den Eijnden} {et~al.}(2018){van den Eijnden}, {Degenaar},
  {Russell}, {Wijnands}, {Miller-Jones}, {Sivakoff}, \& {Hern{\'a}ndez
  Santisteban}}]{Eijnden2018}
{van den Eijnden}, J., {Degenaar}, N., {Russell}, T.~D., {et~al.} 2018, \nat,
  562, 233, \dodoi{10.1038/s41586-018-0524-1}

\bibitem[{{van den Eijnden} {et~al.}(2020){van den Eijnden}, {Degenaar},
  {Wijnands}, {Russell}, {Sivakoff}, {Miller-Jones}, {Rouco Escorial},
  {Her{\'a}dez Santisteban}, \& {Reynolds}}]{Eijnden2020}
{van den Eijnden}, J., {Degenaar}, N., {Wijnands}, R., {et~al.} 2020, The
  Astronomer's Telegram, 14193, 1

\bibitem[{{van der Swaluw} {et~al.}(2001){van der Swaluw}, {Achterberg},
  {Gallant}, \& {T{\'o}th}}]{Swaluw2001}
{van der Swaluw}, E., {Achterberg}, A., {Gallant}, Y.~A., \& {T{\'o}th}, G.
  2001, \aap, 380, 309, \dodoi{10.1051/0004-6361:20011437}

\bibitem[{{Veledina} {et~al.}(2019){Veledina}, {N{\"a}ttil{\"a}}, \&
  {Beloborodov}}]{Veledina2019}
{Veledina}, A., {N{\"a}ttil{\"a}}, J., \& {Beloborodov}, A.~M. 2019, \apj, 884,
  144, \dodoi{10.3847/1538-4357/ab44c6}

\bibitem[{{Venter} \& {de Jager}(2008)}]{Venter2008}
{Venter}, C., \& {de Jager}, O.~C. 2008, \apjl, 680, L125,
  \dodoi{10.1086/589996}

\bibitem[{{Venter} {et~al.}(2009){Venter}, {De Jager}, \&
  {Clapson}}]{Venter2009}
{Venter}, C., {De Jager}, O.~C., \& {Clapson}, A.~C. 2009, \apjl, 696, L52,
  \dodoi{10.1088/0004-637X/696/1/L52}

\bibitem[{{Verbunt} \& {Freire}(2014)}]{Verbunt_Freire2014}
{Verbunt}, F., \& {Freire}, P. C.~C. 2014, \aap, 561, A11,
  \dodoi{10.1051/0004-6361/201321177}

\bibitem[{{Vigan{\`o}} {et~al.}(2015{\natexlab{a}}){Vigan{\`o}}, {Torres},
  {Hirotani}, \& {Pessah}}]{Vigano2015p1}
{Vigan{\`o}}, D., {Torres}, D.~F., {Hirotani}, K., \& {Pessah}, M.~E.
  2015{\natexlab{a}}, \mnras, 447, 2631, \dodoi{10.1093/mnras/stu2564}

\bibitem[{{Vigan{\`o}} {et~al.}(2015{\natexlab{b}}){Vigan{\`o}}, {Torres},
  {Hirotani}, \& {Pessah}}]{Vigano2015p2}
---. 2015{\natexlab{b}}, \mnras, 447, 2649, \dodoi{10.1093/mnras/stu2565}

\bibitem[{{Vigan{\`o}} {et~al.}(2015{\natexlab{c}}){Vigan{\`o}}, {Torres},
  {Hirotani}, \& {Pessah}}]{Vigano2015}
---. 2015{\natexlab{c}}, \mnras, 447, 1164, \dodoi{10.1093/mnras/stu2456}

\bibitem[{{Vigan{\`o}} {et~al.}(2015{\natexlab{d}}){Vigan{\`o}}, {Torres}, \&
  {Mart{\'\i}n}}]{Vigano2015p3}
{Vigan{\`o}}, D., {Torres}, D.~F., \& {Mart{\'\i}n}, J. 2015{\natexlab{d}},
  \mnras, 453, 2599, \dodoi{10.1093/mnras/stv1582}

\bibitem[{{Vorster} {et~al.}(2013){Vorster}, {Tibolla}, {Ferreira}, \&
  {Kaufmann}}]{Vorster2013}
{Vorster}, M.~J., {Tibolla}, O., {Ferreira}, S.~E.~S., \& {Kaufmann}, S. 2013,
  \apj, 773, 139, \dodoi{10.1088/0004-637X/773/2/139}

\bibitem[{{Wang} {et~al.}(2022){Wang}, {Kong}, {Zhang}, {Doroshenko},
  {Santangelo}, {Ji}, {Yorgancioglu}, {Chen}, {Zhang}, {Qu}, {Ge}, {Li},
  {Chang}, {Tao}, {Peng}, \& {Shui}}]{Wang2022}
{Wang}, P.~J., {Kong}, L.~D., {Zhang}, S., {et~al.} 2022, \apj, 935, 125,
  \dodoi{10.3847/1538-4357/ac8230}

\bibitem[{Wasserstein {et~al.}(2019)Wasserstein, Schirm, \& Lazar}]{Ronald2019}
Wasserstein, R.~L., Schirm, A.~L., \& Lazar, N.~A. 2019, Moving to a world
  beyond “p< 0.05”,  Taylor \& Francis

\bibitem[{{Weiler} \& {Panagia}(1978)}]{Weiler1978}
{Weiler}, K.~W., \& {Panagia}, N. 1978, \aap, 70, 419

\bibitem[{{Weiler} \& {Seielstad}(1971)}]{Weiler1971}
{Weiler}, K.~W., \& {Seielstad}, G.~A. 1971, \apj, 163, 455,
  \dodoi{10.1086/150791}

\bibitem[{{Weng} {et~al.}(2022){Weng}, {Qian}, {Wang}, {Torres}, {Papitto},
  {Jiang}, {Xu}, {Li}, {Yan}, {Liu}, {Ge}, \& {Yuan}}]{Weng2022}
{Weng}, S.-S., {Qian}, L., {Wang}, B.-J., {et~al.} 2022, Nature Astronomy, 6,
  698, \dodoi{10.1038/s41550-022-01630-1}

\bibitem[{{Wilson}(1986)}]{wilson1986x}
{Wilson}, A.~S. 1986, \apj, 302, 718, \dodoi{10.1086/164033}

\bibitem[{{Woo} {et~al.}(2023){Woo}, {An}, {Gelfand}, {Hailey}, {Mori},
  {Mukherjee}, {Safi-Harb}, \& {Temim}}]{woo2023}
{Woo}, J., {An}, H., {Gelfand}, J.~D., {et~al.} 2023, \apj, 954, 9,
  \dodoi{10.3847/1538-4357/acdd5e}

\bibitem[{{Wood} {et~al.}(2017){Wood}, {Caputo}, {Charles}, {Di Mauro},
  {Magill}, {Perkins}, \& {Fermi-LAT Collaboration}}]{Wood2017}
{Wood}, M., {Caputo}, R., {Charles}, E., {et~al.} 2017, in International Cosmic
  Ray Conference, Vol. 301, 35th International Cosmic Ray Conference
  (ICRC2017), 824, \dodoi{10.22323/1.301.0824}

\bibitem[{{Wu} {et~al.}(2014){Wu}, {Hui}, {Kong}, {Tam}, {Cheng}, \&
  {Dogiel}}]{Wu2014}
{Wu}, E.~M.~H., {Hui}, C.~Y., {Kong}, A.~K.~H., {et~al.} 2014, \apjl, 788, L40,
  \dodoi{10.1088/2041-8205/788/2/L40}

\bibitem[{{Wu} {et~al.}(2018){Wu}, {Clark}, {Pletsch}, {Guillemot}, {Johnson},
  {Torne}, {Champion}, {Deneva}, {Ray}, {Salvetti}, {Kramer}, {Aulbert},
  {Beer}, {Bhattacharyya}, {Bock}, {Camilo}, {Cognard}, {Cu{\'e}llar},
  {Eggenstein}, {Fehrmann}, {Ferrara}, {Kerr}, {Machenschalk}, {Ransom},
  {Sanpa-Arsa}, \& {Wood}}]{wu2018einstein}
{Wu}, J., {Clark}, C.~J., {Pletsch}, H.~J., {et~al.} 2018, \apj, 854, 99,
  \dodoi{10.3847/1538-4357/aaa411}

\bibitem[{{Wu} {et~al.}(2013){Wu}, {Hui}, {Wu}, {Kong}, {Huang}, {Tam},
  {Takata}, \& {Cheng}}]{Wu2013}
{Wu}, J.~H.~K., {Hui}, C.~Y., {Wu}, E.~M.~H., {et~al.} 2013, \apjl, 765, L47,
  \dodoi{10.1088/2041-8205/765/2/L47}

\bibitem[{Wu {et~al.}(2021)Wu, Qiao, \& Xu}]{Wu2021}
Wu, X., Qiao, G., \& Xu, R. 2021, Pulsar Physics (BEIJING BOOK CO. INC.)

\bibitem[{{Xia} {et~al.}(2025){Xia}, {Lv}, {Fang}, \& {Liu}}]{Xia2025}
{Xia}, J., {Lv}, X., {Fang}, K., \& {Liu}, S. 2025, arXiv e-prints,
  arXiv:2503.15052, \dodoi{10.48550/arXiv.2503.15052}

\bibitem[{{Xia} {et~al.}(2023){Xia}, {Zhou}, \& {Fang}}]{xia2023}
{Xia}, Q., {Zhou}, L.-C., \& {Fang}, J. 2023, Research in Astronomy and
  Astrophysics, 23, 105003, \dodoi{10.1088/1674-4527/ace51d}

\bibitem[{Xiang(2008)}]{Xiang2008}
Xiang, S. 2008, Introduction to Astrophysics (USTC Press)

\bibitem[{{Xin} {et~al.}(2019){Xin}, {Zeng}, {Liu}, {Fan}, \& {Wei}}]{xin19}
{Xin}, Y., {Zeng}, H., {Liu}, S., {Fan}, Y., \& {Wei}, D. 2019, \apj, 885, 162,
  \dodoi{10.3847/1538-4357/ab48ee}

\bibitem[{{Xing} {et~al.}(2019){Xing}, {Wang}, {Zhang}, {Chen}, \&
  {Jithesh}}]{Xing2019a}
{Xing}, Y., {Wang}, Z., {Zhang}, X., {Chen}, Y., \& {Jithesh}, V. 2019, \apj,
  872, 25, \dodoi{10.3847/1538-4357/aafc60}

\bibitem[{{Yan} {et~al.}(2020){Yan}, {Manchester}, {Wang}, {Wen}, {Yuan},
  {Lee}, \& {Chen}}]{yan2020periodic}
{Yan}, W.~M., {Manchester}, R.~N., {Wang}, N., {et~al.} 2020, \mnras, 491,
  4634, \dodoi{10.1093/mnras/stz3399}

\bibitem[{{Yao} {et~al.}(2017){Yao}, {Manchester}, \& {Wang}}]{Yao2017}
{Yao}, J.~M., {Manchester}, R.~N., \& {Wang}, N. 2017, \apj, 835, 29,
  \dodoi{10.3847/1538-4357/835/1/29}

\bibitem[{Yu {et~al.}(2022)Yu, Wu, Wen, \& Fang}]{yu22}
Yu, H., Wu, K., Wen, L., \& Fang, J. 2022, New Astronomy, 90, 101669,
  \dodoi{https://doi.org/10.1016/j.newast.2021.101669}

\bibitem[{{Zanin} {et~al.}(2016){Zanin}, {Fern{\'a}ndez-Barral}, {de O{\~n}a
  Wilhelmi}, {Aharonian}, {Blanch}, {Bosch-Ramon}, \& {Galindo}}]{Zanin2016}
{Zanin}, R., {Fern{\'a}ndez-Barral}, A., {de O{\~n}a Wilhelmi}, E., {et~al.}
  2016, \aap, 596, A55, \dodoi{10.1051/0004-6361/201628917}

\bibitem[{{Zhang} {et~al.}(2008){Zhang}, {Chen}, \& {Fang}}]{zhang2008}
{Zhang}, L., {Chen}, S.~B., \& {Fang}, J. 2008, \apj, 676, 1210,
  \dodoi{10.1086/527466}

\bibitem[{{Zhang} \& {Cheng}(1997)}]{Zhang1997psrmodel}
{Zhang}, L., \& {Cheng}, K.~S. 1997, \apj, 487, 370, \dodoi{10.1086/304589}

\bibitem[{Zhang {et~al.}(2023)Zhang, Freire, Ridolfi, Pan, Zhao, Heinke, \&
  Chen}]{Zhangl2023}
Zhang, L., Freire, P.~C.~C., Ridolfi, A., {et~al.} 2023, \apjs, 269,
  \dodoi{10.3847/1538-4365/acfb03}

\bibitem[{{Zhang} {et~al.}(2022){Zhang}, {Xing}, \& {Wang}}]{Zhangp2022}
{Zhang}, P., {Xing}, Y., \& {Wang}, Z. 2022, \apjl, 935, L36,
  \dodoi{10.3847/2041-8213/ac88bf}

\bibitem[{{Zhang} {et~al.}(2016){Zhang}, {Xin}, {Fu}, {Zhou}, {Yan}, {Liu}, \&
  {Zhang}}]{Zhang2016}
{Zhang}, P.~F., {Xin}, Y.~L., {Fu}, L., {et~al.} 2016, \mnras, 459, 99,
  \dodoi{10.1093/mnras/stw567}

\bibitem[{{Zhang} {et~al.}(2024){Zhang}, {Torres}, {Garc{\'\i}a}, {Li}, \&
  {Mestre}}]{Zhang2024}
{Zhang}, W., {Torres}, D.~F., {Garc{\'\i}a}, C.~R., {Li}, J., \& {Mestre}, E.
  2024, \aap, 691, A332, \dodoi{10.1051/0004-6361/202348741}

\bibitem[{{Zhang} \& {Bordas}(2019)}]{zhang2019chandra}
{Zhang}, X., \& {Bordas}, P. 2019, in High Energy Phenomena in Relativistic
  Outflows VII, 90, \dodoi{10.22323/1.354.0090}

\bibitem[{{Zhou} {et~al.}(2015){Zhou}, {Zhang}, {Huang}, {Li}, {Liang}, {Fu},
  {Yan}, \& {Liu}}]{Zhou2015}
{Zhou}, J.~N., {Zhang}, P.~F., {Huang}, X.~Y., {et~al.} 2015, \mnras, 448,
  3215, \dodoi{10.1093/mnras/stv185}

\bibitem[{{Zhu} {et~al.}(2021){Zhu}, {Lu}, {Zhou}, \& {Zhang}}]{zhu2021}
{Zhu}, B.-T., {Lu}, F.-W., {Zhou}, B., \& {Zhang}, L. 2021, \aap, 655, A41,
  \dodoi{10.1051/0004-6361/202141042}

\bibitem[{{Zyuzin} {et~al.}(2018){Zyuzin}, {Karpova}, \&
  {Shibanov}}]{zyuzin2018x}
{Zyuzin}, D.~A., {Karpova}, A.~V., \& {Shibanov}, Y.~A. 2018, \mnras, 476,
  2177, \dodoi{10.1093/mnras/sty359}

\end{thebibliography}

\end{document}